%% file: main.tex
\pdfoutput=1

\documentclass[12pt,a4paper]{article}

\usepackage{ifthen} 
\newboolean{pdflatex}
\setboolean{pdflatex}{true} 

\newboolean{articletitles}
\setboolean{articletitles}{true} 

\newboolean{uprightparticles}
\setboolean{uprightparticles}{false} 

\def\paperauthors{LHCb collaboration} 
\def\paperasciititle{Amplitude analysis of B+ -> K+ pi+ pi- decays} 
\def\papertitle{Amplitude analysis of $B^+ \!\to K^+ \pi^+ \pi^-$ decays} 
\def\paperkeywords{{High Energy Physics}, {LHCb}} 
\def\papercopyright{\the\year\ CERN for the benefit of the LHCb collaboration} 
\def\paperlicence{CC BY 4.0 licence}
\def\paperlicenceurl{https://creativecommons.org/licenses/by/4.0/}

\newif\ifEnableSectionTOCLinks
\EnableSectionTOCLinksfalse 

\input{LHCb/preamble}

\newcommand{\mKpipi}{{\ensuremath{m_{\Kp\pip\pim}}}\xspace}
\newcommand{\mKpi}{{\ensuremath{m_{\Kp\pim}}}\xspace}
\newcommand{\mpipi}{{\ensuremath{m_{\pip\pim}}}\xspace}

\begin{document}

\renewcommand{\thefootnote}{\fnsymbol{footnote}}
\setcounter{footnote}{1}

\input{title-LHCb-PAPER}

\renewcommand{\thefootnote}{\arabic{footnote}}
\setcounter{footnote}{0}


\cleardoublepage


\pagestyle{plain} 
\setcounter{page}{1}
\pagenumbering{arabic}


\input{body}
\input{LHCb/acknowledgements}

\clearpage
\input{appendix}
\clearpage

\input{supplemental}

\addcontentsline{toc}{section}{References}
\bibliographystyle{LHCb/LHCb}
\bibliography{main,LHCb/standard,LHCb/LHCb-PAPER,LHCb/LHCb-CONF,LHCb/LHCb-DP,LHCb/LHCb-TDR}

\newpage
\input{Authorship_LHCb-PAPER-2025-067}

\end{document}

%% file: LHCb/preamble.tex
\usepackage[top=1in, bottom=1.25in, left=1in, right=1in]{geometry}

\usepackage{microtype}
\usepackage{lineno}  
\usepackage{xspace} 
\usepackage{caption} 

\usepackage{graphicx}  
\usepackage{color}
\usepackage{colortbl}
\graphicspath{{./figs/}} 

\usepackage{amsmath} 
\usepackage{amssymb}
\usepackage{amsfonts}
\usepackage{upgreek} 

\usepackage[normalem]{ulem} 

\newcommand*\patchAmsMathEnvironmentForLineno[1]{%
\expandafter\let\csname old#1\expandafter\endcsname\csname #1\endcsname
\expandafter\let\csname oldend#1\expandafter\endcsname\csname
end#1\endcsname
 \renewenvironment{#1}%
   {\linenomath\csname old#1\endcsname}%
   {\csname oldend#1\endcsname\endlinenomath}%
}
\newcommand*\patchBothAmsMathEnvironmentsForLineno[1]{%
  \patchAmsMathEnvironmentForLineno{#1}%
  \patchAmsMathEnvironmentForLineno{#1*}%
}
\AtBeginDocument{%
\patchBothAmsMathEnvironmentsForLineno{equation}%
\patchBothAmsMathEnvironmentsForLineno{align}%
\patchBothAmsMathEnvironmentsForLineno{flalign}%
\patchBothAmsMathEnvironmentsForLineno{alignat}%
\patchBothAmsMathEnvironmentsForLineno{gather}%
\patchBothAmsMathEnvironmentsForLineno{multline}%
\patchBothAmsMathEnvironmentsForLineno{eqnarray}%
}

\usepackage[pdftex,
            pdfauthor={\paperauthors},
            pdftitle={\paperasciititle},
            pdfkeywords={\paperkeywords}]{hyperref}
\usepackage{hyperxmp}
\hypersetup{
    pdfcopyright={Copyright (C) \papercopyright},
    pdflicenseurl={\paperlicenceurl}
}

\usepackage[colorinlistoftodos,textsize=scriptsize]{todonotes}

\usepackage[bottom,flushmargin,hang,multiple]{footmisc}

\usepackage[all]{hypcap} 

\input{LHCb/lhcb-symbols-def} 

\hypersetup{
  colorlinks   = true, 
  urlcolor     = blue, 
  linkcolor    = blue, 
  citecolor    = red   
}

\ifEnableSectionTOCLinks
    \usepackage[explicit]{titlesec} 
    
    \let\oldcontentsline\contentsline
    \renewcommand

    \titleformat{\section}{\normalfont\Large\bf}{\hyperlink{tocsection.\thesection}{{\thesection} \parbox[t]{\dimexpr\textwidth-1pc}{#1}}}{1pc}{}

    \titleformat{\subsection}{\normalfont\bf}{\hyperlink{tocsubsection.\thesubsection}{{\thesubsection} \parbox[t]{\dimexpr\textwidth-1pc}{#1}}}{1pc}{}

    \titleformat{name=\section,numberless}[display]{}{}{0pt}{\normalfont\Huge\bfseries #1}
\fi

\usepackage{cite} 
\usepackage{LHCb/mciteplus}
\makeatletter
\g@addto@macro\bfseries{\boldmath}
\makeatother

%% file: LHCb/lhcb-symbols-def.tex
\usepackage{xspace} 
\usepackage{upgreek}

\def\lhcb   {\mbox{LHCb}\xspace}

\def\babar  {\mbox{BaBar}\xspace}
\def\belle  {\mbox{Belle}\xspace}

\def\besiii {\mbox{BESIII}\xspace}

\def\MagUp {\mbox{\em Mag\kern -0.05em Up}\xspace}

\ifdefined\ifuprightparticles
\else
\newboolean{uprightparticles}
\setboolean{uprightparticles}{false} 
\fi

\ifthenelse{\boolean{uprightparticles}}%
{

 \def\Peta        {\ensuremath{\upeta}\xspace}

 \def\Pmu         {\ensuremath{\upmu}\xspace}

 \def\Ppi         {\ensuremath{\uppi}\xspace}

 \def\Pchi        {\ensuremath{\upchi}\xspace}                 
 \def\Ppsi        {\ensuremath{\uppsi}\xspace}

 \def\PDelta      {\ensuremath{\Delta}\xspace}                 
 \def\PXi         {\ensuremath{\Xi}\xspace}                 
 \def\PLambda     {\ensuremath{\Lambda}\xspace}                 
 \def\PSigma      {\ensuremath{\Sigma}\xspace}                 
 \def\POmega      {\ensuremath{\Omega}\xspace}                 
 \def\PUpsilon    {\ensuremath{\Upsilon}\xspace}
 \let\oldPi\Pi
 \def\PPi         {\ensuremath{\oldPi}\xspace}

 \def\PB      {\ensuremath{\mathrm{B}}\xspace}                 
 \def\PD      {\ensuremath{\mathrm{D}}\xspace}                 
 \def\PJ      {\ensuremath{\mathrm{J}}\xspace}                 
 \def\PK      {\ensuremath{\mathrm{K}}\xspace}                 
 \def\Pb      {\ensuremath{\mathrm{b}}\xspace}                 
 \def\Pc      {\ensuremath{\mathrm{c}}\xspace}                 
 \def\Pd      {\ensuremath{\mathrm{d}}\xspace}                 

 \def\Pp      {\ensuremath{\mathrm{p}}\xspace}                 

 \def\Ps      {\ensuremath{\mathrm{s}}\xspace}                 
 \def\Pt      {\ensuremath{\mathrm{t}}\xspace}                 
 \def\Pu      {\ensuremath{\mathrm{u}}\xspace}                 
 \def\thebaroffset{0.0em}
}
{

 \def\Peta        {\ensuremath{\eta}\xspace}

 \def\Pmu         {\ensuremath{\mu}\xspace}

 \def\Ppi         {\ensuremath{\pi}\xspace}

 \def\Pchi        {\ensuremath{\chi}\xspace}                 
 \def\Ppsi        {\ensuremath{\psi}\xspace}                 
                  
 \mathchardef\PDelta="7101
 \mathchardef\PXi="7104
 \mathchardef\PLambda="7103
 \mathchardef\PSigma="7106
 \mathchardef\POmega="710A
 \mathchardef\PUpsilon="7107
 \mathchardef\PPi="7105
 \def\PB      {\ensuremath{B}\xspace}                 
 \def\PD      {\ensuremath{D}\xspace}                 
 \def\PJ      {\ensuremath{J}\xspace}                 
 \def\PK      {\ensuremath{K}\xspace}                 
 \def\Pb      {\ensuremath{b}\xspace}                 
 \def\Pc      {\ensuremath{c}\xspace}                 
 \def\Pd      {\ensuremath{d}\xspace}                 

 \def\Pp      {\ensuremath{p}\xspace}                 

 \def\Ps      {\ensuremath{s}\xspace}                 
 \def\Pt      {\ensuremath{t}\xspace}                 
 \def\Pu      {\ensuremath{u}\xspace}                 
 \def\thebaroffset{0.18em}
}
\newcommand{\offsetoverline}[2][\thebaroffset]{\kern #1\overline{\kern -#1 #2}}%

\makeatletter
\ifcase \@ptsize \relax
  \newcommand{\miniscule}{\@setfontsize\miniscule{4}{5}}
\or
  \newcommand{\miniscule}{\@setfontsize\miniscule{5}{6}}
\or
  \newcommand{\miniscule}{\@setfontsize\miniscule{5}{6}}
\fi
\makeatother

\DeclareRobustCommand{\optbar}[1]{\shortstack{{\miniscule (\rule[.5ex]{1.25em}{.18mm})}
  \\ [-.7ex] $#1$}}

\def\mup        {{\ensuremath{\Pmu^+}}\xspace}
\def\mun        {{\ensuremath{\Pmu^-}}\xspace} 

\def\mumu       {{\ensuremath{\Pmu^+\Pmu^-}}\xspace}

\def\uquark    {{\ensuremath{\Pu}}\xspace}

\def\dquark    {{\ensuremath{\Pd}}\xspace}

\def\squark    {{\ensuremath{\Ps}}\xspace}

\def\cquark    {{\ensuremath{\Pc}}\xspace}
\def\cquarkbar {{\ensuremath{\overline \cquark}}\xspace}

\def\bquark    {{\ensuremath{\Pb}}\xspace}

\def\tquark    {{\ensuremath{\Pt}}\xspace}

\def\pion   {{\ensuremath{\Ppi}}\xspace}
\def\piz    {{\ensuremath{\pion^0}}\xspace}
\def\pip    {{\ensuremath{\pion^+}}\xspace}
\def\pim    {{\ensuremath{\pion^-}}\xspace}

\def\kaon    {{\ensuremath{\PK}}\xspace}
\def\Kbar    {{\ensuremath{\offsetoverline{\PK}}}\xspace}
\def\Kb      {{\ensuremath{\Kbar}}\xspace}
\def\KorKbar {\kern \thebaroffset\optbar{\kern -\thebaroffset \PK}{}\xspace}
\def\Kz      {{\ensuremath{\kaon^0}}\xspace}

\def\Kp      {{\ensuremath{\kaon^+}}\xspace}
\def\Km      {{\ensuremath{\kaon^-}}\xspace}

\newcommand{\etapr}{\ensuremath{\Peta^{\prime}}\xspace}

\def\Dbar    {{\ensuremath{\offsetoverline{\PD}}}\xspace}
\def\D       {{\ensuremath{\PD}}\xspace}
\def\Db      {{\ensuremath{\Dbar}}\xspace}
\def\DorDbar {\kern \thebaroffset\optbar{\kern -\thebaroffset \PD}\xspace}
\def\Dz      {{\ensuremath{\D^0}}\xspace}
\def\Dzb     {{\ensuremath{\Dbar{}^0}}\xspace}
\def\Dp      {{\ensuremath{\D^+}}\xspace}
\def\Dm      {{\ensuremath{\D^-}}\xspace}

\def\DpDm    {\ensuremath{\Dp {\kern -0.16em \Dm}}\xspace}

\def\Dstarp  {{\ensuremath{\D^{*+}}}\xspace}

\def\Dsp     {{\ensuremath{\D^+_\squark}}\xspace}
\def\Dsm     {{\ensuremath{\D^-_\squark}}\xspace}

\def\B       {{\ensuremath{\PB}}\xspace}
\def\Bbar    {{\ensuremath{\offsetoverline{\PB}}}\xspace}

\def\BorBbar {\kern \thebaroffset\optbar{\kern -\thebaroffset \PB}\xspace}
\def\Bz      {{\ensuremath{\B^0}}\xspace}
\def\Bzb     {{\ensuremath{\Bbar{}^0}}\xspace}
\def\Bd      {{\ensuremath{\B^0}}\xspace}

\def\BdorBdbar {\kern \thebaroffset\optbar{\kern -\thebaroffset \Bd}\xspace}
\def\Bu      {{\ensuremath{\B^+}}\xspace}
\def\Bub     {{\ensuremath{\B^-}}\xspace}
\def\Bp      {{\ensuremath{\Bu}}\xspace}
\def\Bm      {{\ensuremath{\Bub}}\xspace}
\def\Bpm     {{\ensuremath{\B^\pm}}\xspace}

\def\Bs      {{\ensuremath{\B^0_\squark}}\xspace}

\def\BsorBsbar {\kern \thebaroffset\optbar{\kern -\thebaroffset \Bs}\xspace}

\def\jpsi     {{\ensuremath{{\PJ\mskip -3mu/\mskip -2mu\Ppsi}}}\xspace}

\def\chiczero {{\ensuremath{\Pchi_{\cquark 0}}}\xspace}

\def\Y#1S{\ensuremath{\PUpsilon{(#1S)}}\xspace}

\def\proton      {{\ensuremath{\Pp}}\xspace}

\def\LorLbar     {\kern \thebaroffset\optbar{\kern -\thebaroffset \PLambda}\xspace}

\def\BF         {{\ensuremath{\mathcal{B}}}\xspace}

\newcommand{\decay}[2]{\ensuremath{\mathinner{#1\!\to #2}}\xspace}

\def\to                 {\ensuremath{\rightarrow}\xspace}

\def\CP                {{\ensuremath{C\!P}}\xspace}

\def\Vud  {{\ensuremath{V_{\uquark\dquark}^{\phantom{\ast}}}}\xspace}
\def\Vcd  {{\ensuremath{V_{\cquark\dquark}^{\phantom{\ast}}}}\xspace}
\def\Vtd  {{\ensuremath{V_{\tquark\dquark}^{\phantom{\ast}}}}\xspace}

\def\Vubs  {{\ensuremath{V_{\uquark\bquark}^\ast}}\xspace}
\def\Vcbs  {{\ensuremath{V_{\cquark\bquark}^\ast}}\xspace}
\def\Vtbs  {{\ensuremath{V_{\tquark\bquark}^\ast}}\xspace}

\def\AT#1     {\ensuremath{A_{\mathrm{T}}^{#1}}\xspace}           

\def\C#1      {\ensuremath{\mathcal{C}_{#1}}\xspace}                       
\def\Cp#1     {\ensuremath{\mathcal{C}_{#1}^{'}}\xspace}                    
\def\Ceff#1   {\ensuremath{\mathcal{C}_{#1}^{\mathrm{(eff)}}}\xspace}        
\def\Cpeff#1  {\ensuremath{\mathcal{C}_{#1}^{'\mathrm{(eff)}}}\xspace}       
\def\Ope#1    {\ensuremath{\mathcal{O}_{#1}}\xspace}                       
\def\Opep#1   {\ensuremath{\mathcal{O}_{#1}^{'}}\xspace}                    

\newcommand{\nospaceunit}[1]{\ensuremath{\text{#1}}}       
\newcommand{\aunit}[1]{\ensuremath{\text{\,#1}}}       

\newcommand{\tev}{\aunit{Te\kern -0.1em V}\xspace}
\newcommand{\gev}{\aunit{Ge\kern -0.1em V}\xspace}
\newcommand{\mev}{\aunit{Me\kern -0.1em V}\xspace}
\newcommand{\kev}{\aunit{ke\kern -0.1em V}\xspace}
\newcommand{\ev}{\aunit{e\kern -0.1em V}\xspace}
 
\newcommand{\mevc}{\ensuremath{\aunit{Me\kern -0.1em V\!/}c}\xspace}
\newcommand{\gevc}{\ensuremath{\aunit{Ge\kern -0.1em V\!/}c}\xspace}
\newcommand{\mevcc}{\ensuremath{\aunit{Me\kern -0.1em V\!/}c^2}\xspace}
\newcommand{\gevcc}{\ensuremath{\aunit{Ge\kern -0.1em V\!/}c^2}\xspace}
\newcommand{\gevgevcccc}{\ensuremath{\gev^2\!/c^4}\xspace} 

\def\m    {\aunit{m}\xspace}

\def\mum  {\ensuremath{\,\upmu\nospaceunit{m}}\xspace}

\def\fm   {\aunit{fm}\xspace}

\def\fb   {\ensuremath{\aunit{fb}}\xspace}
\def\invfb   {\ensuremath{\fb^{-1}}\xspace}

\newcommand{\stat}{\aunit{(stat)}\xspace}
\newcommand{\syst}{\aunit{(syst)}\xspace}

\def\gsim{{~\raise.15em\hbox{$>$}\kern-.85em
          \lower.35em\hbox{$\sim$}~}\xspace}
\def\lsim{{~\raise.15em\hbox{$<$}\kern-.85em
          \lower.35em\hbox{$\sim$}~}\xspace}

\newcommand{\Real}{\ensuremath{\mathcal{R}e}\xspace}

\def\pt         {\ensuremath{p_{\mathrm{T}}}\xspace}

\def\ptot       {\ensuremath{p}\xspace}

\def\evtgen     {\mbox{\textsc{EvtGen}}\xspace}

\def\geant      {\mbox{\textsc{Geant4}}\xspace}

\def\photos     {\mbox{\textsc{Photos}}\xspace}

\def\pythia     {\mbox{\textsc{Pythia}}\xspace}

\def\tell1  {TELL1\xspace}
\def\ukl1   {UKL1\xspace}

\newcommand{\eg}{\mbox{\itshape e.g.}\xspace}
\newcommand{\ie}{\mbox{\itshape i.e.}\xspace}

\newcommand{\lhcborcid}[1]{\href{https://orcid.org/#1}{\hspace*{0.1em}\raisebox{-0.45ex}{\includegraphics[width=1em]{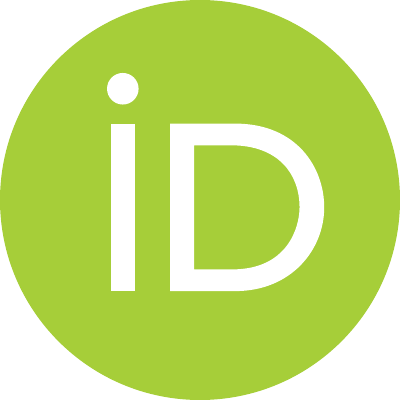}}}}


%% file: title-LHCb-PAPER.tex
\begin{titlepage}
\pagenumbering{roman}

\vspace*{-1.5cm}
\centerline{\large EUROPEAN ORGANIZATION FOR NUCLEAR RESEARCH (CERN)}
\vspace*{1.5cm}
\noindent
\begin{tabular*}{\linewidth}{lc@{\extracolsep{\fill}}r@{\extracolsep{0pt}}}
\ifthenelse{\boolean{pdflatex}}
{\vspace*{-1.5cm}\mbox{\!\!\!\includegraphics[width=.14\textwidth]{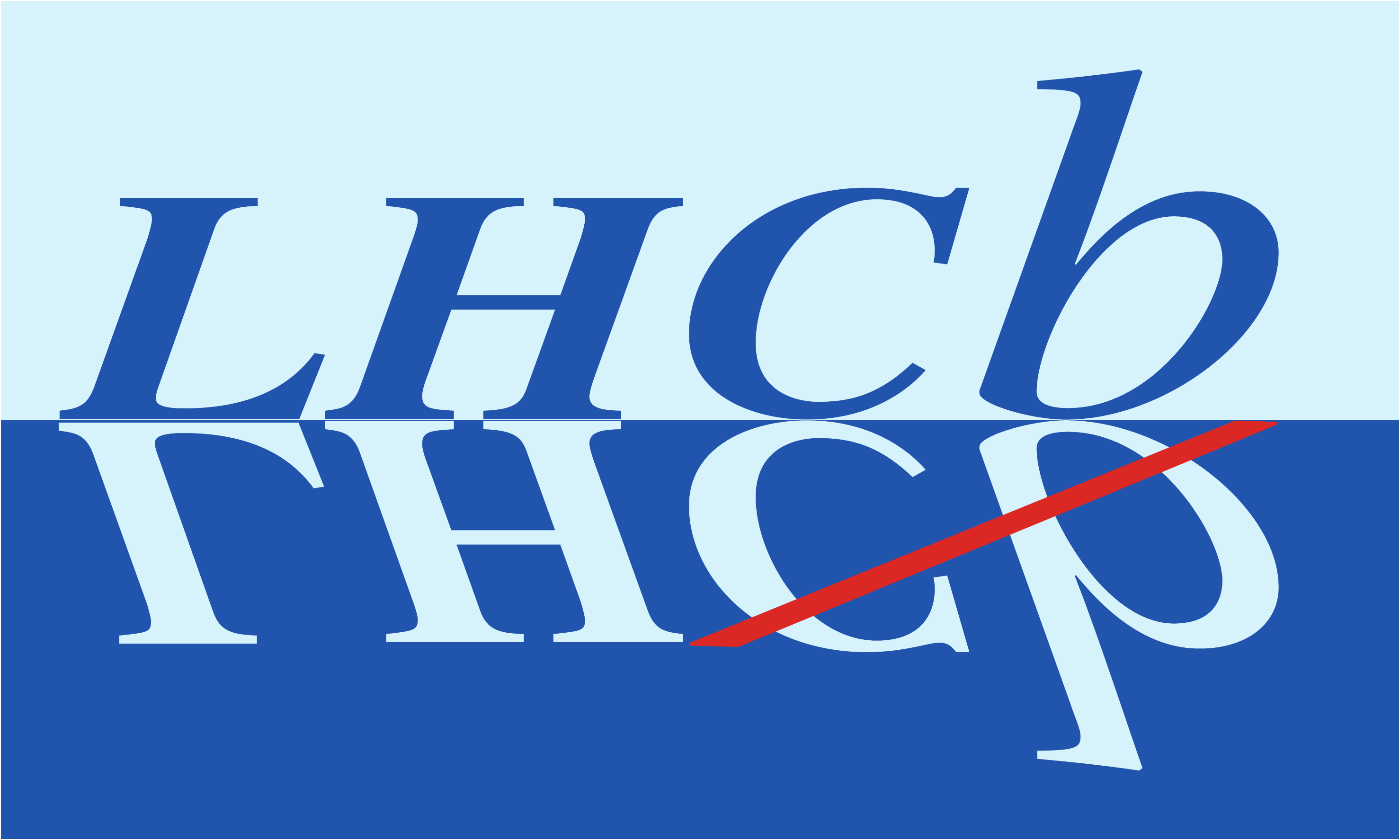}} & &}%
{\vspace*{-1.2cm}\mbox{\!\!\!\includegraphics[width=.12\textwidth]{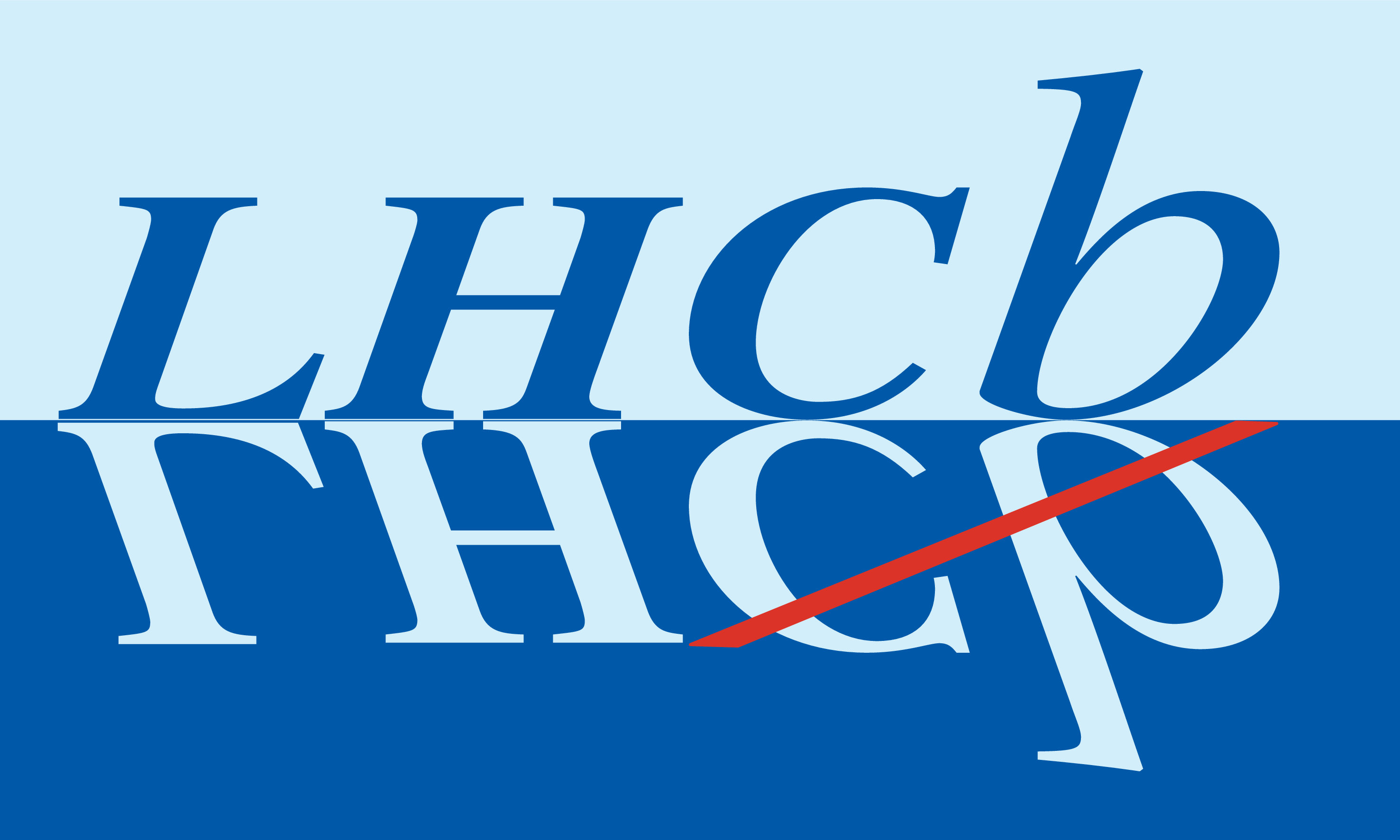}} & &}%
\\
 & & CERN-EP-2026-216 \\  
 & & LHCb-PAPER-2025-067 \\  
 & & 17 August 2026 \\ 
 & & \\
\end{tabular*}

\vspace*{1.0cm}

{\normalfont\bfseries\boldmath\huge
\begin{center}
  \papertitle 
\end{center}
}

\vspace*{0.5cm}

\begin{center}
\paperauthors\footnote{Authors are listed at the end of this paper.}
\end{center}

\vspace{\fill}

\begin{abstract}
    \noindent
The branching fractions and quasi-two-body $C\!P$-violating asymmetries of intermediate states obtained through an amplitude analysis of the charmless three-body decay $B^+ \!\to K^+ \pi^+ \pi^-$ are reported. The analysis is based on $pp$ collision data at centre-of-mass energies $\sqrt{s}=7$ and $8\,\text{TeV}$ recorded with the LHCb detector, corresponding to an integrated luminosity of $3\,\text{fb}^{-1}$. The most challenging aspect of the amplitude modelling lies in the description of the dominant $K^+ \pi^-$ and $\pi^+ \pi^-$ S-wave contributions. This is achieved by three complementary approaches based on a physically motivated analytic model built on the isobar approximation, the K-matrix formalism, and a quasi-model-independent procedure in which overlapping crossing partial waves are simultaneously studied. In addition, alternative sets of results are presented, considering the $\pi^+ \pi^-$ final state to manifest either through direct $\omega(782)$ decays or $\rho(770)^0\textrm{--}\omega(782)$ mixing. The most precise measurements of branching fractions and $C\!P$ asymmetries are obtained for the vast majority of intermediate states, establishing firmer reference points against which to cleanly probe model-independent physics beyond the Standard Model. The results from all three approaches agree and provide new insight into strong dynamics and the origin of $C\!P$-violation effects in $B^+ \!\to K^+ \pi^+ \pi^-$ decays.
\end{abstract}

\vspace*{0.5cm}

\begin{center}
  Submitted to
  Phys.~Rev.~D
\end{center}

\vspace{\fill}

{\footnotesize 
\centerline{\copyright~\papercopyright. \href{\paperlicenceurl}{\paperlicence}.}}
\vspace*{2mm}

\end{titlepage}


\newpage
\setcounter{page}{2}
\mbox{~}

%% file: body.tex
\section{Introduction}
\label{sec:Introduction}

In the Standard Model of particle physics (SM), \CP violation originates from a single irreducible complex phase in the Cabibbo--Kobayashi--Maskawa (CKM) matrix~\cite{Cabibbo:1963yz,Kobayashi:1973fv}.
Thus far, all measurements of \CP\ violation in particle decays are consistent with this explanation.
Nevertheless, the degree of \CP\ violation permitted in the SM is inconsistent with the macroscopic matter-antimatter asymmetry observed in the Universe~\cite{Shaposhnikov:1991cu}, motivating further studies and searches for sources of \CP\ violation beyond the SM.

For the manifestation of \CP\ violation in decay, at least two interfering amplitudes with different strong and weak phases are required.
In the SM, weak phases are associated with the complex elements of the CKM matrix and have opposite sign between charge-conjugate processes, while strong phases are associated with hadronic final-state effects and do not change sign under \CP\ conjugation.
In decays of \bquark hadrons to charmless hadronic final states, contributions from both tree-level and loop (so-called ``penguin'') diagrams, which can provide the relative weak phase that is necessary for \CP violation to manifest, are possible with comparable magnitudes.

In multibody decays, variation of strong phases across the phase space, caused by hadronic effects, allows for a richer tapestry of \CP\ violation effects and phenomenology compared to two-body decays.
Large \CP asymmetries localised in regions of phase space of charmless three-body $B$-meson decays have been observed in model-independent analyses~\cite{LHCb-PAPER-2013-027,LHCb-PAPER-2013-051,LHCb-PAPER-2014-044,LHCb-PAPER-2021-049,LHCb-PAPER-2021-050}, however a description of these effects with an accurate model of the contributing resonances has only become available for \decay{\Bp}{\pip\Kp\Km}~\cite{LHCb-PAPER-2018-051} and \decay{\Bp}{\pip\pip\pim}~\cite{LHCb-PAPER-2019-018,LHCb-PAPER-2019-017} decays.

This paper documents an analysis of the \decay{\Bp}{\Kp\pip\pim} decay amplitude in the two-dimensional phase space known as the Dalitz plot~\cite{Dalitz:1953cp, Fabri:1954zz}.
The inclusion of charge-conjugate processes is implied, except where asymmetries are discussed.
Apart from the extraction of meaningful information on the intermediate branching fractions and \CP-violating asymmetries in this decay, it is also of paramount importance to improve the handling of the underlying strong interaction. 
Focussed discussions can be found in two companion Letters dedicated to the \CP-conserving~\cite{BuToKpPipPimPRLStrong} and \CP-violating~\cite{BuToKpPipPimPRLCP} implications facilitated by the improved understanding of strong-interaction dynamics discussed herein.

The primary quasi-two-body results reported have several uses beyond providing essential feedback for the myriad predictions on offer from several effective-theory frameworks such as QCD factorisation~\cite{Beneke:2003zv,Cheng:2009cn,Cheng:2016shb,Cheng:2010yd,Cheng:2013fba,Krankl:2015fha}, perturbative QCD~\cite{Li:2006jv,Chai:2022ptk,Li:2018lbd,Wang:2006ria,Shen:2006ms,Chen:2002th,Wang:2014ira}, soft-collinear effective theory~\cite{Wang:2008rk,Bell:2015koa} and SU(3) flavour symmetry~\cite{Cheng:2014rfa}. These extend to model-independent tests for new physics in loop-mediated processes via isospin sum rules~\cite{Lipkin:1991st,Gronau:2005kz,Gronau:2010dd} and in measurements of the effective weak phase $\gamma = \phi_3 \equiv \arg[-\Vud \Vubs/\Vcd \Vcbs]$~\cite{Bhattacharya:2013cla,Bhattacharya:2014eca,Bertholet:2018tmx,Bhattacharya:2023pef}. Furthermore, results for charmonium~(\cquark \cquarkbar) intermediates, such as the $\chi_{c0}(1P)$ resonance, open the possibility~\cite{Jung:2012mp,DeBruyn:2014oga} to correct measurements of the \Bz--\Bzb mixing phase $\beta = \phi_1 \equiv \arg[-\Vcd \Vcbs/\Vtd \Vtbs]$, from their related \decay{\Bz}{(\cquark \cquarkbar) \Kz} channels arising from penguin contamination, thereby improving the fidelity of constraints to the CKM matrix.

The present analysis is performed on data corresponding to an integrated luminosity of 3\invfb collected by the LHCb experiment, of which $1\invfb$ was collected in 2011 with a \proton\proton collision centre-of-mass energy of $\sqrt{s} = 7\tev$ and $2\invfb$ was collected in 2012 with $\sqrt{s} = 8\tev$.
Models of the Dalitz-plot distribution are constructed in terms of intermediate resonant and nonresonant structures.
Due to their magnitude and potential importance to the observed \CP violation in $\decay{\Bp}{\Kp\pip\pim}$ decay, particular attention is given to the \Kp\pim and \pip\pim S-wave contributions, which are known to consist of numerous overlapping
resonances with structures related to decay-threshold openings~\cite{Garcia-Martin:2011iqs,Pelaez:2020gnd}.
Three different state-of-the-art approaches to the modelling of the S-wave are used to ensure that any inaccuracies in the description of this part of the amplitude do not impact the interpretation of the physical quantities reported.

This paper is organised as follows: 
Section~\ref{sec:Detector} gives a brief description of the LHCb detector, the event reconstruction and simulation software; 
the signal candidate selection procedure is described in Section~\ref{sec:Selection}; 
Section~\ref{sec:Mass_fit} describes the procedure for estimating the signal and background yields that enter into the amplitude fit; 
Section~\ref{sec:DPformalism} outlines the formalism used for the construction of the amplitude models, as well as a description of the mass lineshapes used to parameterise the intermediate structures; 
Section~\ref{sec:systematics} describes the systematic uncertainties associated with the analysis procedure; 
Section~\ref{sec:results} documents the physics parameters of interest obtained from the amplitude models and presents comparisons between projections of the data and the fit models; 
these results are then discussed in Section~\ref{sec:interpretation}; 
and a summary of the work can be found in Section~\ref{sec:conclusions}.

\section{Detector and simulation}
\label{sec:Detector}

The \lhcb detector~\cite{LHCb-DP-2008-001,LHCb-DP-2014-002} is a single-arm forward
spectrometer covering the \mbox{pseudorapidity} range $2<\eta <5$,
designed for the study of particles containing \bquark or \cquark
quarks. The detector used to collect the data analysed in this paper includes a high-precision tracking system
consisting of a silicon-strip vertex detector surrounding the $pp$
interaction region~\cite{LHCb-DP-2014-001}, a large-area silicon-strip detector located
upstream of a dipole magnet with a bending power of about
$4{\mathrm{\,T}}\m$, and three stations of silicon-strip detectors and straw
drift tubes~\cite{LHCb-DP-2013-003} placed downstream of the magnet.
The tracking system provides a measurement of the momentum \ptot of charged particles with
relative uncertainty that varies from 0.5\% at low momentum to 1.0\% at 200\gevc.
The minimum distance of a track to a primary vertex (PV), or impact parameter (IP),
is measured with a resolution of $(15+29/\pt)\mum$,
where \pt is the component of the momentum transverse to the beam (in\,\gevc).
Different types of charged hadrons are distinguished using information
from two ring-imaging Cherenkov detectors~\cite{LHCb-DP-2012-003}.
Photons, electrons and hadrons are identified by a calorimeter system consisting of
scintillating-pad and preshower detectors, an electromagnetic
and a hadronic calorimeter. Muons are identified by a
system composed of alternating layers of iron and multiwire
proportional chambers~\cite{LHCb-DP-2012-002}.
The magnetic field deflects oppositely charged particles in opposite directions and this can lead to detection asymmetries. Periodically reversing the magnetic field polarity throughout the data-taking reduces this effect to a negligible level.
Approximately $60\%$ of 2011 data and $52\%$ of 2012 data were collected in the ``down'' polarity configuration, and the remainder in the ``up'' configuration.

The online event selection is performed by a trigger~\cite{LHCb-DP-2012-004}
which consists of a hardware stage followed by a software stage. The hardware stage is based on information from the calorimeter and muon systems in which events are required to contain a muon with high \pt, or a hadron, photon or electron with high transverse energy in the calorimeters.
The software trigger requires a two- or three-track secondary vertex with significant displacement from all primary $pp$ interaction vertices. 
All charged particles with \mbox{$\pt>500\,(300)\mevc$}  are reconstructed, for data collected in 2011\,(2012), in events where at least one charged particle has transverse momentum $\pt > 1.7\,(1.6)\gevc$ and is inconsistent with originating from a PV. 
A multivariate algorithm~\cite{BBDT} is used for the identification of secondary vertices consistent with \bquark-hadron decays. Triggered data further undergo a centralised, offline processing step
to deliver physics-analysis-ready data across the entire \lhcb physics programme~\cite{Stripping}.

Simulated samples are used to investigate backgrounds from other
\bquark-hadron decays and also to study the detection and reconstruction efficiency of the signal.
In the simulation, $pp$ collisions are generated using
\pythia~\cite{Sjostrand:2006za,Sjostrand:2007gs} with a specific \lhcb
configuration~\cite{LHCb-PROC-2010-056}.
Decays of unstable particles are described by \evtgen~\cite{Lange:2001uf},
in which final-state radiation is generated using \photos~\cite{Golonka:2005pn}.
The interaction of the generated particles with the detector and its
response are implemented using the \geant toolkit~\cite{Allison:2006ve,
Agostinelli:2002hh} as described in Ref.~\cite{LHCb-PROC-2011-006}.

\section{\texorpdfstring{\boldmath \Bp-candidate selection}{B+ candidate selection}}
\label{sec:Selection}

The selection of \Bp signal candidates follows well-established procedures developed for the measurement of the relative branching fractions of \decay{\Bp}{h^+ h^{\prime+} h^{\prime-}} decays, where $h$ and $h^\prime$ represent either a pion or kaon, performed on the same data sample~\cite{LHCb-PAPER-2020-031}. 
The most important difference is in the approach to the suppression of random associations of tracks, commonly referred to as combinatorial background, which is based on a multivariate boosted decision tree~(BDT) classifier~\cite{Breiman}.
In the relative branching fractions measurement of \decay{\Bp}{h^+ h^{\prime+} h^{\prime-}} decays, this algorithm was trained with the \pip\Kp\Km final state, but in this analysis the training is based instead on the \Kp\pip\pim final state.
The signal training sample is taken from simulation, while the background training sample is taken from a region of the \Kp\pip\pim mass in data where combinatorial background dominates, \mbox{$5.65 < \mKpipi < 6.10\gevcc$}.
The variables that enter this classifier are \Bp and decay product kinematic properties, quantities based on the quality of the reconstructed tracks and decay vertices, as well as the \Bp displacement from the PV. 
The requirement on the output of the classifier is optimised to maximise the expected approximate signal significance, $N_{\rm s}/\sqrt{N_{\rm s} + N_{\rm b}}$, where $N_{\rm s}$ is the expected signal yield within the range \mbox{$5.24 < \mKpipi < 5.32\gevcc$}, corresponding to approximately $\pm3$~standard deviations in mass resolution around the known $\Bp$ mass~\cite{PDG2024}, and $N_{\rm b}$ is the corresponding combinatorial background yield within the same region. 
These yield inputs to the aforementioned figure of merit are obtained from rudimentary fits to \mKpipi data.

To remove the bulk of \Dzb decays into two oppositely charged mesons, the following two-body mass regions are vetoed:
\mbox{$1.75 < \mKpi < 1.95 \gevcc$} and
\mbox{$1.75 < \mpipi < 1.90 \gevcc$}.
These vetoed windows are relatively wide so as to also suppress backgrounds with particle misidentification.
Furthermore, the requirements placed on the particle-identification~(PID) information associated with each final-state track are more stringent than those in Ref.~\cite{LHCb-PAPER-2020-031}. This suppresses further backgrounds that still arise when any number of kaons or pions are misidentified and ensures that systematic uncertainties related to such backgrounds remain negligible in the amplitude analysis. 
Tighter PID requirements are imposed on the \Kp candidate in the region
\mbox{$1.95 < \mKpi < 2.00 \gevcc$} and on the \pim candidate in
\mbox{$1.65 < \mKpi < 1.75 \gevcc$}, while tighter PID requirements are applied to each pion candidate in
\mbox{$1.65 < \mpipi < 1.75 \gevcc$}. 
These requirements reduce most misidentified \Dzb backgrounds to negligible levels. Finally, another veto, \mbox{$3.05 < \mpipi < 3.15 \gevcc$}, is imposed to remove the narrow \jpsi structure. 
The $\chiczero(1P)$ resonance is retained as the only charmonium structure visible in phase space. 
Misidentified contributions from the dimuon and dielectron \jpsi decay channels are known to be negligible in the \Kp\pip\pim final state~\cite{LHCb-PAPER-2020-031}. 
Other potential contributions from decays of broader charmonia states are included in the amplitude model.

Two-body charmless \Bz decays can inhabit the background in \Kp \pip \pim phase space when derived from the sideband above the \Bp mass. 
Candidates containing such contributions are removed by vetoing the two-body mass ranges $[5.22, 5.32]\gevcc$. 
Approximately $0.2\%$ of selected events contain multiple \Bp decay candidates following the aforementioned selection procedure; all such candidates are included in the analysis. 
The Dalitz-plot variables are calculated applying a kinematic mass constraint, fixing the \Bp candidate mass to the known value~\cite{PDG2020} to improve resolution and to ensure that all events remain within the Dalitz-plot boundary.

\section{\texorpdfstring{\boldmath \Bp-candidate mass fit}{B+ candidate mass fit}}
\label{sec:Mass_fit}

An extended, unbinned, maximum-likelihood fit is performed to the \mKpipi mass spectra to extract yields and global charge asymmetries of the \decay{\Bp}{\Kp\pip\pim} signal and various contributing backgrounds.
The fit is performed to candidates in the range \mbox{$5.1 < \mKpipi <6.0 \gevcc$}, and its results are used to obtain signal and background yields in the signal region, \mbox{$5.24 < \mKpipi < 5.32 \gevcc$}, in which the subsequent Dalitz-plot fit is performed.
All shape parameters of the probability density functions~(PDFs) comprising the fit model are shared between \Bp and \Bm candidates; the only differences between the two final states, which are fitted simultaneously, are due to signal and combinatorial-background yield asymmetries.
At this stage, the data are also subdivided by data-taking year, and whether the hardware trigger decision is due to hadronic calorimeter deposits associated with the signal candidate, or due to other particles in the \proton\proton collision, to permit correction for possible differences in efficiency between subsamples (see Section~\ref{sec:efficiencies}).

The mass distribution of the \decay{\Bp}{\Kp\pip\pim} signal decay is parameterised by the sum of an underlying broad Gaussian and a core Crystal Ball function~\cite{Skwarnicki:1986xj}, with power-law tails on both sides of the peak in order to describe asymmetric non-Gaussian effects due to detector resolution and final-state radiation. 
The parameters describing these tails are determined from simulation relative to the peak position and width, which themselves are free to vary in the data fit to account for small differences between simulation and data.
All remaining parameters, apart from the total yield, are obtained from a fit to simulated decays.
The global raw charge asymmetry for the signal, while free to vary in the mass fit, is not propagated to the subsequent amplitude analysis. 
Rather, the fitted amplitude model accounts for the global charge asymmetry as well as local asymmetries between the \Bp and \Bm data.

The distribution of the combinatorial background is modelled with a falling exponential function. Partially reconstructed backgrounds, which predominantly arise from four-body \mbox{\B-meson} decays where a charged hadron or neutral particle is not reconstructed, are modelled with an ARGUS function~\cite{ARGUS:1990jet} convolved with a Gaussian resolution function. 
Separate components are included in the fit to account for such partially reconstructed backgrounds from \Bp, \Bz and \Bs decays.
Two significant sources of peaking cross-feed background arise from misidentified \decay{\Bp}{\pip\pip\pim} and \decay{\Bp}{\pip\Kp\Km} decays.
To obtain an accurate model for these backgrounds when reconstructed in the \decay{\Bp}{\Kp\pip\pim} final state, simulated decays are weighted according to their respective amplitude models obtained by the \lhcb collaboration~\cite{LHCb-PAPER-2018-051, LHCb-PAPER-2019-017, LHCb-PAPER-2019-018}. 
Further corrections for the differences between data and simulation in trigger, tracking and PID efficiencies are applied accounting for dependence on kinematics to obtain more reliable shapes.
These peaking-background shapes are modelled as Crystal Ball functions, where all parameters are determined from fits to the calibrated simulation.
Furthermore, the yields of these components are constrained to the \decay{\Bp}{\Kp\pip\pim} signal yield multiplied by the product of the relative branching fractions of these decays and the relative overall reconstruction and selection efficiencies, which are described in
Section~\ref{sec:efficiencies}.
Finally, a component for \decay{\Bp}{\Kp\etapr(\to \pip\pim\gamma)} decays is described using simulated events, corrected to match the \decay{\etapr}{\pip\pim\gamma} amplitude model from the \besiii collaboration~\cite{BESIII:2017kyd} and further calibrated to represent data. 
Here and throughout the paper, the symbol \etapr\ is used to denote the $\etapr(958)$ meson.
The shape for this component is modelled in the same way as that of the partially reconstructed backgrounds, while its yield is fixed relative to signal in the same way as for the misidentified peaking backgrounds.

The mass fit results are shown in Fig.~\ref{fig:massFit} with a logarithmic $y$-axis scale (corresponding plots with a linear scale are provided in Appendix~\ref{app:bmassfit}),
while Table~\ref{tab:massFit} quotes the component yields and phase-space-integrated background asymmetries in the \Bp\ signal region. 
These values are subsequently used in the Dalitz-plot fit and to estimate the associated systematic uncertainties outlined in Section~\ref{sec:systematics}.

\begin{figure}[!tb]
    \centering
    \includegraphics[width=0.5\linewidth]{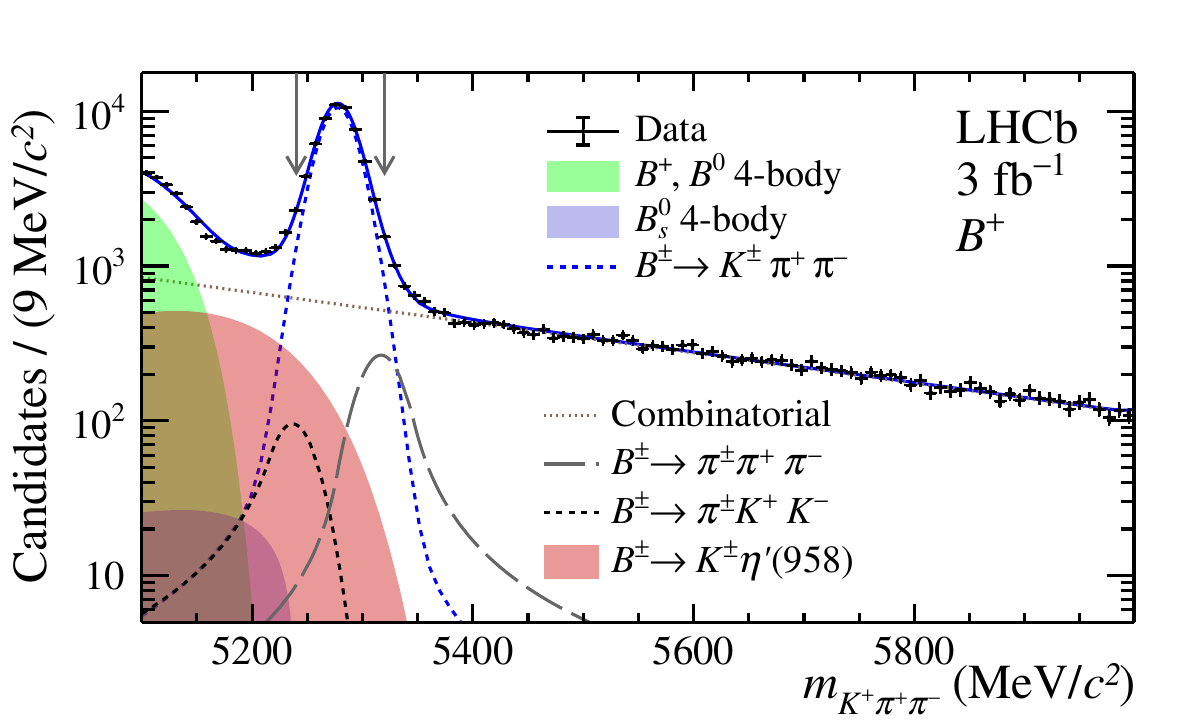}%
    \includegraphics[width=0.5\linewidth]{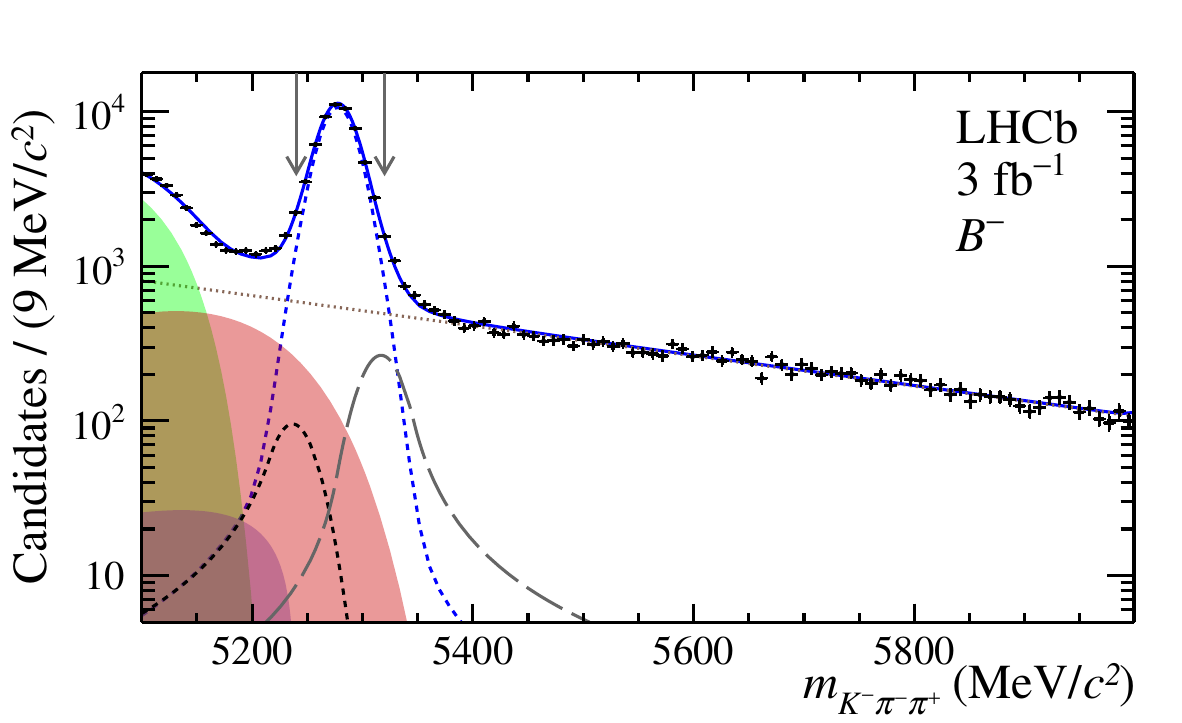}

    \caption{
        Mass distributions for (left)~$B^+$ and (right)~$B^-$ candidates reconstructed in the $K^+ \pi^+ \pi^-$ and $K^- \pi^- \pi^+$ final state, respectively, for the combined 2011 and 2012 data taking samples, along with the results of the fit. The signal region is indicated by arrows.
    }
    \label{fig:massFit}
\end{figure}

\begin{table}[!tb]
    \centering
    \caption{\label{tab:massFit}
        Component \CP-integrated yields and phase-space-integrated background asymmetries in the signal region, calculated from the results of the mass fit, where the uncertainties are statistical and systematic, respectively. For the parameters marked as being fixed, the uncertainties derive from either limited simulation sample sizes or from an external source as indicated by the reference. The raw asymmetry of signal is withheld as it does not propagate to the amplitude analysis.
    }
\resizebox{\linewidth}{!}{
    \renewcommand{\arraystretch}{1.1}
    \begin{tabular}
    {l@{\hspace{0.25cm}}
      @{\hspace{0.25cm}}r@{\hspace{0.25cm}}
      @{\hspace{0.25cm}}l}\hline
    Component & \multicolumn{1}{c}{Yield scaled to signal region} & \multicolumn{1}{c}{Asymmetry\phantom{000}} \\ \hline

Signal & $102\,073 \pm 311 \pm 560$\phantom{---} & \multicolumn{1}{c}{---\phantom{00}} \\
Combinatorial background& $9\,404 \pm 451 \pm 287$\phantom{---} & $-0.022 \pm 0.006 \pm 0.001$ \\
\decay{\Bp}{\pip\pip\pim} & $1\,358 \pm \phantom{0}51 \pm 658$\phantom{---} & $\phantom{-}0.000  \pm 0.006$ (fixed) \\
\decay{\Bp}{\pip\Kp\Km} & $471 \pm \phantom{0}39 \pm \phantom{0}43$\phantom{---} & $\phantom{-}0.000 \pm 0.008$ (fixed) \\
\decay{\Bp}{\Kp\etapr} & $1\,776 \pm 200 \pm \phantom{0}68$\phantom{---} & $\phantom{-}0.000 \pm 0.011$ (fixed)~\cite{HFLAV21} \\

    \hline
    \end{tabular}
    }
\end{table}

\section{Dalitz-plot fit model}
\label{sec:DPformalism}

\begin{figure}[!tb]
    \centering
    \includegraphics[width=0.5\linewidth]{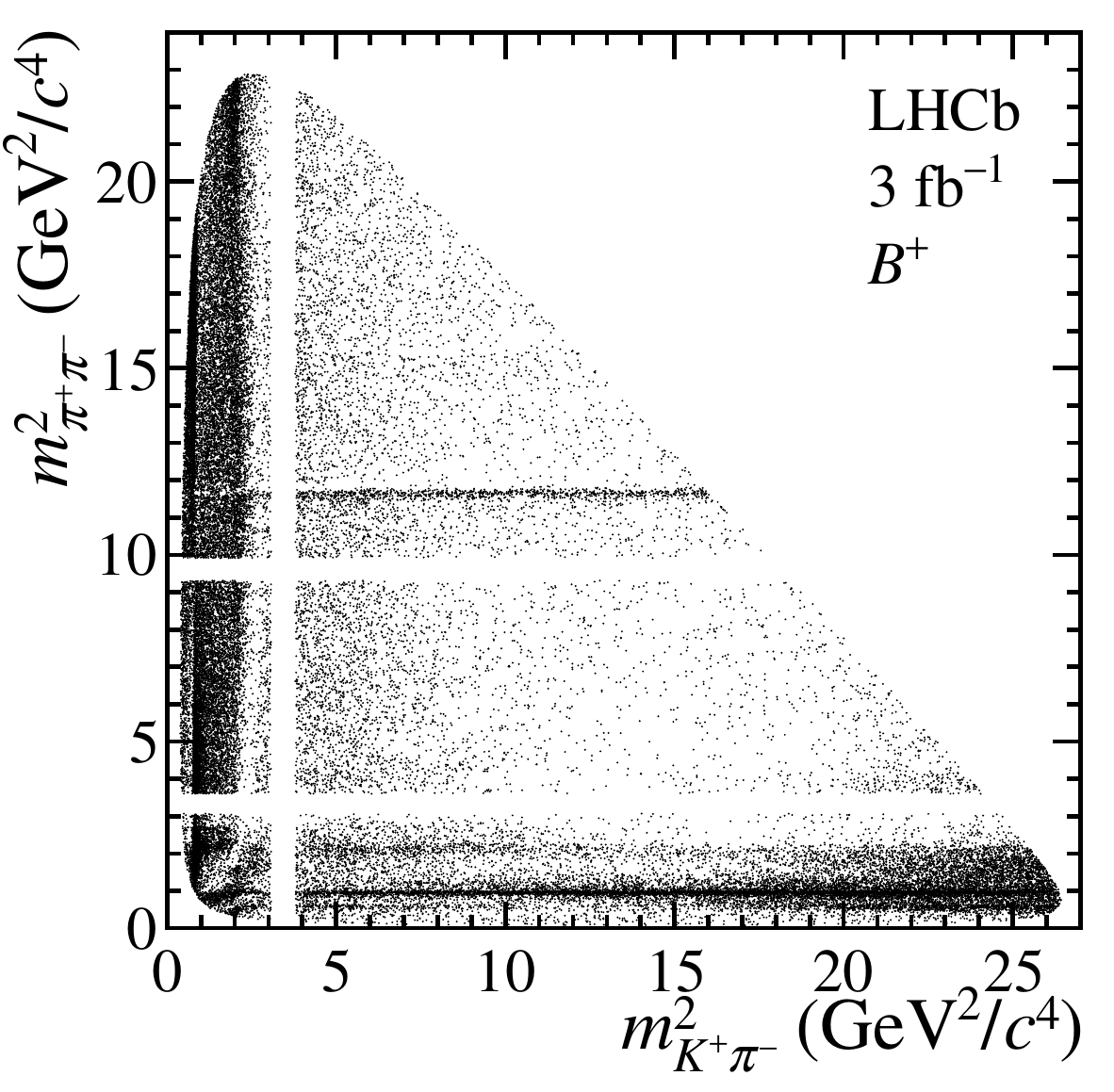}%
    \includegraphics[width=0.5\linewidth]{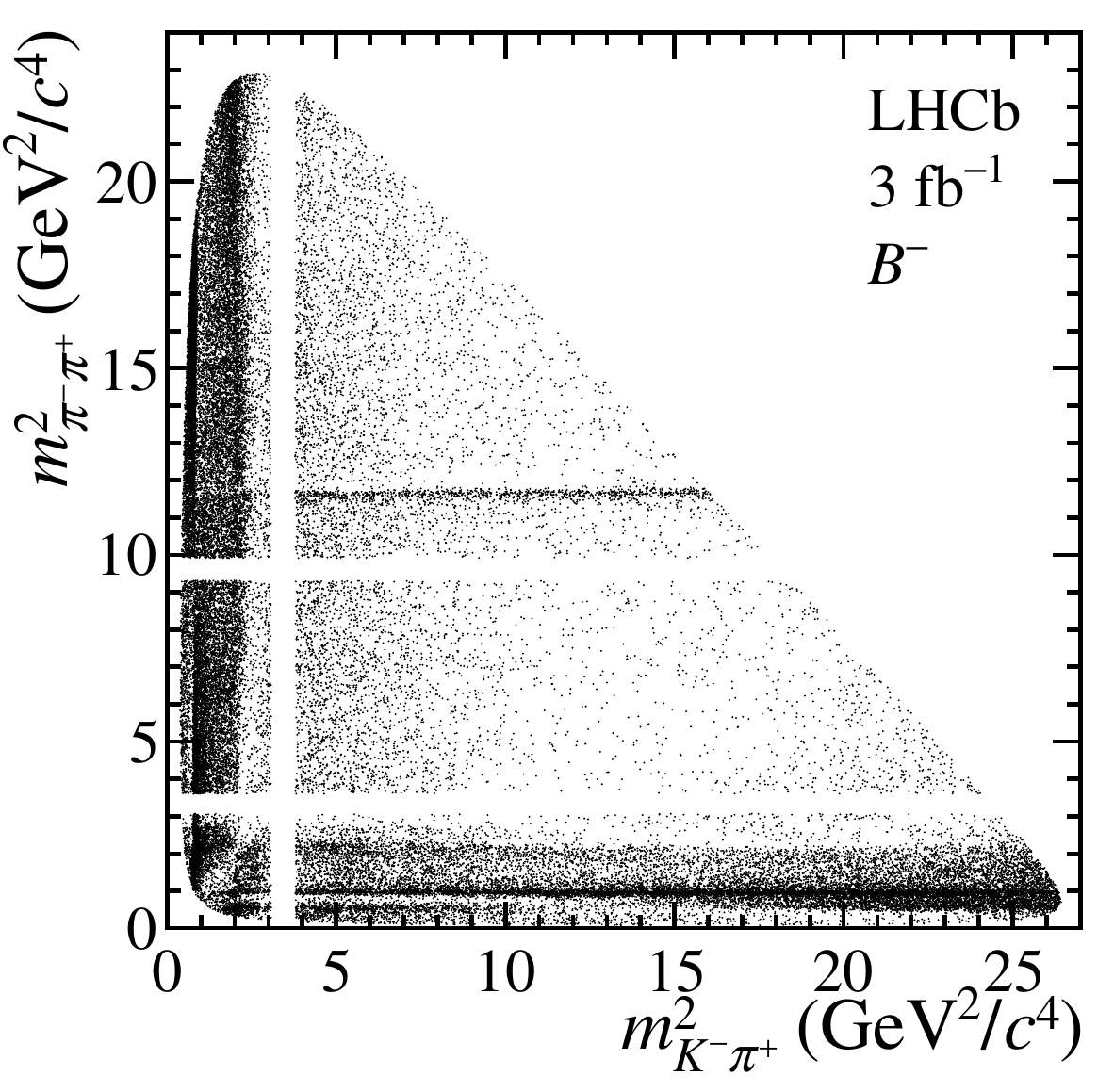}

    \caption{
        Dalitz-plot distributions for (left)~$B^+$ and (right)~$B^-$ candidate decays to the $K^+ \pi^+ \pi^-$ and $K^- \pi^- \pi^+$ final state, respectively. Depleted regions are due to vetoes of $\kern 0.18em \overline{\kern -0.18em D}{}^{0} \!\to K^+ \pi^-$, $\kern 0.18em \overline{\kern -0.18em D}{}^{0} \!\to \pi^+ \pi^-$ and $J\mskip -3mu/\mskip -2mu\psi \! \to \pi^+ \pi^-$ decays.
    }
    \label{fig:dataDP}
\end{figure}

The \decay{\Bp}{\Kp \pip \pim} decay amplitude can be expressed fully in terms of the mass-squared of two pairs of the decay products $m^2_{\Kp\pim}$ and $m^2_{\pip\pim}$, collectively denoted as the position in three-body phase space, $\Phi_3$. The Dalitz-plot distributions of the selected candidates can be seen in Fig.~\ref{fig:dataDP}.

\subsection{Amplitude analysis formalism}
\label{sec:DPformalism:amplitude}

In general, the total amplitude for \Bpm decays is taken as a coherent sum over $N$ components, each described by a decay amplitude $A_j$ that is parameterised with two-body intermediate processes~\cite{isobar1,isobar2,isobar3},
\begin{equation}
    \label{eq:isobar}
    A^\pm(\Phi_3) \equiv \sum_j^N a^\pm_j A_j(\Phi_3)\,,
\end{equation}
where the complex coefficients $a^\pm_j$ represent the relative contribution of component $j$ in the relevant final state. For statistical stability in function minimisation, these are expressed in the ``Cartesian'' parametrisation of \CP violation,
\begin{equation}
    \label{eq:cartesian}
    a_j^{\pm} \equiv (x_j \pm \Delta x_j) + i(y_j \pm \Delta y_j)\,.
\end{equation}
The individual amplitude $A_j$ contains only \CP-conserving dynamics, and is generically parameterised, removing the component index $j$ for brevity, as
\begin{equation}
    A(\Phi_3) \equiv B_{\rm P}(\Phi_3) B_{\rm D}(\Phi_3) S(\Phi_3) T(\Phi_3)\,,
\end{equation}
where the production and decay factors $B_{\rm P}$ and $B_{\rm D}$, respectively, account for deviations from pointlike interactions. 
These are taken to be the normalised Blatt--Weisskopf penetration factors~\cite{Blatt:1952ije,BlattWeisskopf}, that depend on the effective interaction radius taken to be $r_{\rm P} = r_{\rm D} = 4.0 \gev^{-1} \approx 0.8 \fm$, the breakup three-momentum of either decay product $q$ defined in the
rest frame of the decaying state, and the orbital angular momentum $L$ between decay products. 
The angular correlation of the final state particles, which arises due to total angular momentum conservation, is encoded in the spin factor $S$ with the Zemach tensor formalism~\cite{Zemach:1963bc, Zemach:1965ycj}. 
Finally, the propagator $T$ describes the mass lineshape of the intermediate two-body state.

\subsection{Mass lineshapes}

Many intermediate resonant contributions $R$ are described by the relativistic Breit--Wigner propagator
\begin{equation}
    T(s) \equiv \frac{1}{m_0^2 + \xi(s) - s - i
    \sum_j g^2_j \rho_j(s)} \,,
\label{BW}
\end{equation}
where $s$ is the square of the two-body mass. 
The imaginary part is constrained by unitarity, giving rise to the optical theorem which relates this component of the forward scattering amplitude to the sum of probabilities of all possible transitions from the initial state to any final state. 
This simplifies to the form given in Eq.~\eqref{BW}, iterating over the $j$ decay channels that $R$ may decay into with coupling $g_j$ and phase-space factor $\rho_j$.
When considering contributions only from decays into two stable particles, this further simplifies to
\begin{equation}
    \sum_j g^2_j \rho_j(s) = m_0 \Gamma(s) \equiv m_0 \Gamma_0 \frac{m_0}{\sqrt{s}} \left( \frac{q}{q_0} \right)^{2L + 1} B_{\rm D} (q, q_0)^2 \,,
\end{equation}
where $q_0$ is the value of the break-up momentum at the resonance mass, $m_0$.
The energy-dependent decay width, $\Gamma(s)$, is normalised to give the resonance width, $\Gamma_0$, when evaluated at the resonance mass.

The energy-dependent dispersive correction $\xi(s)$ is often neglected with notable exceptions. 
In particular, for contributions involving the broad $\rho$-resonance states, an analytic dispersive term, known as the Gounaris--Sakurai model~\cite{GS}, is included to ensure unitarity far from the pole mass.
Other forms for $\xi(s)$ appear in specific S-wave models to be described in the following.

\subsection{S-wave models}
\label{sec:DPformalism:S-wave}

The $\Kp\pim$ and $\pip\pim$ S-wave ($L = 0$) components of the \decay{\Bp}{\Kp\pip\pim} amplitude are both large in magnitude and contain many overlapping resonances.
In addition, the kinematic ranges of these components include several relevant decay channel thresholds, where increasing two-body mass opens additional decay channels that at lower mass are kinematically inaccessible, which can modulate the intensity.
This analysis includes three distinct treatments of the \mbox{S-wave} components in \decay{\Bp}{\Kp\pip\pim} decays in an attempt to better understand its behaviour. 
The use of multiple approaches also increases confidence that parameters reported for the non-S-wave contributions are robust, and provides additional information for further study.

As such, three sets of results are presented here, corresponding to the cases where the $\Kp\pim$ and $\pip\pim$ S-waves are described by:
(i)~a coherent sum of specific two-body contributions (this model is referred to subsequently as ``Isobar'');
(ii)~substitution of the Isobar S-wave in $\pip\pim$ with a monolithic, two-body unitarity-preserving model informed by historical scattering data \mbox{(``K-matrix'');}
and (iii)~a quasi-model-independent binned approach (``QMI'').
All approaches contain the same contributions from higher-order partial waves, where $L > 0$.

\subsubsection{Isobar model}
\label{sec:Isobar}

In addition to the $K_0^*(1950)^0$, $f_0(1500)$ and $f_0(1710)$ states described by the Breit--Wigner function, the Isobar model S-wave amplitude includes dedicated lineshapes tuned to various phenomena. 
The low \Kp \pim mass region is modelled with the so-called Generalised LASS (GLASS) lineshape~\cite{BaBar:2008inr}, an extension to the unitarity-preserving LASS function~\cite{Estabrooks:1978de,Aston:1987ir} that breaks two-body unitarity in scattering by allowing the complex coupling of the $K_0^*(1430)^0$ resonance to vary relative to an underlying elastic $\pi K$--$\pi K$ scattering contribution. 
This is justified as additional phase motion may be induced by contributions from $K\eta^{(\prime)}$ coupled channels, from the $I=3/2$ amplitude, from interaction between the companion \pip and the \Kp\pim S-wave, and from penguin amplitudes.
The first three of these have been considered as potential sources of unitarity violation in the $K\pi$ S-wave obtained from \decay{\Dp}{\Km\pip\pip} decays~\cite{E791:2005gev}; the fourth is an additional possibility in \decay{\Bp}{\Kp\pip\pim} decays.

As a function of the \Kp\pim mass squared, $s$, the GLASS lineshape is given by~\cite{BaBar:2008inr}
\begin{equation}
    A_{\rm GLASS}(s) \equiv A_{\rm Res} + A_{\rm ER} = a_{\rm GLASS} \left(\sin \delta_{\rm Res} e^{i\delta_{\rm Res}}e^{2i\delta_{\rm ER}} + a_{\rm ER} \sin \delta_{\rm ER} e^{i\delta_{\rm ER}}\right) \,,
    \label{eq:app:glass}
\end{equation}
with
\begin{align}
    \delta_{\rm Res}(s) &= \phi_{\rm Nonres} + \arctan\biggl[\frac{m_{K^*_0}\Gamma(s)}{m^2_{K^*_0} - s}\biggr] \,, \nonumber\\
    \delta_{\rm ER}(s) &= \phi_{\rm ER} + {\rm arccot}\biggl[\frac{1}{aq} + \frac{rq}{2}\biggr] \,, 
    \label{eq:app:glasssplit}
\end{align}
where $a_{\rm GLASS}$ and $a_{\rm ER}$ are free complex couplings of the model determined separately for each \B-meson charge. 
In the resonance-sensitive term $A_{\rm Res}$, $m_{K^*_0}$ is the $K^*_0(1430)^0$ mass and the usual energy-dependent width $\Gamma(s)$ contains the $K^*_0(1430)^0$ resonance width. 
In the effective-range-like term $A_{\rm ER}$, the parameters $a$ and $r$, multiplying the \Kp\pim breakup momentum $q$, represent the scattering
length and effective interaction length, respectively. 
The parameters $\phi_{\rm Nonres}$ and $\phi_{\rm ER}$ are offsets to the scattering strong phases $\delta_{\rm Res}$ and $\delta_{\rm ER}$, respectively. 
In the $\delta_{\rm Res}$ term, $\phi_{\rm Nonres}$ sets the strength of nonresonant entanglement; this term approaches purely resonant behaviour when $\phi_{\rm Nonres}$ tends to zero. 
All parameters of the GLASS lineshape are determined from data in this analysis.

For the broad scalar $f_0(500)$ resonance at low \pip \pim mass, the modified Breit--Wigner model from Bugg is used~\cite{Bugg:2006gc}. 
Besides $f_0(500) \to \pi\pi$ decays, it includes contributions from the $f_0(500) \to K\Kbar$, $\eta \eta$ and $4\pi$ decay modes as well as dispersive effects in the elastic channel. 
The baseline Isobar model has parameters fixed to the values from fit~(iii) of Ref.~\cite{Bugg:2006gc}. 
The Flatt\'{e} function~\cite{Flatte:1976xu} is used to parameterise the $f_0(980)$ resonance, whose pole is close to the $K\Kbar$ threshold. 
The couplings to the $\pi\pi$ and $K\Kbar$ channels governing the total width and the $f_0(980)$ resonance mass are left as free parameters of the model. 
Concerning the $f_0(1370)$ state, where the physical lineshape is not clear at this time, a simple $T$-matrix pole expression is used,
\begin{equation}
    T_{\rm Pole}(s) = \frac{1}{s_0 - s},
\end{equation}
where the pole position, $\sqrt{s_0} = m_0 - i\Gamma_0/2$, is determined from dispersive phenomenological studies with meson-meson scattering data~\cite{Pelaez:2022qby}. 
The pole extracted from $\pi\pi \to K \Kbar $ data is taken, since that from $\pi\pi$ scattering in the elastic channel is less robust due to the need to separate more contributions with different isospin. 
Moreover, trials with the pole from the elastic channel give a much worse $-2\log\mathcal{L}$ value, as defined in Section~\ref{sec:aman:fit}, in the fit to data.

For the lineshapes discussed above, only the direct production of the final state undergoing no further interaction is accounted for in the isobaric sum of S-wave resonant states.
However, rescattering effects need also be considered for a more complete description of final-state interactions. 
Assuming the additional particle produced in the $B$ decay does not interact with the two-meson intermediate state, Watson's theorem guarantees the phase motion in decay will be identical to that of two-meson scattering~\cite{Watson:1952ji}.  
Precise phenomenological amplitude parameterisations based on $\pi\pi$--$\pi\pi$ scattering data are available, from which the parameters of the dispersively constrained~(``CFD'') model are fixed in this analysis~\cite{Garcia-Martin:2011iqs}.

This $\pi\pi$--$\pi\pi$ scattering amplitude, $|\eta^0_0(s)| \exp(i\delta^0_0(s))$, contains a noninteractive term, which is already represented in production by the S-wave isobaric sum, and therefore must first be subtracted. 
An additional phenomenological form factor is typically included to mimic a mild $s$-dependence of the scattering terms for the initial-state two-meson pair. 
It comes from the partonic decay amplitude that produces the three-meson final state, in which the relative momentum between the pair of mesons is distributed among the quarks in the momentum loop within the microscopic amplitude, \eg\ the tree and penguin diagrams, and probes the internal structure of the mesons involved in the initial and final states~\cite{AlvarengaNogueira:2015wpj,Bediaga:2013ela}.

The overall $\pi\pi$--$\pi\pi$ rescattering amplitude is then given by
\begin{equation}
    T_{\pi\pi\textrm{--}\pi\pi}(s) \equiv \frac{|\eta^0_0(s)| e^{i\delta^0_0(s)}-1}{1+s/\Lambda^2_\pi},
    \label{eq:pipipipi-rescattering}
\end{equation}
where $\Lambda_\pi$ would be the only free parameter of the model.
Studies with this rescattering model, however, indicate that $\Lambda_\pi$, which is typically taken to be of order $1\gev$~\cite{LHCb-PAPER-2018-051,LHCb-PAPER-2019-018,LHCb-PAPER-2019-017}, tends to infinity, indicating that the partonic form factor is not essential to the model. 
Therefore, the denominator of the expression in Eq.~\eqref{eq:pipipipi-rescattering} is set to unity.

While the concept of $\pi\pi$--$K \Kbar$ rescattering was originally developed within the context of two-body interactions, for three-body decays rescattering means that a pair of mesons produced in one channel will appear in the final state of a coupled channel. 
Theoretically motivated amplitude models have been derived from scattering data with constraints from $\pi K$--$\pi K$ and $\pi\pi$--$K \Kbar$ dispersion relations~\cite{Pelaez:2020gnd}, from which the parameters referred to as the ``$\textrm{CFD}_{B}$'' set are taken for this analysis. 
The alternative constrained~(``$\textrm{CFD}_{C}$'') set includes spurious experimental data that are in conflict with Watson's theorem near the $K \Kbar$ threshold, and hence is not considered in this analysis.

This $\pi\pi\textrm{--}K \Kbar$ scattering amplitude is presented as unnormalised, so in order to facilitate comparison with the data, it is multiplied by the phase-space factor $4 \sqrt{q_\pi q_K / s}$, where $q$ is the usual breakup momenta for the indicated two-meson final state.\footnote{This analysis is based on a preliminary $\pi\pi\textrm{-}K \Kbar$ scattering model adapted from Ref.~\cite{Garrote:2022uub} in which the definition of $q$ was incorrect by a factor of $2/\sqrt{s}$, with $s$ given in\gev. The difference with the correct version shown in Eq.~\eqref{eq:pipiKK-rescattering} is included as a model systematic uncertainty.} 
The overall $\pi\pi\textrm{--}K \Kbar$ rescattering amplitude is then given by~\cite{Pelaez:2020gnd}
\begin{equation}
    T_{\pi\pi\textrm{--}K \Kbar}(s) \equiv i \frac{4 \sqrt{q_\pi(s) q_K(s)}}{\sqrt{s}} \frac{|g^0_0(s)| e^{i\phi^0_0(s)}}{1+s/\Lambda^2_K},
    \label{eq:pipiKK-rescattering}
\end{equation}
where the partonic form factor that modulates the unknown production magnitude is similarly applied as in Eq.~\eqref{eq:pipipipi-rescattering}, within which $\Lambda_K$ would be the only free parameter of the model. 
However, studies with this rescattering model indicate the partonic form factor can also be discarded from the overall model, and hence the denominator of the right-most term in Eq.~\eqref{eq:pipiKK-rescattering} is set to unity.

\subsubsection{K-matrix model}
\label{sec:Kmatrix}

The coherent sum of resonant contributions modelled with Breit--Wigner lineshapes can be used to describe the dynamics of three-body decays when the quasi-two-body resonances are relatively narrow and isolated.
However, when there are broad, overlapping resonances (with the same isospin and spin-parity quantum numbers) or structures that are near open decay channels, this
model does not satisfy $S$-matrix unitarity, thereby violating the conservation of quantum mechanical probability current.

Assuming that the dynamics is dominated by two-body processes (\ie\ that the \mbox{S-wave} does not interact with other decay products in the final state),
then two-body unitarity is naturally conserved within the K-matrix approach~\cite{Chung:1995dx}.
This approach was originally developed for two-body scattering~\cite{Dalitz:1960du} and the study of resonances in nuclear reactions~\cite{Wigner:1946zz,Wigner:1947zz}, but was extended
to describe resonance production and $n$-body decays in a more general way~\cite{Aitchison:1972ay}. 
In this analysis, the K-matrix approach is applied exclusively to the \pip\pim S-wave, while the \Kp\pim S-wave retains the lineshape model of the Isobar approach.

The K-matrix is parameterised as a sum over the $n$ coupled channels, \mbox{$u \in \left[ \pi\pi, 4\pi, K\Kbar, \eta\eta, \eta\etapr \right]$}, as
\begin{equation}
    T_u(s) \equiv \sum_{v=1}^{n} [(\hat{I} - i \hat{K}(s) \hat{\rho} (s))^{-1}]_{uv} \, \hat{P}_{v}(s) \,,
\end{equation}
where the diagonal matrix $\hat{\rho}$ accounts for the amount of phase space available to each channel. 
The scattering matrix, $\hat{K}$, is taken as the combination of the sum over the $f_0(500)$, $f_0(980)$, $f_0(1370)$, $f_0(1500)$ and $f_0(1710)$ bare mass poles together with nonresonant, or ``slowly varying'', parts and is fixed from a global analysis of $\pi\pi$ scattering data~\cite{Anisovich:2002ij,BaBar:2008inr}. 
Differences in the initial-state environment between particle scattering and decay are accounted for by splicing the $\hat{P}$ production vector with the scattering propagator. 
The parameterisation of $\hat{P}$ is similar in structure to $\hat{K}$ with the exception that the channel couplings are complex and different for \Bp and \Bm decays to allow for \CP violation. 
These couplings are free parameters of the model. However, there is no sensitivity to the parameter governing the shape of the slowly varying part in production, $s_0^{\rm{prod}}$, which is a part of $\hat{P}$. 
Therefore, this parameter is fixed from an external measurement~\cite{FOCUS:2003tdy}.

\subsubsection{Quasi-model-independent analysis}
\label{sec:qmi}

The QMI approach simply describes each of the $K\pi$ and $\pi\pi$ S-waves by binning in mass and allowing a $\CP$-violating complex parameter for each bin with no other structure considered. This set of parameters, varied in the fit, thus represents the average amplitude in each bin. 
This approach exploits the distinctive flat angular structure of the S-wave amplitude to disentangle this component from other contributions to the phase space, assuming the higher-order waves to be well modelled by the isobar approximation. 
Building upon the technical implementation of the \decay{\Bp}{\pip \pip \pim} amplitude analysis from the LHCb collaboration~\cite{LHCb-PAPER-2019-018,LHCb-PAPER-2019-017}, which in turn developed methods pioneered by previous works such as Ref.~\cite{E791:2005gev}, both the overlapping $K^+ \pi^-$ and $\pi^+ \pi^-$ S-wave contributions are modelled.
This marks the first attempt with such an approach in which multiple partial waves produced by particle decay are simultaneously modelled. 
The binning is largely {\it ad hoc}, except that bin edges are tuned to align with the kinematic openings of coupled channels, since cusps in the amplitude are always expected at these thresholds from scattering theory. 
In $K^+ \pi^-$ the relevant thresholds are for the $K 3\pi$, $K\eta$, $K\etapr$, $K \eta_c$ and $D_sD$ channels, while in $\pi^+ \pi^-$ the $4\pi$, $K \Kbar$, $\eta\eta$, $\eta\etapr$, $\eta\eta_c$, $\D\Db$ and $\Dsp\Dsm/\etapr\eta_c$ thresholds are considered. Each S-wave is assigned 50 bins. 
In $K^+ \pi^-$ 35 of these bins are allocated below the open charm threshold, while in $\pi^+ \pi^-$ this number is set to 40. 
Overlapping crossing binned S-waves contain an inherent mathematical ambiguity in the expression of their sum that is not present when a singular partial wave is modelled~\cite{Krinner:2017dba}. For fit convergence, it is therefore essential to fix a value for each component of the couplings, \ie $x$, $y$, $\Delta x$ and $\Delta y$, in one QMI bin. This bin need not necessarily be the same for each coupling component. 
As far as the $-2\log\mathcal{L}$ value (see Eq.~\eqref{eq:likelihood} below) is concerned, the value in each of the fixed bins may take on any finite number. However, these should be as close as possible to their true physical values in the interest of obtaining meaningful separation between the crossing partial S-waves. 
For this purpose, one bin in each coupling component is fixed from the measured Isobar amplitude, by means of a one-dimensional $\chi^2$ scan, ultimately selecting the bin that leads to the closest agreement between the Isobar prediction and the QMI S-wave fit results. 
Due to the potential for bias in the derived one-dimensional QMI S-wave amplitudes, it is also necessary to include an additional systematic uncertainty on these projections, accounting for the full spread of the amplitude value in the fixed bins coming from the Isobar result. 
The two-dimensional sum of the $K^+ \pi^-$ and $\pi^+ \pi^-$ QMI S-waves is unaffected by this effect.

\subsection{Measurement quantities}

The primary outputs of the Dalitz-plot fit are the complex coefficients $a_j^\pm$ defined in Eq.~\eqref{eq:isobar}. 
However, since these are convention-dependent, they have limited physical meaning apart from their phase values, $\arg(a_j^\pm)$, which are reported. 
It is often useful to compare \CP-averaged fit fractions for each intermediate component, $j$, defined as ${\cal F}_{j} = {\cal F}^-_{j} + {\cal F}^+_{j}$, where
\begin{equation}
    {\cal F}^\pm_j \equiv \frac{ \int |a_j^\pm A_j(\Phi_{3})|^2~{\rm d}\Phi_3}{\int (|\sum_j a_j^- A_j(\Phi_3)|^2 + |\sum_j a_j^+ A_j(\Phi_3)|^2)~{\rm d}\Phi_3}\,,
\label{equation:fitFracs}
\end{equation}
for charge-separated \Bpm decays. 
These fit fractions alone will not sum to unity if there is net constructive or destructive interference. 
Such effects are described by the interference fit fractions, given by
\begin{equation}
    {\cal I}^\pm_{j\,<\,k} \equiv \frac{\int 2\Real [a_j^\pm a_k^{\pm *} A_j(\Phi_3) A_k^*(\Phi_3)]~{\rm d}\Phi_3}{\int (|\sum_j a_j^- A_j(\Phi_3)|^2 + |\sum_j a_j^+ A_j(\Phi_3)|^2)~{\rm d}\Phi_3}\,,
    \label{equation:intFracs}
\end{equation}
where the sum over all fit and interference fractions is unity by definition. 
Another important physical quantity is the quasi-two-body parameter of \CP violation in decay associated with a particular intermediate contribution,
\begin{equation}
    \mathcal{A}_{\CP}^j \equiv \frac{{\cal F}^-_{j} - {\cal F}^+_{j}}{{\cal F}^-_{j} + {\cal F}^+_{j}}\,.
    \label{eq:cpAsy}
\end{equation}

The violation of \CP originating from long-distance interference effects cannot be quantified through an asymmetry parameter defined analogously to Eq.~\eqref{eq:cpAsy} by replacing $\mathcal{F}^\pm$ with the corresponding $\mathcal{I}^\pm$ terms of Eq.~\eqref{equation:intFracs}, as the denominators could become null or negative. 
Instead, this effect is described within the quasi-two-body asymmetry of the combined components,
\begin{equation}
    \mathcal{A}^{jk}_{\CP} \equiv \frac{{\cal F}^-_{jk} - {\cal F}^+_{jk}}{{\cal F}^-_{jk} + {\cal F}^+_{jk}} =
    \mathcal{A}^j_{\CP}\frac{\mathcal{F}_j}{\mathcal{F}_{jk}} + 
    \mathcal{A}^k_{\CP}\frac{\mathcal{F}_k}{\mathcal{F}_{jk}} +
    2\mathcal{A}_\mathcal{I}^{jk} \,,
    \label{eq:ACP-IFFdef}
\end{equation}
where $\mathcal{A}_\mathcal{I}^{jk} = (\mathcal{I}^-_{jk} - \mathcal{I}^+_{jk})/\mathcal{F}_{jk}$.
The compound fit fractions with two indices, ${\cal F}^-_{jk}$ and ${\cal F}^+_{jk}$, are determined using Eq.~\eqref{equation:fitFracs}, with coherent sums of the $j$ and $k$ components instead comprising the numerator, and ${\cal F}_{jk} = {\cal F}^-_{jk} + {\cal F}^+_{jk}$. In the decomposition of Eq.~\eqref{eq:ACP-IFFdef}, the first two terms arise from quasi-two-body \CP asymmetry in their respective components, while the third term, $\mathcal{A}_\mathcal{I}^{jk}$, quantifies \CP asymmetry in their interference.

\subsection{Square Dalitz plot}
\label{sec:sqDP}

Since resonances tend to populate the edges of the conventional Dalitz plot in charmless \B decays, it is useful to define the so-called ``square'' Dalitz plot~\cite{BaBar:2005jqu}, which provides improved resolution in these critical regions when using uniform binning, for example when modelling experimental effects.  Furthermore, the mapping to a square space aligns the constant bin boundaries to the kinematic phase-space boundaries. 
The square Dalitz plot is defined in terms of the variables $m^\prime$ and $\theta^\prime$
\begin{equation}
m^\prime \equiv \frac{1}{\pi} \arccos \left( 2 \, \frac{m_{\pip\pim} - m_{\pip\pim}^{\rm min}}{m_{\pip\pim}^{\rm max} - m_{\pip\pim}^{\rm min}} - 1 \right), \ \ \ \ \theta^\prime \equiv \frac{1}{\pi} \theta_{\pip\pim},
\end{equation}
where $m_{\pip\pim}^{\rm max} = m_{\Bp} - m_{\Kp}$ and $m_{\pip\pim}^{\rm min} = 2 m_{\pip}$ represent the kinematic limits of the $\pip\pim$ mass 
permitted in the \decay{\Bp}{\Kp\pip\pim} decay, and $\theta_{\pip\pim}$ is the angle between the \Kp and \pim three-momentum vectors in the $\pip\pim$ rest frame.
Transition between phase-space bases, ${\rm d}\Phi_3 \to |\!\det J|{\rm d}m^\prime {\rm d}\theta^\prime$, is facilitated by the Jacobian of the transformation,
\begin{equation}
    |\!\det J| = 2\pi^2 |\vec p_\pim| |\vec p_\Kp| m_{\pip\pim} (m_{\pip\pim}^{\rm max} - m_{\pip\pim}^{\rm min}) \sin (\pi m^\prime) \sin (\pi \theta^\prime) \,,
\end{equation}
where the three-momenta $\vec p$ are also evaluated in the $\pip\pim$ rest frame.

All efficiencies and backgrounds described in Sections~\ref{sec:efficiencies} and~\ref{sec:backgrounds} are determined as functions of the square Dalitz-plot variables, modelled as histograms with a uniform $26\times25$ binning scheme in $m^\prime$ and $\theta^\prime$. 
These histograms are determined separately for \Bp and \Bm decays and further smoothed by a two-dimensional cubic spline to mitigate effects of discontinuity at the bin edges, with bins abutting kinematic boundaries reflected to ensure good behaviour at the edge of the phase space.
The \Dz\ and \jpsi\ vetoes are not included in these maps, as they cover very narrow regions of phase space where the subsequent binning and smoothing of the histograms would otherwise lead to biased results.

\subsection{Efficiency model}
\label{sec:efficiencies}

The efficiency of selecting a signal decay is parameterised in the two-dimensional square Dalitz plot and determined separately for \Bp and \Bm decays.
Nonuniformities in the efficiency of selecting a signal decay arise as a result of the detector geometry, trigger and reconstruction algorithms, particle identification selections, production and detection asymmetries, and other background rejection requirements such as that imposed by the BDT classifier to discriminate against combinatorial background.
The efficiency map is first obtained using simulated decays, however differences between data and simulation arising in the aforementioned effects are each calibrated with data control samples. 
These corrections are applied to the baseline model except for those related to the combinatorial background suppression BDT, which are considered in the systematic uncertainties. 

The effect of an asymmetry between the production rates of $\Bm$ and $\Bp$ mesons is indistinguishable in this analysis from a global detection efficiency asymmetry.
Therefore, the \Bp production asymmetry, as measured within the LHCb acceptance~\cite{LHCb-PAPER-2016-062}, is taken into
account by introducing a global asymmetry of approximately $-0.6\%$ into the efficiency maps. This is obtained as an average of the measured production asymmetries, weighted by the relative integrated luminosity obtained in 2011 and 2012.

The hardware-trigger efficiency correction is calculated using pions from \decay{\Dz}{\Km\pip} decays, arising from promptly produced \decay{\Dstarp}{\Dz(\to\Km\pip)\pip} decays, and affects two disjoint subsets of the selected candidates: those where the trigger requirements were satisfied by hadronic calorimeter deposits as a result of the signal decay and those where the requirements were satisfied only by deposits from the rest of the event.
In the first case, the probability to satisfy the trigger requirements is calculated using calibration data as a function of the transverse energy of each final-state particle of a given species, the dipole-magnet polarity, and the hadronic calorimeter region.
In the second subset, a smaller correction is applied following the same procedure in order to account for the requirement that these tracks did not fire the hardware trigger.
These corrections are combined according to the relative abundance of each category in data.

\begin{figure}[!tb]
    \centering
    \includegraphics[width=0.5\linewidth]{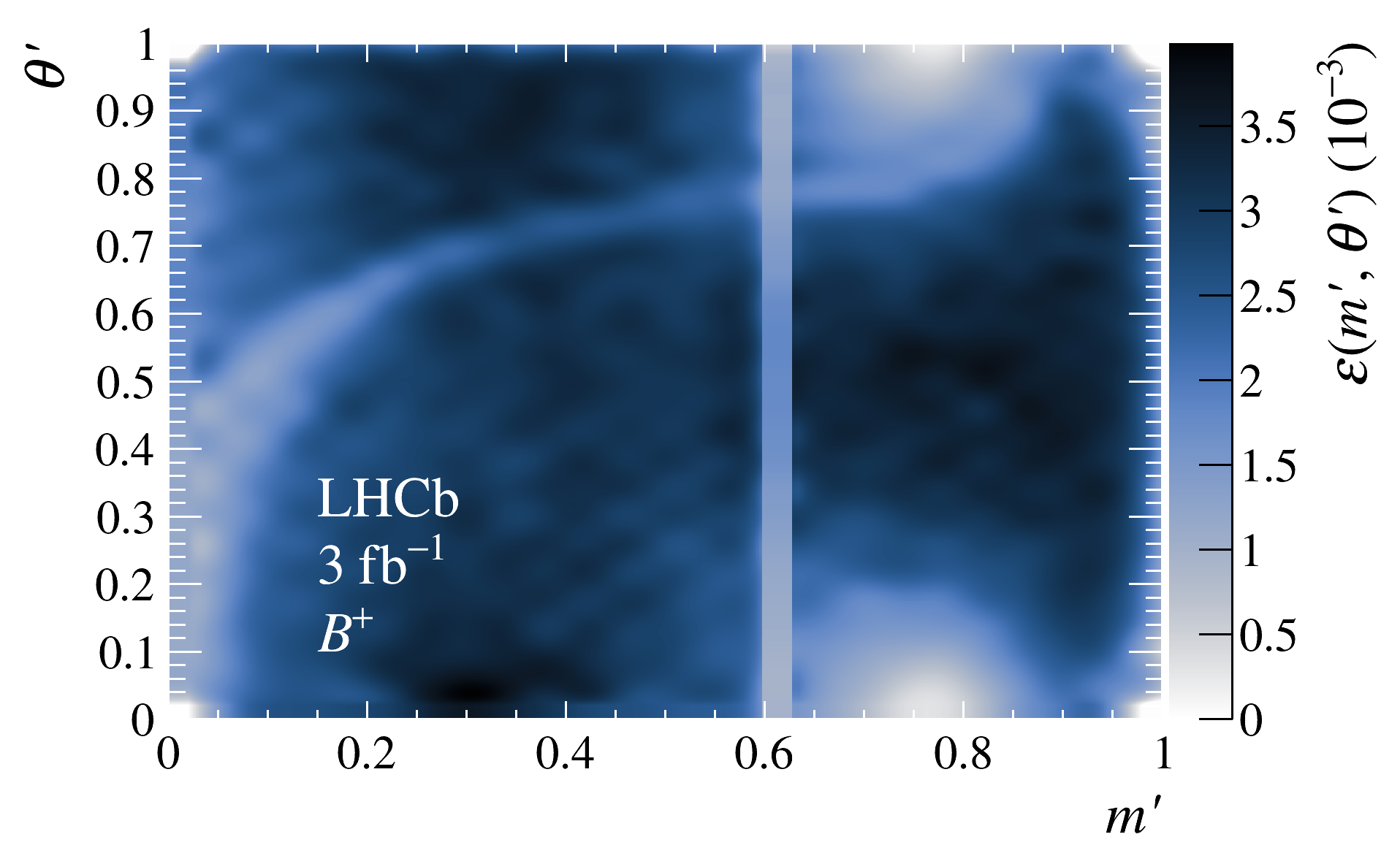}%
    \includegraphics[width=0.5\linewidth]{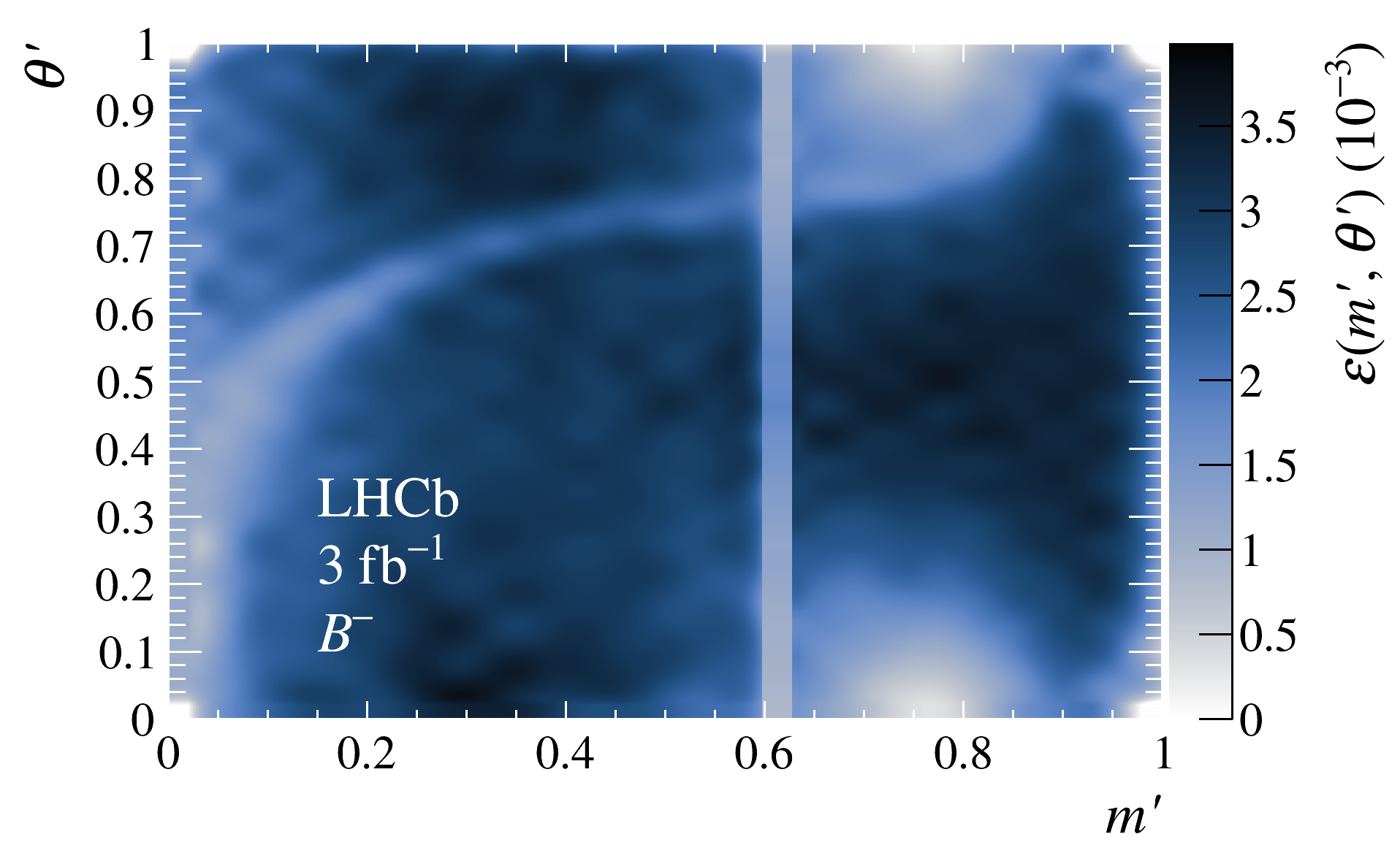}

    \caption{
        Square Dalitz-plot distributions for the (left)~$B^+$ and (right)~$B^-$ signal efficiency models, $\epsilon(m^\prime,\theta^\prime)$, smoothed using a two-dimensional cubic spline. Depleted regions that traverse the plots arise where more stringent PID requirements are applied near the $\kern 0.18em \overline{\kern -0.18em D}{}^{0}$ meson.
    }
    \label{fig:effSDP}
\end{figure}

Prompt \decay{\jpsi}{\mup \mun} decays provide a calibration of the tracking efficiency~\cite{LHCb-DP-2013-002}, by comparing the performance of the default track-finding algorithm to that obtained when dropping information from at least one subdetector. 
This calibration is determined as a function of the track momentum and pseudorapidity, and of the multiplicity of the event, and is assumed to factorise with respect to the final-state tracks so that the efficiency for each track is multiplied to form the overall efficiency.
The particle identification efficiency, also assumed to factorise, is calculated from calibration data corresponding to the
\decay{\Dstarp}{\Dz(\to \Km\pip)\pip} decay, where pions and kaons can be identified without the use of the LHCb particle identification system~\cite{LHCb-PUB-2016-021}. 
The particle identification efficiencies for the background-subtracted pions and kaons are parameterised in terms of their total and transverse momentum, and the number of tracks in the event.
Finally, with the approach devised for the $b$-hadron production asymmetry measurements~\cite{LHCb-PAPER-2016-062}, residual detection asymmetries not already accounted for are determined from prompt charm control samples in bins of momentum.

The overall efficiency, as a function of square Dalitz-plot position, can be seen in Fig.~\ref{fig:effSDP} for \Bp and \Bm decays separately. 
Here, and in similar plots, depleted regions 
associated to the more stringent PID requirements placed around the \Dzb meson are visible.

\subsection{Resolution effects}

Due to narrow \pip\pim mass structures, particularly the $\omega(782)$ and $\chi_{c0}(1P)$ resonances, the signal PDF is convolved with a one-dimensional Gaussian function representing the detector resolution,~$R$. 
Since there are no structures in the $m_{\Kp\pim}$ spectrum that are narrow compared to the detector resolution, only a one-dimensional smearing is required.
The resolution itself is taken from simulated decays depending on the mass square $s_{\pip\pim}$.

The signal PDF for \Bp or \Bm decays is then given by
\begin{equation}
  {\cal P}^{\pm}_{\rm Sig}(\Phi_3) = \frac{\epsilon^{\pm}(m^\prime, \theta^\prime) |A^{\pm}(\Phi_3)|^2 \otimes R(s_{\pip\pim})}{\displaystyle\int (\epsilon^{-}(m^\prime, \theta^\prime) |A^{-}(\Phi_3)|^2 + \epsilon^{+}(m^\prime, \theta^\prime) |A^{+}(\Phi_3)|^2)~{\rm d}\Phi_3} \,,
\end{equation}
where $\epsilon^\pm$ represents the Dalitz-plot dependent efficiency for the $B^\pm$ decay. 
Furthermore, the convolution is not applied to the denominator as this would amount to a trivial additive constant in the $-2\log\mathcal{L}$ value (see Eq.~\eqref{eq:likelihood} below) due to Fubini's theorem.

\subsection{Background model}
\label{sec:backgrounds}

\begin{figure}[!tb]
    \centering
    \includegraphics[width=0.5\linewidth]{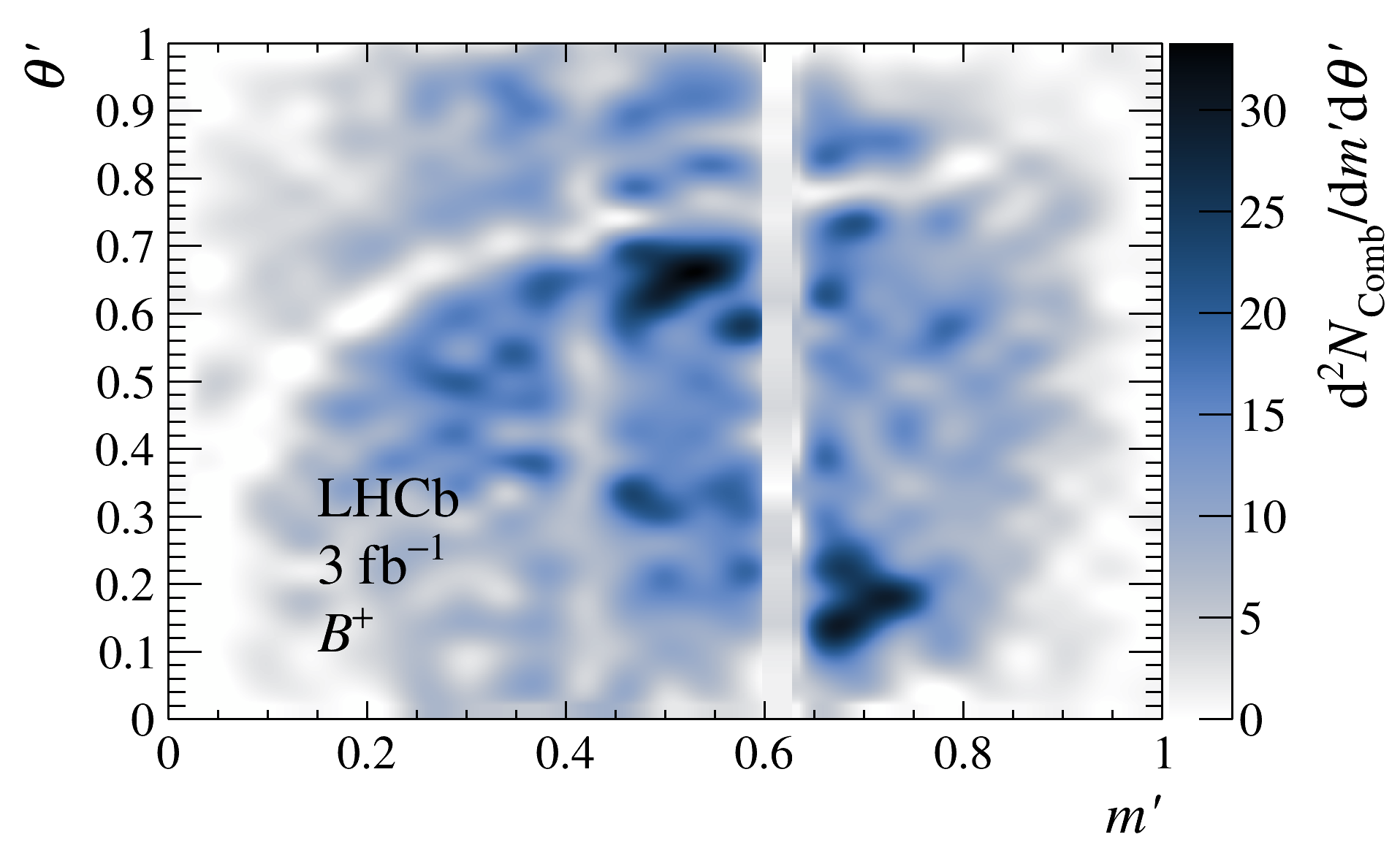}%
    \includegraphics[width=0.5\linewidth]{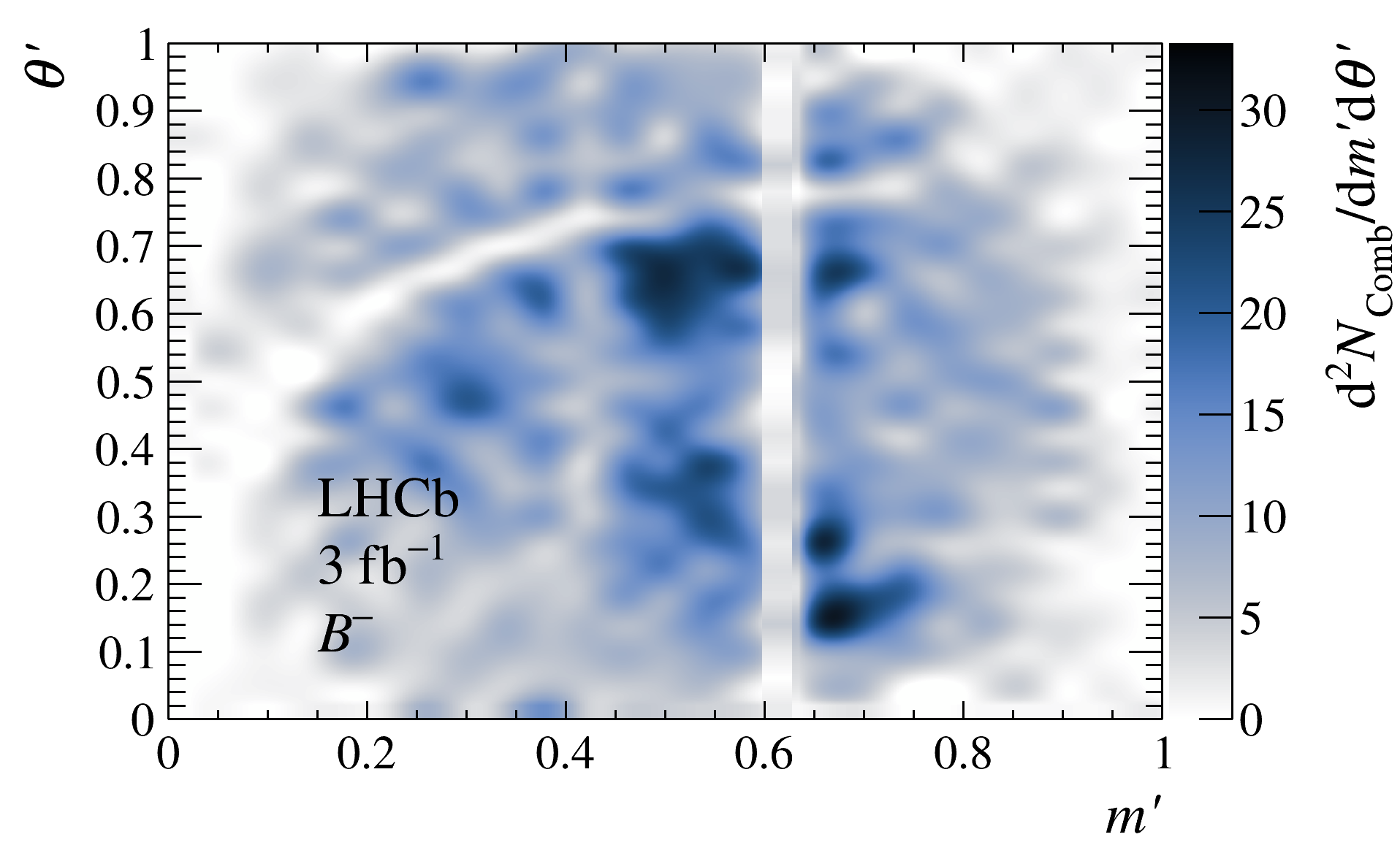}

    \caption{
        Square Dalitz-plot distributions for the (left)~$B^+$ and (right)~$B^-$ combinatorial background models, scaled to represent their respective yields in the signal region.
        }
    \label{fig:combinatorialBkg}
\end{figure}

The dominant source of background in the signal region is combinatorial in nature, modelled using upper-sideband data from the region $5.65 < \mKpipi < 6.10\gevcc$. 
The corresponding combinatorial background distributions can be seen in Fig.~\ref{fig:combinatorialBkg}.
In the Dalitz-plot fit, the charge asymmetry in the combinatorial background yield is fixed to that obtained in the \Bp mass fit described in Section~\ref{sec:Mass_fit}.

\begin{figure}[!tb]
    \centering
    \includegraphics[width=0.5\linewidth]{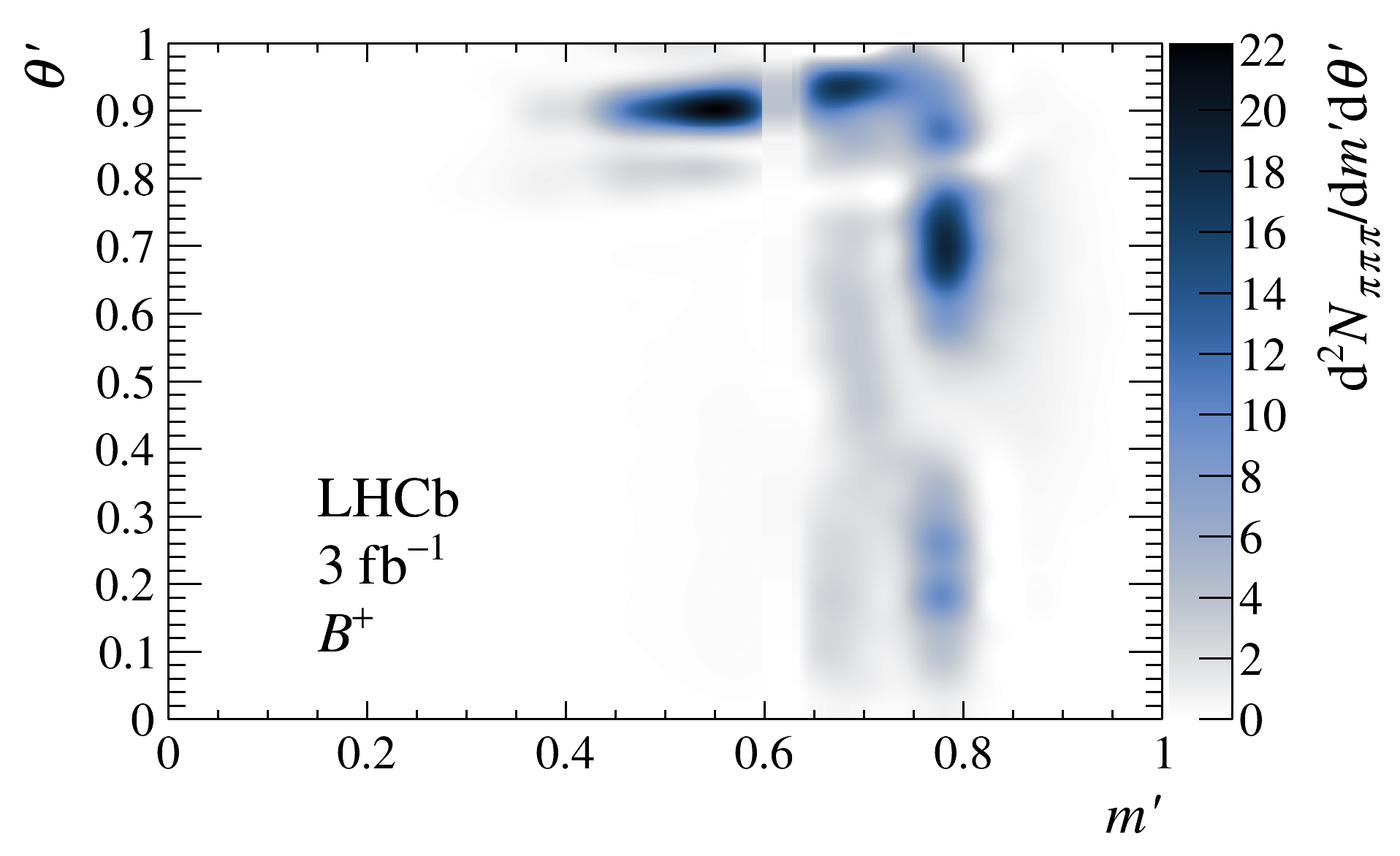}%
    \includegraphics[width=0.5\linewidth]{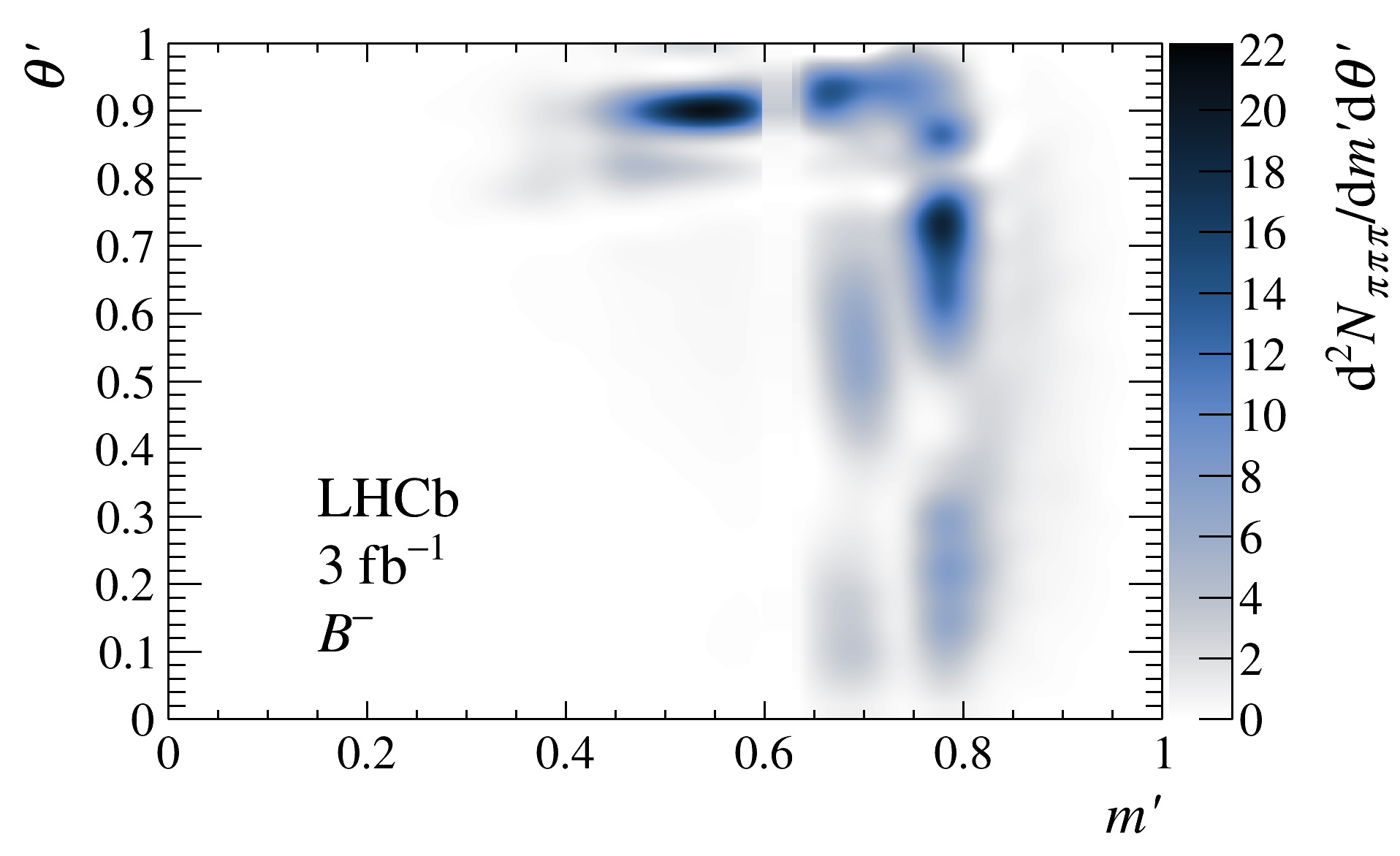}

    \caption{
        Square Dalitz-plot distributions for the misidentified $B^+ \!\to \pi^+ \pi^+ \pi^-$ background models in (left)~$B^+$ and (right)~$B^-$ samples, scaled to represent their respective yields in the signal region.
        }
    \label{fig:crossFeedBkgB2pipipi}
\end{figure}

\begin{figure}[!tb]
    \centering
    \includegraphics[width=0.5\linewidth]{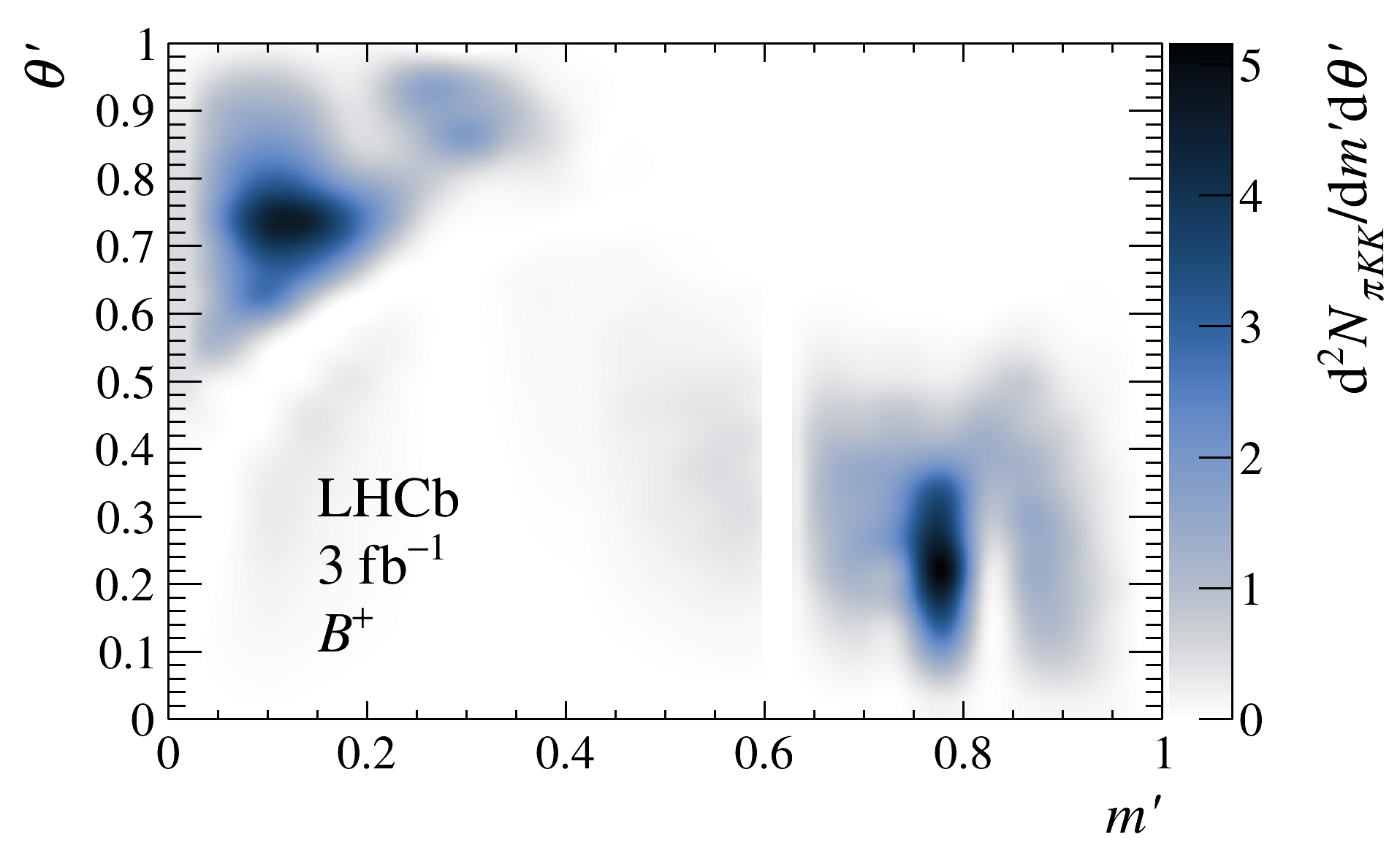}%
    \includegraphics[width=0.5\linewidth]{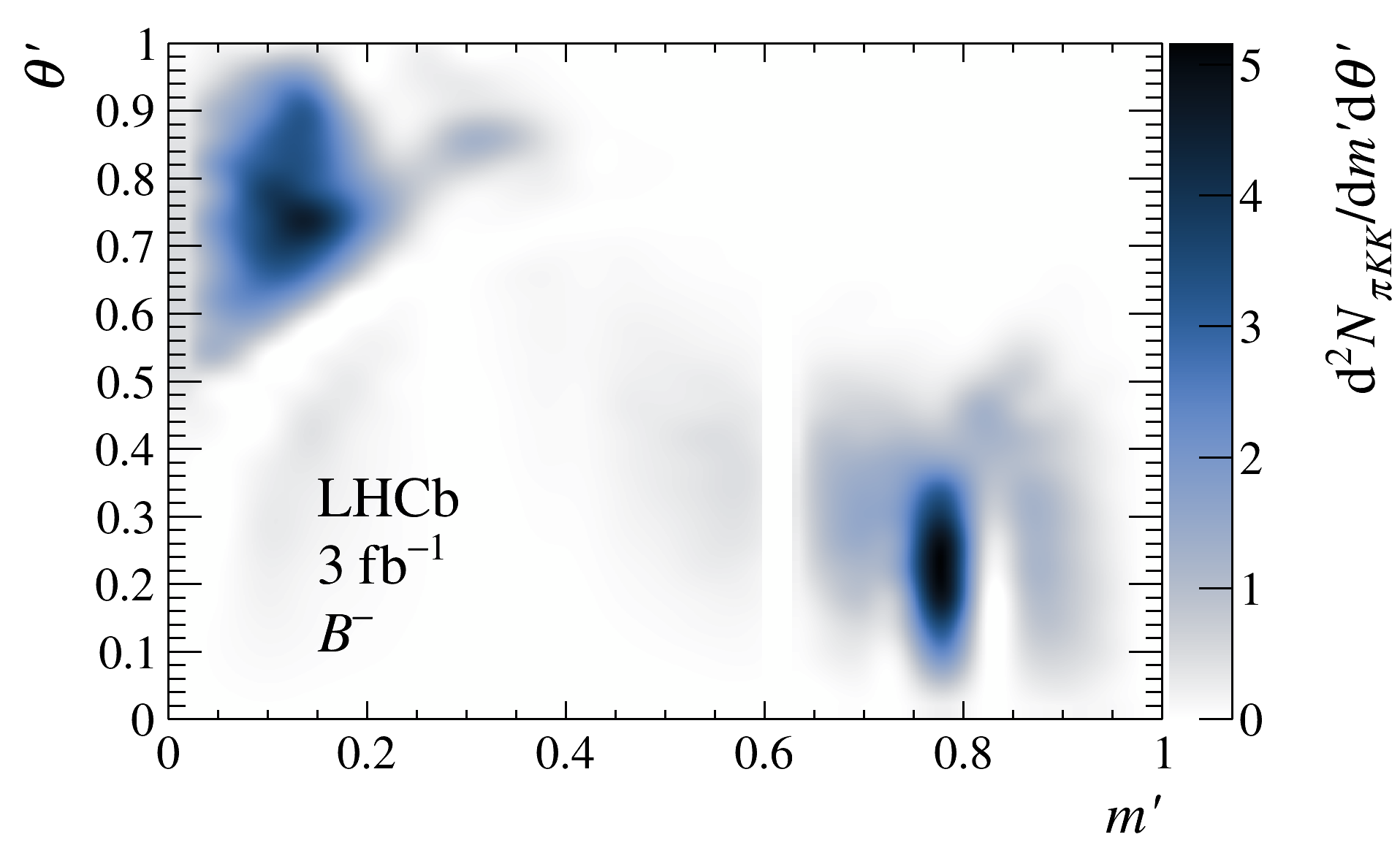}

    \caption{
        Square Dalitz-plot distributions for the misidentified $B^+ \!\to \pi^+ K^+ K^-$ background models in (left)~$B^+$ and (right)~$B^-$ samples, scaled to represent their respective yields in the signal region.
        }
    \label{fig:crossFeedBkgB2pikk}
\end{figure}

Sources of peaking background arise from misidentified decays reconstructed under the \decay{\Bp}{\Kp\pip\pim} decay hypothesis. These backgrounds, including \decay{\Bp}{\pip\pip\pim} and \decay{\Bp}{\pip\Kp\Km} decays, are modelled using simulated events with corrections to account for differences between data and simulation applied in the same way as for the signal efficiency model. 
To account for the phase-space distributions of these backgrounds, the events are further weighted according to their respective amplitude models obtained by the \lhcb collaboration~\cite{LHCb-PAPER-2019-017,LHCb-PAPER-2019-018,LHCb-PAPER-2018-051}. 
The corresponding distributions can be seen in Figs.~\ref{fig:crossFeedBkgB2pipipi} and~\ref{fig:crossFeedBkgB2pikk}.

\begin{figure}[!tb]
    \centering
    \includegraphics[width=0.5\linewidth]{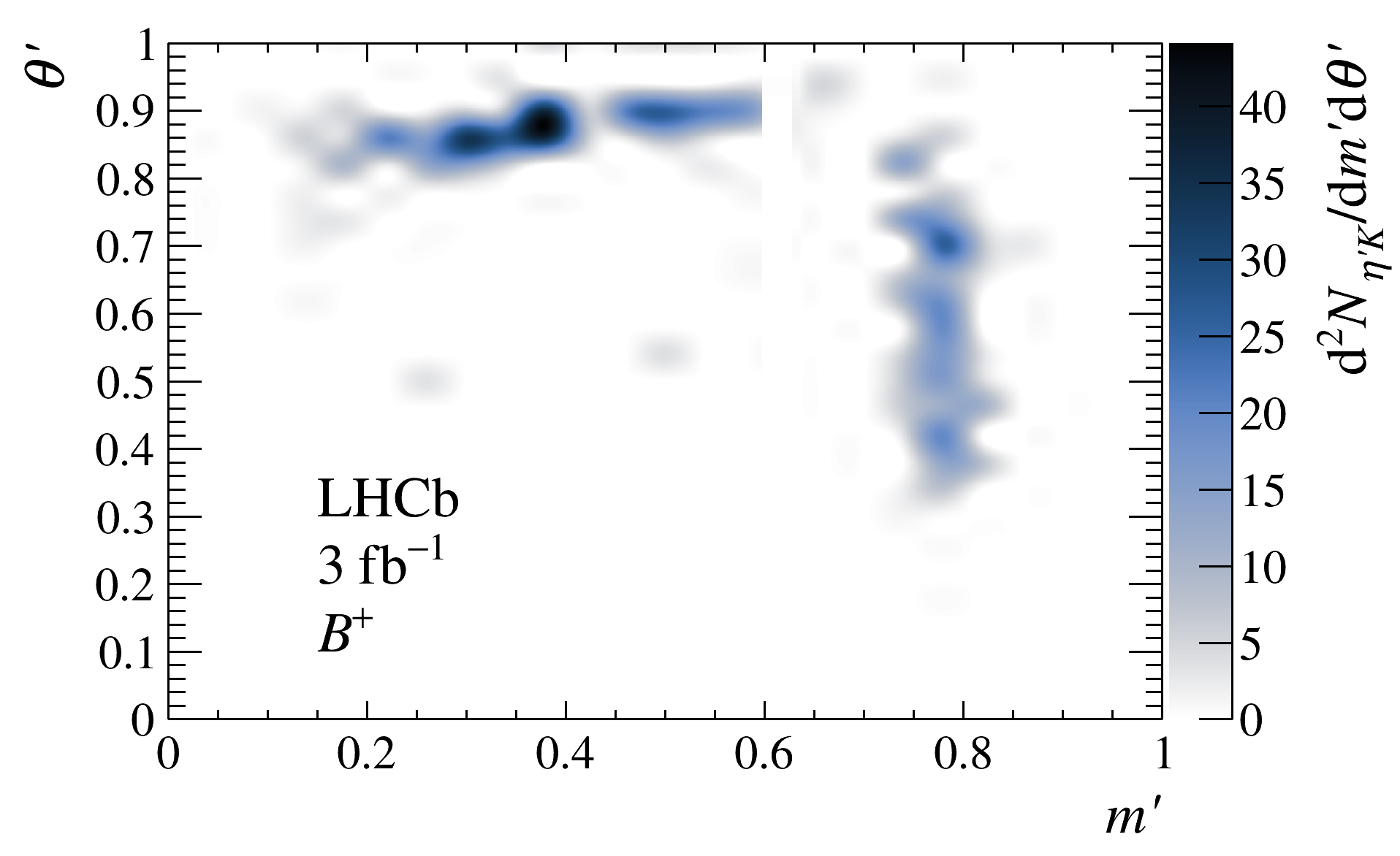}%
    \includegraphics[width=0.5\linewidth]{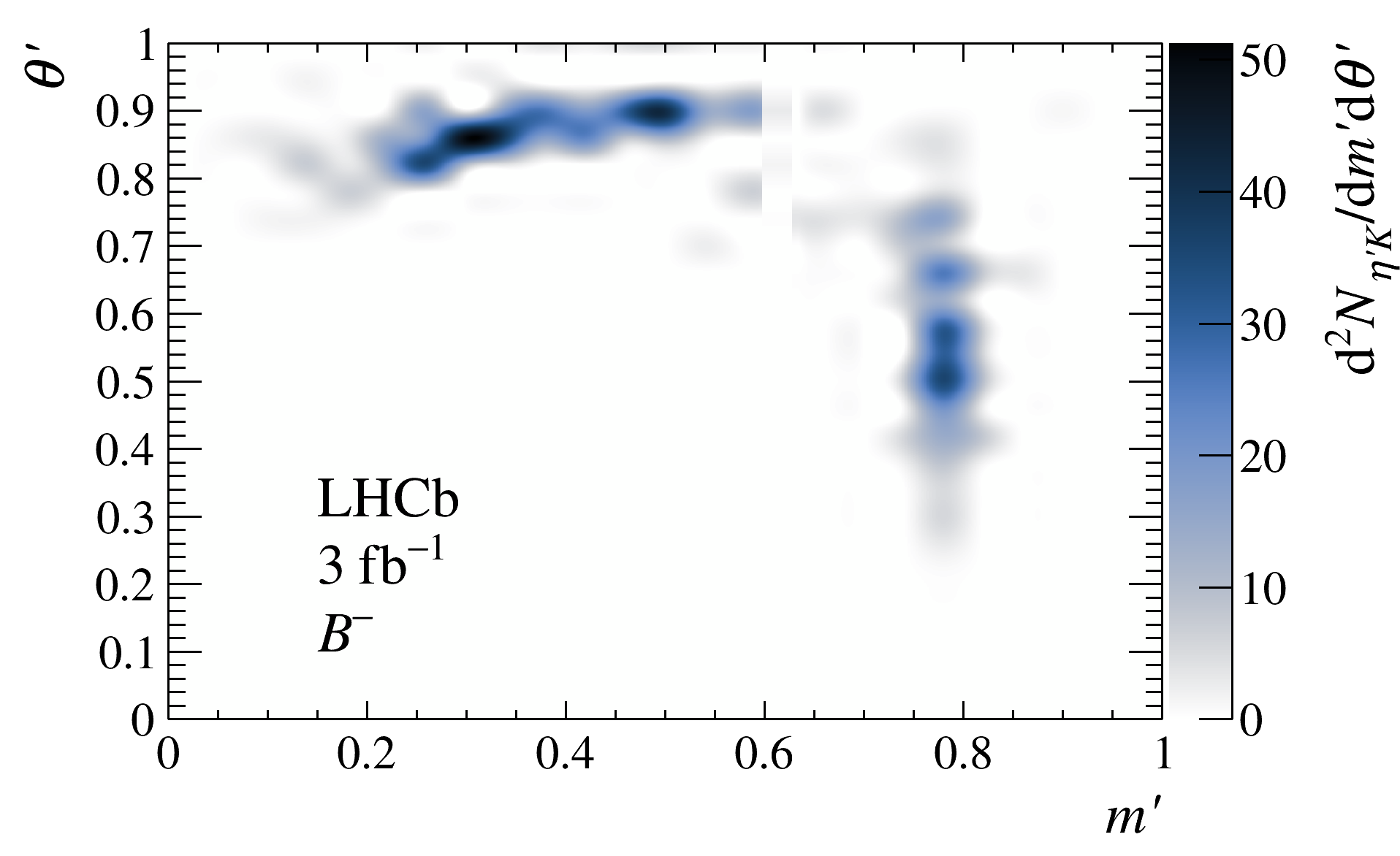}

    \caption{Square Dalitz-plot distributions for the partially reconstructed $B^+ \!\to K^+ \eta^{\prime}(\pi^+ \pi^- \gamma)$ background models in (left)~$B^+$ and (right)~$B^-$, scaled to represent their respective yields in the signal region.
    }
    \label{fig:crossFeedBkgB2etapk}
\end{figure}

A final category of partially reconstructed peaking background originates from \decay{\Bp}{\Kp\etapr(\to\pip\pim\gamma)} decays. These are treated in the same way as the misidentified peaking backgrounds except that the corrected simulation is reweighted by the \decay{\etapr}{\pip\pim\gamma} amplitude model from the \besiii collaboration~\cite{BESIII:2017kyd}.
The background distributions are shown in Fig.~\ref{fig:crossFeedBkgB2etapk}. All background models transformed back to the traditional Dalitz plot can be found in Appendix~\ref{app:background}.

\subsection{Fit procedure}
\label{sec:aman:fit}

Of the three approaches to the S-wave, the Isobar and K-matrix fits are performed using the \textsc|Laura++| Dalitz-plot
fitter package \verb|v3r5p3|~\cite{Laura++}, which interfaces to the \verb|MINUIT| function minimisation
algorithm~\cite{James:1975dr,Brun:1997pa}.
In contrast, the QMI approach relies on
the \textsc|Mint2|~\cite{Mint2} amplitude-analysis interface to \verb|Minuit2|~\cite{Brun:1997pa}.
The fundamental difference between these amplitude-analysis software packages is in the handling of the normalisation.
The former approximates the definite integral by employing a Gaussian quadrature approach, while the latter invokes a stochastic sampling technique.
Additionally, due to the size of its parameter space, the QMI approach greatly benefits from the use of GPU-accelerated solutions.

In all cases, the $K^*(892)^0$ component is set to be the ``reference'' amplitude.
In practice, this means that the average magnitude of the \Bp and \Bm coefficients for this component is set to unity, \ie $x=1$ in Eq.~\eqref{eq:cartesian}, while the $\Delta x$ parameter is left free to vary to allow for \CP violation.
Since there is no sensitivity to the phase difference between the \Bp and \Bm amplitudes, the imaginary part of the $K^*(892)^0$ component is set to zero for both \Bp and \Bm ($y = \Delta y = 0$), which means that the coefficients of all other contributions to the model are measured relative to this component.

The extended likelihood function that is optimised is of the form
\begin{equation}
    \label{eq:likelihood}
	\mathcal{L} = \frac{e^{-\sum_j N_j}}{N!} \prod_{i=1}^{N_{\rm Cand}} \left[\sum_j N_j \mathcal{P}_j(\Phi^i_3)\right]\,,
\end{equation}
where $N_j$ is the yield for the candidate category $j$ (fixed from Table~\ref{tab:massFit}), $N_{\rm Cand}$ is the total number of candidates in the signal region, and $\mathcal{P}_j$ is the probability density function for candidates in category $j$ in terms of the Dalitz-plot coordinates, $\Phi^i_3$.
The \Bp and \Bm data are modelled simultaneously and the optimal values of the fitted parameters are found by minimising twice the negative log-likelihood, $-2\log \mathcal{L}$.

\begin{table}[!tb]
    \centering
    \caption{\label{tab:aman:swave}
        Components of the \Kp\pim and \pip\pim S-waves in each of the three approaches considered in the analysis, together with their fixed lineshape parameters.
    }
    \resizebox{\textwidth}{!}{
    \renewcommand{\arraystretch}{1.1}
    \begin{tabular}
    {l@{\hspace{0.125cm}}
     @{\hspace{0.125cm}}l@{\hspace{0.125cm}} 
     @{\hspace{0.125cm}}l@{\hspace{0.125cm}}
     @{\hspace{0.125cm}}l} \hline
    Component & Approach & Mass lineshape, $T(s)$ & Fixed parameters\\ \hline

    $(\Kp \pim)_{\rm Res}$ & Isobar/K-matrix & GLASS~\cite{BaBar:2008inr,Aston:1987ir} & ---\\
    $(\Kp \pim)_{\rm ER}$\\ \hline
    $K_0^*(1950)^0$ & Isobar/K-matrix & Breit--Wigner & $m_0 = 1.945 \pm 0.022 \gevcc$~\cite{PDG2020}\\
    & & & $\Gamma_0~ =  0.201 \pm 0.086 \gev$~\cite{PDG2020}\\ \hline
    $f_0(500)$ & Isobar & Bugg & fit (iii) of Ref.~\cite{Bugg:2006gc}\\ \hline
    $f_0(980)$ & Isobar & Flatt\'{e}~\cite{Flatte:1976xu} & ---\\ \hline
    $f_0(1370)$ & Isobar & T-matrix pole & $m_0 = 1.38 \pm 0.07 \gevcc$~\cite{Pelaez:2022qby}\\
    & & & $\Gamma_0/2 ~=  0.22 \pm 0.08 \gev$~\cite{Pelaez:2022qby}\\ \hline
    $f_0(1500)$ & Isobar & Breit--Wigner & $m_0 = 1.506 \pm 0.006 \gevcc$~\cite{PDG2020}\\
    & & & $\Gamma_0~ =  0.112 \pm 0.009 \gev$~\cite{PDG2020}\\ \hline
    $f_0(1710)$ & Isobar & Breit--Wigner & $m_0 = 1.704 \pm 0.012 \gevcc$~\cite{PDG2020}\\
    & & & $\Gamma_0~ =  0.123 \pm 0.018 \gev$~\cite{PDG2020}\\ \hline
    $\pi\pi$--$\pi\pi$ & Isobar & Rescattering & CFD of Ref.~\cite{Garcia-Martin:2011iqs}\\ \hline
    $\pi\pi$--$K \Kbar$ & Isobar & Rescattering & $\textrm{CFD}_B$ of Ref.~\cite{Pelaez:2020gnd}\\ \hline
    $\pip\pim$ S-wave & K-matrix & K-matrix~\cite{Chung:1995dx,Dalitz:1960du,Wigner:1946zz,Wigner:1947zz,Aitchison:1972ay} & Refs.~\cite{Anisovich:2002ij,BaBar:2008inr,FOCUS:2003tdy} \\ \hline
    $\Kp\pim$ and & QMI & Constant per bin & ---\\
    $\pip\pim$ S-waves\\
    \hline
\end{tabular}
}
\end{table}

\begin{table}[!tb]
    \centering
    \caption{\label{tab:aman:nonswave}
        Resonances common to each S-wave approach, together with their default lineshapes and fixed lineshape parameters.
    }
    \renewcommand{\arraystretch}{1.1}
    \begin{tabular}
    {l@{\hspace{0.25cm}}
     @{\hspace{0.25cm}}c@{\hspace{0.25cm}} 
     @{\hspace{0.25cm}}l@{\hspace{0.25cm}}
     @{\hspace{0.25cm}}l} \hline
    Resonance & Spin & Mass lineshape, $T(s)$ & Fixed parameters~\cite{PDG2020}\\ \hline

    $\chiczero(1P)$ & 0 & Breit--Wigner & $m_0 = 3.41471 \pm 0.00030 \gevcc$\\
    & & & $\Gamma_0 ~=  0.0105 \pm 0.0008 \gev$\\ \hline
    $K^{*}(892)^0$ & 1 & Breit--Wigner & $m_0 = 0.89555 \pm 0.00020 \gevcc$\\
    & & & $\Gamma_0 ~= 0.0473 \pm 0.0005 \gev$\\ \hline
    $K^{*}(1410)^0$ & 1 & Breit--Wigner & $m_0 = 1.414 \pm 0.015 \gevcc$\\
    & & & $\Gamma_0 ~= 0.232 \pm 0.021 \gev$\\ \hline
    $K^{*}(1680)^0$ & 1 & Breit--Wigner & $m_0 = 1.718 \pm 0.018 \gevcc$\\
    & & & $\Gamma_0 ~= 0.322 \pm 0.110 \gev$\\ \hline
    $\rho(770)^0$~${}^\dagger$& 1 & Gounaris--Sakurai~\cite{GS} & $m_0 = 0.7690 \pm 0.0009 \gevcc$\\
    & & & $\Gamma_0 ~= 0.1509 \pm 0.0017 \gev$\\ \hline
    $\omega(782)$ & 1 & Breit--Wigner & $m_0 = 0.78266 \pm 0.00013 \gevcc$\\
    & & & $\Gamma_0 ~=  0.00868 \pm 0.00013 \gev$\\ \hline
    $\rho(1450)^0$ & 1 & Gounaris--Sakurai~\cite{GS} & $m_0 = 1.465 \pm 0.025 \gevcc$\\
    & & & $\Gamma_0 ~= 0.400 \pm 0.060 \gev$\\ \hline
    $\rho(1700)^0$ & 1 & Gounaris--Sakurai~\cite{GS} & $m_0 = 1.720 \pm 0.020 \gevcc$\\
    & & & $\Gamma_0~ = 0.250 \pm 0.100 \gev$\\ \hline
    $K^*_2(1430)^0$ & 2 & Breit--Wigner & $m_0 = 1.4324 \pm 0.0013 \gevcc$\\
    & & & $\Gamma_0 ~=  0.109 \pm 0.005 \gev$\\ \hline
    $f_2(1270)$ & 2 & Breit--Wigner & $m_0 = 1.2755 \pm 0.0008 \gevcc$\\
    & & & $\Gamma_0 ~= 0.1859 \pm 0.0028 \gev$\\ \hline
    $f_2^\prime(1525)$ & 2 & Breit--Wigner & $m_0 = 1.5174 \pm 0.0025 \gevcc$\\
    & & & $\Gamma_0 ~= 0.0869 \pm 0.0023 \gev$\\ \hline
    $K^*_3(1780)^0$ & 3 & Breit--Wigner & $m_0 = 1.779 \pm 0.008 \gevcc$\\
    & & & $\Gamma_0 ~= 0.161 \pm 0.017 \gev$\\ \hline
    $\rho_3(1690)^0$ & 3 & Breit--Wigner & $m_0 = 1.686 \pm 0.004 \gevcc$\\
    & & & $\Gamma_0 ~= 0.186 \pm 0.014 \gev$\\
    \hline
\end{tabular} \\
{\footnotesize
    ${}^\dagger$ The ``neutral only, other reactions'' values from Ref.~\cite{PDG2020} are used.
}
\end{table}

Contrary to the majority of amplitude analyses, there is no formal model building procedure conducted to optimise which resonant contributions to include {\it per se}. 
This is because every resonance that is considered as established in Ref.~\cite{PDG2020}, and which plausibly decays to either \Kp\pim or \pip\pim in the charmless region, is found to be individually significant considering statistical uncertainties only, based on the reduction in $-2\log \mathcal{L}$ for the additional degrees of freedom. Since Dalitz-plot analyses are susceptible to multiple solutions, depending upon the initial parameter values the results may correspond to a local, rather than global, minimum of the $-2\log \mathcal{L}$ function.
To attempt to find the global minimum, a large number of fits are performed where the initial values of the complex coefficients are randomised.
The fit result with the smallest $-2\log \mathcal{L}$ value out of this ensemble is then taken to be the baseline result for each S-wave method. 
No secondary solutions are found within 25 units of $-2\log \mathcal{L}$ of this baseline. 
A summary of the components in each of the three models considered in this analysis is given in Tables~\ref{tab:aman:swave} and~\ref{tab:aman:nonswave}.

\section{Systematic uncertainties}
\label{sec:systematics}

Sources of systematic uncertainty are separated into two major categories: those that arise from experimental effects and those from the inherent lack of knowledge of the amplitude models.
The sources are described below, with the numerical systematic uncertainties for the \CP-averaged fit fractions, quasi-two-body \CP asymmetries, and relative phases for \Bp and \Bm decays summarised in Appendix~\ref{app:systematic}. 
Throughout this section, labels in brackets are assigned to facilitate matching with each source listed in the systematic-uncertainty tables.

Beginning with the experimental systematic uncertainties, the uncertainties on the signal yield and the background yields and asymmetries, given in Table~\ref{tab:massFit}, comprise a statistical component as well as systematic effects due to the \B-candidate mass fit procedure. 
Apart from varying parameters of the baseline model within one standard deviation, such experimentally motivated models are themselves changed in order to mitigate imperfections in the \mKpipi model.
The uncertainty arising from assumptions regarding the signal parameterisation is found by releasing the underlying Gaussian of its restrictions relative to the core Crystal Ball function. 
Similarly, the model for the combinatorial background is replaced with a second-order polynomial.
The relative fractions between the remaining peaking backgrounds are varied within two standard deviations as they are derived from simulation, while their charge asymmetries are varied within their world-average expectations~\cite{HFLAV21}.
The combined statistical and systematic uncertainties on the signal and background yields and asymmetries are then propagated to the amplitude analysis, where those variations causing the largest upward and downward deviations with respect to the baseline yield and charge asymmetry values are taken to form a covariance matrix between these parameters. Values sampled from this matrix are then fixed in the amplitude-analysis fit to assign the systematic uncertainty relating to the three-body \B-candidate mass fit~(``\B~mass'').

To account for the statistical uncertainty on the efficiency description, an ensemble of efficiency maps is created by sampling bin-by-bin from the baseline description, according to uncorrelated Gaussian distributions with means corresponding to the calibrated central value of the efficiency in each bin, and widths corresponding to the uncertainty derived from data. Sources of these systematic uncertainties include the \Bp production asymmetry~\cite{LHCb-PAPER-2016-062}, tracking~\cite{LHCb-DP-2013-002}, particle identification~\cite{LHCb-PUB-2016-021} and residual detection asymmetries~\cite{LHCb-PAPER-2016-062}. 
To account for extraneous potential biases in the method used to correct the hardware trigger efficiency, an alternative method using \decay{\Bz}{\jpsi(\to\mumu)\Kp\pim} decays in the region of the $K^*(892)^0$ resonance is employed. Corrections to the simulation are determined from events first triggered by muons from the \jpsi decay, with the trigger decision then recomputed for the kaon and pion tracks~\cite{LHCb-PAPER-2017-010, LHCb-PAPER-2018-045}. 
Additionally, to account for potential variation of the efficiency within a square Dalitz-plot bin, the efficiency map is constructed using a finer binning scheme. 
The standard deviation of the distribution of resulting Dalitz-plot fit parameters obtained with these variations are then taken to be the associated systematic uncertainty~(``Eff'').

As the detector resolution in $m^2_{\pip\pim}$ convolved into the signal model is determined from simulation, this candidate-dependent resolution is increased by a global scale factor determined from data. 
This is found from a one-dimensional fit to $m_{\pip\pim}$ in the immediate vicinity of the prominent $\chi_{c0}(1P)$ contribution, which is clearly visible Fig.~\ref{fig:dataDP}. 
Adopting a Voigtian function for the signal and a polynomial for the remainder, the difference from the baseline fit parameters with the scaled resolution is taken as their associated systematic uncertainties~(``Resol'').

The uncertainties on the combinatorial and peaking background distributions are propagated to the Dalitz-plot fit results with an ensemble procedure similar to that for the efficiency maps. 
To account for a scenario in which the combinatorial background phase-space structure varies with the \B-candidate mass, an alternative combinatorial map much closer to the signal region $5400 < \mKpipi < 5650 \mevcc$ is produced, where contamination from \Bp decays is still minimal. 
As peaking-background models derive from simulation, the systematic effects considered for the efficiency maps are also accounted for here. 
Furthermore, the amplitude-model weights in phase space are varied within their known uncertainties~\cite{LHCb-PAPER-2019-017,LHCb-PAPER-2019-018,LHCb-PAPER-2018-051,BESIII:2017kyd}. 
The standard deviation in the variation of the subsequent Dalitz-plot fit results is taken to be the systematic uncertainty due to these effects~(``Bkg'') and generally constitutes the dominant systematic uncertainty in this analysis.

Systematic uncertainties related to possible intrinsic fit bias are investigated using an ensemble of pseudoexperiments based on the baseline fit result~(``Fit~bias'').
Differences between the input and fitted values from the ensemble for the fit parameters are generally found to be small.
However, for the fits involving the QMI approach, an additional bias arises from the intrinsic limitations in the capacity of the approach to reproduce the underlying analytic S-wave. 
The cause of such a bias is primarily due to the approximation of the analytic lineshape by a constant amplitude in each bin; this is by far the dominant systematic uncertainty in this approach.
Its evaluation involves reusing the ensemble of pseudoexperiments generated for estimating the Isobar fit bias, fitting them with the QMI model, and determining the difference between the obtained and true bin-averaged values of the S-wave amplitude~(``QMI~bias''). Systematic uncertainties are assigned as the sum in quadrature of the difference between the input and output values and the uncertainty on the mean of the output value determined from a fit to the ensemble.

The remaining sources of uncertainty are endemic to the amplitude model and are presented as a separate major category because their improvement typically lies outside the realm of experiment. Lack of knowledge on the radius parameter of the {\it ad hoc} Blatt--Weisskopf barrier factors is accounted for by modifying its value between $3$ and $5 \gev^{-1}$ using a flat prior, with uncertainties calculated from an ensemble~(``Barrier'').

The systematic uncertainty due to fixed partial-wave shape parameters~(``Lineshape'') such as masses and widths is assigned with an ensemble technique, where their values for the resonances common to each S-wave approach are fluctuated according to the uncertainties listed in the Particle Data Group tables~\cite{PDG2020}. 
Specific to the Isobar approach, the S-wave lineshape variations are made according to the uncertainties listed in Table~\ref{tab:aman:swave}, whereas for the K-matrix an alternative fit is performed releasing the slowly-varying production parameter, $s_0^{\rm{prod}}$. 
Use of the corrected $\pi\pi\textrm{--}K \Kbar$ rescattering lineshape described in Eq.~\eqref{eq:pipiKK-rescattering} is also included in this category for the Isobar model. 
Systematic uncertainties due to the fixed values of the parameters reported in Tables~\ref{tab:aman:swave} and~\ref{tab:aman:nonswave} are negligible, and the impact of updated measurements of these parameters since Ref.~\cite{PDG2020} was published is also neglected. 

Finally, potential additional resonant contributions to the amplitude are considered through the inclusion of the $K^*_4(2045)^0$ and $f_2(1430)$ states into the model, with the systematic impact of each summed in quadrature~(``Extra'').

In general, the largest sources of systematic uncertainty for the Isobar and K-matrix approaches are due to the combinatorial-background distribution across phase space. 
For the \CP-averaged fit fractions this source dominates the uncertainty, while for the \CP asymmetries, which are more robust against such effects, the statistical and systematic uncertainties are somewhat comparable. 
For the relative phases between components, the background and efficiency models comprise the dominant systematic uncertainty, while the impact of additional resonances is the largest contributor to the model uncertainty. 
The total systematic uncertainty combines the effects of each category in quadrature as they are considered to be uncorrelated.

\section{Results}
\label{sec:results}

Information on the relative global minima of the $-2\log\mathcal{L}$ function between the baseline results in each of the three S-wave approaches can be found in Table~\ref{tab:res:min}. 
Naturally, the QMI approach gives the best absolute minimum, however the additional number of free parameters by which this is achieved compared to the other approaches indicates comparable fit quality and hence that the Isobar and K-matrix S-wave models are fairly reliable.     
It is also interesting to note the comparison between the Isobar and K-matrix minima. The improvement in $-2\log\mathcal{L}$ for fewer degrees of freedom supports the use of more physically motivated dispersive S-wave lineshapes, compared to what has been done previously in similar analyses, as this is the first incidence of the K-matrix being surpassed in this fit-quality metric by an isobaric approach in which all lineshapes have physical meaning.

\begin{table}[!tb]
    \centering
    \caption{\label{tab:res:min}
        Properties of the global $-2\log\mathcal{L}$ minima for each of the three S-wave approaches with respect to the K-matrix result.
    }
    \renewcommand{\arraystretch}{1.1}
    \begin{tabular}
    {l@{\hspace{0.25cm}}
     @{\hspace{0.25cm}}r@{\hspace{0.25cm}}
     @{\hspace{0.25cm}}c@{\hspace{0.25cm}}
     @{\hspace{0.25cm}}r} \hline
     & Isobar & K-matrix & QMI\\ \hline

    $\Delta(-2\log\mathcal{L})$ & $-24$\phantom{--} & 0 & $-291$ \\
    $\Delta$(ndf) & $-9$\phantom{--} & 0 & $+338$ \\

    \hline
    \end{tabular}
\end{table}

The goodness of fit is assessed by comparing the fit model with the data in square Dalitz-plot bins that can be seen in Appendix~\ref{app:gof} and determining an associated $\chi^2$ value. 
A binning common to \Bp and \Bm decays is chosen through an adaptive procedure that requires an approximately constant number of candidates from the combined \Bp and \Bm data samples in each bin. For various values of the required number of candidates per bin, the ratio of the $\chi^2$ (in the range 5000--5800 for 3600 bins) to the number of bins is approximately 1.6 accounting for statistical uncertainties only. 
Given the impact of the systematic uncertainties on the results discussed in Section~\ref{sec:systematics}, the agreement of the fit models with the data is reasonable. 
Smaller $\chi^2$ values tend to be obtained for the S-wave models with larger numbers of free parameters, such that all three approaches have comparable goodness-of-fit overall. 
The exception is for \Bp decays, where the $\chi^2$ metric favours the K-matrix model over the Isobar. 
The distribution in the square Dalitz plot of bins that contribute significantly to the $\chi^2$ does not reveal any clear source of mismodelling.

Due to the number of numerical results involved in this analysis, it is not practical to tabulate every measured value and uncertainty necessary to replicate a complete expression of the amplitudes. 
Therefore, these are recorded by electronic means as discussed in the Supplemental Material~\cite{supplemental}. 
In particular, these contain additional tables of interference fractions and their \CP-violating asymmetries as defined in Eq.~\eqref{eq:ACP-IFFdef} that are not further discussed here in the text. 
Instead, the focus in the main body of this paper is placed on the measured quasi-two-body quantities and relative phases, while the free lineshape parameters appear in Appendices~\ref{app:isobar} and~\ref{app:kmatrix}.

\subsection{Fit fractions}

\begin{figure}[tb]
    \centering
    \includegraphics[width=0.5\linewidth]{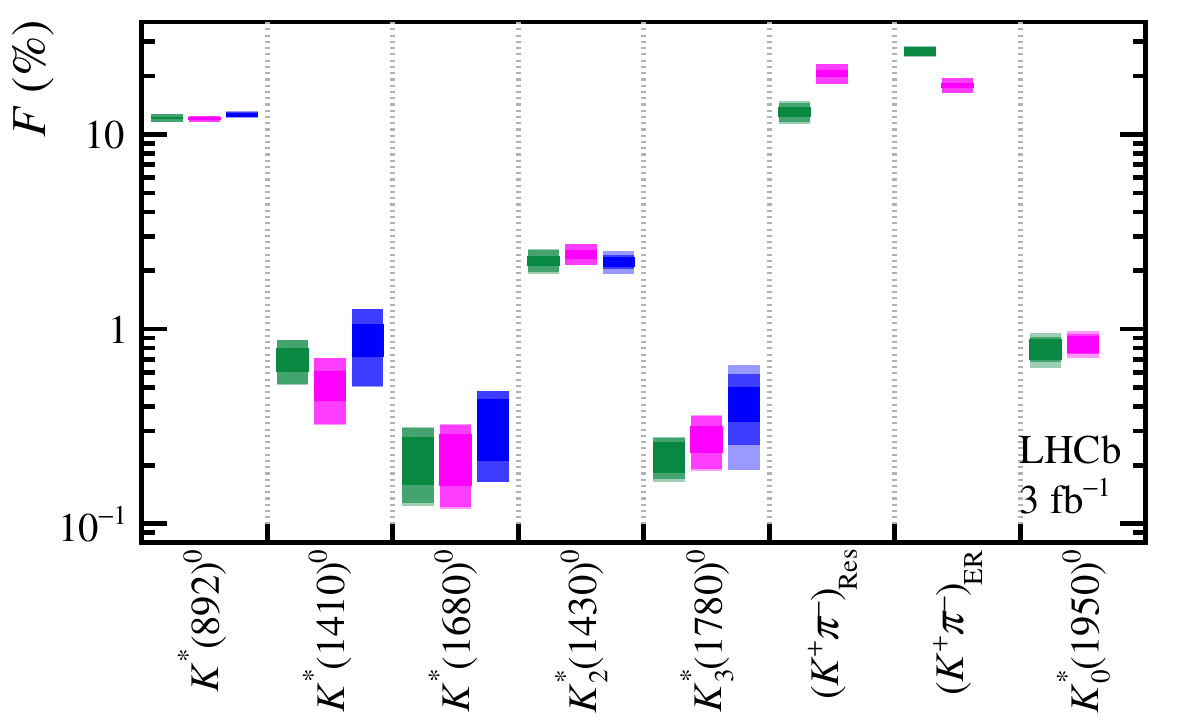}%
    \includegraphics[width=0.5\linewidth]{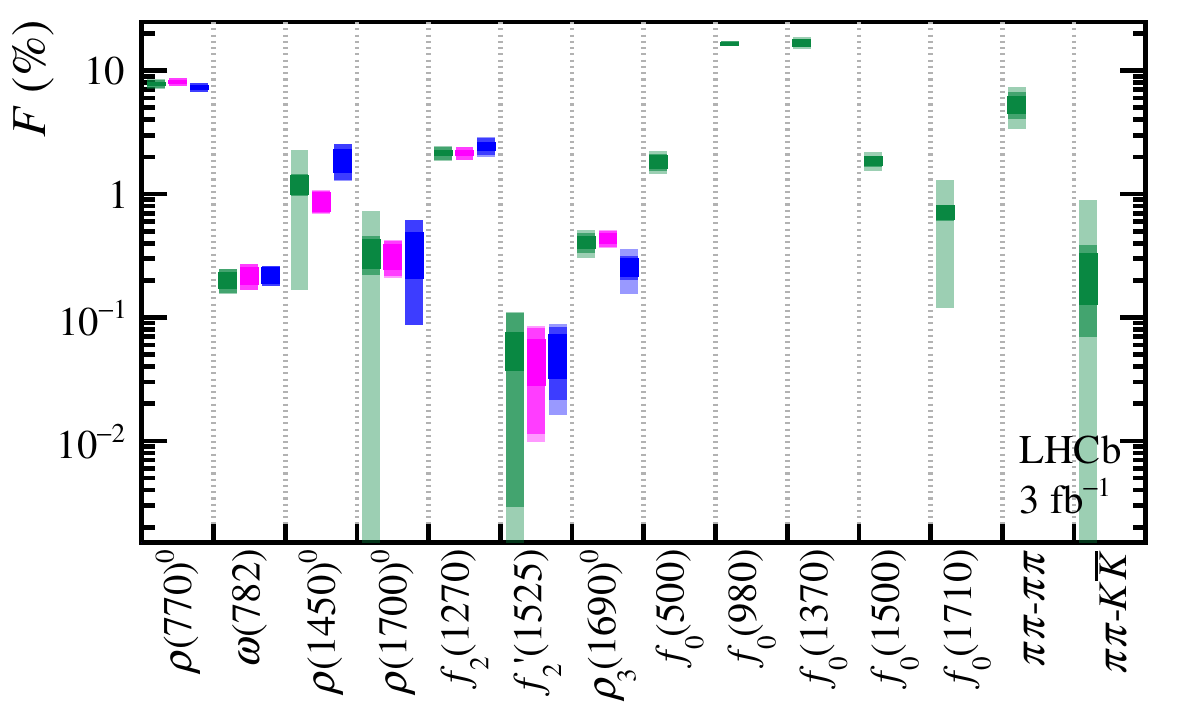}

    \caption{
        Comparison between Isobar~(green), K-matrix~(magenta) and QMI~(blue) fit fractions for states decaying to (left)~$K^+ \pi^-$ and (right)~$\pi^+ \pi^-$. Solid bands show the statistical spread, with progressively lighter shading indicating the combined statistical and systematic uncertainties, followed by the total uncertainty including that from the model.
    }
    \label{fig:res:ff}
\end{figure}

The \CP-averaged fit fractions are given in Table~\ref{tab:res:ff} for all three S-wave approaches and are visually represented in Fig.~\ref{fig:res:ff} to ease comparison. 
In all cases, statistical uncertainties are calculated using $68\%$ confidence intervals obtained from the results of fits performed to data sets sampled from the nominal fit models. 
Generally, the statistical uncertainty is lowest for the model with the fewest parameters (Isobar), and highest for the model with the largest number of parameters (QMI). 
For the prominent $K^*(892)^0$ and $\rho(770)^0$ vector states, and the $K_2^*(1430)^0$ and $f_2(1270)$ tensor states, the systematic uncertainties overwhelm the statistical error, whereas for the remaining contributions, the tendency is for these to be at comparable levels. 
Another trend is that the model uncertainty dominates if the associated state has low statistical significance. 
Including systematic and amplitude-model sources of uncertainty, several intermediate states are observed for the first time as discussed in a companion Letter~\cite{BuToKpPipPimPRLStrong}.

While the total fit fractions given in Table~\ref{tab:res:ff} may appear to be in disagreement between the S-wave approaches, this quantity can be deceiving. 
As composite objects, the internal interference-fraction contributions between terms comprising the K-matrix and QMI S-waves are included in their total fit fractions, while in the Isobar approach, these are absent. 
To facilitate a meaningful comparison between approaches, the total S-wave is taken as a single amplitude in each approach. 
Given in Table~\ref{tab:res:ffsum}, the S-wave sum and total fit fractions for the \Bp and \Bm amplitudes are found to be in excellent agreement between approaches. When split into individual overall \Kp\pim and \pip\pim S-waves, the Isobar and K-matrix results are not in agreement, whereas the QMI results are in agreement with both, showing a slight preference for the Isobar results.

\begin{table}[tb]
    \centering
    \caption{\label{tab:res:ff}
        \CP-averaged fit fractions for each approach, where the first uncertainty is statistical, the second the experimental systematic and the third is the model systematic. 
    }

    \resizebox{\textwidth}{!}{
    \renewcommand{\arraystretch}{1.1}
    \begin{tabular}
    {@{\hspace{0.0cm}}l@{\hspace{0.125cm}}
     @{\hspace{0.125cm}}c@{\hspace{0.125cm}}
     @{\hspace{0.125cm}}c@{\hspace{0.125cm}}
     @{\hspace{0.125cm}}c@{\hspace{0.0cm}}} \hline
    & \multicolumn{3}{c}{Fit fraction  $(\%)$}\\ \hline
    Component & Isobar & K-matrix & QMI\\ \hline

$K^*(892)^0$ & $12.167 \pm 0.128 \pm 0.494 \pm 0.157$ & $12.071 \pm 0.118 \pm 0.397 \pm 0.047$ & $12.669 \pm 0.223 \pm 0.304 \pm 0.286$\\
$K^*(1410)^0$ & $\phantom{0}0.700 \pm 0.092 \pm 0.149 \pm 0.053$ & $\phantom{0}0.517 \pm 0.088 \pm 0.169 \pm 0.037$ & $\phantom{0}0.891 \pm 0.166 \pm 0.343 \pm 0.050$\\
$K^*(1680)^0$ & $\phantom{0}0.219 \pm 0.060 \pm 0.069 \pm 0.029$ & $\phantom{0}0.223 \pm 0.065 \pm 0.077 \pm 0.024$ & $\phantom{0}0.322 \pm 0.111 \pm 0.111 \pm 0.018$\\
$\rho(770)^0$ & $\phantom{0}7.834 \pm 0.192 \pm 0.579 \pm 0.301$ & $\phantom{0}8.122 \pm 0.194 \pm 0.543 \pm 0.101$ & $\phantom{0}7.326 \pm 0.259 \pm 0.543 \pm 0.158$\\
$\omega(782)$ & $\phantom{0}0.202 \pm 0.030 \pm 0.033 \pm 0.009$ & $\phantom{0}0.219 \pm 0.035 \pm 0.037 \pm 0.002$ & $\phantom{0}0.221 \pm 0.032 \pm 0.027 \pm 0.004$\\
$\rho(1450)^0$ & $\phantom{0}1.213 \pm 0.212 \pm 0.123 \pm 1.016$ & $\phantom{0}0.881 \pm 0.153 \pm 0.095 \pm 0.081$ & $\phantom{0}1.908 \pm 0.409 \pm 0.460 \pm 0.102$\\
$\rho(1700)^0$ & $\phantom{0}0.340 \pm 0.088 \pm 0.076 \pm 0.372$ & $\phantom{0}0.317 \pm 0.073 \pm 0.070 \pm 0.035$ & $\phantom{0}0.351 \pm 0.142 \pm 0.223 \pm 0.009$\\
$K_2^*(1430)^0$ & $\phantom{0}2.252 \pm 0.125 \pm 0.259 \pm 0.157$ & $\phantom{0}2.434 \pm 0.117 \pm 0.265 \pm 0.083$ & $\phantom{0}2.220 \pm 0.119 \pm 0.130 \pm 0.229$\\
$f_2(1270)$ & $\phantom{0}2.152 \pm 0.110 \pm 0.242 \pm 0.136$ & $\phantom{0}2.145 \pm 0.094 \pm 0.224 \pm 0.087$ & $\phantom{0}2.442 \pm 0.174 \pm 0.337 \pm 0.227$\\
$f_2^\prime(1525)$ & $\phantom{0}0.056 \pm 0.019 \pm 0.050 \pm 0.015$ & $\phantom{0}0.047 \pm 0.019 \pm 0.030 \pm 0.011$ & $\phantom{0}0.053 \pm 0.021 \pm 0.023 \pm 0.019$\\
$K_3^*(1780)^0$ & $\phantom{0}0.222 \pm 0.039 \pm 0.036 \pm 0.024$ & $\phantom{0}0.274 \pm 0.042 \pm 0.071 \pm 0.029$ & $\phantom{0}0.420 \pm 0.084 \pm 0.145 \pm 0.158$\\
$\rho_3(1690)^0$ & $\phantom{0}0.408 \pm 0.047 \pm 0.056 \pm 0.076$ & $\phantom{0}0.439 \pm 0.042 \pm 0.050 \pm 0.029$ & $\phantom{0}0.258 \pm 0.043 \pm 0.038 \pm 0.085$\\
$\chi_{c0}(1P)$ & $\phantom{0}2.000 \pm 0.065 \pm 0.093 \pm 0.028$ & $\phantom{0}1.927 \pm 0.052 \pm 0.079 \pm 0.007$ & $\phantom{0}1.823 \pm 0.077 \pm 0.146 \pm 0.014$\\\hline
$(K^+\pi^-)_{\rm Res}$ & $13.076 \pm 0.633 \pm 1.293 \pm 0.958$ & $20.610 \pm 0.765 \pm 2.280 \pm 0.582$ & ---\\
$(K^+\pi^-)_{\rm ER}$ & $26.852 \pm 1.281 \pm 0.664 \pm 1.031$ & $17.857 \pm 0.460 \pm 1.491 \pm 0.304$ & ---\\
$K_0^*(1950)^0$ & $\phantom{0}0.793 \pm 0.092 \pm 0.076 \pm 0.110$ & $\phantom{0}0.842 \pm 0.079 \pm 0.058 \pm 0.093$ & ---\\\hline
$f_0(500)$ & $\phantom{0}1.830 \pm 0.214 \pm 0.205 \pm 0.245$ & --- & ---\\
$f_0(980)$ & $16.582 \pm 0.440 \pm 0.467 \pm 0.429$ & --- & ---\\
$f_0(1370)$ & $16.786 \pm 0.951 \pm 0.615 \pm 1.566$ & --- & ---\\
$f_0(1500)$ & $\phantom{0}1.857 \pm 0.150 \pm 0.110 \pm 0.264$ & --- & ---\\
$f_0(1710)$ & $\phantom{0}0.714 \pm 0.094 \pm 0.052 \pm 0.586$ & --- & ---\\
$\pi\pi$--$\pi\pi$ & $\phantom{0}5.365 \pm 0.833 \pm 1.023 \pm 1.467$ & --- & ---\\
$\pi\pi$--$K\Kb$ & $\phantom{0}0.227 \pm 0.101 \pm 0.121 \pm 0.656$ & --- & ---\\\hline
\pip\pim S-wave & --- & $30.901 \pm 0.303 \pm 1.018 \pm 0.124$ & --- \\\hline
\Kp\pim and & --- & --- & $67.547 \pm 0.606 \pm 0.570 \pm 0.291$ \\
\pip\pim S-waves \\ \hline
Total & $113.847 \pm 2.593 \pm 3.381 \pm 1.548$ & $99.825 \pm 0.749 \pm 2.244 \pm 0.858$ & $98.451 \pm 0.431 \pm 0.545 \pm 0.158$\\

    \hline
    \end{tabular}
    }
\end{table}

\begin{table}[tb]
    \centering
    \caption{\label{tab:res:ffsum}
        Sum of fit fractions in percent deriving from the indicated overall S-waves instead of individual components comprising the S-wave in each of the three approaches, and their impact on the total fit fractions. The first uncertainty is statistical, the second the experimental systematic and the third is the model systematic. The fourth uncertainty arises from the approach adopted to lift the mathematical ambiguity inherent to the one-dimensional QMI S-waves.
    }

    \resizebox{\textwidth}{!}{
    \renewcommand{\arraystretch}{1.1}
    \begin{tabular}
    {@{\hspace{0.0cm}}l@{\hspace{0.125cm}}
     @{\hspace{0.125cm}}c@{\hspace{0.125cm}}
     @{\hspace{0.125cm}}c@{\hspace{0.125cm}}
     @{\hspace{0.125cm}}c@{\hspace{0.0cm}}} \hline
    & \multicolumn{3}{c}{Fit fraction  $(\%)$}\\ \hline
    Component & Isobar & K-matrix & QMI\\ \hline

    \Kp\pim S-wave & $63.606 \pm 2.019 \pm 3.490 \pm 3.264$ & $52.010 \pm 0.489 \pm 1.524 \pm 0.298$ & $66.196 \pm 3.341 \pm 8.468 \pm 2.099 \pm 2.267$ \\\hline
    \pip\pim S-wave & $38.800 \pm 1.201 \pm 1.733 \pm 4.405$ & $30.901 \pm 0.303 \pm 1.018 \pm 0.124$ & $36.905 \pm 2.231 \pm 6.692 \pm 3.512 \pm 2.991$ \\\hline
    \Kp\pim and & $68.248 \pm 0.387 \pm 0.252 \pm 0.747$ & $68.475 \pm 0.329 \pm 0.389 \pm 0.190$ & $67.547 \pm 0.606 \pm 0.570 \pm 0.291 \, \phantom{\pm 0.000} \,\,$ \\
    \pip\pim S-waves \\ \hline
    Total & $98.433 \pm 0.301 \pm 0.316 \pm 0.221$ & $98.090 \pm 0.377 \pm 0.294 \pm 0.169$ & $98.451 \pm 0.431 \pm 0.545 \pm 0.158 \, \phantom{\pm 0.000} \,\,$\\

    \hline
    \end{tabular}
    }
\end{table}

\subsection{\texorpdfstring{\boldmath{\CP} asymmetries}{CP asymmetries}} 

\begin{figure}[tb]
    \centering
    \includegraphics[width=0.5\linewidth]{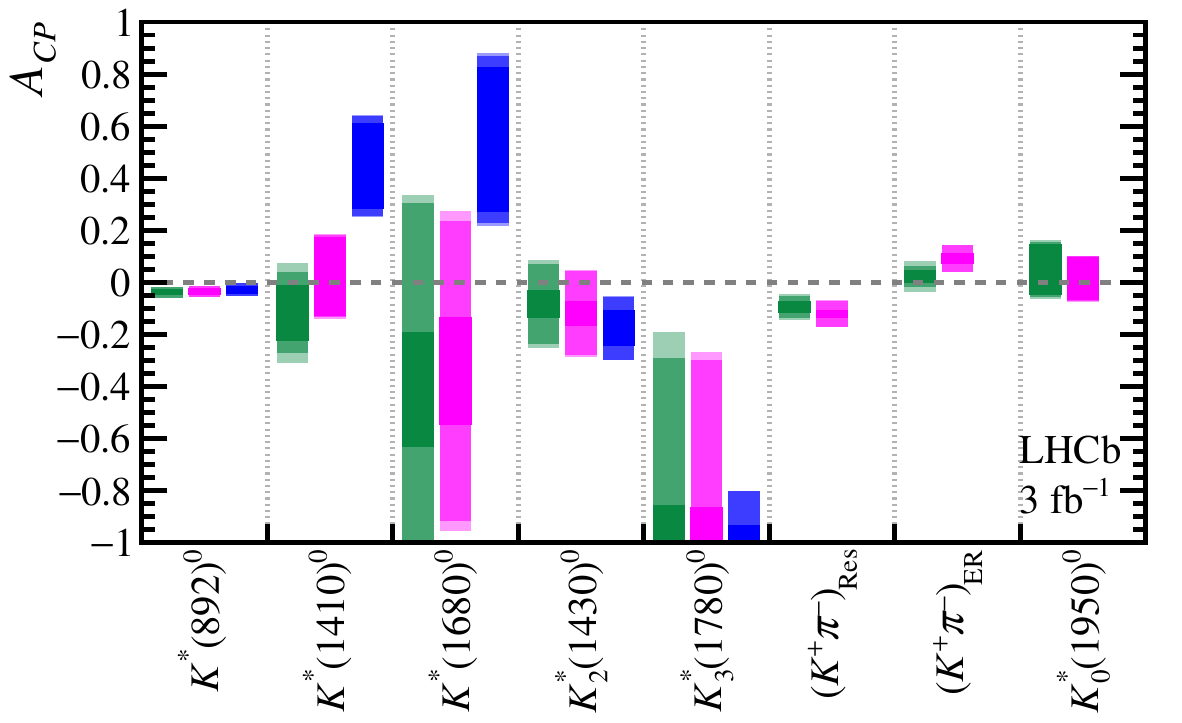}%
    \includegraphics[width=0.5\linewidth]{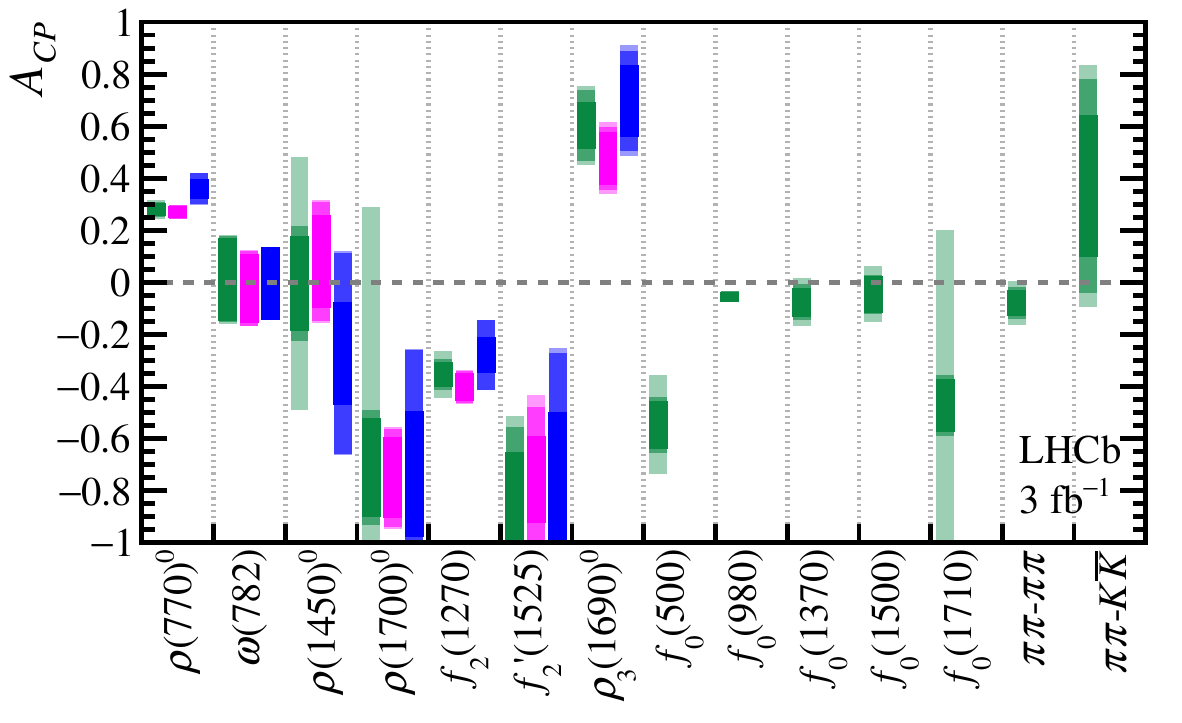}

    \caption{
        Comparison between Isobar~(green), K-matrix~(magenta) and QMI~(blue) $\mathcal{A}_{C\!P}$ for states decaying to (left)~$K^+ \pi^-$ and (right)~$\pi^+ \pi^-$. Solid bands show the statistical spread, with progressively lighter shading indicating the combined statistical and systematic uncertainties, followed by the total uncertainty including that from the model.
    }
    \label{fig:res:acp}
\end{figure}

Quasi-two-body \CP asymmetries associated to each component are shown in Table~\ref{tab:res:acp} and Fig.~\ref{fig:res:acp}, also demonstrating good agreement between the S-wave approaches. 
Several significant measurements of \CP violation can be noted, however an in-depth discussion is left to another companion Letter~\cite{BuToKpPipPimPRLCP}. 
As far as the uncertainties are concerned, many of the points discussed for the fit fractions remain relevant. However, because of the propensity for partial cancellations in asymmetry parameters, the systematic uncertainties tend to be at the same level or smaller than the statistical errors, leaving space for future improvements.

As with the naive total fit fractions from Table~\ref{tab:res:ff}, the total \CP asymmetries associated to the S-wave components in Table~\ref{tab:res:acp} also show an apparent discrepancy when calculated from these fit fractions. 
When considering the S-wave either as a whole or as individual overall \Kp\pim and \pip\pim contributions, their \CP asymmetries and that of the total model are found to be in excellent agreement between approaches, as shown in Table~\ref{tab:res:acpsum}.

\begin{table}[tb]
    \centering
    \caption{\label{tab:res:acp}
        Quasi-two-body \CP asymmetries for each approach, where the first uncertainty is statistical, the second the experimental systematic and the third is the model systematic. 
    }
    \renewcommand{\arraystretch}{1.1}
    \resizebox{\textwidth}{!}
    {\begin{tabular}
    {@{\hspace{0.0cm}}l@{\hspace{0.125cm}}
     @{\hspace{0.125cm}}c@{\hspace{0.125cm}}
     @{\hspace{0.125cm}}c@{\hspace{0.125cm}}
     @{\hspace{0.125cm}}c@{\hspace{0.0cm}}} \hline &
    \multicolumn{3}{c}{Quasi-two-body \CP asymmetry}\\ \hline
    Component & Isobar & K-matrix & QMI \\ \hline

$K^*(892)^0$ & $-0.038 \pm 0.010 \pm 0.018 \pm 0.002$ & $-0.035 \pm 0.012 \pm 0.017 \pm 0.001$ & $-0.027 \pm 0.014 \pm 0.022 \pm 0.002$\\
$K^*(1410)^0$ & $-0.116 \pm 0.106 \pm 0.113 \pm 0.113$ & $+0.024 \pm 0.150 \pm 0.048 \pm 0.040$ & $+0.447 \pm 0.163 \pm 0.102 \pm 0.045$\\
$K^*(1680)^0$ & $-0.412 \pm 0.220 \pm 0.681 \pm 0.221$ & $-0.340 \pm 0.206 \pm 0.539 \pm 0.217$ & $+0.549 \pm 0.277 \pm 0.165 \pm 0.077$\\
$\rho(770)^0$ & $+0.280 \pm 0.022 \pm 0.016 \pm 0.024$ & $+0.271 \pm 0.022 \pm 0.014 \pm 0.003$ & $+0.359 \pm 0.037 \pm 0.047 \pm 0.009$\\
$\omega(782)$ & $+0.011 \pm 0.158 \pm 0.048 \pm 0.043$ & $-0.023 \pm 0.131 \pm 0.061 \pm 0.015$ & $-0.004 \pm 0.139 \pm 0.014 \pm 0.004$\\
$\rho(1450)^0$ & $-0.005 \pm 0.180 \pm 0.131 \pm 0.433$ & $+0.081 \pm 0.178 \pm 0.143 \pm 0.062$ & $-0.273 \pm 0.197 \pm 0.332 \pm 0.068$\\
$\rho(1700)^0$ & $-0.711 \pm 0.189 \pm 0.113 \pm 0.975$ & $-0.750 \pm 0.153 \pm 0.111 \pm 0.051$ & $-0.735 \pm 0.240 \pm 0.408 \pm 0.058$\\
$K_2^*(1430)^0$ & $-0.084 \pm 0.052 \pm 0.144 \pm 0.073$ & $-0.119 \pm 0.046 \pm 0.155 \pm 0.044$ & $-0.176 \pm 0.066 \pm 0.102 \pm 0.007$\\
$f_2(1270)$ & $-0.354 \pm 0.047 \pm 0.036 \pm 0.066$ & $-0.402 \pm 0.051 \pm 0.036 \pm 0.020$ & $-0.279 \pm 0.067 \pm 0.115 \pm 0.006$\\
$f_2^\prime(1525)$ & $-0.827 \pm 0.175 \pm 0.206 \pm 0.157$ & $-0.757 \pm 0.167 \pm 0.224 \pm 0.169$ & $-0.764 \pm 0.263 \pm 0.419 \pm 0.128$\\
$K_3^*(1780)^0$ & $-0.936 \pm 0.080 \pm 0.639 \pm 0.372$ & $-0.934 \pm 0.069 \pm 0.632 \pm 0.195$ & $-0.971 \pm 0.036 \pm 0.165 \pm 0.004$\\
$\rho_3(1690)^0$ & $+0.603 \pm 0.090 \pm 0.103 \pm 0.064$ & $+0.477 \pm 0.100 \pm 0.067 \pm 0.069$ & $+0.698 \pm 0.135 \pm 0.136 \pm 0.091$\\
$\chi_{c0}(1P)$ & $+0.043 \pm 0.028 \pm 0.013 \pm 0.005$ & $+0.027 \pm 0.028 \pm 0.011 \pm 0.004$ & $+0.012 \pm 0.038 \pm 0.020 \pm 0.008$\\\hline
$(K^+\pi^-)_{\rm Res}$ & $-0.094 \pm 0.020 \pm 0.038 \pm 0.027$ & $-0.121 \pm 0.015 \pm 0.048 \pm 0.013$ & ---\\
$(K^+\pi^-)_{\rm ER}$ & $+0.023 \pm 0.022 \pm 0.033 \pm 0.042$ & $+0.092 \pm 0.021 \pm 0.048 \pm 0.009$ & ---\\
$K_0^*(1950)^0$ & $+0.050 \pm 0.097 \pm 0.042 \pm 0.041$ & $+0.014 \pm 0.081 \pm 0.031 \pm 0.014$ & ---\\\hline
$f_0(500)$ & $-0.547 \pm 0.091 \pm 0.058 \pm 0.157$ & --- & ---\\
$f_0(980)$ & $-0.054 \pm 0.017 \pm 0.011 \pm 0.010$ & --- & ---\\
$f_0(1370)$ & $-0.075 \pm 0.054 \pm 0.045 \pm 0.062$ & --- & ---\\
$f_0(1500)$ & $-0.046 \pm 0.069 \pm 0.030 \pm 0.077$ & --- & ---\\
$f_0(1710)$ & $-0.474 \pm 0.099 \pm 0.064 \pm 0.664$ & --- & ---\\
$\pi\pi$--$\pi\pi$ & $-0.079 \pm 0.049 \pm 0.039 \pm 0.057$ & --- & ---\\
$\pi\pi$--$K\Kb$ & $+0.371 \pm 0.272 \pm 0.309 \pm 0.213$ & --- & ---\\\hline
\pip\pim S-wave & --- & $+0.052 \pm 0.033 \pm 0.108 \pm 0.014$ & --- \\\hline
\Kp\pim and  & --- & --- & $+0.033 \pm 0.010 \pm 0.018 \pm 0.002$ \\
\pip\pim S-waves \\\hline
Total & $-0.036 \pm 0.010 \pm 0.008 \pm 0.012$ & $+0.011 \pm 0.004 \pm 0.010 \pm 0.003$ & $+0.030 \pm 0.007 \pm 0.010 \pm 0.003$\\

    \hline
    \end{tabular}
    }
\end{table}

\begin{table}[tb]
    \centering
    \caption{\label{tab:res:acpsum}
        Quasi-two-body \CP asymmetries deriving from the indicated overall S-waves instead of individual components comprising the S-wave in each of the three approaches, and their impact on the total \CP asymmetry. The first uncertainty is statistical, the second the experimental systematic and the third is the model systematic. The fourth uncertainty arises from the approach adopted to lift the mathematical ambiguity inherent to the one-dimensional QMI S-waves.
    }

    \resizebox{\textwidth}{!}{
    \renewcommand{\arraystretch}{1.1}
    \begin{tabular}
    {@{\hspace{0.0cm}}l@{\hspace{0.125cm}}
     @{\hspace{0.125cm}}c@{\hspace{0.125cm}}
     @{\hspace{0.125cm}}c@{\hspace{0.125cm}}
     @{\hspace{0.125cm}}c@{\hspace{0.0cm}}} \hline
    & \multicolumn{3}{c}{Quasi-two-body \CP asymmetry}\\ \hline
    Component & Isobar & K-matrix & QMI\\ \hline

    \Kp\pim S-wave & $-0.016 \pm 0.011 \pm 0.014 \pm 0.019$ & $-0.025 \pm 0.008 \pm 0.016 \pm 0.003$ & $-0.035 \pm 0.077 \pm 0.049 \pm 0.010 \pm 0.061$ \\\hline
    \pip\pim S-wave & $+0.080 \pm 0.012 \pm 0.007 \pm 0.060$ & $+0.052 \pm 0.033 \pm 0.108 \pm 0.014$ & $+0.022 \pm 0.072 \pm 0.032 \pm 0.023 \pm 0.038$ \\\hline
    \Kp\pim and & $+0.023 \pm 0.006 \pm 0.013 \pm 0.025$ & $+0.024 \pm 0.006 \pm 0.014 \pm 0.003$ & $+0.033 \pm 0.010 \pm 0.018 \pm 0.002 \, \phantom{\pm 0.000} \,\,$ \\
    \pip\pim S-waves \\ \hline
    Total & $+0.021 \pm 0.004 \pm 0.010 \pm 0.016$ & $+0.020 \pm 0.004 \pm 0.010 \pm 0.001$ & $+0.030 \pm 0.007 \pm 0.010 \pm 0.003 \, \phantom{\pm 0.000} \,\,$\\

    \hline
    \end{tabular}
    }
\end{table}

\subsection{Relative phases}

\begin{figure}[tb]
    \centering
    \includegraphics[width=0.5\textwidth]{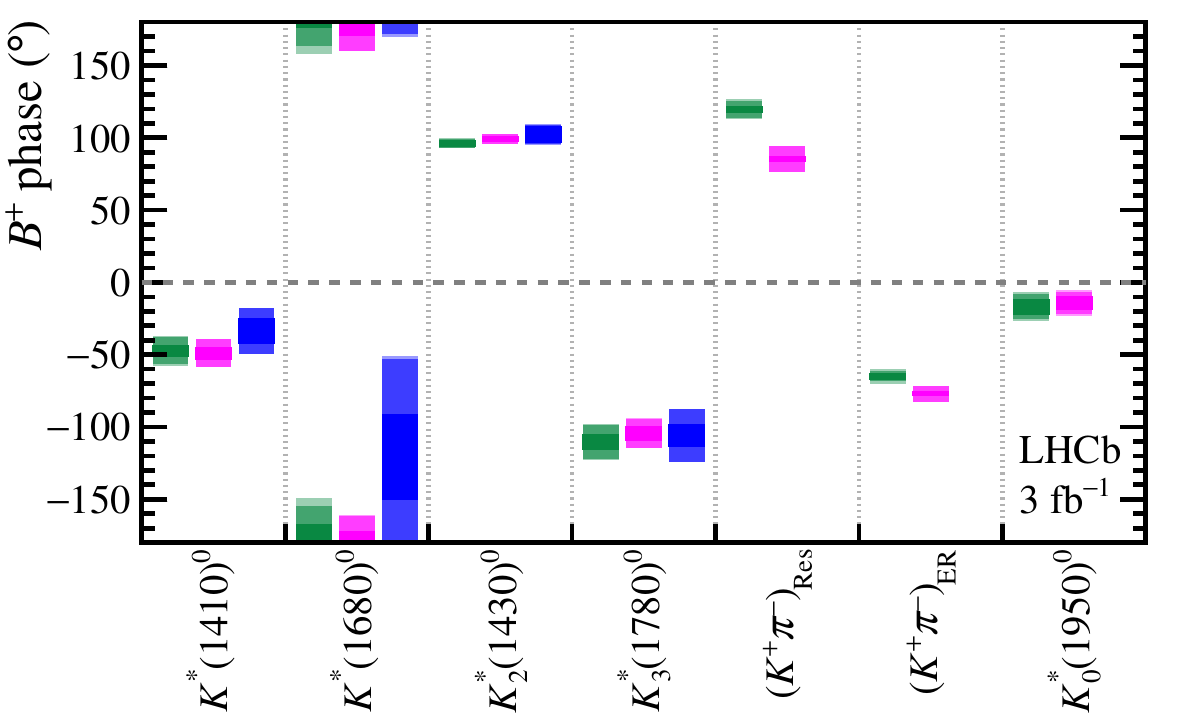}%
    \includegraphics[width=0.5\textwidth]{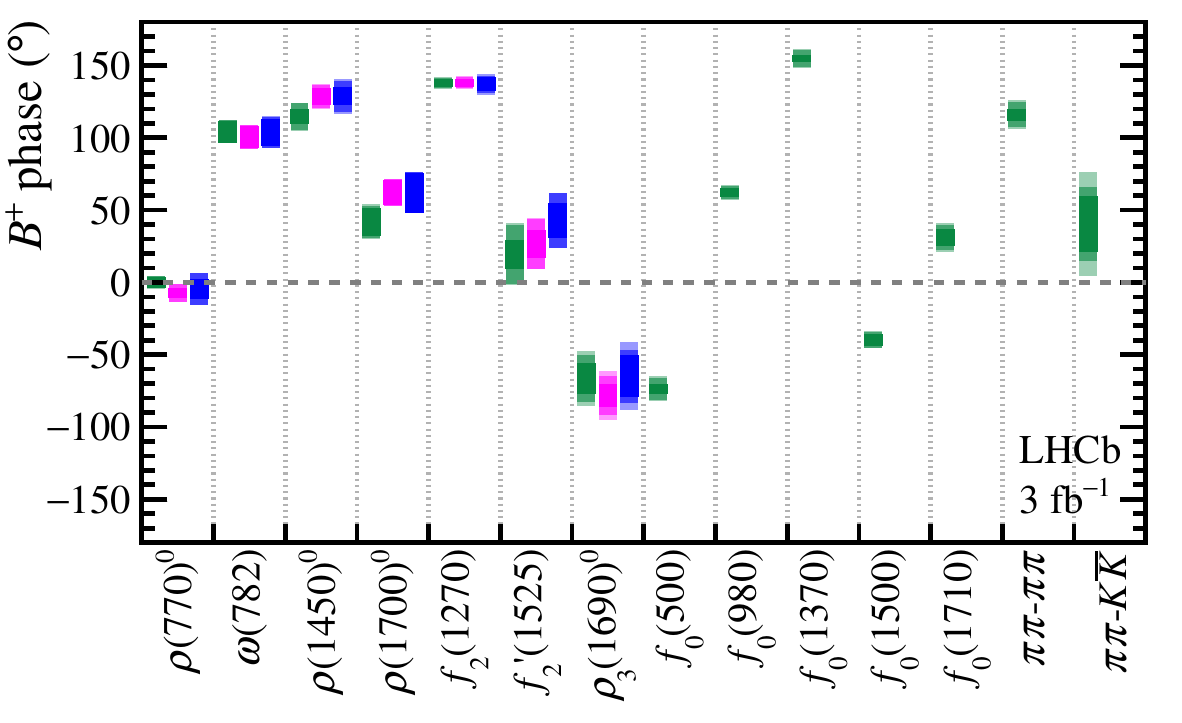}

    \includegraphics[width=0.5\textwidth]{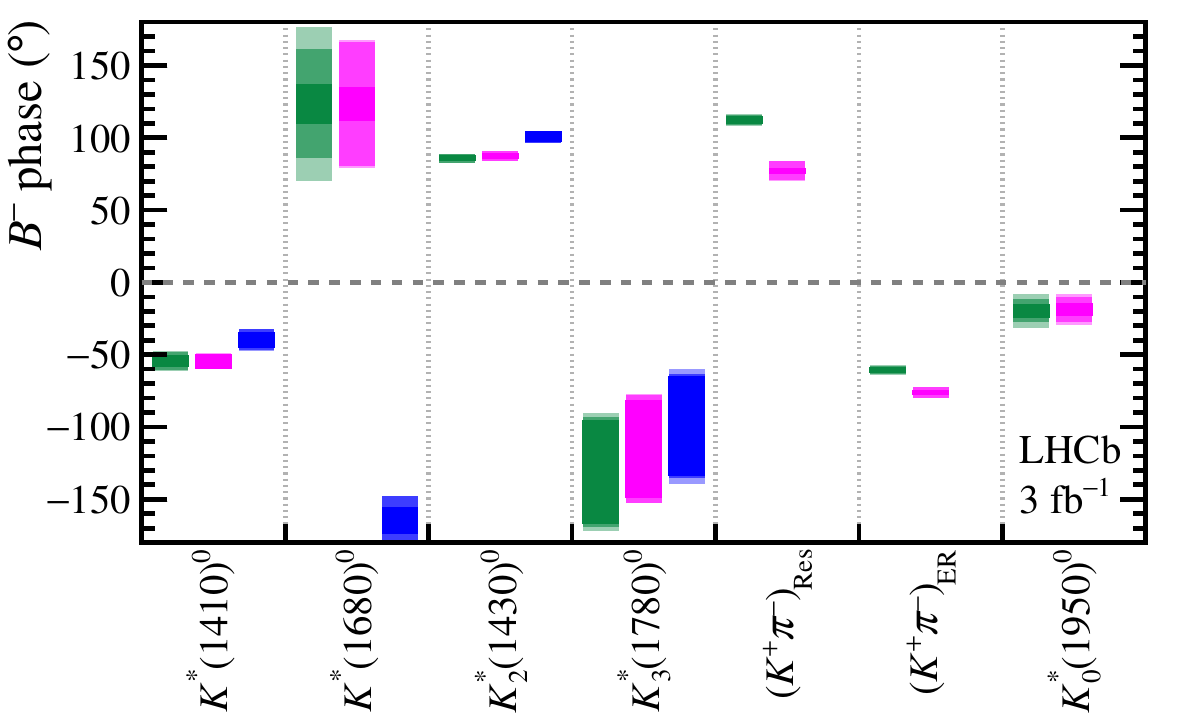}%
    \includegraphics[width=0.5\textwidth]{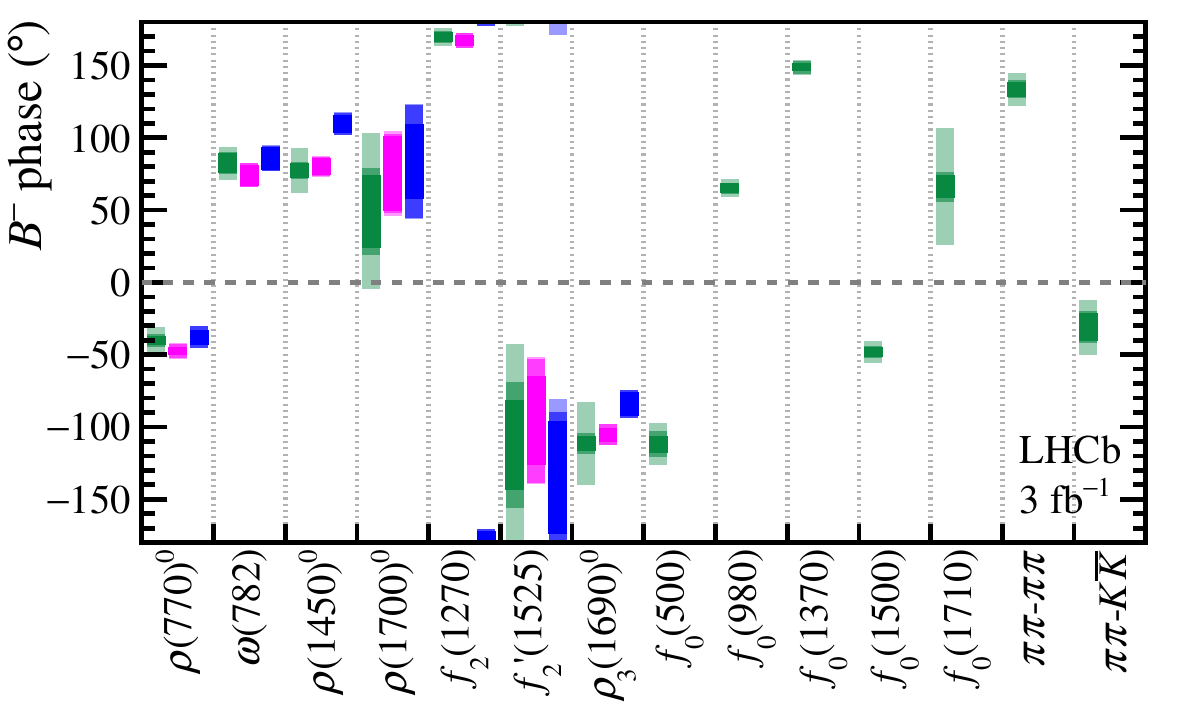}

    \caption{
        Comparison between Isobar~(green), K-matrix~(magenta) and QMI~(blue) amplitude phases of (top)~$B^+$ and (bottom)~$B^-$ decays for states decaying to (left)~$K^+ \pi^-$ and (right)~$\pi^+ \pi^-$. Solid bands show the statistical spread, with progressively lighter shading indicating the combined statistical and systematic uncertainties, followed by the total uncertainty including that from the model. 
        The $K^*(892)^0$ phase is fixed to zero to serve as a reference, so does not appear here.
    }
    \label{fig:res:phases}
\end{figure}

For completeness, the relative phases for each contribution in \Bp~(\Bm) decays are reported in Table~\ref{tab:res:phasep}~(\ref{tab:res:phasem}) and Fig.~\ref{fig:res:phases}, showing broad agreement between the S-wave approaches. 
The results for these amplitude-level quantities tend to be dominated by the systematic uncertainty.

\begin{table}[tb]
  \centering
  \caption{\label{tab:res:phasep}
        Phase comparison for \Bp decay amplitudes between the three S-wave approaches where the first uncertainty is statistical, the second systematic and the third from the model. 
    }

    \resizebox{\textwidth}{!}{
    \renewcommand{\arraystretch}{1.1}
    \begin{tabular}
    {@{\hspace{0.0cm}}l@{\hspace{0.125cm}}
     @{\hspace{0.125cm}}c@{\hspace{0.125cm}}
     @{\hspace{0.125cm}}c@{\hspace{0.125cm}}
     @{\hspace{0.125cm}}c@{\hspace{0.0cm}}} \hline & 
    \multicolumn{3}{c}{Phase of \Bp decay amplitude $(^{\circ})$}\\ \hline
    Component & Isobar & K-matrix & QMI\\ \hline

$K^*(892)^0$ & 0 (fixed) & 0 (fixed) & 0 (fixed) \\
$K^*(1410)^0$ & $\phantom{0}{-47.4} \pm \phantom{0}4.1 \pm \phantom{0}8.4 \pm \phantom{0}4.5$ & $\phantom{0}{-49.0} \pm \phantom{0}4.4 \pm \phantom{0}8.6 \pm \phantom{0}1.5$ & $\phantom{0}{-33.6} \pm \phantom{0}8.8 \pm 13.1 \pm \phantom{0}1.5$\\
$K^*(1680)^0$ & ${-175.5} \pm \phantom{0}8.2 \pm 19.2 \pm 16.0$ & $+179.3 \pm \phantom{0}8.5 \pm 17.2 \pm \phantom{0}3.8$ & $-120.7 \pm 29.3 \pm 60.8 \pm 17.7$\\
$\rho(770)^0$ & $\phantom{00}{+0.1} \pm \phantom{0}3.3 \pm \phantom{0}2.4 \pm \phantom{0}1.5$ & $\phantom{00}{-7.2} \pm \phantom{0}3.2 \pm \phantom{0}5.4 \pm \phantom{0}1.1$ & $\phantom{00}{-4.3} \pm \phantom{0}6.4 \pm \phantom{0}9.0 \pm \phantom{0}1.3$\\
$\omega(782)$ & $+104.3 \pm \phantom{0}7.2 \pm \phantom{0}2.8 \pm \phantom{0}1.4$ & $+100.8 \pm \phantom{0}7.4 \pm \phantom{0}3.7 \pm \phantom{0}1.5$ & $+103.8 \pm \phantom{0}9.0 \pm \phantom{0}5.9 \pm \phantom{0}2.1$\\
$\rho(1450)^0$ & $+114.5 \pm \phantom{0}4.8 \pm \phantom{0}7.8 \pm \phantom{0}2.4$ & $+128.5 \pm \phantom{0}5.7 \pm \phantom{0}5.8 \pm \phantom{0}2.3$ & $+128.8 \pm \phantom{0}5.9 \pm \phantom{0}8.9 \pm \phantom{0}5.5$\\
$\rho(1700)^0$ & $\phantom{0}{+42.1} \pm \phantom{0}9.3 \pm \phantom{0}6.1 \pm \phantom{0}4.7$ & $\phantom{0}{+62.2} \pm \phantom{0}8.3 \pm \phantom{0}3.8 \pm \phantom{0}2.8$ & $\phantom{0}{+62.1} \pm 13.6 \pm \phantom{0}3.8 \pm \phantom{0}2.3$\\
$K_2^*(1430)^0$ & $\phantom{0}{+96.1} \pm \phantom{0}1.8 \pm \phantom{0}2.4 \pm \phantom{0}1.9$ & $\phantom{0}{+99.2} \pm \phantom{0}1.5 \pm \phantom{0}2.7 \pm \phantom{0}0.6$ & $+102.3 \pm \phantom{0}5.8 \pm \phantom{0}3.5 \pm \phantom{0}2.0$\\
$f_2(1270)$ & $+137.8 \pm \phantom{0}2.2 \pm \phantom{0}2.8 \pm \phantom{0}2.6$ & $+138.1 \pm \phantom{0}2.3 \pm \phantom{0}2.8 \pm \phantom{0}2.5$ & $+137.0 \pm \phantom{0}4.5 \pm \phantom{0}4.0 \pm \phantom{0}3.6$\\
$f_2^\prime(1525)$ & $\phantom{0}{+19.5} \pm \phantom{0}9.8 \pm 17.8 \pm \phantom{0}6.4$ & $\phantom{0}{+26.7} \pm \phantom{0}9.3 \pm 14.6 \pm \phantom{0}2.7$ & $\phantom{0}{+42.7} \pm 11.6 \pm 15.2 \pm \phantom{0}1.6$\\
$K_3^*(1780)^0$ & $-110.4 \pm \phantom{0}5.4 \pm 10.2 \pm \phantom{0}3.7$ & $-104.4 \pm \phantom{0}5.1 \pm \phantom{0}8.6 \pm \phantom{0}2.5$ & $-106.0 \pm \phantom{0}7.5 \pm 16.5 \pm \phantom{0}3.6$\\
$\rho_3(1690)^0$ & $\phantom{0}{-66.4} \pm 10.3 \pm 12.2 \pm 10.1$ & $\phantom{0}{-78.3} \pm \phantom{0}7.7 \pm 10.7 \pm 10.2$ & $\phantom{0}{-64.9} \pm 14.3 \pm 11.3 \pm 14.9$\\
$\chi_{c0}(1P)$ & $+115.6 \pm \phantom{0}3.8 \pm \phantom{0}1.6 \pm \phantom{0}1.6$ & $+116.1 \pm \phantom{0}3.1 \pm \phantom{0}2.0 \pm \phantom{0}0.3$ & $+119.3 \pm \phantom{0}5.3 \pm \phantom{0}2.6 \pm \phantom{0}0.7$\\\hline
$(K^+\pi^-)_{\rm Res}$ & $+119.7 \pm \phantom{0}2.0 \pm \phantom{0}5.5 \pm \phantom{0}3.8$ & $\phantom{0}{+85.5} \pm \phantom{0}1.6 \pm \phantom{0}8.7 \pm \phantom{0}1.3$ & ---\\
$(K^+\pi^-)_{\rm ER}$ & $\phantom{0}{-64.9} \pm \phantom{0}1.9 \pm \phantom{0}2.8 \pm \phantom{0}3.7$ & $\phantom{0}{-76.9} \pm \phantom{0}1.5 \pm \phantom{0}5.3 \pm \phantom{0}0.5$ & ---\\
$K_0^*(1950)^0$ & $\phantom{0}{-16.7} \pm \phantom{0}5.2 \pm \phantom{0}6.7 \pm \phantom{0}5.2$ & $\phantom{0}{-14.2} \pm \phantom{0}4.6 \pm \phantom{0}6.0 \pm \phantom{0}4.8$ & ---\\\hline
$f_0(500)$ & $\phantom{0}{-73.6} \pm \phantom{0}3.3 \pm \phantom{0}6.8 \pm \phantom{0}3.8$ & --- & ---\\
$f_0(980)$ & $\phantom{0}{+62.1} \pm \phantom{0}2.8 \pm \phantom{0}3.4 \pm \phantom{0}2.2$ & --- & ---\\
$f_0(1370)$ & $+154.9 \pm \phantom{0}2.3 \pm \phantom{0}5.5 \pm \phantom{0}2.2$ & --- & ---\\
$f_0(1500)$ & $\phantom{0}{-39.6} \pm \phantom{0}3.7 \pm \phantom{0}4.1 \pm \phantom{0}2.1$ & --- & ---\\
$f_0(1710)$ & $\phantom{0}{+30.8} \pm \phantom{0}5.6 \pm \phantom{0}6.8 \pm \phantom{0}4.9$ & --- & ---\\
$\pi\pi$--$\pi\pi$ & $+116.0 \pm \phantom{0}3.9 \pm \phantom{0}7.9 \pm \phantom{0}4.2$ & --- & ---\\
$\pi\pi$--$K\Kb$ & $\phantom{0}{+40.4} \pm 19.0 \pm 17.5 \pm 25.3$ & --- & ---\\

    \hline
    \end{tabular}
    }
\end{table}

\begin{table}[tb]
  \centering
  \caption{\label{tab:res:phasem}
        Phase comparison for \Bm decay amplitudes between the three S-wave approaches where the first uncertainty is statistical, the second systematic and the third from the model. 
    }

    \resizebox{\textwidth}{!}{
    \renewcommand{\arraystretch}{1.1}
    \begin{tabular}
    {@{\hspace{0.0cm}}l@{\hspace{0.125cm}}
     @{\hspace{0.125cm}}c@{\hspace{0.125cm}}
     @{\hspace{0.125cm}}c@{\hspace{0.125cm}}
     @{\hspace{0.125cm}}c@{\hspace{0.0cm}}} \hline
    & \multicolumn{3}{c}{Phase of \Bm decay amplitude $(^{\circ})$}\\ \hline
    Component & Isobar & K-matrix & QMI\\ \hline

$K^*(892)^0$ & 0 (fixed) & 0 (fixed) & 0 (fixed) \\
$K^*(1410)^0$ & $\phantom{0}{-54.4} \pm \phantom{0}4.0 \pm \phantom{0}4.4 \pm \phantom{0}3.7$ & $\phantom{0}{-54.4} \pm \phantom{0}4.9 \pm \phantom{0}1.8 \pm \phantom{0}0.8$ & $\phantom{0}{-39.8} \pm \phantom{0}5.1 \pm \phantom{0}5.2 \pm \phantom{0}0.8$\\
$K^*(1680)^0$ & $+123.5 \pm 13.5 \pm 35.1 \pm 37.7$ & $+123.4 \pm 11.1 \pm 41.2 \pm 11.8$ & $-164.5 \pm \phantom{0}9.0 \pm 14.0 \pm \phantom{0}1.5$\\
$\rho(770)^0$ & $\phantom{0}{-40.2} \pm \phantom{0}2.8 \pm \phantom{0}3.3 \pm \phantom{0}8.0$ & $\phantom{0}{-47.5} \pm \phantom{0}2.7 \pm \phantom{0}4.4 \pm \phantom{0}2.0$ & $\phantom{0}{-38.0} \pm \phantom{0}4.8 \pm \phantom{0}5.9 \pm \phantom{0}0.6$\\
$\omega(782)$ & $\phantom{0}{+82.4} \pm \phantom{0}6.5 \pm \phantom{0}4.0 \pm \phantom{0}8.4$ & $\phantom{0}{+74.2} \pm \phantom{0}6.9 \pm \phantom{0}4.6 \pm \phantom{0}2.0$ & $\phantom{0}{+85.9} \pm \phantom{0}7.6 \pm \phantom{0}4.4 \pm \phantom{0}0.8$\\
$\rho(1450)^0$ & $\phantom{0}{+77.4} \pm \phantom{0}5.0 \pm \phantom{0}3.4 \pm 14.4$ & $\phantom{0}{+80.3} \pm \phantom{0}5.8 \pm \phantom{0}3.2 \pm \phantom{0}2.6$ & $+109.7 \pm \phantom{0}5.9 \pm \phantom{0}5.1 \pm \phantom{0}1.9$\\
$\rho(1700)^0$ & $\phantom{0}{+49.1} \pm 25.0 \pm 16.3 \pm 44.9$ & $\phantom{0}{+75.2} \pm 25.7 \pm 10.0 \pm \phantom{0}9.8$ & $\phantom{0}{+83.8} \pm 25.4 \pm 29.8 \pm \phantom{0}7.8$\\
$K_2^*(1430)^0$ & $\phantom{0}{+85.9} \pm \phantom{0}1.6 \pm \phantom{0}2.6 \pm \phantom{0}0.5$ & $\phantom{0}{+87.4} \pm \phantom{0}1.6 \pm \phantom{0}2.8 \pm \phantom{0}1.0$ & $+100.8 \pm \phantom{0}3.5 \pm \phantom{0}2.3 \pm \phantom{0}0.4$\\
$f_2(1270)$ & $+169.6 \pm \phantom{0}3.0 \pm \phantom{0}2.8 \pm \phantom{0}4.4$ & $+167.3 \pm \phantom{0}3.5 \pm \phantom{0}3.4 \pm \phantom{0}2.0$ & $-176.5 \pm \phantom{0}4.4 \pm \phantom{0}4.0 \pm \phantom{0}1.6$\\
$f_2^\prime(1525)$ & $-112.5 \pm 31.0 \pm 30.6 \pm 55.0$ & $\phantom{0}{-95.6} \pm 30.5 \pm 30.1 \pm \phantom{0}8.8$ & $-135.0 \pm 38.8 \pm 23.0 \pm 30.0$\\
$K_3^*(1780)^0$ & $-131.1 \pm 35.8 \pm 13.7 \pm 13.9$ & $-115.1 \pm 33.7 \pm 16.5 \pm \phantom{0}4.1$ & $\phantom{0}{-99.5} \pm 34.2 \pm 10.8 \pm 17.1$\\
$\rho_3(1690)^0$ & $-111.5 \pm \phantom{0}4.8 \pm \phantom{0}5.7 \pm 27.9$ & $-105.4 \pm \phantom{0}4.6 \pm \phantom{0}5.5 \pm \phantom{0}1.3$ & $\phantom{0}{-84.1} \pm \phantom{0}8.3 \pm \phantom{0}5.3 \pm \phantom{0}1.3$\\
$\chi_{c0}(1P)$ & $+115.2 \pm \phantom{0}3.2 \pm \phantom{0}2.0 \pm \phantom{0}0.9$ & $+116.0 \pm \phantom{0}3.0 \pm \phantom{0}1.7 \pm \phantom{0}0.2$ & $+126.7 \pm \phantom{0}5.0 \pm \phantom{0}2.1 \pm \phantom{0}0.2$\\\hline
$(K^+\pi^-)_{\rm Res}$ & $+112.4 \pm \phantom{0}2.3 \pm \phantom{0}2.9 \pm \phantom{0}1.8$ & $\phantom{0}{+77.2} \pm \phantom{0}1.7 \pm \phantom{0}6.3 \pm \phantom{0}1.6$ & ---\\
$(K^+\pi^-)_{\rm ER}$ & $\phantom{0}{-60.7} \pm \phantom{0}1.9 \pm \phantom{0}1.8 \pm \phantom{0}2.6$ & $\phantom{0}{-75.9} \pm \phantom{0}1.3 \pm \phantom{0}3.6 \pm \phantom{0}0.6$ & ---\\
$K_0^*(1950)^0$ & $\phantom{0}{-19.7} \pm \phantom{0}4.5 \pm \phantom{0}6.4 \pm \phantom{0}8.6$ & $\phantom{0}{-18.8} \pm \phantom{0}4.1 \pm \phantom{0}7.6 \pm \phantom{0}6.7$ & ---\\\hline
$f_0(500)$ & $-111.9 \pm \phantom{0}5.6 \pm \phantom{0}6.8 \pm 11.7$ & --- & ---\\
$f_0(980)$ & $\phantom{0}{+65.2} \pm \phantom{0}3.1 \pm \phantom{0}2.2 \pm \phantom{0}4.8$ & --- & ---\\
$f_0(1370)$ & $+148.8 \pm \phantom{0}2.4 \pm \phantom{0}3.8 \pm \phantom{0}2.9$ & --- & ---\\
$f_0(1500)$ & $\phantom{0}{-48.1} \pm \phantom{0}3.4 \pm \phantom{0}2.3 \pm \phantom{0}6.4$ & --- & ---\\
$f_0(1710)$ & $\phantom{0}{+66.3} \pm \phantom{0}7.4 \pm \phantom{0}7.2 \pm 39.1$ & --- & ---\\
$\pi\pi$--$\pi\pi$ & $+133.5 \pm \phantom{0}5.1 \pm \phantom{0}3.7 \pm \phantom{0}9.2$ & --- & ---\\
$\pi\pi$--$K\Kb$ & $\phantom{0}{-31.1} \pm \phantom{0}9.3 \pm \phantom{0}5.8 \pm 15.3$ & --- & ---\\

    \hline
    \end{tabular}
    }
\end{table}

\clearpage

\subsection{Fit projections}

\begin{figure}[tb]
    \centering
    \includegraphics[width=0.5\linewidth]{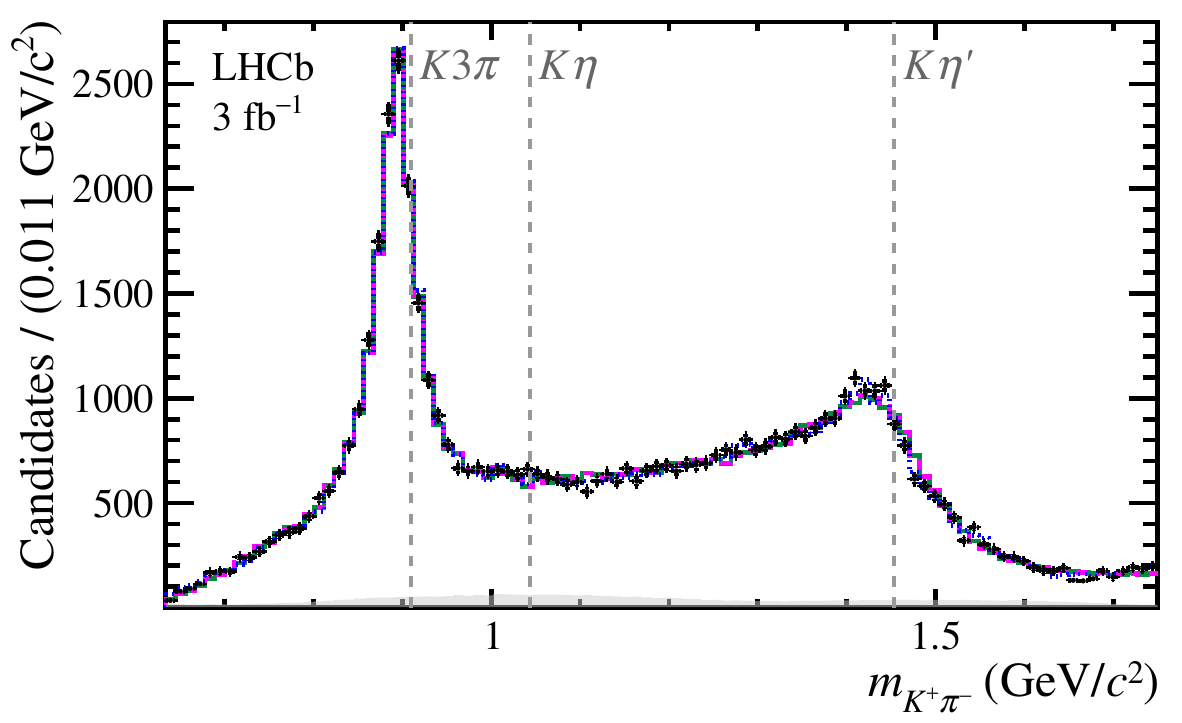}%
    \includegraphics[width=0.5\linewidth]{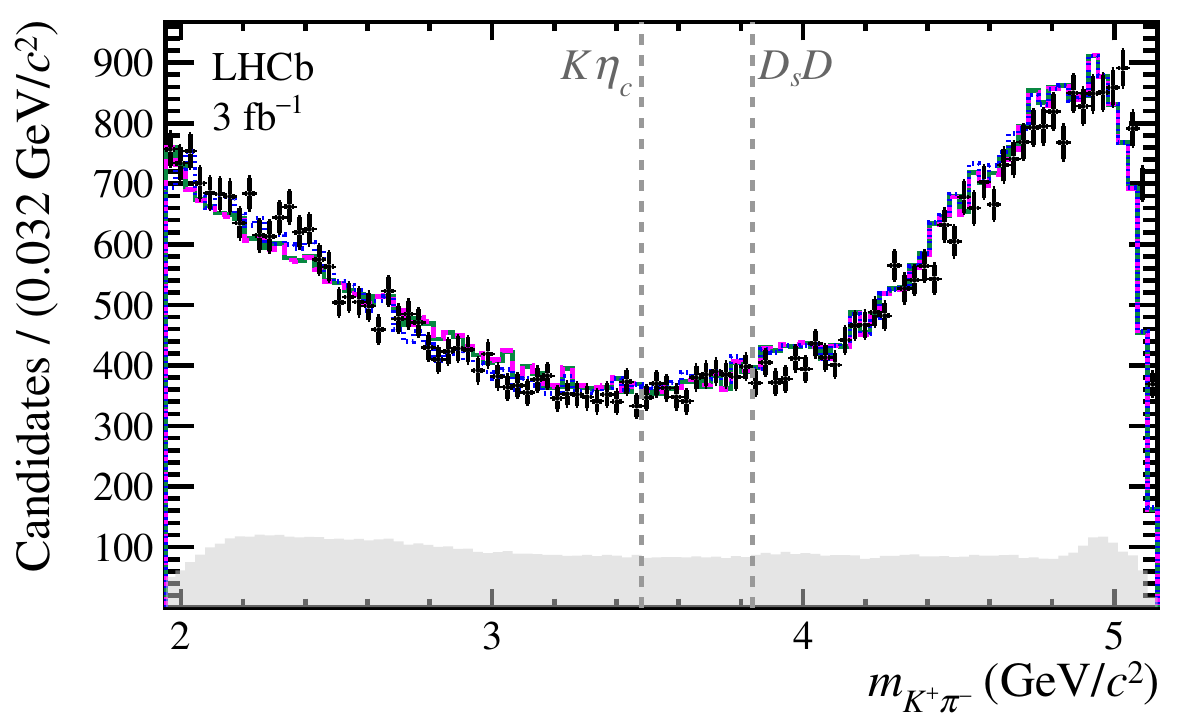}

    \includegraphics[width=0.5\linewidth]{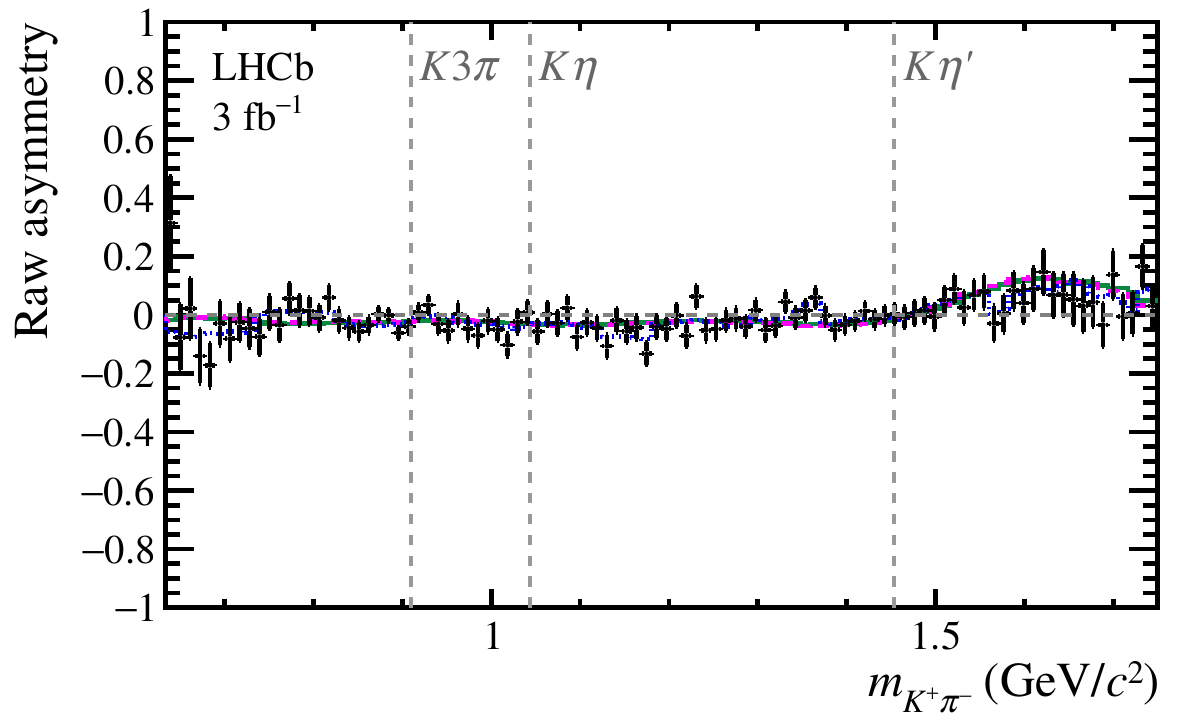}%
    \includegraphics[width=0.5\linewidth]{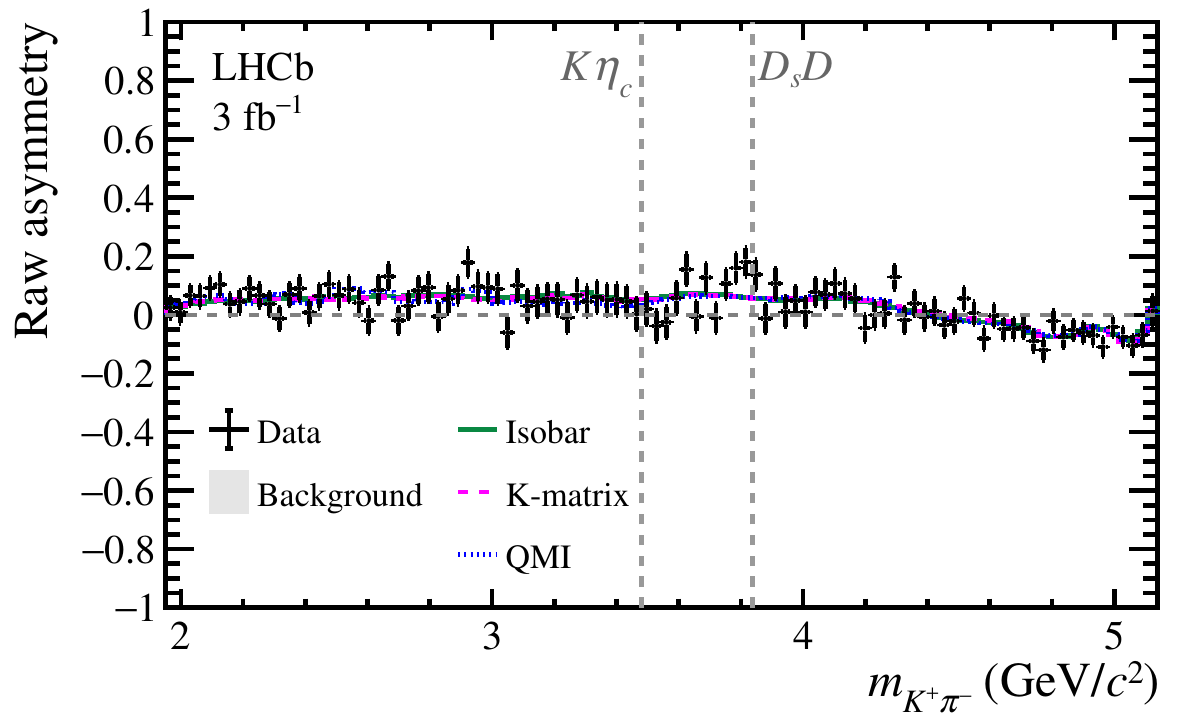}
    \caption{
        Fit projections of each model~(top) in the $m_{K^+ \pi^-}$ region (left)~below and (right)~above the open charm threshold, with (bottom)~the corresponding asymmetries. 
        Vertical dashed lines correspond to the opening thresholds of the indicated coupled channels.
    }
    \label{fig:res:kpi}
\end{figure}

\begin{figure}[tb]
    \centering
    \includegraphics[width=0.5\linewidth]{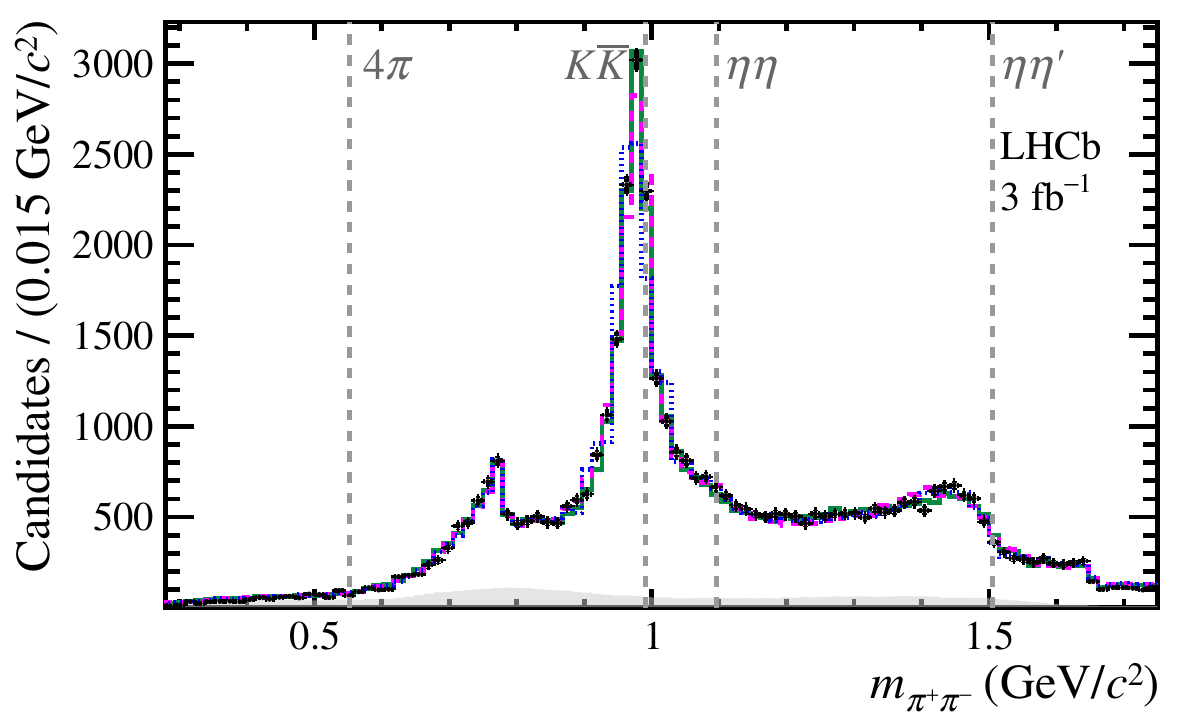}%
    \includegraphics[width=0.5\linewidth]{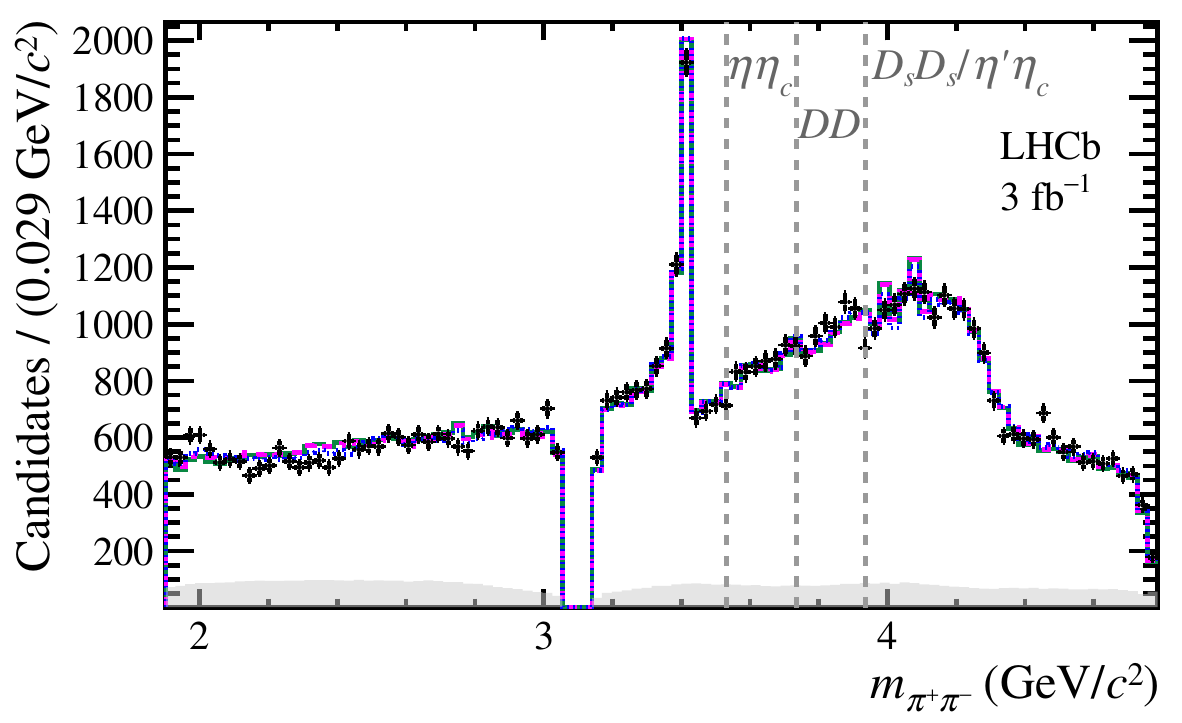}

    \includegraphics[width=0.5\linewidth]{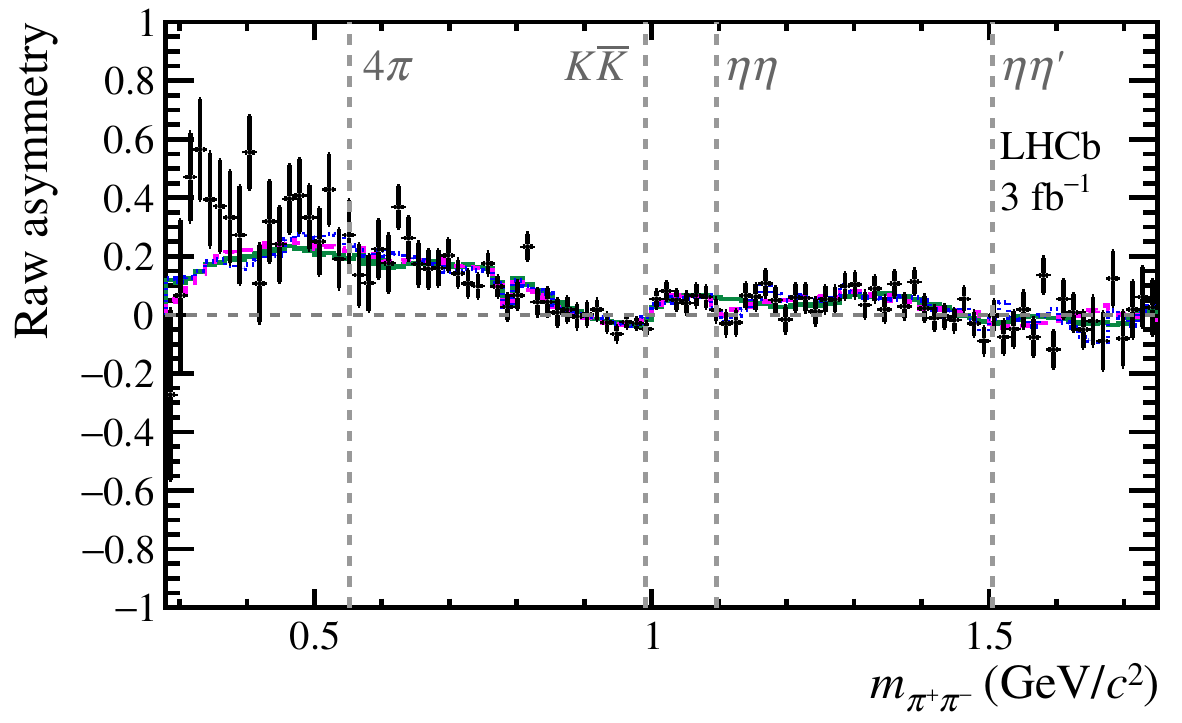}%
    \includegraphics[width=0.5\linewidth]{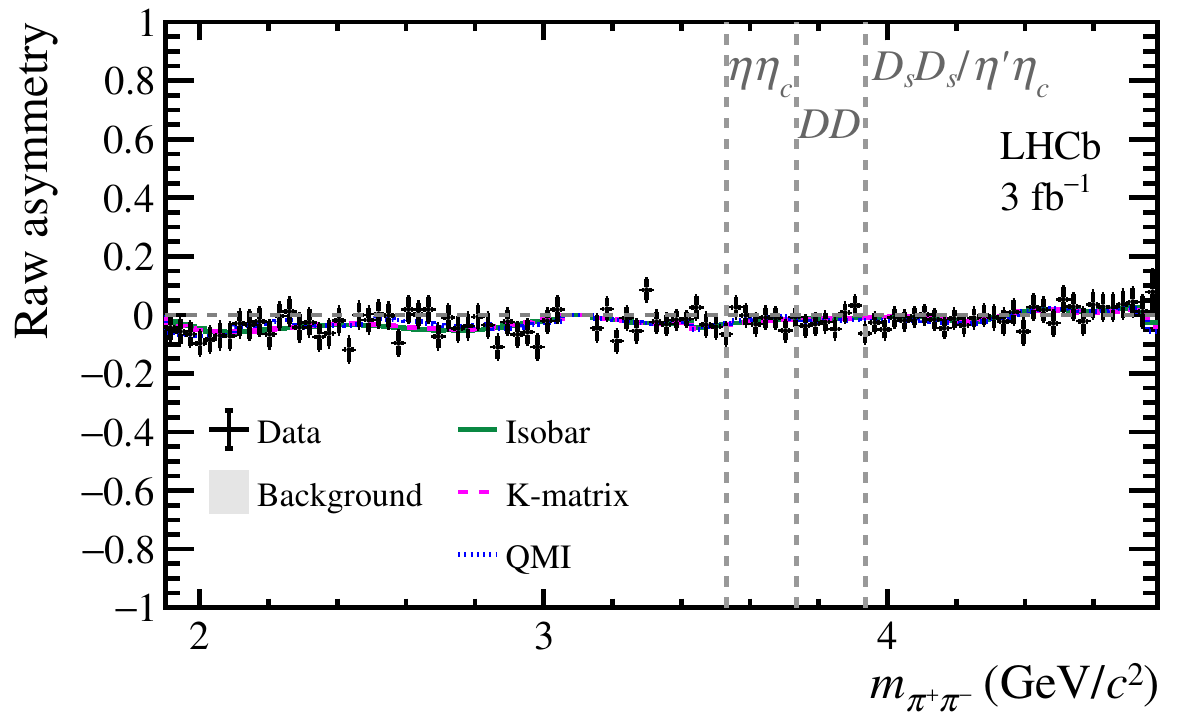}

    \caption{
        Fit projections of each model~(top) in the $m_{\pi^+ \pi^-}$ region (left)~below and (right)~above the open charm threshold, with (bottom)~the corresponding asymmetries. 
        Vertical dashed lines correspond to the opening thresholds of the indicated coupled channels.
    }
    \label{fig:res:pipi}
\end{figure}

\begin{figure}[tb]
    \centering
    \includegraphics[width=0.5\linewidth]{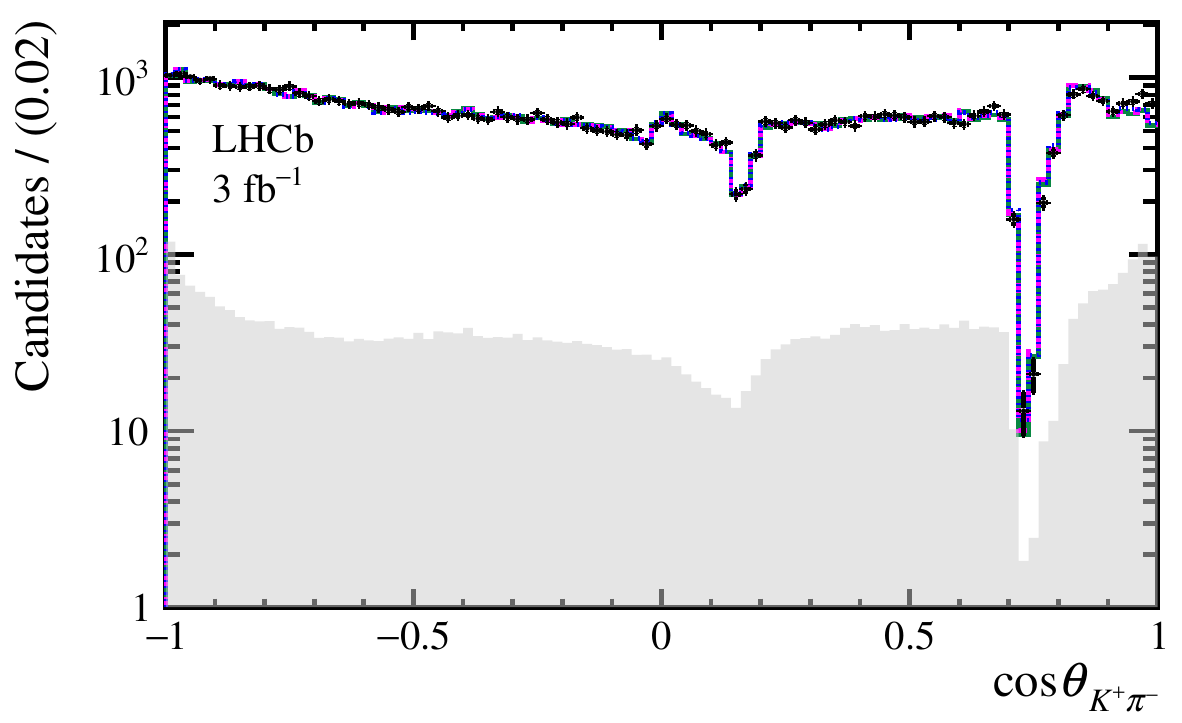}%
    \includegraphics[width=0.5\linewidth]{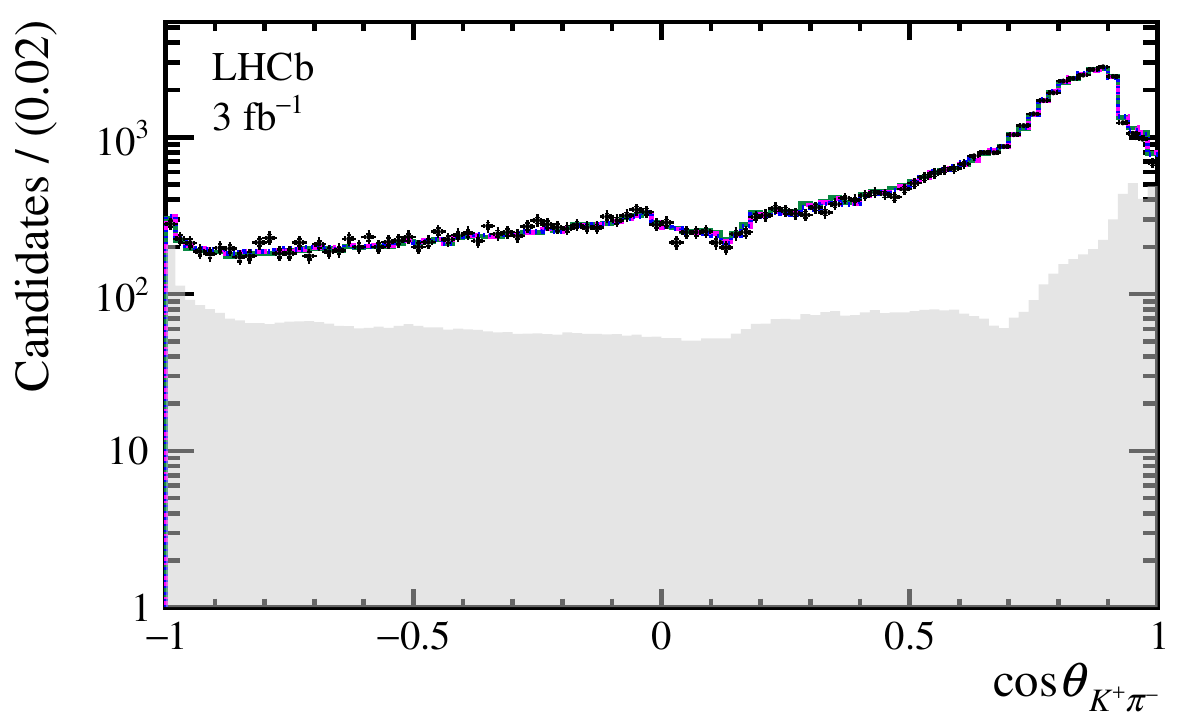}

    \includegraphics[width=0.5\linewidth]{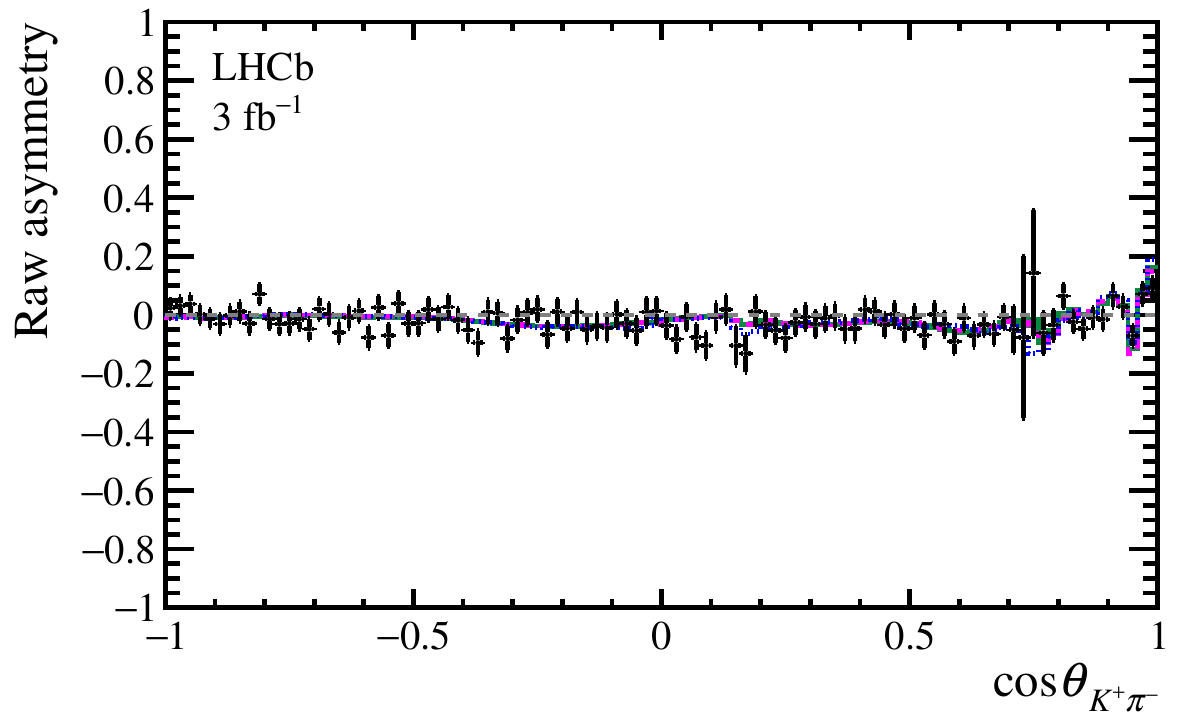}%
    \includegraphics[width=0.5\linewidth]{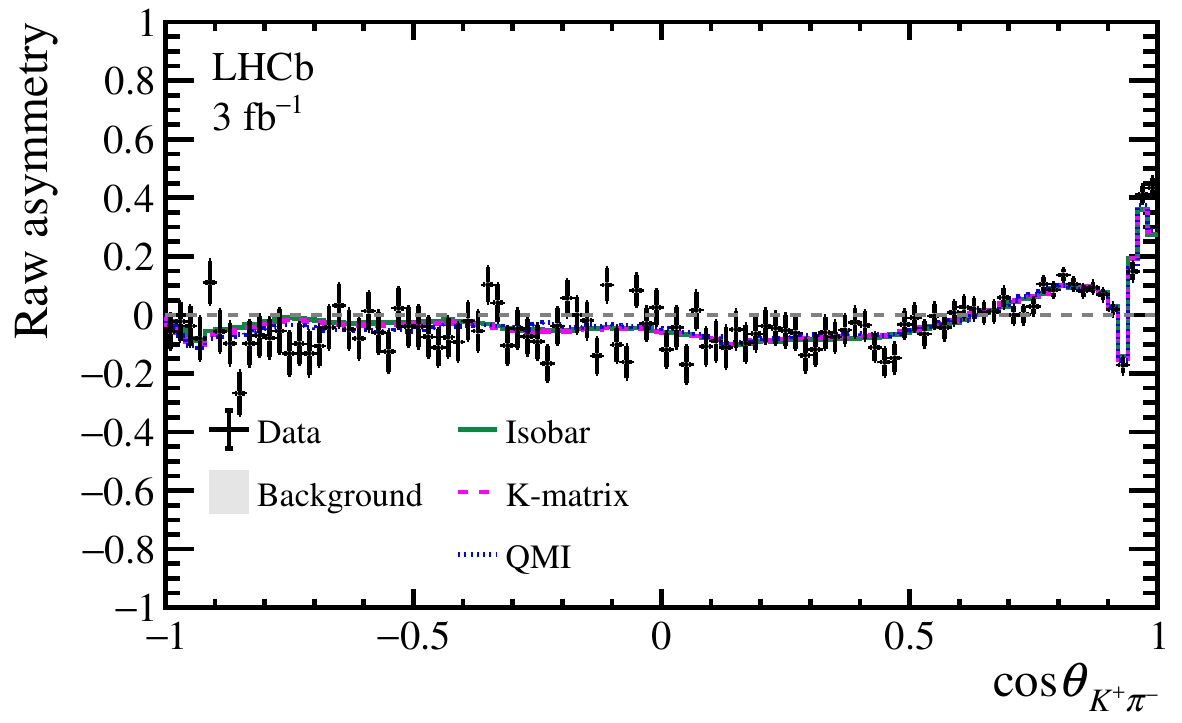}
    \caption{
        Fit projections of each model~(top) in the $\cos\theta_{K^+ \pi^-}$ region (left)~below and (right)~above the open charm threshold, with (bottom)~the corresponding asymmetries.
    }
    \label{fig:res:cthkpi}
\end{figure}

\begin{figure}[tb]
    \centering
    \includegraphics[width=0.5\linewidth]{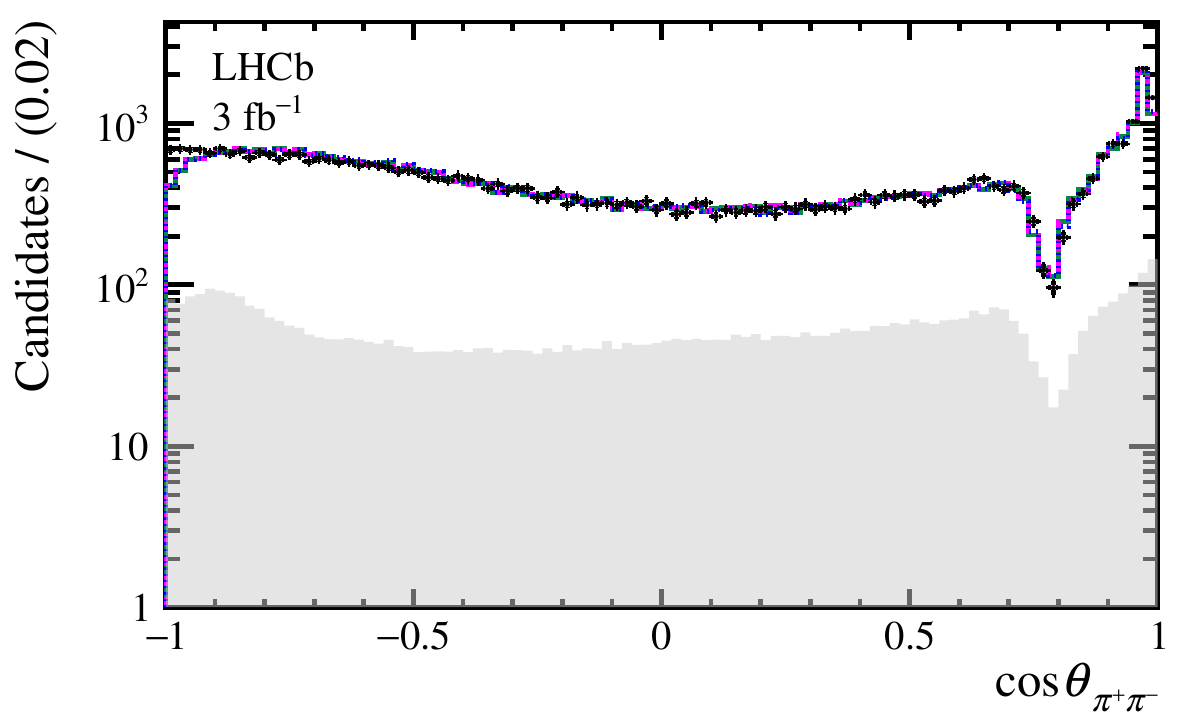}%
    \includegraphics[width=0.5\linewidth]{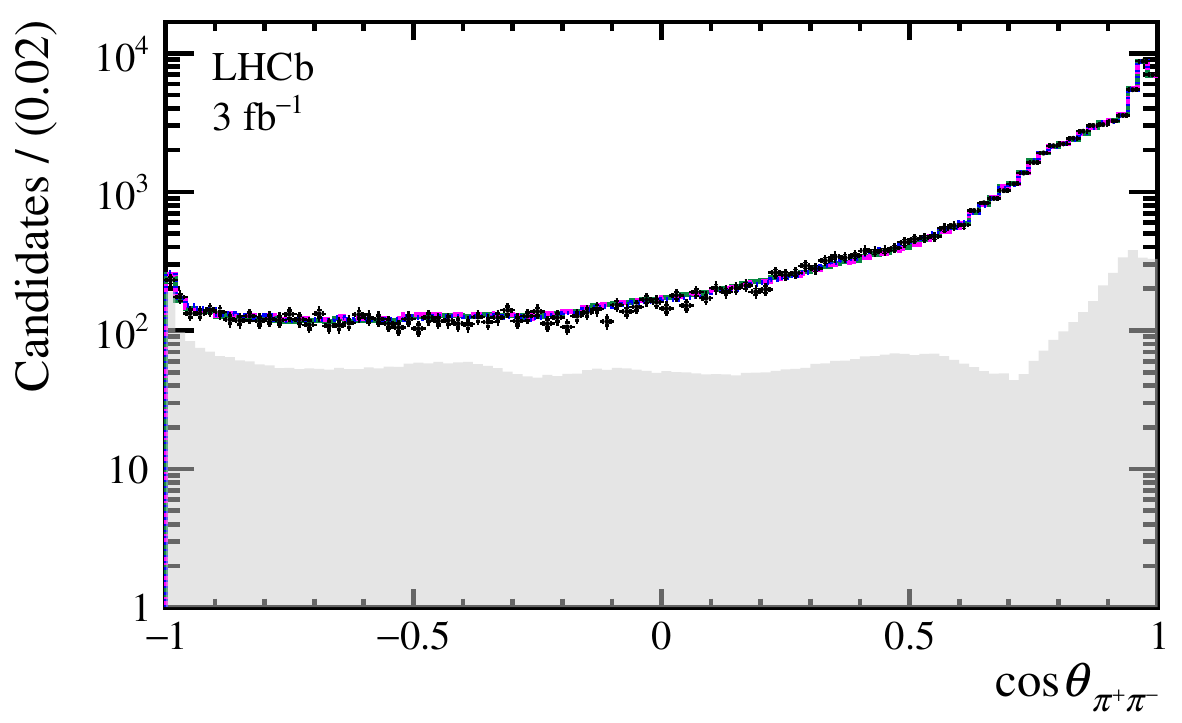}

    \includegraphics[width=0.5\linewidth]{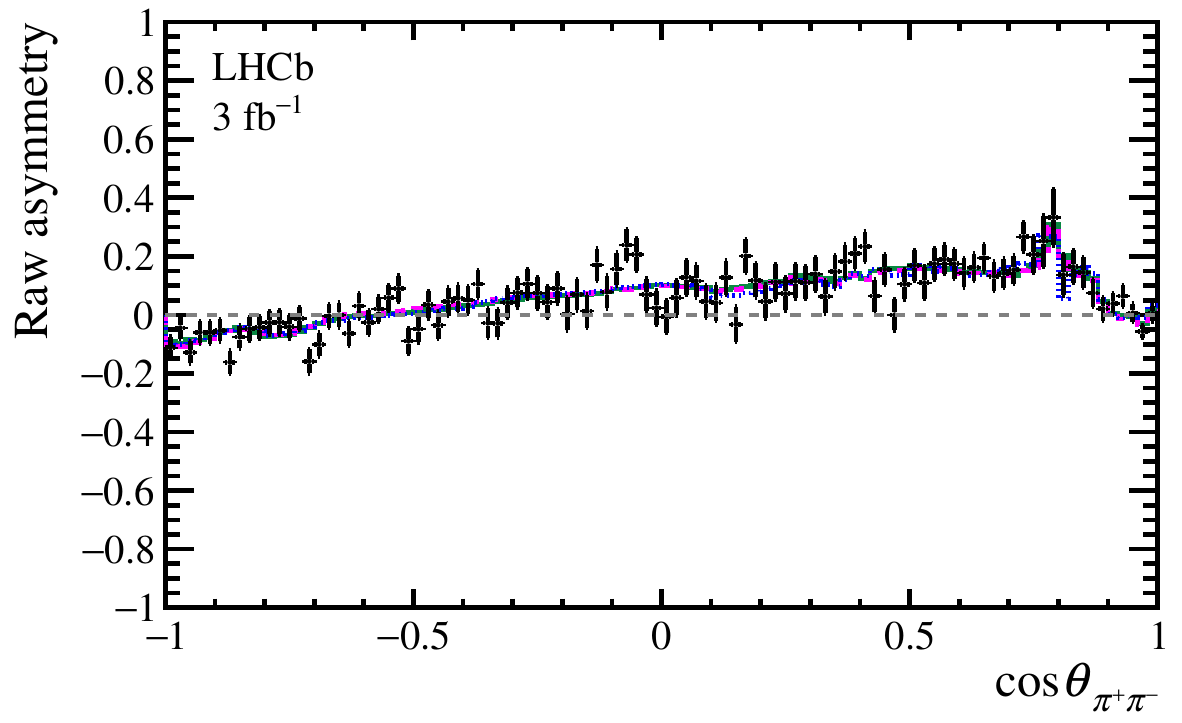}%
    \includegraphics[width=0.5\linewidth]{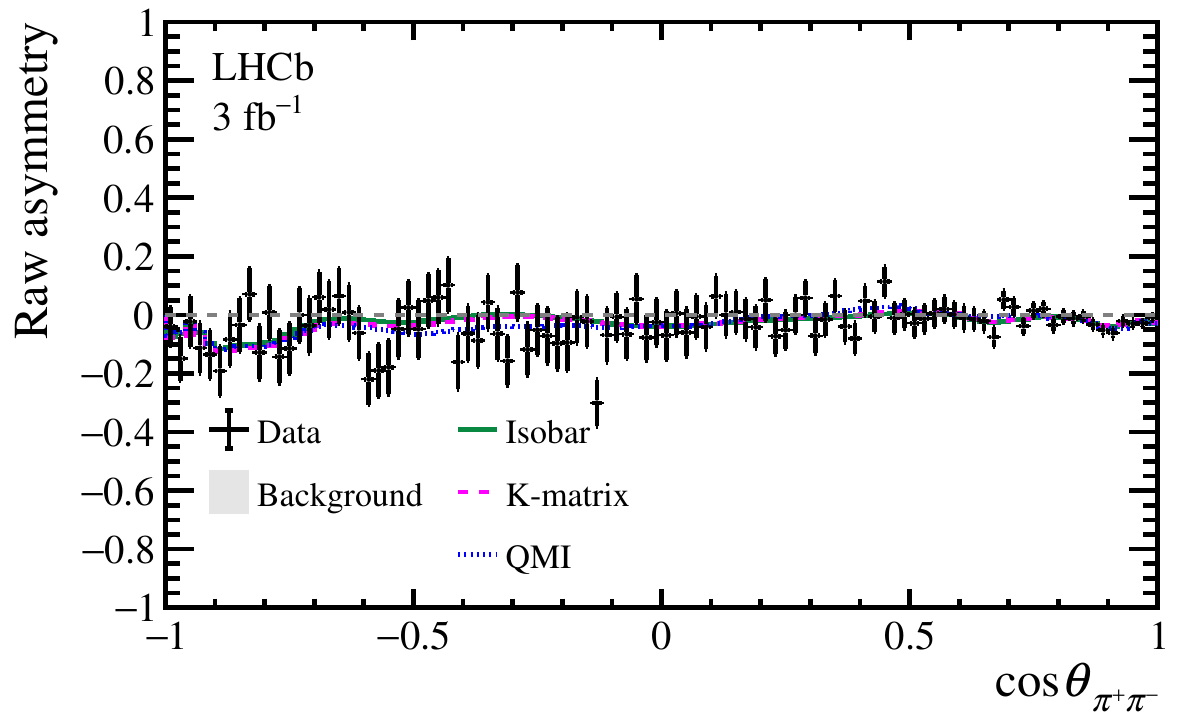}
    \caption{
        Fit projections of each model~(top) in the $\cos\theta_{\pi^+ \pi^-}$ region (left)~below and (right)~above the open charm threshold, with (bottom)~the corresponding asymmetries.
    }
    \label{fig:res:cthpipi}
\end{figure}

Comparisons of the data and all three fit models, projected onto $m_{\Kp\pim}$, $m_{\pip\pim}$ and the cosines of their associated helicity angles, along with the raw asymmetries between $\Bm$ and $\Bp$ decays, can be seen in Figs.~\ref{fig:res:kpi}--\ref{fig:res:cthpipi}. These models appear to be in good overall agreement with the data and with each other, both in \CP-averaged projections and in the variation of the asymmetries across the phase space. Additional projections separating the contributions for various components of the amplitude model are shown in
Appendices~\ref{app:isobar},~\ref{app:kmatrix} and~\ref{app:qmi} for the Isobar, K-matrix and QMI approaches, respectively.
Further comparisons of the different S-wave models are given in Appendix~\ref{app:swave}, and comparison of the data and fit models using weights to enhance components with particular orbital angular momenta are shown in Appendix~\ref{app:moments}.

\clearpage

\subsection{\texorpdfstring{\boldmath Evaluation of the \decay{\Bp}{K^*_0(1430)^0 \pip} decay properties}{Evaluation of the B+ -> K*0(1430)0 pi+ decay properties}}
\label{sec:results:Kstarz1430FF}

\begin{figure}[tb]
    \centering
    \includegraphics[width=0.5\linewidth]{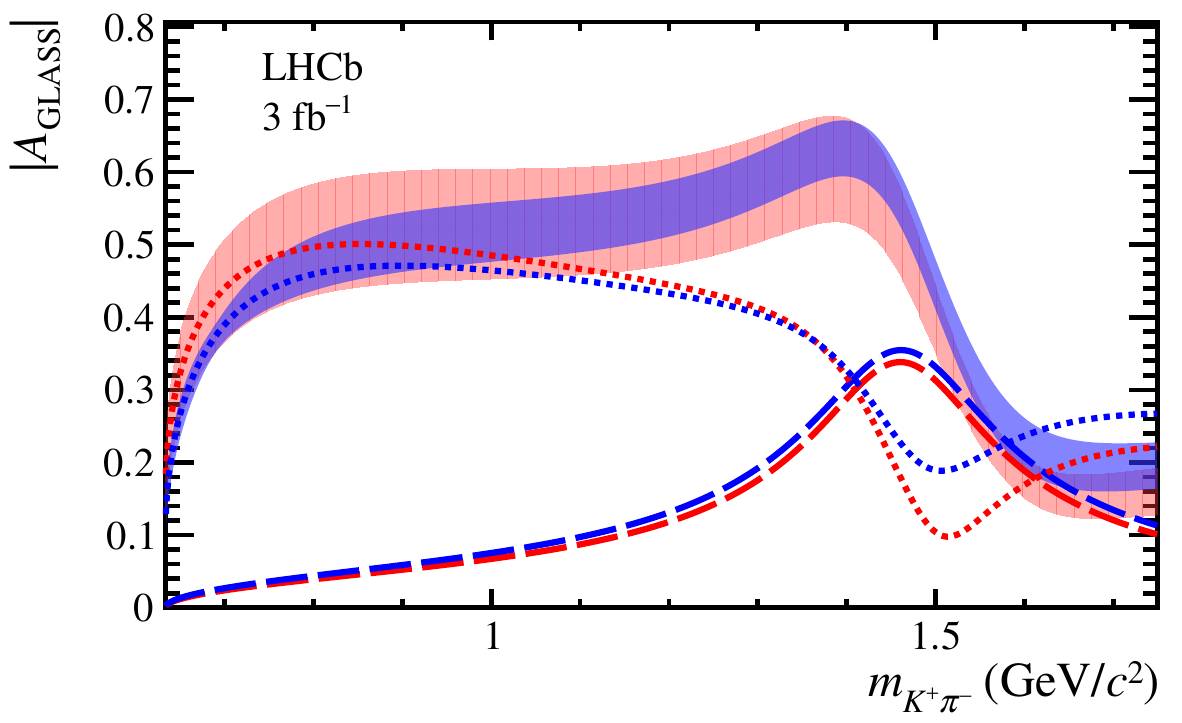}%
    \includegraphics[width=0.5\linewidth]{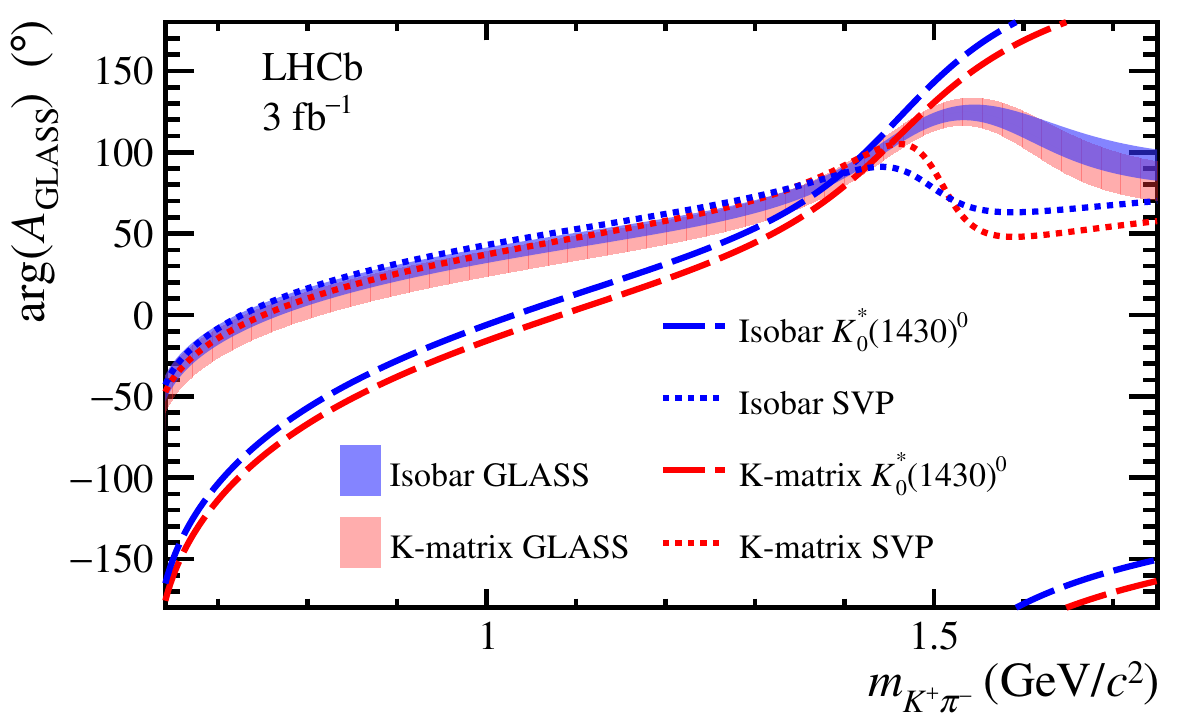}

    \caption{(left)~Amplitude and (right)~phase of the GLASS lineshape, where the band includes the total uncertainty. Individual contributions from the $K^*_0(1430)^0$ and SVP components are also shown.}
    \label{fig:GLASS}
\end{figure}

The GLASS lineshape as determined from data can be seen in Fig.~\ref{fig:GLASS}, for the Isobar and K-matrix approaches. 
The corresponding overall fit fractions are \mbox{$(58.9 \pm 2.0\stat \pm 3.2\syst \pm 3.0\textrm{\,(model)})\%$} and \mbox{$(47.7 \pm 0.5\stat \pm 1.4\syst \pm 0.3\textrm{\,(model)})\%$} for the Isobar and K-matrix approaches, respectively. 
This difference is attributed to interference with each S-wave approach in \pip\pim as their total S-wave fit fractions agree to within 1\%, as does that obtained with the QMI approach, reported in Table~\ref{tab:res:ffsum}.

The measured resonance parameters of the $K^*_0(1430)^0$ resonance, presented in Appendices~\ref{app:isobar} and~\ref{app:kmatrix} for the Isobar and K-matrix approaches respectively, are practically identical. 
The $K^*_0(1430)^0$ fit fraction is likewise expected to be in agreement; however calculation of the fit fraction from the resonance-sensitive term in Eq.~\eqref{eq:app:glass} (as was done by the \babar collaboration), shows a significant disagreement in Table~\ref{tab:res:ff} between the Isobar and K-matrix approaches. 
This arises from the $\phi_{\rm Nonres}$ term, defined in Eq.~\eqref{eq:app:glasssplit}, in the GLASS lineshape and impact of the differing values obtained for this parameter in the Isobar and K-matrix approaches.
Considering the $K^*_0(1430)^0$ resonance phase to have purely a Breit--Wigner form,
\begin{equation}
    \phi_{K^*_0}(s) \equiv \arctan\biggl[\frac{m_{K^*_0}\Gamma(s)}{(m^2_{K^*_0} - s)}\biggr] \,,
\end{equation}
a double-angle expansion of $\sin\delta_{\rm Res}$ leads to the peaking amplitude best identified with the $K^*_0(1430)^0$ resonance, which evaluates to
\begin{equation}
    A_{K^*_0}(s) \equiv \biggl[\frac{1}{2}\cos 2\phi_{\rm Nonres}\sin 2\phi_{K^*_0}(s) + i\cos^2 \phi_{\rm Nonres}\sin^2 \phi_{K^*_0}(s)\biggr] e^{2i\delta_{\rm ER}} \,,
\end{equation}
with the remainder of $A_{\rm GLASS}$ comprising the slowly varying part~(SVP). 
With this definition, the \decay{\Bp}{K^*_0(1430)^0 \pip} decay properties are given in Table~\ref{tab:res:kst1430}. 
The $K^*_0(1430)^0$ and SVP components of the GLASS lineshape are shown in Fig.~\ref{fig:GLASS} as the dashed and dotted curves, respectively.
 
\begin{table}[tb]
    \centering
    \caption{\label{tab:res:kst1430}
        Results obtained in the Isobar and K-matrix approaches for the \decay{\Bp}{K^*_0(1430)^0 \pip} decay properties where the first uncertainty is statistical, the second systematic and the third from the model.
    }

    \begin{tabular}
    {@{\hspace{0.25cm}}l@{\hspace{-0.7cm}}
     @{\hspace{-0.7cm}}l@{\hspace{1.0cm}}
     @{\hspace{0.25cm}}c@{\hspace{0.25cm}}
     @{\hspace{0.25cm}}c@{\hspace{0.25cm}}}\hline
    Observable & &  Isobar & K-matrix \\ \hline

$\mathcal{F}$ & $(\%)$ & $\phantom{+}7.912 \pm 0.270 \pm 0.267 \pm 0.381$ & $\phantom{+}6.736 \pm 0.208 \pm 0.551 \pm 0.148$\\
$\mathcal{A}_{\CP}$ & & $-0.094 \pm 0.021 \pm 0.038 \pm 0.027$ & $-0.121 \pm 0.015 \pm 0.048 \pm 0.013$\\
$\delta^+$ & $({}^\circ)$ & $+119.7 \pm \phantom{00}2.6 \pm \phantom{00}5.5 \pm \phantom{00}3.8$ & $\phantom{0}{+85.5} \pm \phantom{00}1.8 \pm \phantom{00}8.7 \pm \phantom{00}1.3$\\
$\delta^-$ & $({}^\circ)$ & $+112.4 \pm \phantom{00}2.6 \pm \phantom{00}2.9 \pm \phantom{00}1.8$ & $\phantom{0}{+77.2} \pm \phantom{00}1.9 \pm \phantom{00}6.3 \pm \phantom{00}1.6$\\\hline

    \hline
    \end{tabular}
\end{table}

\subsection{\texorpdfstring{\boldmath Alternative $\rho^0$--$\omega$ transition results}{Alternative rho0--omega transition results}}

In the baseline result, $\rho^0$--$\omega$ mixing is considered to manifest through direct \decay{\omega}{\pip \pim} decay. A commonly held alternate view is that the $\omega(782)$ meson first mixes into the $\rho(770)^0$ state, which subsequently decays into the \pip\pim final state. 
In order to model the latter, the direct $\omega$ contribution is replaced by its product with the $\rho^0$ lineshape as
\begin{equation}
    T_\omega(s) \to T_{\rho\omega}(s) = T_\rho(s) T_\omega(s).
\end{equation}
The functional difference between these two paradigms is that the magnitude of $T_{\rho\omega}$ closely resembles $|T_\omega|$ while the phase motion shares more similarities with $\arg(T_\rho)$. 
The main difference is that the phase of $T_{\rho\omega}$ instead rotates by $2\pi$ in the same range, $[m_0-\Gamma_0, m_0+\Gamma_0]$, where $T_\rho$ will rotate by $\pi$.

The $-2\log\mathcal{L}$ value is found to be practically unchanged between these two views of $\rho^0$--$\omega$ mixing, as shown in Table~\ref{tab:res:alt}, indicating that there is little prospect of separating them even in future analyses. 
Therefore, only the physics parameters that are impacted by these lineshapes are reported in Table~\ref{tab:res:alt}. 
Due to the similarity of fit fractions between the alternate and baseline models, the systematic and model uncertainties can be directly imported from the baseline. 
As expected, the only meaningful differences are in the $\rho^0$--$\omega$ phases which experience a shift of $\sim -90^\circ$ from the baseline $\omega$ phases.

\begin{table}[b]
    \centering
    \caption{\label{tab:res:alt}
        Results obtained with the alternative $\rho^0$--$\omega$ transition model.
        Only parameters impacted by the change of model are shown, and uncertainties are statistical only.
        The results for the $\rho^0$--$\omega$ component should be compared to those for the $\omega$ component in the baseline model.
    }

    \begin{tabular}
    {l@{\hspace{0.25cm}}
     @{\hspace{0.125cm}}l@{\hspace{-0.8cm}}
     @{\hspace{-0.8cm}}l@{\hspace{1.0cm}}
     @{\hspace{0.25cm}}c@{\hspace{0.25cm}}
     @{\hspace{0.25cm}}c@{\hspace{0.25cm}}
     @{\hspace{0.25cm}}c}\hline
    Component & Observable & &  Isobar & K-matrix & QMI \\ \hline

$\rho(770)^0$ & $\mathcal{F}$ & $(\%)$ & $\phantom{0}{\,7.654} \pm 0.150$ & $\phantom{0}{\,7.922} \pm 0.177$ & $\phantom{0}{\,7.147} \pm 0.248$\\
& $\mathcal{A}_{\CP}$ & & $+0.274 \pm 0.019$ &$+0.268 \pm 0.021$ & $+0.357 \pm 0.036$\\
& $\delta^+$ & $({}^\circ)$ & $\phantom{00}{+2.5} \pm \phantom{00}3.0$ &  $\phantom{00}{-4.8} \pm \phantom{00}3.2$& $\phantom{00}{-1.8} \pm \phantom{00}5.4$\\
& $\delta^-$ & $({}^\circ)$ &  $\phantom{0}{-38.7} \pm \phantom{00}2.1$& $\phantom{0}{-45.9} \pm \phantom{00}2.7$ & $\phantom{0}{-36.4} \pm \phantom{00}5.2$\\\hline

$\rho^0$--$\omega$ & $\mathcal{F}$ & $(\%)$ & $\phantom{0}{\,0.175} \pm 0.026$ & $\phantom{0}{\,0.190} \pm 0.029$ & $\phantom{0}{\,0.193} \pm 0.029$\\
& $\mathcal{A}_{\CP}$ & & $+0.005\pm 0.129$ & $-0.028 \pm0.136$ & $-0.013 \pm 0.168$\\
& $\delta^+$ & $({}^\circ)$ & $\phantom{00}{+2.1} \pm \phantom{00}7.5$ & $\phantom{00}{-1.3} \pm \phantom{00}7.4$ & $\phantom{00}{+2.3} \pm \phantom{00}8.2$\\
& $\delta^-$ & $({}^\circ)$ & $\phantom{0}{-20.0} \pm \phantom{00}6.1$  &$\phantom{0}{-28.1} \pm \phantom{00}5.9$ & $\phantom{0}{-15.6} \pm \phantom{00}8.0$\\\hline

& $\Delta(-2\log\mathcal{L})$ & & $+0.2$ & $-0.3$ & $-0.2$\\

    \hline
    \end{tabular}
\end{table}

\section{Interpretation}
\label{sec:interpretation}

\begin{table}[tb]
    \centering
    \caption{\label{tab:int:bf}
        Intermediate absolute branching fractions for averaging purposes and upper limits in cases where the significance of the underlying fit fraction does not exceed three standard deviations. 
        Branching fractions of the resonance to $\Kp\pim$ or $\pip\pim$ have been accounted for, except in cases labelled $\dagger$ where the measured quantity is the product branching fraction. 
        The first uncertainty is the total obtained from the Isobar approach, while the final uncertainty is always from the external inclusive \decay{\Bp}{\Kp \pip \pim} branching fractions and any intermediate branching fractions. A middle uncertainty represents the maximum deviation from the K-matrix and QMI results.
        The branching fraction for the $\chi_{c0}(1P)$ component is divided by 100 to have the same units as the other entries.  
    }
    \renewcommand{\arraystretch}{1.1}
    \begin{tabular}
    {l@{\hspace{0.25cm}}
     @{\hspace{0.25cm}}r@{\hspace{0.25cm}}
     @{\hspace{0.25cm}}c} \hline
    Component & Branching fraction ($\times 10^{-6}$) & Upper limit \\
    & & at 90\% (95\%) CL\\ \hline

\decay{\Bp}{K^*(892)^0 \pip} & $10.54 \pm 0.46 \pm 0.44 \pm 0.26$ & ---\\
\decay{\Bp}{K^*(1410)^0 \pip} & $9.17 \pm 2.39 \pm 2.49 \pm 1.82$ & ---\\
\decay{\Bp}{K^*(1680)^0 \pip} & $0.49 \pm 0.21 \pm 0.23 \pm 0.03$ & $< 0.91$ $\phantom{0}(1.02)$\\
\decay{\Bp}{\rho(770)^0 \Kp} & $4.51 \pm 0.39 \pm 0.29 \pm 0.11$ & ---\\
\decay{\Bp}{\omega(782) \Kp} & $7.59 \pm 1.71 \pm 0.73 \pm 0.67$ & ---\\
\decay{\Bp}{\rho(1450)^0 \Kp} $\dagger$ & $0.70 \pm 0.60 \pm 0.40 \pm 0.02$ & ---\\
\decay{\Bp}{\rho(1700)^0 \Kp} $\dagger$ & $0.20 \pm 0.22 \pm 0.01 \pm 0.00$ & $< 0.51$ $\phantom{0}(0.59)$\\
\decay{\Bp}{K_2^*(1430)^0 \pip} & $3.90 \pm 0.57 \pm 0.31 \pm 0.13$ & ---\\
\decay{\Bp}{f_2(1270) \Kp} & $2.21 \pm 0.31 \pm 0.30 \pm 0.09$ & ---\\
\decay{\Bp}{f_2^\prime(1525) \Kp} & $3.90 \pm 3.84 \pm 0.63 \pm 0.51$ & $< 9.32$ $(10.69)$\\
\decay{\Bp}{K_3^*(1780)^0 \pip} & $1.02 \pm 0.27 \pm 0.91 \pm 0.06$ & $< 2.32$ $\phantom{0}(2.65)$\\
\decay{\Bp}{\rho_3(1690)^0 \Kp} & $1.00 \pm 0.26 \pm 0.37 \pm 0.06$ & ---\\
\decay{\Bp}{\chi_{c0}(1P) \Kp} $(\times 10^{-2})$ & $2.03 \pm 0.12 \pm 0.18 \pm 0.09$ & ---\\
\hline
\decay{\Bp}{K^*_0(1430)^0 \pip} & $7.35 \pm 0.50 \pm 1.09 \pm 0.81$ & ---\\
\decay{\Bp}{(K^+\pi^-)_{\rm SVP} \pip} $\dagger$ & $28.69 \pm 2.66 \pm 6.81 \pm 0.70$ & ---\\
\decay{\Bp}{K_0^*(1950)^0 \pip} & $1.32 \pm 0.27 \pm 0.08 \pm 0.36$ & ---\\\hline
\decay{\Bp}{f_0(500) \Kp} $\dagger$ & $1.06 \pm 0.22 \pm 0.03 \phantom{{}\pm 0.00}$ & ---\\
\decay{\Bp}{f_0(980) \Kp} $\dagger$ & $9.55 \pm 0.44 \pm 0.23 \phantom{{}\pm 0.00}$ & ---\\
\decay{\Bp}{f_0(1370) \Kp} $\dagger$ & $9.67 \pm 1.11 \pm 0.23 \phantom{{}\pm 0.00}$ & ---\\
\decay{\Bp}{f_0(1500) \Kp} & $4.65 \pm 0.81 \pm 0.32 \phantom{{}\pm 0.00}$ & ---\\
\decay{\Bp}{f_0(1710) \Kp} $\dagger$ & $0.41 \pm 0.34 \pm 0.01 \phantom{{}\pm 0.00}$ & ---\\
\decay{\Bp}{(\pi\pi\textrm{--}\pi\pi) \Kp} $\dagger$ & $3.09 \pm 1.14 \pm 0.08 \phantom{{}\pm 0.00}$ & ---\\
\decay{\Bp}{(\pi\pi\textrm{--}K \Kbar) \Kp} $\dagger$ & $0.13 \pm 0.39 \pm 0.00 \phantom{{}\pm 0.00}$ & $< 0.73$ $\phantom{0}(0.85)$\\

    \hline
    \end{tabular}
\end{table}

Whilst all S-wave approaches are considered to be equal in this analysis, the Isobar results are recommended for averaging purposes as every component can be associated to a physical state. 
Where applicable, the maximum deviation from the K-matrix and QMI approaches is assigned as an additional systematic uncertainty.
Fit fractions are first converted to branching fraction measurements as shown in Table~\ref{tab:int:bf}, which is achieved through multiplying by the inclusive absolute branching fraction, $\BF(\decay{\Bp}{\Kp \pip \pim}) = (5.76 \pm 0.14) \times 10^{-5}$~\cite{HFLAV21}. 
Intermediate branching fractions, assuming isospin symmetry where appropriate, are also taken into account wherever possible~\cite{PDG2020}, otherwise the product of branching fractions is reported. 
Thus, for an intermediate resonant decay \decay{\Bp}{R h}, the relation between its fit and branching fraction is given by
\begin{equation}
    \BF(\decay{\Bp}{R h})\BF(\decay{R}{h^{\prime} h^{(\prime)}}) = \mathcal{F}(\decay{\Bp}{R h})\BF(\decay{\Bp}{\Kp \pip \pim}) \,.
\end{equation}
Upper limits based on the Gaussian profiles of the fit fractions are reported for states not exceeding three standard deviations in significance. 
Similarly, the quasi-two-body \CP asymmetries are reported in Table~\ref{tab:int:acp}. 
Information on the total uncertainty correlations between all of these measurements can be found in the Supplemental Material~\cite{supplemental}.

Being based on a much larger data sample, with better separation of signal from background and improved modelling, these results represent a significant improvement in precision and in understanding of the \decay{\Bp}{\Kp\pip\pim} decay.
Due to the importance of using the GLASS lineshape as discussed in a companion article~\cite{BuToKpPipPimPRLStrong}, these results should not be averaged with those from previous amplitude analyses of the same decay process that relied on unitarity-preserving arguments ultimately originating with LASS data~\cite{Belle:2005rpz,BaBar:2008lpx}. 
As a case in point, the result
\begin{equation*}
    \BF(\decay{\Bp}{\chi_{c0}(1P) \Kp}) = (2.03 \pm 0.12 \textrm{ (Isobar)} \pm 0.18 \textrm{ (S-wave)} \pm 0.09 \textrm{ (External)}) \times 10^{-4}
\end{equation*}
is in slight tension with the world average \mbox{$\BF(\decay{\Bp}{\chi_{c0}(1P) \Kp}) =\left(1.51\,^{+0.15}_{-0.13}\right)\times 10^{-4}$}~\cite{PDG2024}, however the two lowest contributions to this average are from previous \decay{\Bp}{\Kp\pip\pim} amplitude analyses; the result of this analysis is in excellent agreement with, and more precise than, all other previous measurements of this quantity. 
The branching fraction and \CP asymmetry measurements for \decay{\Bp}{\omega(782) \Kp} are also in agreement with previous measurements~\cite{BaBar:2007nku,Belle:2013nby}, however they are not as precise as those with the $\decay{\omega(782)}{\pip\pim\piz}$ final state on account of the larger yields available. 
First upper limits are provided for decays involving the $\rho(1700)^0$, $K_3^*(1780)^0$ and $\pi\pi$--$K \Kbar$ intermediate states.

\begin{table}[tb]
    \centering
    \caption{\label{tab:int:acp}
        Quasi-two-body \CP asymmetries for averaging purposes, where the first uncertainty is the total obtained from the Isobar approach and a second uncertainty represents the maximum deviation from the K-matrix and QMI results.
    }
    \renewcommand{\arraystretch}{1.1}
    \begin{tabular}
    {l@{\hspace{0.25cm}}
     @{\hspace{0.25cm}}c} \hline
    Component & $\mathcal{A}_{\CP}$\\ \hline

\decay{\Bp}{K^*(892)^0 \pip} & $-0.038 \pm 0.021 \pm 0.010$\\
\decay{\Bp}{K^*(1410)^0 \pip} & $-0.116 \pm 0.192 \pm 0.563$\\
\decay{\Bp}{K^*(1680)^0 \pip} & $-0.412 \pm 0.749 \pm 0.960$\\
\decay{\Bp}{\rho(770)^0 \Kp} & $+0.280 \pm 0.036 \pm 0.080$\\
\decay{\Bp}{\omega(782) \Kp} & $+0.011 \pm 0.171 \pm 0.034$\\
\decay{\Bp}{\rho(1450)^0 \Kp} & $-0.005 \pm 0.486 \pm 0.268$\\
\decay{\Bp}{\rho(1700)^0 \Kp} & $-0.711 \pm 0.999 \pm 0.039$\\
\decay{\Bp}{K_2^*(1430)^0 \pip} & $-0.084 \pm 0.169 \pm 0.092$\\
\decay{\Bp}{f_2(1270) \Kp} & $-0.354 \pm 0.089 \pm 0.075$\\
\decay{\Bp}{f_2^\prime(1525) \Kp} & $-0.827 \pm 0.312 \pm 0.069$\\
\decay{\Bp}{K_3^*(1780)^0 \pip} & $-0.936 \pm 0.744 \pm 0.035$\\
\decay{\Bp}{\rho_3(1690)^0 \Kp} & $+0.603 \pm 0.151 \pm 0.126$\\
\decay{\Bp}{\chi_{c0}(1P) \Kp} & $+0.043 \pm 0.032 \pm 0.031$\\
\hline
\decay{\Bp}{K^*_0(1430)^0 \pip} & $-0.094 \pm 0.051 \pm 0.026$\\
\decay{\Bp}{(K^+\pi^-)_{\rm SVP}} & $-0.026 \pm 0.031 \pm 0.002$\\
\decay{\Bp}{K_0^*(1950)^0 \pip} & $+0.050 \pm 0.113 \pm 0.036$\\\hline
\decay{\Bp}{f_0(500) \Kp} & $-0.547 \pm 0.191 \phantom{{}\pm 0.000}$\\
\decay{\Bp}{f_0(980) \Kp} & $-0.054 \pm 0.023 \phantom{{}\pm 0.000}$\\
\decay{\Bp}{f_0(1370) \Kp} & $-0.075 \pm 0.093 \phantom{{}\pm 0.000}$\\
\decay{\Bp}{f_0(1500) \Kp} & $-0.046 \pm 0.108 \phantom{{}\pm 0.000}$\\
\decay{\Bp}{f_0(1710) \Kp} & $-0.474 \pm 0.674 \phantom{{}\pm 0.000}$\\
\decay{\Bp}{(\pi\pi\textrm{--}\pi\pi) \Kp} & $-0.079 \pm 0.084 \phantom{{}\pm 0.000}$\\
\decay{\Bp}{(\pi\pi\textrm{--}K \Kbar) \Kp} & $+0.371 \pm 0.464 \phantom{{}\pm 0.000}$\\

    \hline
    \end{tabular}
\end{table}

For the penguin-dominated \decay{\Bp}{K^*(892)^0 \pip} decay, the measured branching fraction and \CP asymmetry in decay are in excellent agreement with results from \babar\ and \belle~\cite{Belle:2005rpz,BaBar:2008lpx} and with theoretical expectations~\cite{Beneke:2003zv,Cheng:2009cn,Cheng:2016shb,Li:2006jv,Chai:2022ptk,Cheng:2014rfa,Wang:2008rk,Bell:2015koa}. 
These results and those for all other excited kaon states comprise particularly crucial input to model-independent tests for physics beyond the SM via isospin sum rules~\cite{Lipkin:1991st,Gronau:2005kz,Gronau:2010dd} adapted to excited intermediate states as they provide the purest constraint for the dominant, sensitive penguin loop amplitude, with the only other contribution coming from weak-annihilation diagrams. 
On the other hand, for the \decay{\Bp}{\rho(770)^0 \Kp} decay, the measured branching fraction is larger with respect to the previous measurements, and close to the central values predicted by various theories~\cite{Beneke:2003zv,Cheng:2009cn,Cheng:2016shb,Li:2006jv,Chai:2022ptk,Cheng:2014rfa,Wang:2008rk,Bell:2015koa}. 
Similarly, the branching fractions of both the \decay{\Bp}{K_2^*(1430)^0 \pip} and \decay{\Bp}{f_2(1270) \Kp} decays move towards their theoretical predictions~\cite{Cheng:2010yd,Li:2018lbd}, compared to previous measurements. 
In the scalar sector, the product branching fraction and \CP asymmetry of \decay{\Bp}{f_0(980) \Kp} is in agreement with previous results and with theory~\cite{Cheng:2013fba,Cheng:2016shb,Wang:2006ria}.

\section{Conclusions}
\label{sec:conclusions}

An amplitude analysis of the \decay{\Bp}{\Kp\pip\pim} decay is performed, based on $pp$ collision data at centre-of-mass energies $\sqrt{s}=7$ and $8\tev$ recorded with the LHCb detector, corresponding to an integrated luminosity of $3\,\text{fb}^{-1}$. 
The detailed description of the analysis procedure presented here complements shorter companion articles focussed on the \CP-conserving implications prompted by an improved understanding of strong interaction dynamics~\cite{BuToKpPipPimPRLStrong}, and the consequent first observations of different sources of \CP-violation effects in this channel~\cite{BuToKpPipPimPRLCP}. 
The results showcase the effectiveness of three distinct yet complementary approaches to describing the large S-wave contribution to this decay: Isobar, K-matrix, and Quasi-Model-Independent. 
Overall, good agreement is found between all three models and the data, indicating significant progress in understanding the intricate dynamics governing charmless hadronic three-body decays.
One important improvement is the modelling of the $K\pi$ S-wave, in the Isobar and K-matrix approaches, with the GLASS lineshape rather than the unitarity-preserving LASS function used in previous amplitude analyses of the same process.
As a consequence, the branching fractions and quasi-two-body \CP asymmetries associated with each intermediate state reported from this analysis should not be averaged with previous measurements. 

Both direct and transition-mediated contributions from the $\omega(782)$ resonance are considered, with results presented using alternative models. 
The data do not provide significant discriminating power between these two cases.
Despite the similarities between the two models, their distinctions are clarified, particularly in the phase shifts seen in the $\rho^0$--$\omega$ coupling. 
As for the results themselves, the precision on the \decay{\Bp}{\chi_{c0}(1P) \Kp} branching fraction and \CP asymmetry is greatly improved, offering the prospect for a tighter constraint on contaminating penguin amplitudes in measurements of the \Bz--\Bzb mixing phase from \decay{\Bz}{\chi_{c0}(1P) \Kz} decays. Similar improvements are noted in most other contributions, each of which comprises fundamental input to model-independent isospin sum-rule tests of new physics in loop-mediated decays. Finally, the first upper-limit measurements are made with regards to contributions from the $\rho(1700)^0$ and $K_3^*(1780)^0$ resonances, and the $\pi\pi$--$K \Kbar$ rescattering contribution.

On the technical front, it is interesting to note that in this analysis, the isobar approach with a sophisticated description of the S-wave contributions, yields a similar, if not better, fit quality while having fewer parameters than the K-matrix approach. 
This is also the first attempt with particle decay at extracting multiple partial waves in a quasi-model-independent manner. Systematic biases induced by fixing the contents of one bin to the Isobar prediction, and arising only in the presentation of individual waves, are found to be under control.

These results establish a solid foundation for future analyses and provide valuable input to phenomenological work on the strong interaction mechanisms underscoring the remarkably large \CP violation observed in charmless hadronic three-body decays of the charged \B meson, and on \B-meson decays in general.

%% file: LHCb/acknowledgements.tex
\section*{Acknowledgements}
%
%
\noindent We express our gratitude to our colleagues in the CERN
accelerator departments for the excellent performance of the LHC. We
thank the technical and administrative staff at the LHCb
institutes.
We acknowledge support from CERN and from the national agencies:
ARC (Australia);
CAPES, CNPq, FAPERJ and FINEP (Brazil); 
MOST and NSFC (China); 
CNRS/IN2P3 and CEA (France);  
BMFTR, DFG and MPG (Germany);
NKFIH (Hungary);              
INFN (Italy); 
NWO (Netherlands); 
MNiSW and NCN (Poland); 
MEC/IFA (Romania); 
MICIU and AEI (Spain);
SNSF and SER (Switzerland); 
NASU (Ukraine); 
STFC (United Kingdom); 
DOE NP and NSF (USA).
We acknowledge the computing resources that are provided by ARDC (Australia), 
CBPF (Brazil),
CERN, 
IHEP and LZU (China),
IN2P3 (France), 
KIT and DESY (Germany), 
INFN (Italy), 
SURF (Netherlands),
Polish WLCG (Poland),
IFIN-HH (Romania), 
PIC (Spain), CSCS (Switzerland), 
GridPP (United Kingdom),
and NSF (USA).  
We are indebted to the communities behind the multiple open-source
software packages on which we depend.
Individual groups or members have received support from
RTP (Australia), 
FWO Odysseus grant G0ASD25N (Belgium), 
Key Research Program of Frontier Sciences of CAS, CAS PIFI, CAS CCEPP (China); 
Minciencias (Colombia);
EPLANET, Marie Sk\l{}odowska-Curie Actions, ERC and NextGenerationEU (European Union);
A*MIDEX, ANR, IPhU and Labex P2IO, and R\'{e}gion Auvergne-Rh\^{o}ne-Alpes (France);
Alexander-von-Humboldt Foundation (Germany);
ICSC (Italy); 
Severo Ochoa and Mar\'ia de Maeztu Units of Excellence, GVA, XuntaGal, GENCAT, InTalent-Inditex and Prog.~Atracci\'on Talento CM (Spain);
the Leverhulme Trust, the Royal Society and UKRI (United Kingdom).

%% file: appendix.tex
\section*{Appendices}

\appendix

\addcontentsline{toc}{section}{Appendices}

\section{\texorpdfstring{\boldmath \Bp-candidate mass fit}{B+ candidate mass fit}}
\label{app:bmassfit}

\begin{figure}[!b]
    \centering
    \includegraphics[width=0.5\linewidth]{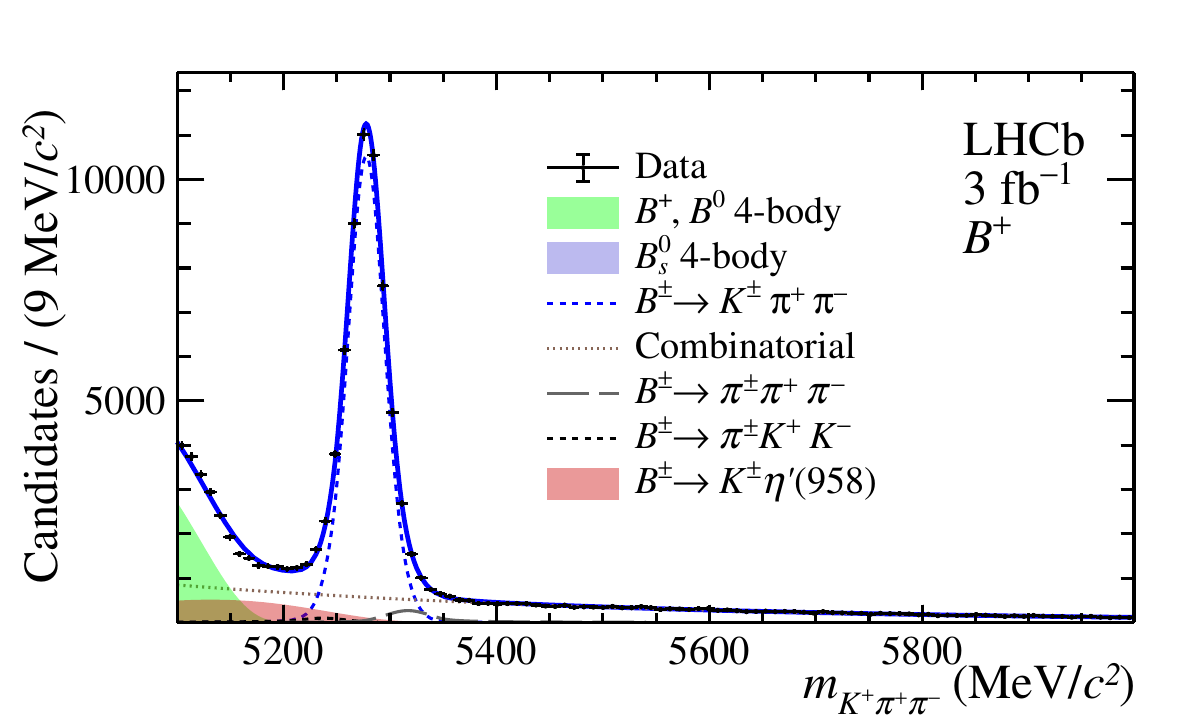}%
    \includegraphics[width=0.5\linewidth]{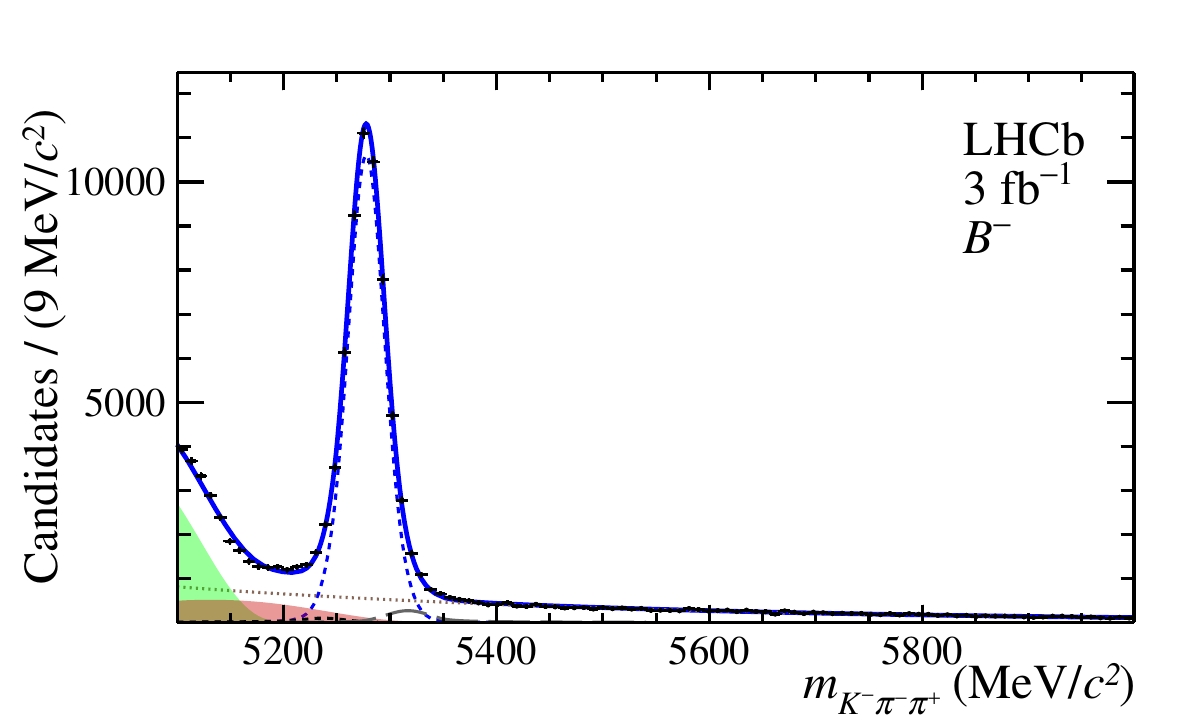}

    \caption{
        Mass distributions for (left)~$B^+$ and (right)~$B^-$ candidates reconstructed in the $K^+ \pi^+ \pi^-$ and $K^- \pi^- \pi^+$ final state, respectively, for the combined 2011 and 2012 data taking samples, along with the results of the fit.
    }
    \label{fig:massFitlinear}
\end{figure}

The mass distributions of the selected \B candidates, together with the results of the fits to these distributions, are shown in Fig.~\ref{fig:massFitlinear}.
These complement the corresponding plots with a logarithmic $y$-axis scale shown in Fig.~\ref{fig:massFit}. 

\clearpage

\section{Background model}
\label{app:background}

\begin{figure}[b]
    \centering
    \includegraphics[width=0.5\linewidth]{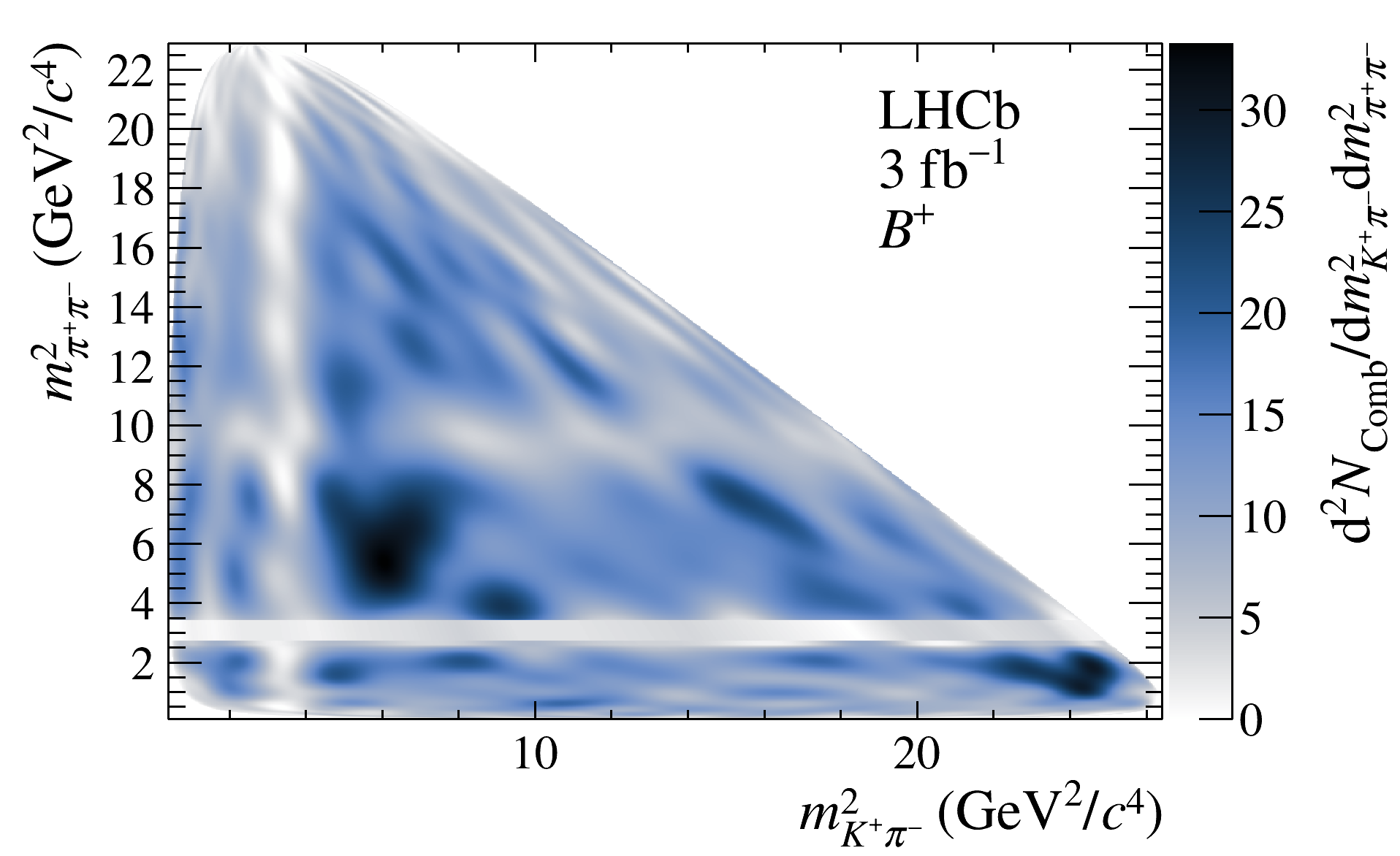}%
    \includegraphics[width=0.5\linewidth]{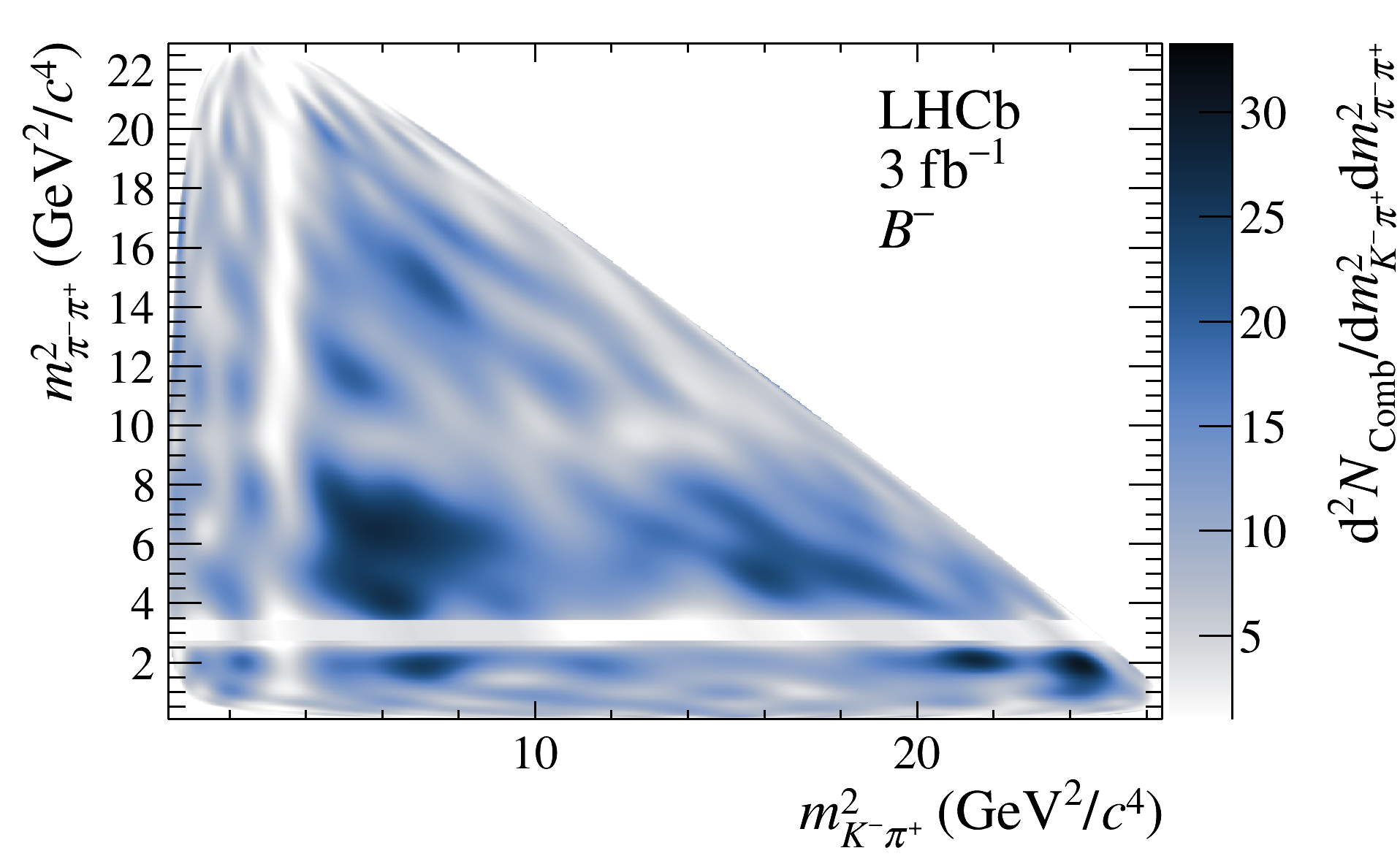}

    \caption{
        Dalitz-plot distributions for the (left)~$B^+$ and (right)~$B^-$ combinatorial background models, scaled to represent their respective yields in the signal region.
        }
    \label{fig:app:combinatorialBkg}
\end{figure}

\begin{figure}[!tb]
    \centering
    \includegraphics[width=0.5\linewidth]{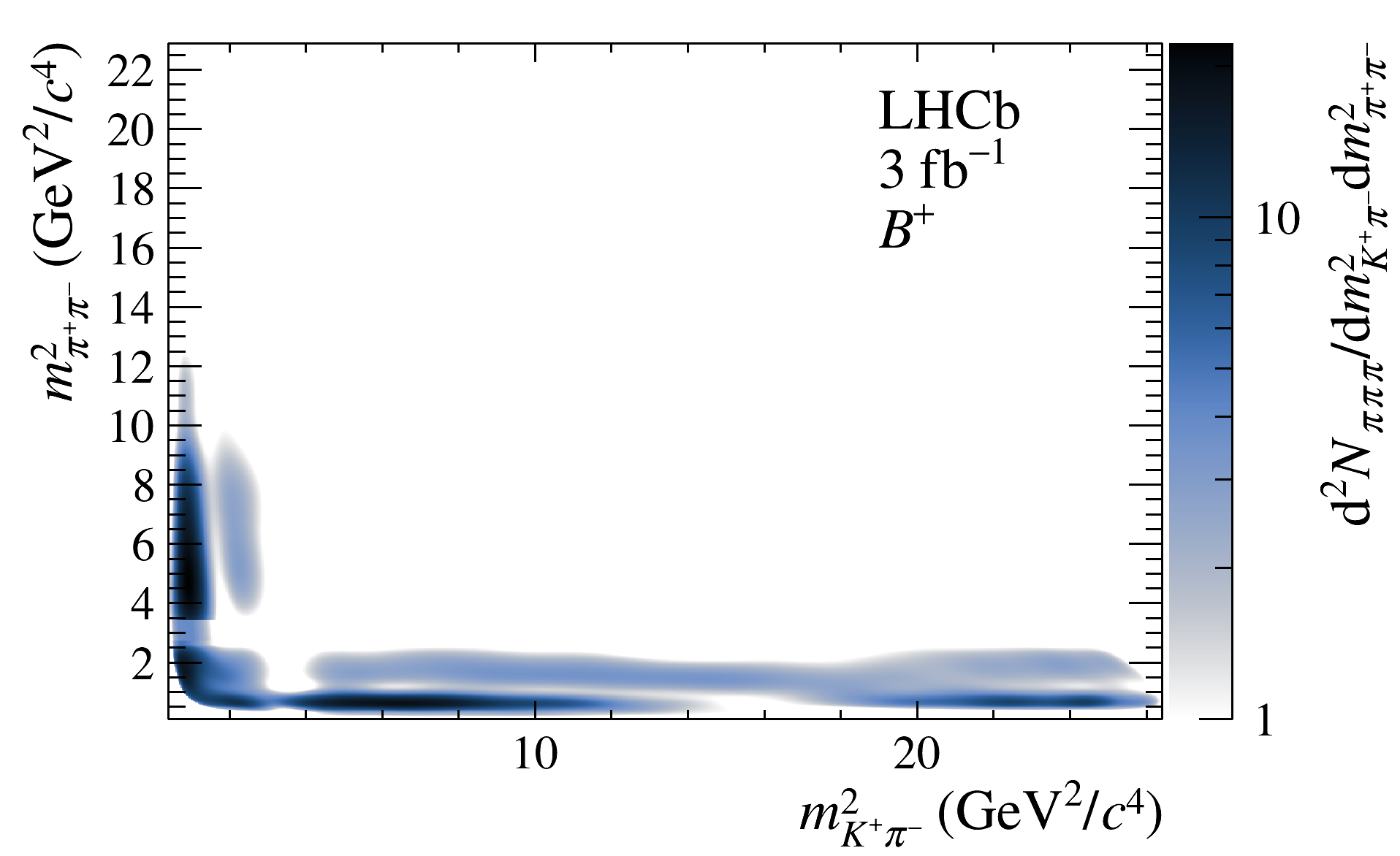}%
    \includegraphics[width=0.5\linewidth]{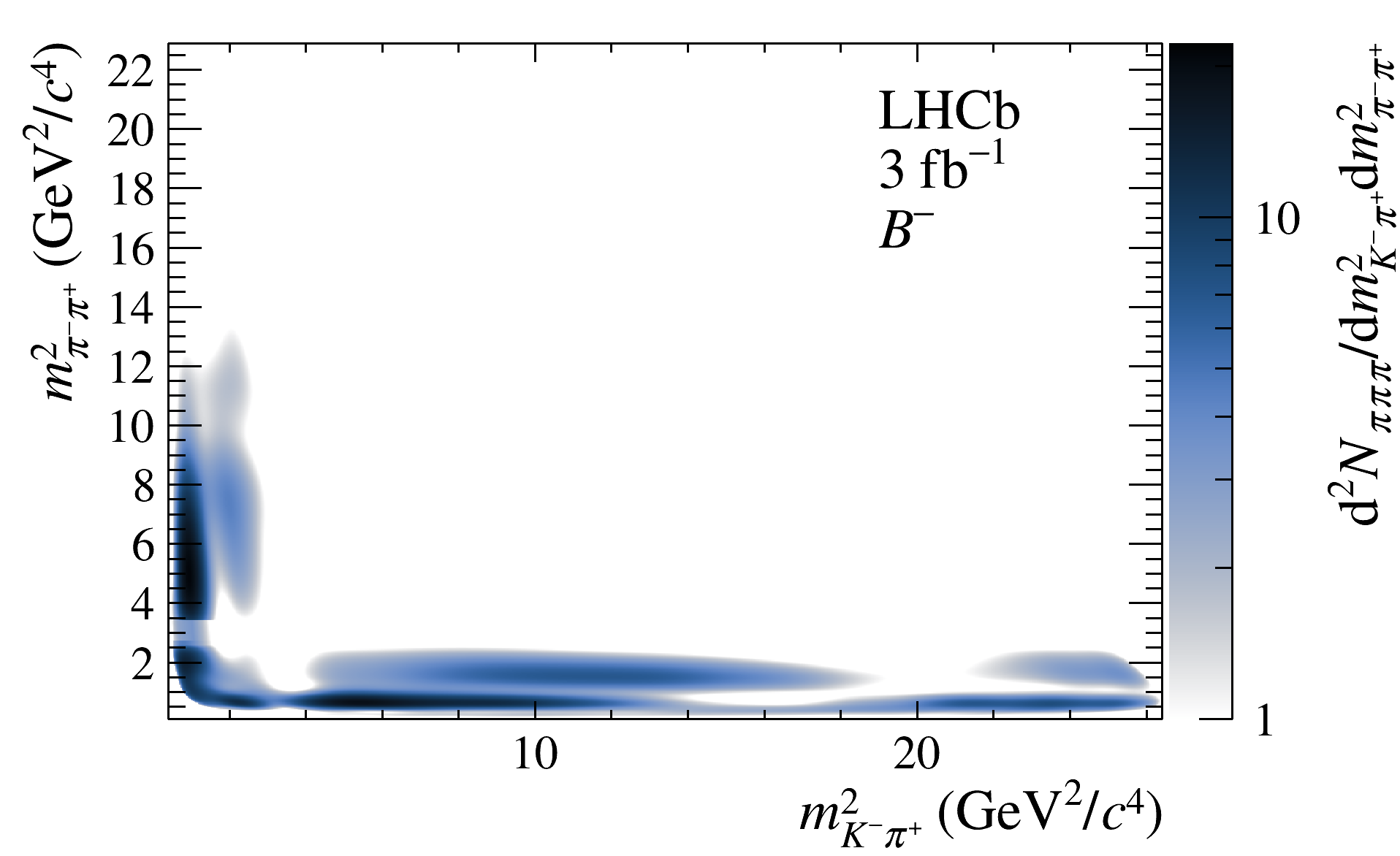}

    \caption{
        Dalitz-plot distributions for the misidentified $B^+ \!\to \pi^+ \pi^+ \pi^-$ background models in (left)~$B^+$ and (right)~$B^-$ samples, scaled to represent their respective yields in the signal region.
        }
    \label{fig:app:crossFeedBkgB2pipipi}
\end{figure}

\begin{figure}[!tb]
    \centering
    \includegraphics[width=0.5\linewidth]{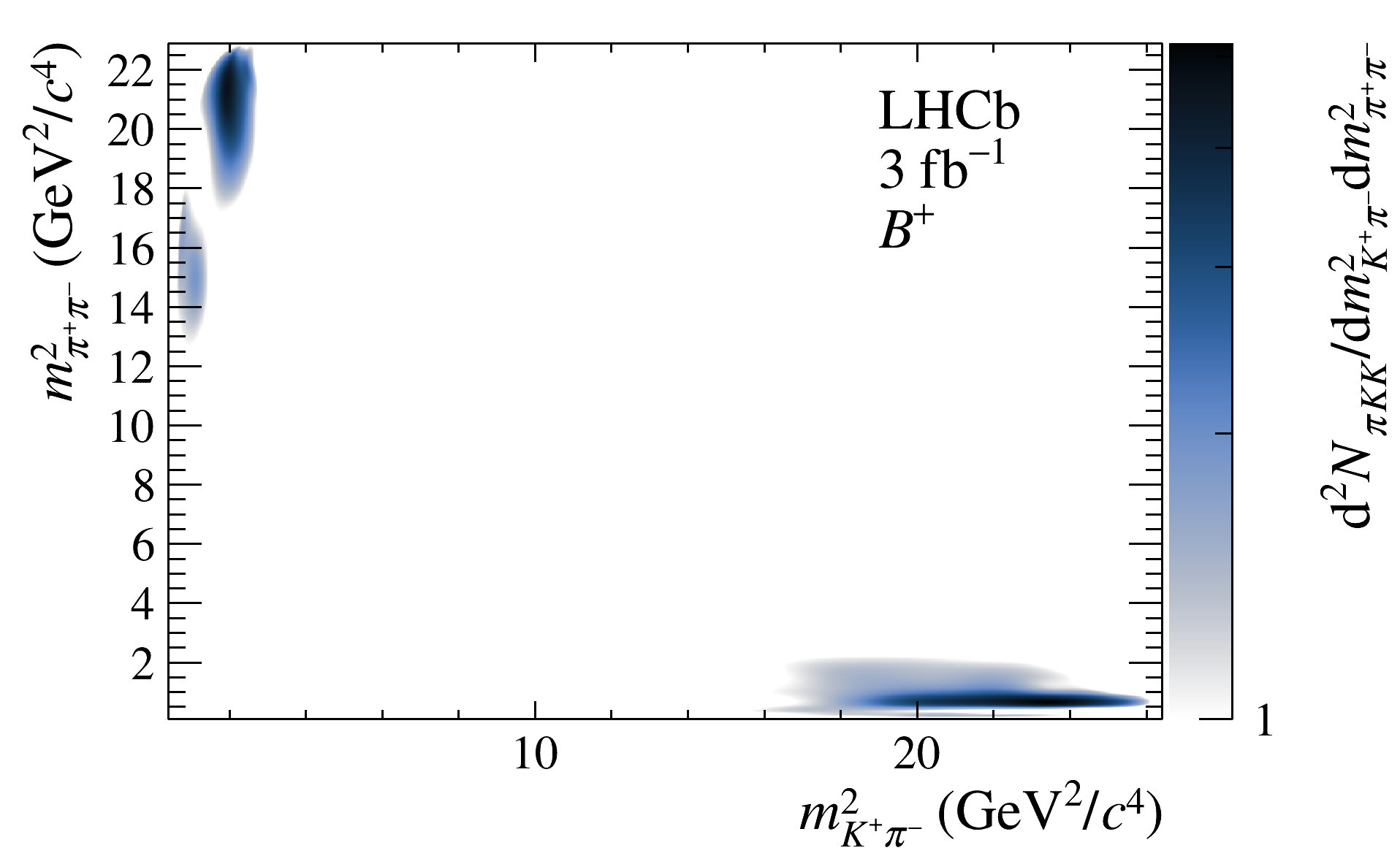}%
    \includegraphics[width=0.5\linewidth]{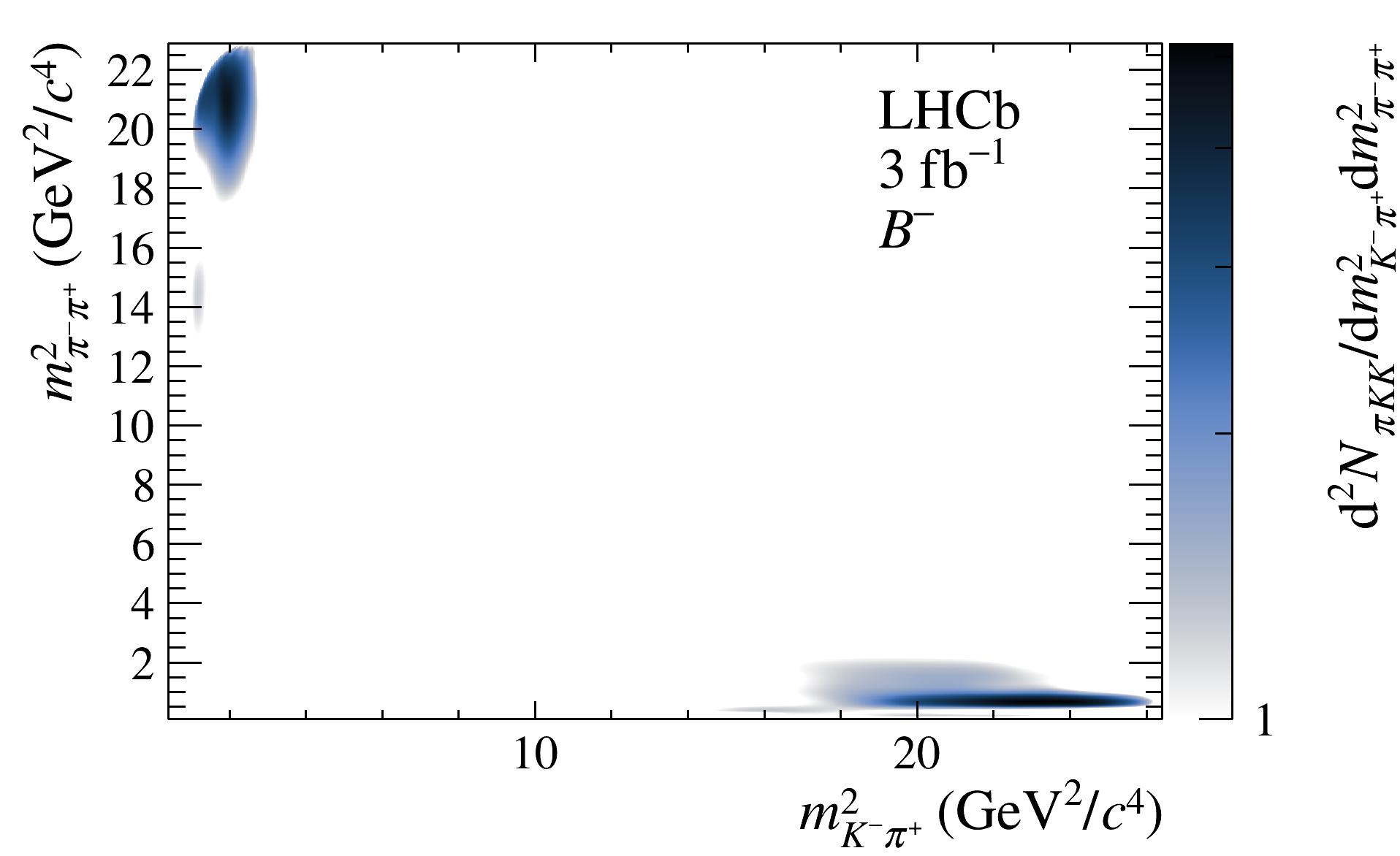}

    \caption{
        Dalitz-plot distributions for the misidentified $B^+ \!\to \pi^+ K^+ K^-$ background models in (left)~$B^+$ and (right)~$B^-$ samples, scaled to represent their respective yields in the signal region.
        }
    \label{fig:app:crossFeedBkgB2pikk}
\end{figure}

\begin{figure}[!tb]
    \centering
    \includegraphics[width=0.5\linewidth]{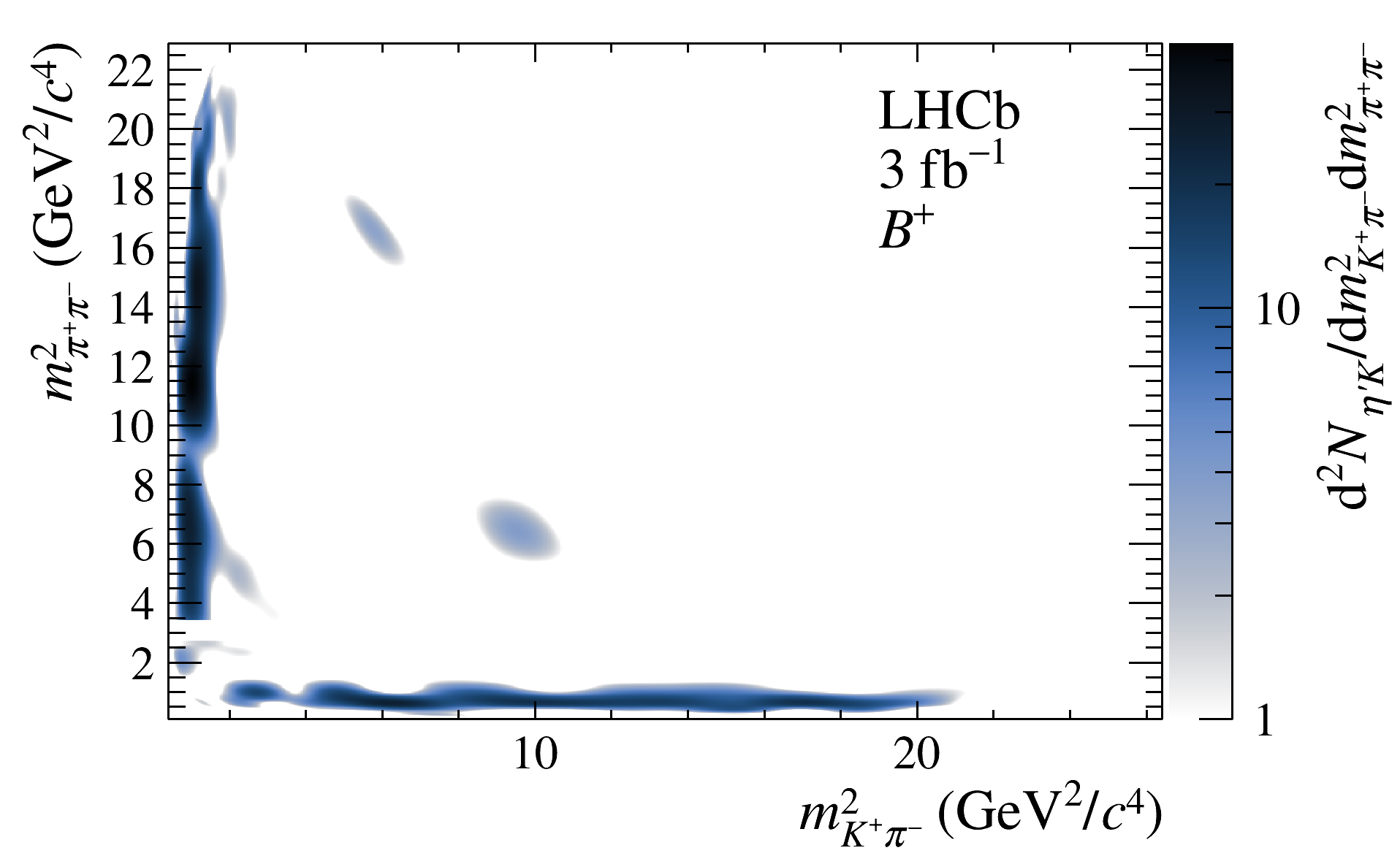}%
    \includegraphics[width=0.5\linewidth]{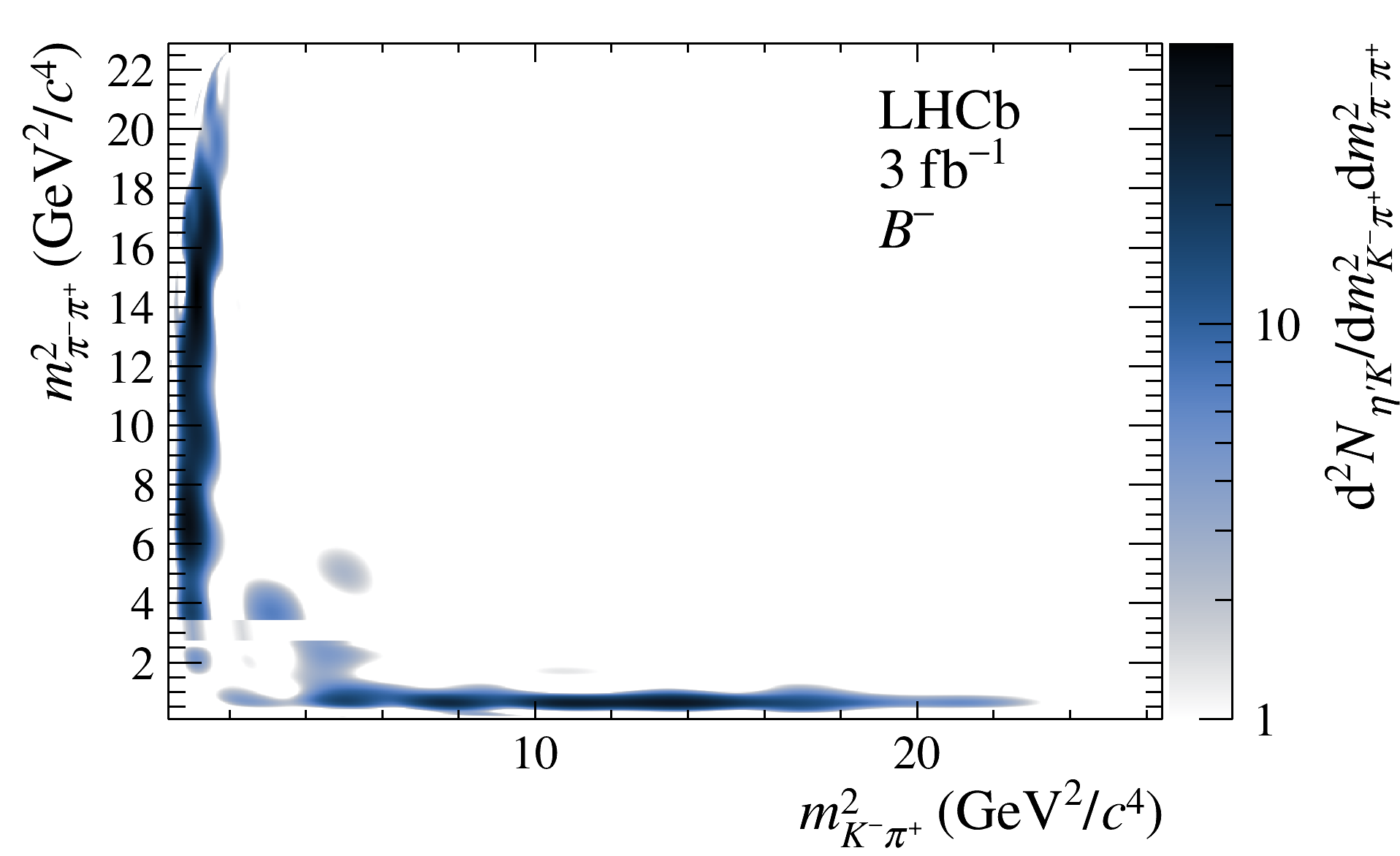}

    \caption{
        Dalitz-plot distributions for the partially reconstructed $B^+ \!\to K^+ \eta^{\prime}(\pi^+ \pi^- \gamma)$ background models in (left)~$B^+$ and (right)~$B^-$, scaled to represent their respective yields in the signal region.
    }
    \label{fig:app:crossFeedBkgB2etapk}
\end{figure}

The background model transformed from the square Dalitz plot to the traditional Dalitz plot are shown. The combinatorial, \decay{\Bp}{\pip\pip\pim}, \decay{\Bp}{\pip\Kp\Km} and \decay{\Bp}{\Kp\etapr(\pip\pim\gamma)} background models corresponding to Figs.~\ref{fig:combinatorialBkg}--\ref{fig:crossFeedBkgB2etapk}, respectively, are shown in Figs.~\ref{fig:app:combinatorialBkg}--\ref{fig:app:crossFeedBkgB2etapk}.

\clearpage

\section{Systematic uncertainties}
\label{app:systematic}

The systematic uncertainties for the \CP-averaged fit fractions, broken down by category, are summarised in Tables~\ref{tab:app:syst:iso:ff}--\ref{tab:app:syst:qmi:ff} for the Isobar, K-matrix and QMI approaches, respectively, while those for the quasi-two-body \CP asymmetries are recorded in Tables~\ref{tab:app:syst:iso:acp}--\ref{tab:app:syst:qmi:acp}. 
For the relative phases of \Bp~(\Bm) decays, the systematic uncertainties are similarly presented in  Tables~\ref{tab:app:syst:iso:phasep}--\ref{tab:app:syst:qmi:phasep}~(\ref{tab:app:syst:iso:phasem}--\ref{tab:app:syst:qmi:phasem}). 
The shorthand labels representing each category of systematic uncertainty are defined throughout Section~\ref{sec:systematics}.

\begin{table}[!htb]
    \centering
    \caption{\label{tab:app:syst:iso:ff}
        Systematic uncertainties on the fit fractions for the Isobar approach given in percent.
    }
    \renewcommand{\arraystretch}{1.1}
    \begin{tabular}
    {l@{\hspace{0.125cm}}
     @{\hspace{0.125cm}}r@{\hspace{0.125cm}}
     @{\hspace{0.125cm}}r@{\hspace{0.125cm}}
     @{\hspace{0.125cm}}r@{\hspace{0.125cm}}
     @{\hspace{0.125cm}}r@{\hspace{0.125cm}}
     @{\hspace{0.125cm}}r@{\hspace{0.125cm}}
     @{\hspace{0.125cm}}r@{\hspace{0.125cm}}
     @{\hspace{0.125cm}}r@{\hspace{0.125cm}}
     @{\hspace{0.125cm}}r}\hline
     & $B$ mass & Eff & Resol & Bkg & Fit bias & Barrier & Lineshape & Extra \\\hline

$K^*(892)^0$ & 0.003 & 0.105 & 0.007 & 0.455 & 0.162 & 0.163 & 0.156 & 0.009 \\
$K^*(1410)^0$ & 0.016 & 0.060 & 0.005 & 0.083 & 0.106 & 0.040 & 0.047 & 0.025 \\
$K^*(1680)^0$ & 0.017 & 0.022 & 0.001 & 0.059 & 0.022 & 0.064 & 0.021 & 0.020 \\
$\rho(770)^0$ & 0.172 & 0.034 & 0.004 & 0.528 & 0.160 & 0.246 & 0.106 & 0.282 \\
$\omega(782)$ & 0.006 & 0.003 & 0.001 & 0.022 & 0.023 & 0.013 & 0.004 & 0.008 \\
$\rho(1450)^0$ & 0.024 & 0.099 & 0.037 & 0.049 & 0.032 & 0.268 & 0.152 & 1.005 \\
$\rho(1700)^0$ & 0.009 & 0.041 & 0.008 & 0.018 & 0.061 & 0.076 & 0.028 & 0.371 \\
$K_2^*(1430)^0$ & $ < 0.001$ & 0.168 & 0.043 & 0.192 & 0.006 & 0.060 & 0.153 & 0.036 \\
$f_2(1270)$ & 0.065 & 0.106 & 0.014 & 0.185 & 0.094 & 0.041 & 0.044 & 0.129 \\
$f_2^\prime(1525)$ & 0.004 & 0.017 & 0.002 & 0.046 & 0.006 & 0.005 & 0.004 & 0.014 \\
$K_3^*(1780)^0$ & $ < 0.001$ & 0.030 & 0.006 & 0.014 & 0.013 & 0.015 & 0.023 & 0.007 \\
$\rho_3(1690)^0$ & 0.017 & 0.018 & 0.011 & 0.037 & 0.032 & 0.022 & 0.038 & 0.066 \\
$\chi_{c0}(1P)$ & 0.005 & 0.053 & 0.005 & 0.061 & 0.045 & 0.005 & 0.028 & 0.004 \\
$(K^+\pi^-)_{\rm Res}$ & 0.107 & 0.383 & 0.050 & 1.030 & 0.672 & 0.254 & 0.272 & 0.919 \\
$(K^+\pi^-)_{\rm ER}$ & 0.098 & 0.403 & 0.105 & 0.395 & 0.318 & 0.479 & 0.965 & 0.364 \\
$K_0^*(1950)^0$ & 0.019 & 0.043 & 0.013 & 0.057 & 0.010 & 0.012 & 0.107 & 0.026 \\
$f_0(500)$ & 0.003 & 0.104 & 0.044 & 0.162 & 0.053 & 0.122 & 0.103 & 0.223 \\
$f_0(980)$ & 0.032 & 0.180 & 0.045 & 0.404 & 0.138 & 0.151 & 0.069 & 0.424 \\
$f_0(1370)$ & 0.074 & 0.403 & 0.045 & 0.342 & 0.303 & 0.480 & 1.524 & 0.363 \\
$f_0(1500)$ & 0.004 & 0.023 & 0.012 & 0.104 & 0.023 & 0.034 & 0.133 & 0.228 \\
$f_0(1710)$ & 0.009 & 0.035 & $ < 0.001$ & 0.033 & 0.017 & 0.019 & 0.078 & 0.580 \\
$\pi\pi$--$\pi\pi$ & 0.193 & 0.285 & 0.021 & 0.908 & 0.320 & 0.307 & 0.847 & 1.198 \\
$\pi\pi$--$K \Kbar$ & 0.028 & 0.034 & 0.010 & 0.112 & 0.008 & 0.023 & 0.656 & 0.014 \\

    \hline
    \end{tabular}
\end{table}

\begin{table}[!htb]
    \centering
    \caption{\label{tab:app:syst:km:ff}
        Systematic uncertainties on the fit fractions for the K-matrix approach given in percent.
    }
    \renewcommand{\arraystretch}{1.1}
    \begin{tabular}
    {l@{\hspace{0.125cm}}
     @{\hspace{0.125cm}}r@{\hspace{0.125cm}}
     @{\hspace{0.125cm}}r@{\hspace{0.125cm}}
     @{\hspace{0.125cm}}r@{\hspace{0.125cm}}
     @{\hspace{0.125cm}}r@{\hspace{0.125cm}}
     @{\hspace{0.125cm}}r@{\hspace{0.125cm}}
     @{\hspace{0.125cm}}r@{\hspace{0.125cm}}
     @{\hspace{0.125cm}}r@{\hspace{0.125cm}}
     @{\hspace{0.125cm}}r@{\hspace{0.125cm}}
     }\hline
     & $B$ mass & Eff & Resol & Bkg & Fit bias & Barrier & Lineshape & Extra \\\hline

$K^*(892)^0$ & 0.062 & 0.094 & 0.070 & 0.354 & 0.121 & 0.588 & 0.046 & 0.012 \\
$K^*(1410)^0$ & 0.010 & 0.069 & 0.012 & 0.018 & 0.153 & 0.033 & 0.036 & 0.010 \\
$K^*(1680)^0$ & 0.008 & 0.041 & 0.009 & 0.059 & 0.024 & 0.092 & 0.024 & 0.002 \\
$\rho(770)^0$ & 0.036 & 0.096 & 0.040 & 0.496 & 0.192 & 0.357 & 0.058 & 0.082 \\
$\omega(782)$ & $ < 0.001$ & 0.003 & $ < 0.001$ & 0.025 & 0.028 & 0.010 & 0.002 & 0.002 \\
$\rho(1450)^0$ & 0.033 & 0.075 & 0.032 & 0.036 & 0.006 & 0.254 & 0.055 & 0.059 \\
$\rho(1700)^0$ & 0.005 & 0.044 & 0.003 & 0.040 & 0.036 & 0.103 & 0.010 & 0.034 \\
$K_2^*(1430)^0$ & 0.050 & 0.235 & 0.047 & 0.098 & 0.031 & 0.026 & 0.077 & 0.030 \\
$f_2(1270)$ & $ < 0.001$ & 0.093 & $ < 0.001$ & 0.184 & 0.086 & 0.131 & 0.021 & 0.085 \\
$f_2^\prime(1525)$ & 0.002 & 0.009 & 0.002 & 0.028 & 0.005 & 0.003 & $ < 0.001$ & 0.011 \\
$K_3^*(1780)^0$ & 0.010 & 0.059 & 0.009 & 0.037 & 0.010 & 0.132 & 0.029 & 0.005 \\
$\rho_3(1690)^0$ & 0.013 & 0.020 & 0.013 & 0.034 & 0.026 & 0.035 & 0.009 & 0.028 \\
$\chi_{c0}(1P)$ & 0.003 & 0.052 & 0.003 & 0.011 & 0.058 & 0.065 & 0.005 & 0.005 \\
$(K^+\pi^-)_{\rm Res}$ & 0.051 & 0.301 & 0.042 & 0.304 & 0.341 & 0.079 & 0.115 & 0.093 \\
$(K^+\pi^-)_{\rm ER}$ & 0.357 & 0.700 & 0.422 & 0.741 & 0.324 & 0.382 & 0.253 & 0.176 \\
$K_0^*(1950)^0$ & 0.011 & 0.052 & 0.011 & 0.019 & 0.013 & 0.304 & 0.093 & 0.008 \\
$\pi^+\pi^-$ S-wave & 0.160 & 0.237 & 0.179 & 0.909 & 0.312 & 1.007 & 0.120 & 0.031 \\

    \hline
    \end{tabular}
\end{table}

\begin{table}[!htb]
    \centering
    \caption{\label{tab:app:syst:qmi:ff}
        Systematic uncertainties on the fit fractions for the QMI approach given in percent.
    }
    \renewcommand{\arraystretch}{1.1}
    \begin{tabular}
    {l@{\hspace{0.125cm}}
     @{\hspace{0.125cm}}r@{\hspace{0.125cm}}
     @{\hspace{0.125cm}}r@{\hspace{0.125cm}}
     @{\hspace{0.125cm}}r@{\hspace{0.125cm}}
     @{\hspace{0.125cm}}r@{\hspace{0.125cm}}
     @{\hspace{0.125cm}}r@{\hspace{0.125cm}}
     @{\hspace{0.125cm}}r@{\hspace{0.125cm}}
     @{\hspace{0.125cm}}r@{\hspace{0.125cm}}
     @{\hspace{0.125cm}}r}\hline
     & $B$ mass & Eff & Resol & Bkg & QMI bias & Barrier & Lineshape & Extra \\ \hline

$K^*(892)^0$ & 0.007 & 0.100 & 0.001 & 0.045 & 0.283 & 0.284 & 0.021 & 0.022 \\
$K^*(1410)^0$ & 0.001 & 0.063 & $<0.001$ & 0.009 & 0.337 & 0.020 & 0.034 & 0.030 \\
$K^*(1680)^0$ & 0.001 & 0.048 & $<0.001$ & 0.022 & 0.098 & 0.002 & 0.013 & 0.013 \\
$\rho(770)^0$ & 0.013 & 0.090 & 0.004 & 0.215 & 0.490 & 0.151 & 0.034 & 0.028 \\
$\omega(782)$ & $<0.001$ & 0.003 & 0.002 & 0.011 & 0.024 & 0.004 & $<0.001$ & 0.002 \\
$\rho(1450)^0$ & 0.041 & 0.136 & 0.025 & 0.089 & 0.427 & 0.003 & 0.090 & 0.049 \\
$\rho(1700)^0$ & 0.014 & 0.044 & 0.006 & 0.010 & 0.218 & 0.009 & 0.003 & 0.002 \\
$K_2^*(1430)^0$ & 0.008 & 0.115 & 0.007 & 0.046 & 0.040 & 0.225 & 0.031 & 0.030 \\
$f_2(1270)$ & 0.020 & 0.095 & 0.003 & 0.312 & 0.080 & 0.223 & 0.037 & 0.021 \\
$f_2^\prime(1525)$ & $<0.001$ & 0.012 & $<0.001$ & 0.014 & 0.015 & 0.004 & 0.001 & 0.018 \\
$K_3^*(1780)^0$ & 0.003 & 0.010 & $<0.001$ & 0.144 & 0.006 & 0.150 & 0.009 & 0.051 \\
$\rho_3(1690)^0$ & 0.001 & 0.018 & $<0.001$ & 0.021 & 0.027 & 0.084 & 0.001 & 0.013 \\
$\chi_{c0}(1P)$ & $<0.001$ & 0.053 & 0.007 & 0.136 & 0.007 & 0.008 & 0.001 & 0.011 \\
S-wave & 0.074 & 0.134 & 0.029 & 0.233 & 0.497 & 0.232 & 0.164 & 0.066 \\

    \hline
    \end{tabular}
\end{table}

\begin{table}[!htb]
    \centering
    \caption{\label{tab:app:syst:iso:acp}
        Systematic uncertainties on the $\mathcal{A}_{\CP}$ values for the Isobar approach.
    }
    \renewcommand{\arraystretch}{1.1}
    \begin{tabular}
    {l@{\hspace{0.125cm}}
     @{\hspace{0.125cm}}r@{\hspace{0.125cm}}
     @{\hspace{0.125cm}}r@{\hspace{0.125cm}}
     @{\hspace{0.125cm}}r@{\hspace{0.125cm}}
     @{\hspace{0.125cm}}r@{\hspace{0.125cm}}
     @{\hspace{0.125cm}}r@{\hspace{0.125cm}}
     @{\hspace{0.125cm}}r@{\hspace{0.125cm}}
     @{\hspace{0.125cm}}r@{\hspace{0.125cm}}
     @{\hspace{0.125cm}}r@{\hspace{0.125cm}}
     }\hline
     & $B$ mass & Eff & Resol & Bkg & Fit bias & Barrier & Lineshape & Extra \\\hline

$K^*(892)^0$ & $ < 0.001$ & 0.015 & $ < 0.001$ & 0.005 & 0.009 & 0.001 & 0.002 & 0.001 \\
$K^*(1410)^0$ & 0.027 & 0.057 & 0.021 & 0.089 & 0.020 & 0.036 & 0.101 & 0.049 \\
$K^*(1680)^0$ & 0.024 & 0.085 & 0.043 & 0.673 & 0.007 & 0.022 & 0.214 & 0.058 \\
$\rho(770)^0$ & $ < 0.001$ & 0.008 & $ < 0.001$ & 0.006 & 0.013 & 0.005 & 0.017 & 0.016 \\
$\omega(782)$ & 0.003 & 0.014 & 0.001 & 0.007 & 0.045 & 0.002 & 0.012 & 0.042 \\
$\rho(1450)^0$ & 0.053 & 0.034 & 0.004 & 0.114 & 0.003 & 0.011 & 0.097 & 0.422 \\
$\rho(1700)^0$ & 0.034 & 0.044 & $ < 0.001$ & 0.071 & 0.067 & 0.061 & 0.077 & 0.972 \\
$K_2^*(1430)^0$ & 0.022 & 0.059 & 0.026 & 0.126 & 0.011 & 0.018 & 0.070 & 0.019 \\
$f_2(1270)$ & 0.005 & 0.019 & 0.006 & 0.029 & 0.005 & 0.015 & 0.003 & 0.066 \\
$f_2^\prime(1525)$ & 0.028 & 0.059 & 0.004 & 0.032 & 0.193 & 0.017 & 0.014 & 0.156 \\
$K_3^*(1780)^0$ & 0.082 & 0.221 & 0.100 & 0.583 & 0.050 & 0.010 & 0.371 & 0.028 \\
$\rho_3(1690)^0$ & 0.005 & 0.019 & $ < 0.001$ & 0.100 & 0.017 & 0.042 & 0.008 & 0.063 \\
$\chi_{c0}(1P)$ & 0.001 & 0.004 & 0.002 & 0.001 & 0.013 & 0.006 & 0.003 & 0.004 \\
$(K^+\pi^-)_{\rm Res}$ & 0.009 & 0.021 & 0.009 & 0.027 & 0.009 & 0.003 & 0.026 & 0.001 \\
$(K^+\pi^-)_{\rm ER}$ & 0.002 & 0.004 & 0.001 & 0.011 & 0.008 & 0.002 & 0.015 & 0.024 \\
$K_0^*(1950)^0$ & 0.007 & 0.028 & 0.006 & 0.026 & 0.013 & 0.010 & 0.037 & 0.019 \\
$f_0(500)$ & 0.008 & 0.031 & 0.026 & 0.037 & 0.018 & 0.067 & 0.136 & 0.079 \\
$f_0(980)$ & 0.002 & 0.009 & 0.002 & 0.004 & 0.005 & 0.002 & 0.007 & 0.008 \\
$f_0(1370)$ & 0.017 & 0.031 & 0.012 & 0.025 & 0.005 & 0.002 & 0.037 & 0.054 \\
$f_0(1500)$ & 0.006 & 0.019 & 0.002 & 0.004 & 0.023 & 0.004 & 0.035 & 0.069 \\
$f_0(1710)$ & 0.008 & 0.036 & $ < 0.001$ & 0.050 & 0.013 & 0.028 & 0.029 & 0.662 \\
$\pi\pi$--$\pi\pi$ & 0.001 & 0.018 & 0.004 & 0.034 & 0.005 & 0.028 & 0.050 & 0.026 \\
$\pi\pi$--$K \Kbar$ & 0.057 & 0.170 & 0.046 & 0.137 & 0.206 & 0.059 & 0.198 & 0.078 \\

    \hline
    \end{tabular}
\end{table}

\begin{table}[!htb]
    \centering
    \caption{\label{tab:app:syst:km:acp}
        Systematic uncertainties on the $\mathcal{A}_{\CP}$ values for the K-matrix approach.
    }
    \renewcommand{\arraystretch}{1.1}
    \begin{tabular}
    {l@{\hspace{0.125cm}}
     @{\hspace{0.125cm}}r@{\hspace{0.125cm}}
     @{\hspace{0.125cm}}r@{\hspace{0.125cm}}
     @{\hspace{0.125cm}}r@{\hspace{0.125cm}}
     @{\hspace{0.125cm}}r@{\hspace{0.125cm}}
     @{\hspace{0.125cm}}r@{\hspace{0.125cm}}
     @{\hspace{0.125cm}}r@{\hspace{0.125cm}}
     @{\hspace{0.125cm}}r@{\hspace{0.125cm}}
     @{\hspace{0.125cm}}r@{\hspace{0.125cm}}
     @{\hspace{0.125cm}}r@{\hspace{0.125cm}}
     @{\hspace{0.125cm}}r}\hline
    & $B$ mass & Eff & Resol & Bkg & Fit bias & Barrier & Lineshape & Extra \\\hline

$K^*(892)^0$ & $ < 0.001$ & 0.015 & $ < 0.001$ & 0.004 & 0.006 & 0.001 & $ < 0.001$ & $ < 0.001$ \\
$K^*(1410)^0$ & 0.010 & 0.040 & 0.010 & 0.022 & 0.005 & 0.061 & 0.025 & 0.032 \\
$K^*(1680)^0$ & 0.066 & 0.202 & 0.066 & 0.489 & 0.030 & 0.040 & 0.205 & 0.070 \\
$\rho(770)^0$ & $ < 0.001$ & 0.006 & $ < 0.001$ & 0.009 & 0.009 & 0.003 & 0.003 & 0.002 \\
$\omega(782)$ & $ < 0.001$ & 0.016 & 0.002 & 0.019 & 0.056 & 0.003 & 0.002 & 0.015 \\
$\rho(1450)^0$ & 0.005 & 0.040 & 0.001 & 0.137 & 0.003 & 0.030 & 0.043 & 0.046 \\
$\rho(1700)^0$ & 0.002 & 0.043 & 0.001 & 0.074 & 0.070 & 0.007 & 0.025 & 0.044 \\
$K_2^*(1430)^0$ & 0.028 & 0.083 & 0.027 & 0.121 & 0.029 & 0.009 & 0.044 & 0.007 \\
$f_2(1270)$ & 0.007 & 0.024 & 0.007 & 0.025 & 0.003 & 0.014 & 0.002 & 0.020 \\
$f_2^\prime(1525)$ & 0.008 & 0.041 & 0.008 & 0.084 & 0.203 & $<0.001$ & 0.010 & 0.169 \\
$K_3^*(1780)^0$ & 0.104 & 0.342 & 0.105 & 0.490 & 0.141 & 0.001 & 0.193 & 0.024 \\
$\rho_3(1690)^0$ & 0.006 & 0.021 & 0.006 & 0.064 & $ < 0.001$ & 0.027 & 0.009 & 0.068 \\
$\chi_{c0}(1P)$ & $ < 0.001$ & 0.004 & 0.001 & 0.002 & 0.010 & 0.004 & $ < 0.001$ & 0.004 \\
$(K^+\pi^-)_{\rm Res}$ & 0.009 & 0.025 & 0.009 & 0.035 & 0.018 & 0.001 & 0.012 & 0.003 \\
$(K^+\pi^-)_{\rm ER}$ & $ < 0.001$ & 0.006 & $ < 0.001$ & 0.007 & 0.013 & 0.001 & 0.002 & 0.002 \\
$K_0^*(1950)^0$ & 0.006 & 0.020 & 0.008 & 0.003 & 0.021 & 0.004 & 0.013 & 0.006 \\
\pip\pim S-wave & 0.018 & 0.025 & 0.020 & 0.100 & 0.031 & $< 0.001$ & 0.013 & 0.003 \\

    \hline
    \end{tabular}
\end{table}

\begin{table}[!htb]
    \centering
    \caption{\label{tab:app:syst:qmi:acp}
    Systematic uncertainties on the $\mathcal{A}_{\CP}$ values for the QMI approach.
    }
    \renewcommand{\arraystretch}{1.1}
    \begin{tabular}
    {l@{\hspace{0.125cm}}
     @{\hspace{0.125cm}}r@{\hspace{0.125cm}}
     @{\hspace{0.125cm}}r@{\hspace{0.125cm}}
     @{\hspace{0.125cm}}r@{\hspace{0.125cm}}
     @{\hspace{0.125cm}}r@{\hspace{0.125cm}}
     @{\hspace{0.125cm}}r@{\hspace{0.125cm}}
     @{\hspace{0.125cm}}r@{\hspace{0.125cm}}
     @{\hspace{0.125cm}}r@{\hspace{0.125cm}}
     @{\hspace{0.125cm}}r}\hline
     & $B$ mass & Eff & Resol & Bkg & QMI bias & Barrier & Lineshape & Extra \\ \hline

$K^*(892)^0$ & 0.001 & 0.017 & $<0.001$ & 0.009 & 0.012 & $<0.001$ & 0.001 & 0.002 \\
$K^*(1410)^0$ & $<0.001$ & 0.033 & 0.002 & 0.049 & 0.083 & 0.003 & 0.024 & 0.038 \\
$K^*(1680)^0$ & 0.020 & 0.082 & 0.005 & 0.134 & 0.047 & 0.002 & 0.003 & 0.077 \\
$\rho(770)^0$ & 0.003 & 0.006 & 0.001 & 0.021 & 0.041 & 0.001 & 0.008 & 0.005 \\
$\omega(782)$ & 0.003 & 0.010 & 0.001 & 0.009 & 0.002 & $<0.001$ & 0.001 & 0.004 \\
$\rho(1450)^0$ & 0.002 & 0.075 & 0.005 & 0.314 & 0.076 & 0.004 & 0.035 & 0.058 \\
$\rho(1700)^0$ & 0.008 & 0.063 & $<0.001$ & 0.395 & 0.083 & 0.001 & 0.004 & 0.058 \\
$K_2^*(1430)^0$ & 0.003 & 0.037 & 0.001 & 0.093 & 0.019 & $<0.001$ & 0.001 & 0.007 \\
$f_2(1270)$ & 0.003 & 0.020 & 0.002 & 0.112 & 0.017 & 0.001 & 0.003 & 0.005 \\
$f_2^\prime(1525)$ & 0.006 & 0.119 & 0.003 & 0.027 & 0.401 & 0.004 & 0.018 & 0.126 \\
$K_3^*(1780)^0$ & 0.001 & 0.027 & $<0.001$ & 0.145 & 0.073 & $<0.001$ & 0.002 & 0.004 \\
$\rho_3(1690)^0$ & 0.004 & 0.024 & 0.004 & 0.131 & 0.030 & 0.005 & 0.003 & 0.091 \\
$\chi_{c0}(1P)$ & 0.001 & 0.006 & 0.001 & 0.017 & 0.008 & $<0.001$ & 0.002 & 0.008 \\
S-wave & 0.001 & 0.002 & $<0.001$ & 0.014 & 0.011 & $<0.001$ & 0.001 & 0.002 \\

    \hline
    \end{tabular}
\end{table}

\begin{table}[!htb]
    \centering
    \caption{\label{tab:app:syst:iso:phasep}
        Systematic uncertainties on the \Bp relative phases for the Isobar approach given in degrees.
    }
    \renewcommand{\arraystretch}{1.1}
    \begin{tabular}
    {l@{\hspace{0.125cm}}
     @{\hspace{0.125cm}}r@{\hspace{0.125cm}}
     @{\hspace{0.125cm}}r@{\hspace{0.125cm}}
     @{\hspace{0.125cm}}r@{\hspace{0.125cm}}
     @{\hspace{0.125cm}}r@{\hspace{0.125cm}}
     @{\hspace{0.125cm}}r@{\hspace{0.125cm}}
     @{\hspace{0.125cm}}r@{\hspace{0.125cm}}
     @{\hspace{0.125cm}}r@{\hspace{0.125cm}}
     @{\hspace{0.125cm}}r@{\hspace{0.125cm}}
     @{\hspace{0.125cm}}r@{\hspace{0.125cm}}
     @{\hspace{0.125cm}}r}\hline
     & $B$ mass & Eff & Resol & Bkg & Fit bias & Barrier & Lineshape & Extra \\\hline

$K^*(1410)^0$ & 0.7 & 2.2 & 0.1 & 7.9 & 1.2 & 0.3 & 4.4 & 1.1 \\
$K^*(1680)^0$ & 0.6 & 4.5 & 0.6 & 18.2 & 4.1 & 3.9 & 15.7 & 3.0 \\
$\rho(770)^0$ & 0.1 & 1.5 & 0.3 & 1.5 & 0.6 & 0.6 & 1.5 & 0.3 \\
$\omega(782)$ & 0.2 & 0.3 & 0.2 & 1.0 & 2.5 & 1.6 & 0.8 & 1.1 \\
$\rho(1450)^0$ & 0.6 & 4.1 & 0.6 & 6.1 & 2.4 & 0.7 & 0.8 & 2.3 \\
$\rho(1700)^0$ & 0.1 & 3.1 & 0.3 & 5.2 & 0.3 & 4.5 & 4.1 & 2.4 \\
$K_2^*(1430)^0$ & 0.5 & 2.1 & 0.4 & 1.1 & 0.3 & 0.5 & 1.9 & 0.4 \\
$f_2(1270)$ & 0.1 & 1.4 & 0.1 & 2.5 & 0.2 & 0.2 & 1.0 & 2.3 \\
$f_2^\prime(1525)$ & 0.8 & 6.5 & 0.7 & 16.4 & 1.8 & 3.0 & 3.1 & 5.5 \\
$K_3^*(1780)^0$ & 0.3 & 1.4 & 0.3 & 10.0 & 1.3 & 3.4 & 1.3 & 3.5 \\
$\rho_3(1690)^0$ & 1.3 & 7.9 & 0.8 & 9.2 & 0.9 & 3.1 & 3.6 & 9.4 \\
$\chi_{c0}(1P)$ & 0.4 & 0.9 & 0.6 & 0.6 & 0.9 & 0.7 & 1.6 & 0.2 \\
$(K^+\pi^-)_{\rm Res}$ & 0.9 & 1.6 & 1.0 & 7.9 & 2.9 & 2.4 & 0.5 & 1.3 \\
$(K^+\pi^-)_{\rm ER}$ & 0.9 & 1.3 & 0.6 & 2.2 & $<0.1$ & 0.6 & 2.9 & 2.3 \\
$K_0^*(1950)^0$ & 0.1 & 3.2 & 0.4 & 5.8 & 0.5 & 1.2 & 4.8 & 2.0 \\
$f_0(500)$ & 1.2 & 1.3 & 0.3 & 6.7 & 0.3 & 0.9 & 3.4 & 1.8 \\
$f_0(980)$ & 0.4 & 0.6 & 0.2 & 2.6 & 2.0 & 2.1 & 2.1 & 0.8 \\
$f_0(1370)$ & 0.3 & 1.2 & 0.1 & 5.0 & 1.7 & 0.6 & 1.6 & 1.5 \\
$f_0(1500)$ & 0.4 & 1.6 & 0.1 & 3.8 & 0.5 & 1.3 & 1.3 & 1.6 \\
$f_0(1710)$ & 0.4 & 1.1 & 0.2 & 6.6 & 0.5 & 0.4 & 1.5 & 4.7 \\
$\pi\pi$--$\pi\pi$ & 1.6 & 3.0 & 1.4 & 6.7 & 2.1 & 0.2 & 3.7 & 1.9 \\
$\pi\pi$--$K \Kbar$ & 1.1 & 5.4 & 0.1 & 3.2 & 15.7 & 7.1 & 24.3 & 7.2 \\

    \hline
    \end{tabular}
\end{table}

\begin{table}[!htb]
    \centering
    \caption{\label{tab:app:syst:km:phasep}
        Systematic uncertainties on the \Bp relative phases for the K-matrix approach given in degrees.
    }
    \renewcommand{\arraystretch}{1.1}
    \begin{tabular}
    {l@{\hspace{0.125cm}}
     @{\hspace{0.125cm}}r@{\hspace{0.125cm}}
     @{\hspace{0.125cm}}r@{\hspace{0.125cm}}
     @{\hspace{0.125cm}}r@{\hspace{0.125cm}}
     @{\hspace{0.125cm}}r@{\hspace{0.125cm}}
     @{\hspace{0.125cm}}r@{\hspace{0.125cm}}
     @{\hspace{0.125cm}}r@{\hspace{0.125cm}}
     @{\hspace{0.125cm}}r@{\hspace{0.125cm}}
     @{\hspace{0.125cm}}r@{\hspace{0.125cm}}
     @{\hspace{0.125cm}}r@{\hspace{0.125cm}}
     @{\hspace{0.125cm}}r}\hline
     & $B$ mass & Eff & Resol & Bkg & Fit bias & Barrier & Lineshape & Extra \\\hline

$K^*(1410)^0$ & 0.6 & 2.1 & 0.7 & 8.1 & 1.6 & 0.6 & 0.9 & 1.1 \\
$K^*(1680)^0$ & 1.7 & 3.2 & 2.0 & 16.0 & 5.0 & 2.5 & 2.2 & 3.1 \\
$\rho(770)^0$ & $<0.1$ & 1.1 & 0.1 & 4.9 & 1.9 & 1.1 & 0.6 & 0.9 \\
$\omega(782)$ & 0.3 & 0.6 & 0.4 & 3.5 & 0.8 & 2.0 & 0.4 & 1.4 \\
$\rho(1450)^0$ & 0.6 & 3.7 & 0.6 & 3.9 & 2.0 & 4.0 & 0.7 & 2.2 \\
$\rho(1700)^0$ & 0.2 & 3.4 & 0.2 & 1.6 & $<0.1$ & 6.0 & 2.5 & 1.2 \\
$K_2^*(1430)^0$ & 0.5 & 1.9 & 0.5 & 1.6 & 0.9 & 0.7 & 0.4 & 0.4 \\
$f_2(1270)$ & 0.1 & 1.5 & 0.1 & 2.3 & 0.7 & 0.6 & 0.6 & 2.4 \\
$f_2^\prime(1525)$ & 0.9 & 5.5 & 0.9 & 13.0 & 3.6 & 3.0 & 1.4 & 2.4 \\
$K_3^*(1780)^0$ & 0.5 & 1.3 & 0.5 & 8.4 & 0.9 & 2.8 & 0.8 & 2.3 \\
$\rho_3(1690)^0$ & 1.1 & 6.9 & 1.0 & 8.0 & 1.3 & 1.5 & 1.5 & 10.1 \\
$\chi_{c0}(1P)$ & 0.4 & 0.9 & 0.6 & 1.0 & 1.2 & 0.7 & 0.3 & 0.2 \\
$(K^+\pi^-)_{\rm Res}$ & 0.9 & 1.6 & 1.0 & 7.9 & 2.9 & 2.4 & 0.5 & 1.3 \\
$(K^+\pi^-)_{\rm ER}$ & 0.4 & 1.8 & 0.3 & 4.6 & 1.8 & 0.8 & 0.3 & 0.3 \\
$K_0^*(1950)^0$ & 0.8 & 3.4 & 0.9 & 4.6 & 1.4 & 0.6 & 4.6 & 1.3 \\

    \hline
    \end{tabular}
\end{table}

\begin{table}[tb]
    \centering
    \caption{\label{tab:app:syst:qmi:phasep}
        Systematic uncertainties on the \Bp relative phases for the QMI approach given in degrees.
    }
    \renewcommand{\arraystretch}{1.1}
    \begin{tabular}
    {l@{\hspace{0.125cm}}
     @{\hspace{0.125cm}}r@{\hspace{0.125cm}}
     @{\hspace{0.125cm}}r@{\hspace{0.125cm}}
     @{\hspace{0.125cm}}r@{\hspace{0.125cm}}
     @{\hspace{0.125cm}}r@{\hspace{0.125cm}}
     @{\hspace{0.125cm}}r@{\hspace{0.125cm}}
     @{\hspace{0.125cm}}r@{\hspace{0.125cm}}
     @{\hspace{0.125cm}}r@{\hspace{0.125cm}}
     @{\hspace{0.125cm}}r@{\hspace{0.125cm}}
     @{\hspace{0.125cm}}r@{\hspace{0.125cm}}
     @{\hspace{0.125cm}}r}\hline
     & $B$ mass & Eff & Resol & Bkg & Fit bias & Barrier & Lineshape & Extra \\\hline

$K^*(1410)^0$ & 1.1 & 3.2 & 0.3 & 12.2 & 3.2 & 0.1 & 1.1 & 1.0 \\
$K^*(1680)^0$ & 0.5 & 8.3 & 0.7 & 60.1 & 3.1 & 1.3 & 4.2 & 17.1 \\
$\rho(770)^0$ & 0.2 & 2.9 & $<0.1$ & 1.2 & 8.4 & $<0.1$ & 0.5 & 1.1 \\
$\omega(782)$ & 0.1 & 2.4 & 0.2 & 2.7 & 4.6 & $<0.1$ & 0.6 & 2.0 \\
$\rho(1450)^0$ & 0.1 & 2.3 & $<0.1$ & 4.4 & 7.4 & 0.2 & 0.2 & 5.5 \\
$\rho(1700)^0$ & 0.7 & 2.1 & 0.6 & 3.0 & 0.2 & 0.1 & 1.0 & 2.1 \\
$K_2^*(1430)^0$ & 0.1 & 2.8 & $<0.1$ & 2.1 & 0.2 & 0.1 & 0.2 & 1.9 \\
$f_2(1270)$ & 0.2 & 2.6 & 0.1 & 0.5 & 3.0 & $<0.1$ & 0.4 & 3.6 \\
$f_2^\prime(1525)$ & 0.2 & 8.0 & 0.2 & 9.2 & 9.0 & 0.3 & 0.5 & 1.5 \\
$K_3^*(1780)^0$ & 0.1 & 1.1 & 0.1 & 16.5 & 0.4 & 0.1 & 0.3 & 3.5 \\
$\rho_3(1690)^0$ & 0.1 & 10.3 & 0.3 & 1.0 & 4.5 & $<0.1$ & 1.5 & 14.8 \\
$\chi_{c0}(1P)$ & 0.1 & 1.9 & $<0.1$ & 1.7 & 0.4 & $<0.1$ & 0.1 & 0.7 \\

    \hline
    \end{tabular}
\end{table}

\begin{table}[!htb]
    \centering
    \caption{\label{tab:app:syst:iso:phasem}
        Systematic uncertainties on the \Bm relative phases for the Isobar approach given in degrees.
    }
    \renewcommand{\arraystretch}{1.1}
    \begin{tabular}
    {l@{\hspace{0.125cm}}
     @{\hspace{0.125cm}}r@{\hspace{0.125cm}}
     @{\hspace{0.125cm}}r@{\hspace{0.125cm}}
     @{\hspace{0.125cm}}r@{\hspace{0.125cm}}
     @{\hspace{0.125cm}}r@{\hspace{0.125cm}}
     @{\hspace{0.125cm}}r@{\hspace{0.125cm}}
     @{\hspace{0.125cm}}r@{\hspace{0.125cm}}
     @{\hspace{0.125cm}}r@{\hspace{0.125cm}}
     @{\hspace{0.125cm}}r@{\hspace{0.125cm}}
     @{\hspace{0.125cm}}r@{\hspace{0.125cm}}
     @{\hspace{0.125cm}}r}\hline
     & $B$ mass & Eff & Resol & Bkg & Fit bias & Barrier & Lineshape & Extra \\\hline

$K^*(1410)^0$ & 1.0 & 1.7 & 0.7 & 1.4 & 2.5 & 1.6 & 3.6 & 0.4 \\
$K^*(1680)^0$ & 9.8 & 19.6 & 10.0 & 25.5 & 5.4 & 2.6 & 37.3 & 5.5 \\
$\rho(770)^0$ & 1.5 & 2.3 & 1.5 & 1.3 & 1.1 & 0.9 & 4.7 & 6.5 \\
$\omega(782)$ & 1.6 & 2.5 & 1.8 & 1.6 & 1.9 & 1.3 & 5.5 & 6.3 \\
$\rho(1450)^0$ & 0.8 & 1.4 & 0.8 & 0.9 & 2.9 & 2.9 & 2.8 & 14.2 \\
$\rho(1700)^0$ & 1.1 & 5.5 & 0.7 & 15.0 & 0.9 & 18.0 & 3.4 & 44.8 \\
$K_2^*(1430)^0$ & 0.1 & 1.4 & 0.2 & 2.1 & 0.1 & 0.4 & 0.3 & 0.4 \\
$f_2(1270)$ & 0.1 & 1.3 & 0.2 & 2.1 & 1.3 & 0.4 & 0.6 & 4.4 \\
$f_2^\prime(1525)$ & 0.4 & 17.1 & 0.3 & 18.5 & 16.3 & 1.1 & 4.2 & 54.9 \\
$K_3^*(1780)^0$ & 2.6 & 9.6 & 3.2 & 8.1 & 3.3 & 20.1 & 11.1 & 8.4 \\
$\rho_3(1690)^0$ & 0.2 & 0.8 & 0.2 & 5.2 & 2.1 & 0.1 & 1.3 & 27.8 \\
$\chi_{c0}(1P)$ & $<0.1$ & 1.2 & 0.8 & 0.3 & 1.4 & $<0.1$ & 0.7 & 0.6 \\
$(K^+\pi^-)_{\rm Res}$ & 0.2 & 0.8 & $<0.1$ & 1.7 & 2.2 & 0.3 & 1.2 & 1.4 \\
$(K^+\pi^-)_{\rm ER}$ & 0.5 & 1.2 & 0.3 & 1.1 & 0.3 & 0.6 & 1.7 & 2.0 \\
$K_0^*(1950)^0$ & 1.1 & 4.6 & 0.9 & 4.3 & 0.4 & 0.1 & 8.6 & 0.4 \\
$f_0(500)$ & 1.3 & 1.7 & 1.0 & 6.5 & $<0.1$ & 0.2 & 5.9 & 10.1 \\
$f_0(980)$ & 0.5 & 0.9 & 0.7 & 0.8 & 1.6 & 2.0 & 2.9 & 3.8 \\
$f_0(1370)$ & 0.5 & 0.7 & 0.3 & 3.2 & 1.8 & 0.7 & 2.3 & 1.7 \\
$f_0(1500)$ & 0.3 & 1.0 & 0.2 & 1.4 & 1.4 & 1.0 & 1.6 & 6.3 \\
$f_0(1710)$ & 0.8 & 3.6 & 0.6 & 6.0 & 0.4 & 1.8 & 5.9 & 38.7 \\
$\pi\pi$--$\pi\pi$ & 1.3 & 2.1 & 1.4 & 2.1 & 1.6 & 2.9 & 2.7 & 8.8 \\
$\pi\pi$--$K \Kbar$ & 0.2 & 3.5 & $<0.1$ & 2.8 & 0.4 & 3.6 & 12.9 & 8.3 \\

    \hline
    \end{tabular}
\end{table}

\begin{table}[!htb]
    \centering
    \caption{\label{tab:app:syst:km:phasem}
        Systematic uncertainties on the \Bm relative phases for the K-matrix approach given in degrees.
    }
    \renewcommand{\arraystretch}{1.1}
    \begin{tabular}
    {l@{\hspace{0.125cm}}
     @{\hspace{0.125cm}}r@{\hspace{0.125cm}}
     @{\hspace{0.125cm}}r@{\hspace{0.125cm}}
     @{\hspace{0.125cm}}r@{\hspace{0.125cm}}
     @{\hspace{0.125cm}}r@{\hspace{0.125cm}}
     @{\hspace{0.125cm}}r@{\hspace{0.125cm}}
     @{\hspace{0.125cm}}r@{\hspace{0.125cm}}
     @{\hspace{0.125cm}}r@{\hspace{0.125cm}}
     @{\hspace{0.125cm}}r@{\hspace{0.125cm}}
     @{\hspace{0.125cm}}r@{\hspace{0.125cm}}
     @{\hspace{0.125cm}}r}\hline
     & $B$ mass & Eff & Resol & Bkg & Fit bias & Barrier & Lineshape & Extra \\\hline

$K^*(1410)^0$ & $<0.1$ & 1.2 & $<0.1$ & 0.5 & 1.2 & 1.5 & 0.8 & 0.1 \\
$K^*(1680)^0$ & 8.9 & 25.8 & 8.8 & 24.7 & 16.4 & 2.2 & 11.7 & 1.9 \\
$\rho(770)^0$ & 1.5 & 3.3 & 1.4 & 1.4 & 1.5 & 1.1 & 1.7 & 1.0 \\
$\omega(782)$ & 1.6 & 3.4 & 1.8 & 1.5 & 1.3 & 1.5 & 1.7 & 1.0 \\
$\rho(1450)^0$ & 1.0 & 2.1 & 0.8 & 0.8 & 1.9 & 2.4 & 1.8 & 1.9 \\
$\rho(1700)^0$ & 0.1 & 6.4 & $<0.1$ & 7.5 & 1.7 & 24.7 & 6.0 & 7.7 \\
$K_2^*(1430)^0$ & $<0.1$ & 1.4 & $<0.1$ & 2.4 & 0.1 & 0.6 & 0.9 & 0.2 \\
$f_2(1270)$ & 0.1 & 2.0 & 0.1 & 2.4 & 1.4 & 0.6 & 0.8 & 1.8 \\
$f_2^\prime(1525)$ & 1.1 & 15.5 & 1.2 & 15.4 & 20.6 & 2.7 & 1.8 & 8.6 \\
$K_3^*(1780)^0$ & 2.4 & 14.4 & 2.4 & 3.9 & 6.0 & 17.6 & 4.0 & 0.8 \\
$\rho_3(1690)^0$ & 0.1 & 1.4 & 0.1 & 4.7 & 2.5 & 0.3 & 0.9 & 0.8 \\
$\chi_{c0}(1P)$ & 0.2 & 1.0 & 0.8 & 0.4 & 1.0 & 0.5 & 0.1 & 0.1 \\
$(K^+\pi^-)_{\rm Res}$ & 0.4 & 2.3 & 0.4 & 4.4 & 3.9 & 2.4 & 0.9 & 1.3 \\
$(K^+\pi^-)_{\rm ER}$  & $<0.1$ & 1.3 & 0.1 & 3.2 & 0.9 & 0.3 & 0.5 & 0.3 \\
$K_0^*(1950)^0$ & 1.5 & 6.2 & 1.6 & 3.4 & 1.8 & 0.1 & 6.7 & 0.5 \\

    \hline
    \end{tabular}
\end{table}

\begin{table}[tb]
    \centering
    \caption{\label{tab:app:syst:qmi:phasem}
        Systematic uncertainties on the \Bm relative phases for the QMI approach given in degrees.
    }
    \renewcommand{\arraystretch}{1.1}
    \begin{tabular}
    {l@{\hspace{0.125cm}}
     @{\hspace{0.125cm}}r@{\hspace{0.125cm}}
     @{\hspace{0.125cm}}r@{\hspace{0.125cm}}
     @{\hspace{0.125cm}}r@{\hspace{0.125cm}}
     @{\hspace{0.125cm}}r@{\hspace{0.125cm}}
     @{\hspace{0.125cm}}r@{\hspace{0.125cm}}
     @{\hspace{0.125cm}}r@{\hspace{0.125cm}}
     @{\hspace{0.125cm}}r@{\hspace{0.125cm}}
     @{\hspace{0.125cm}}r@{\hspace{0.125cm}}
     @{\hspace{0.125cm}}r@{\hspace{0.125cm}}
     @{\hspace{0.125cm}}r}\hline
     & $B$ mass & Eff & Resol & Bkg & Fit bias & Barrier & Lineshape & Extra \\\hline

$K^*(1410)^0$ & 0.3 & 0.5 & 0.1 & 3.1 & 4.1 & 0.1 & 0.6 & 0.5 \\
$K^*(1680)^0$ & 0.5 & 1.7 & 0.1 & 2.5 & 13.7 & $<0.1$ & 0.7 & 1.3 \\
$\rho(770)^0$ & 0.4 & 2.0 & 0.3 & 2.9 & 4.7 & $<0.1$ & 0.2 & 0.6 \\
$\omega(782)$ & 0.4 & 1.8 & 0.4 & 2.9 & 2.7 & $<0.1$ & 0.4 & 0.7 \\
$\rho(1450)^0$ & 0.2 & 1.7 & 0.1 & 1.2 & 4.6 & $<0.1$ & 1.6 & 0.9 \\
$\rho(1700)^0$ & 0.1 & 9.2 & 0.9 & 27.3 & 7.8 & 0.4 & 2.0 & 7.5 \\
$K_2^*(1430)^0$ & $<0.1$ & 2.1 & $<0.1$ & 0.9 & 0.1 & $<0.1$ & 0.2 & 0.4 \\
$f_2(1270)$ & $<0.1$ & 1.2 & 0.1 & 0.8 & 3.8 & $<0.1$ & $<0.1$ & 1.6 \\
$f_2^\prime(1525)$ & 0.2 & 9.2 & 0.6 & 8.4 & 19.4 & 0.2 & 1.7 & 30.0 \\
$K_3^*(1780)^0$ & 1.3 & 8.8 & 0.3 & 6.2 & 0.9 & $<0.1$ & 3.3 & 16.8 \\
$\rho_3(1690)^0$ & 0.1 & 2.9 & 0.1 & 0.4 & 4.4 & $<0.1$ & 0.8 & 1.0 \\
$\chi_{c0}(1P)$ & $<0.1$ & 1.1 & 0.2 & 1.6 & 0.7 & 0.1 & 0.2 & 0.1 \\

    \hline
    \end{tabular}
\end{table}
\clearpage

\section{Goodness-of-fit comparison}
\label{app:gof}

\begin{figure}[tb]
    \centering
    \includegraphics[width=0.5\linewidth]{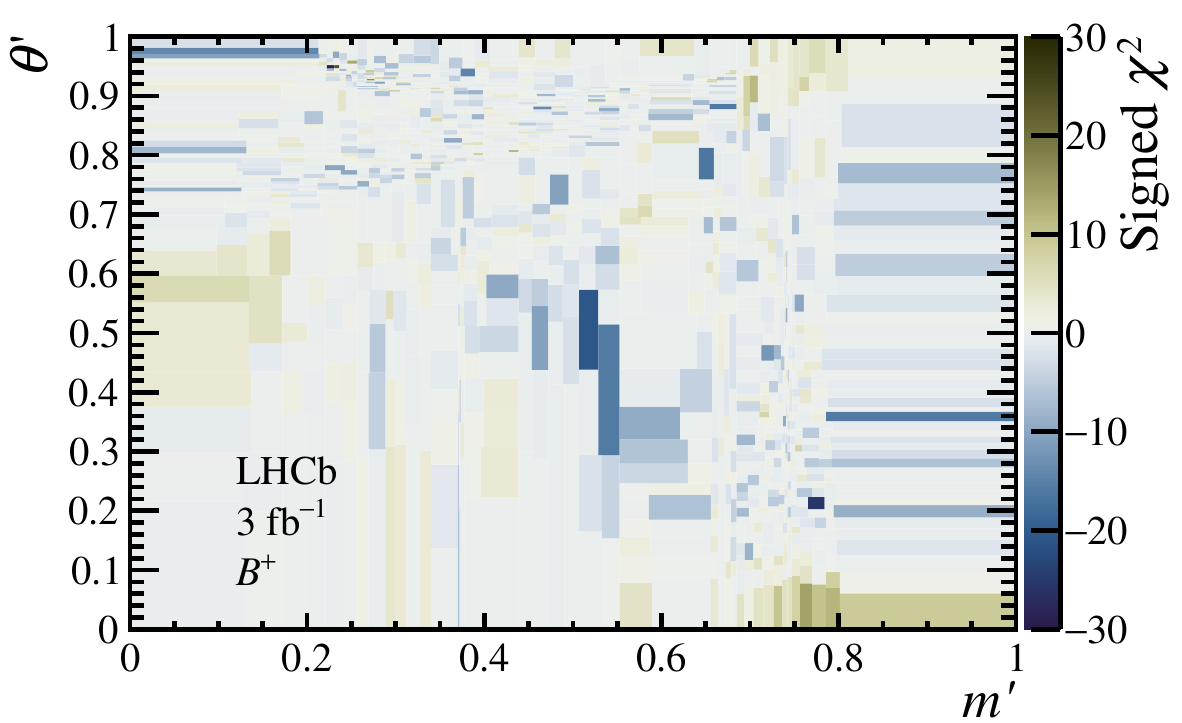}%
    \includegraphics[width=0.5\linewidth]{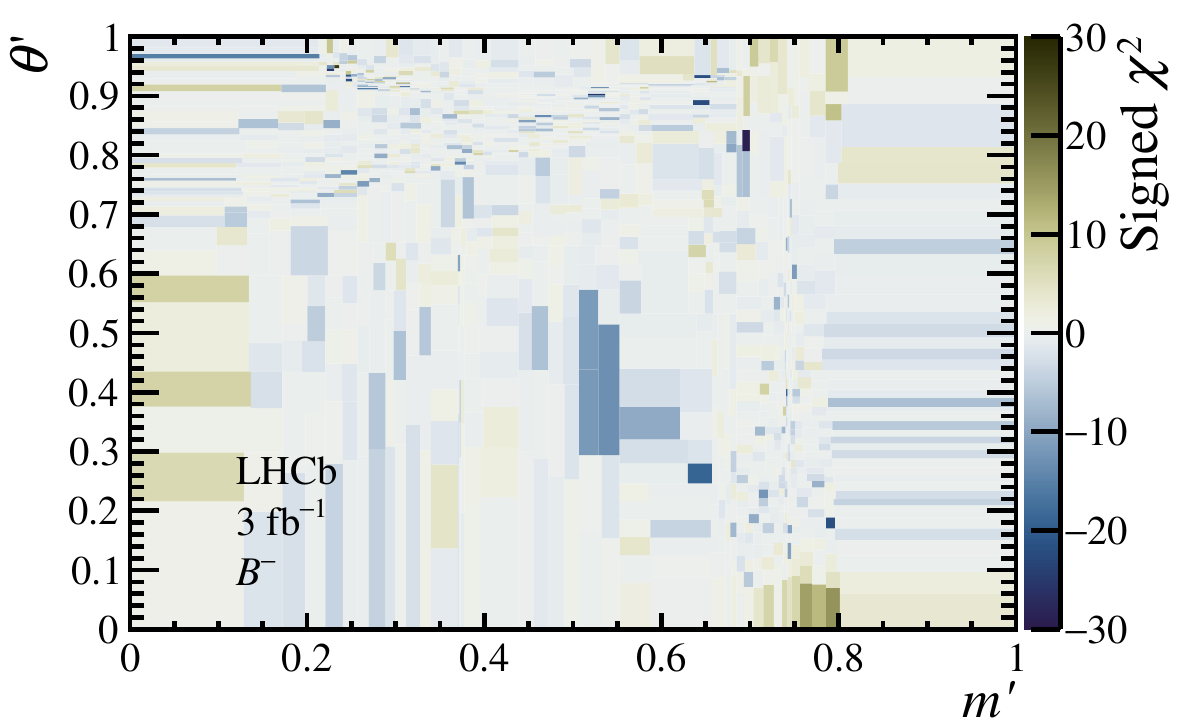}

    \includegraphics[width=0.5\linewidth]{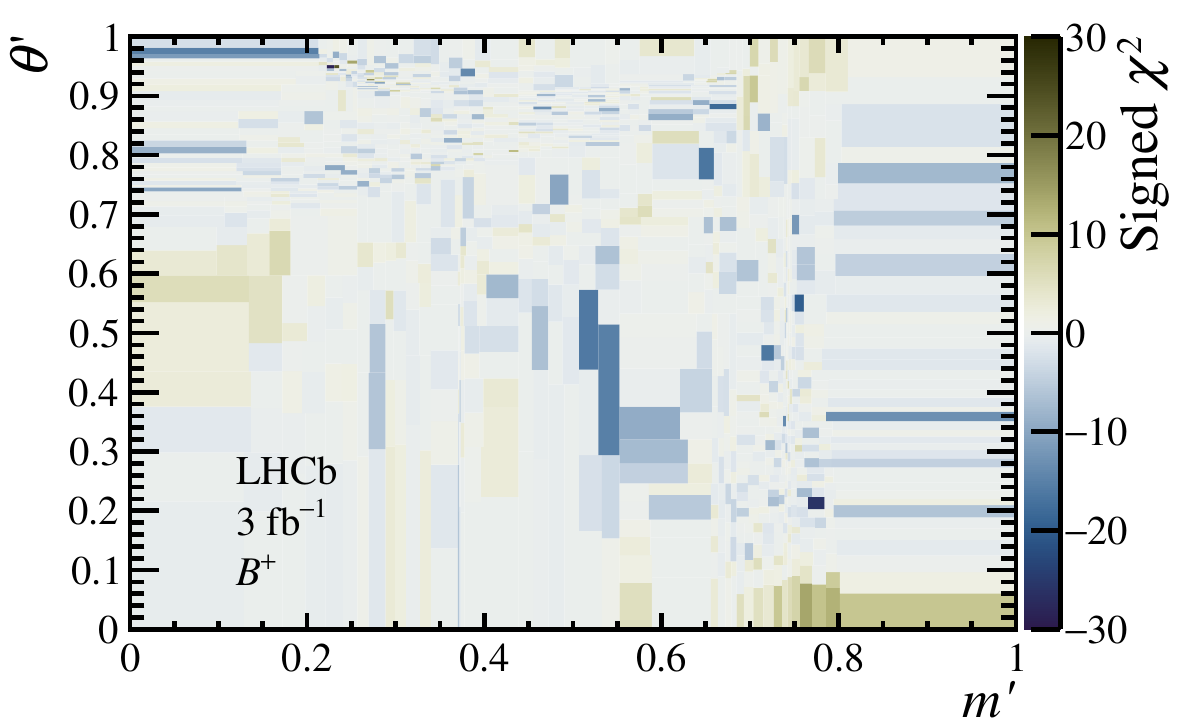}%
    \includegraphics[width=0.5\linewidth]{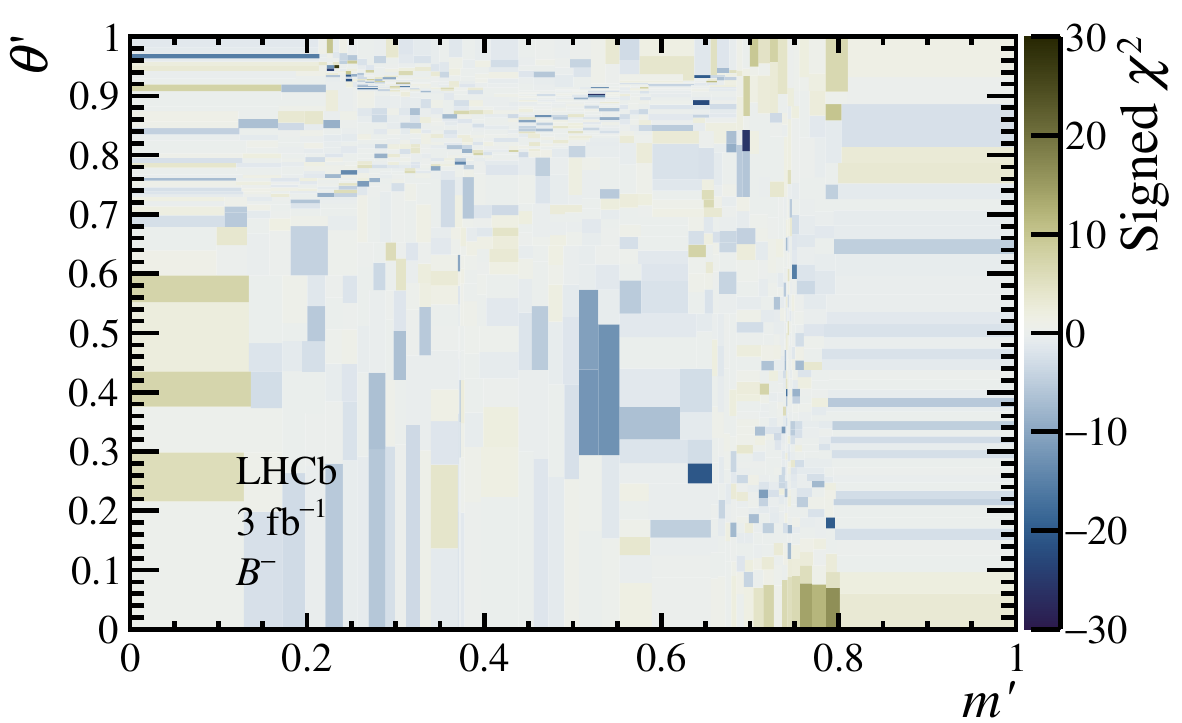}

    \includegraphics[width=0.5\linewidth]{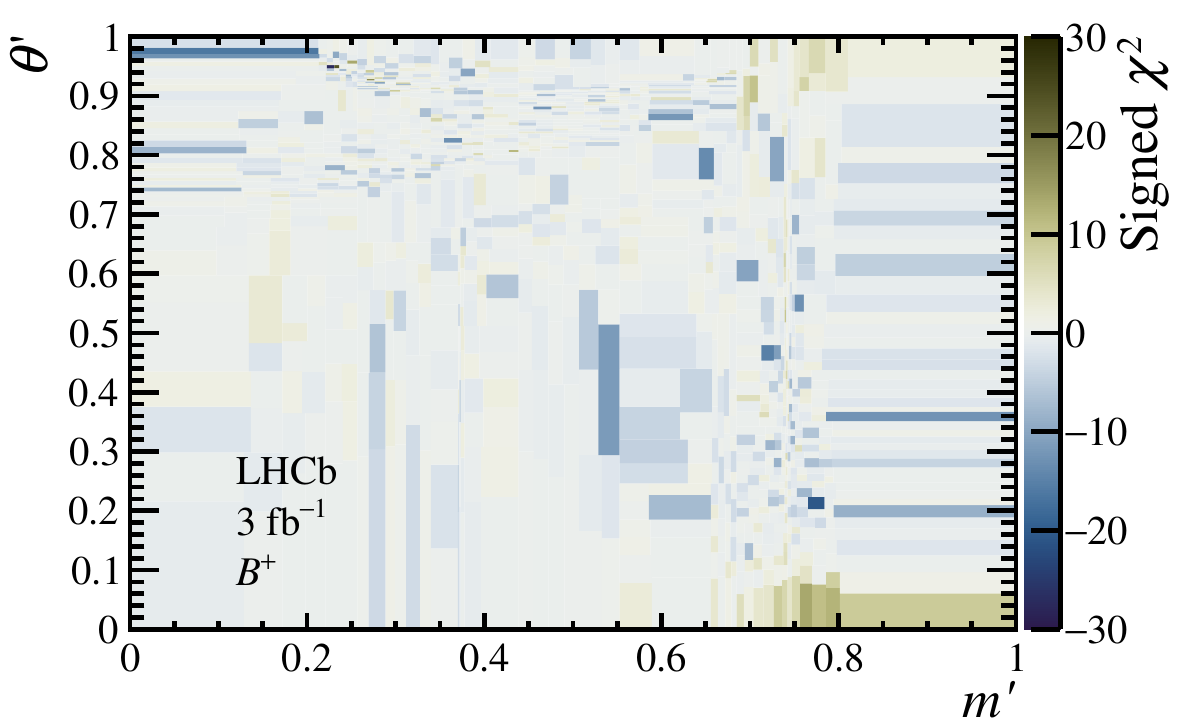}%
    \includegraphics[width=0.5\linewidth]{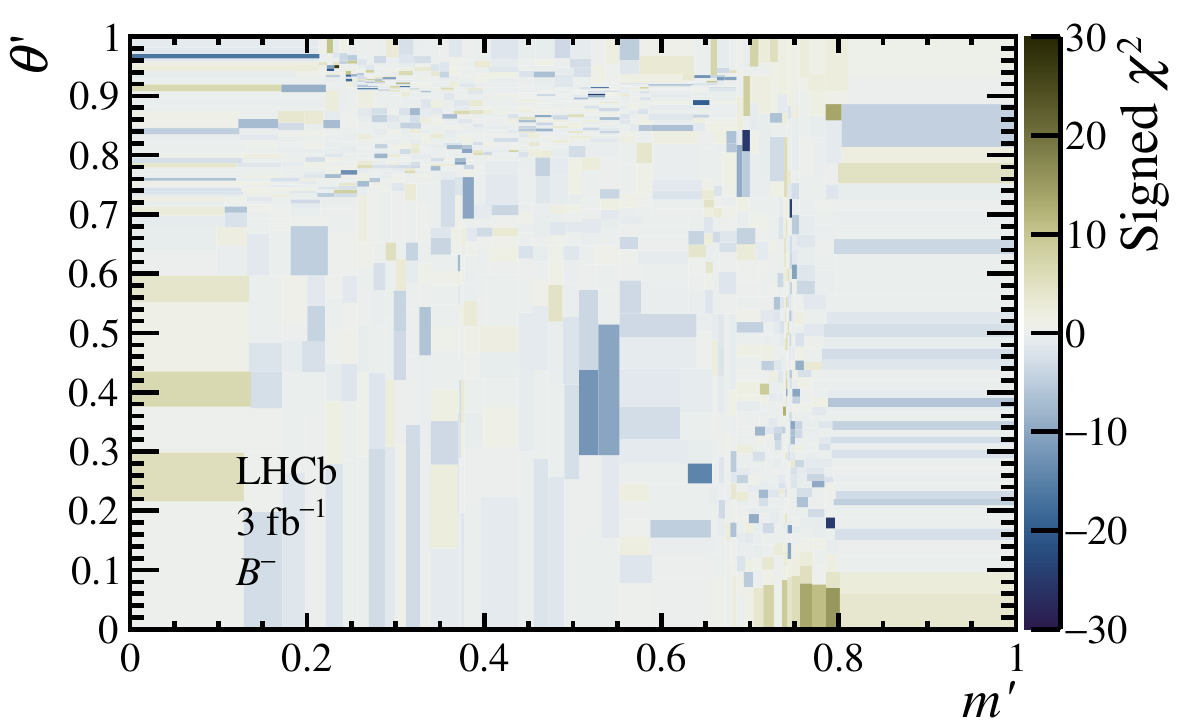}

     \caption{
       Signed $\chi^2$ distributions indicating the agreement between the (top)~Isobar, (middle)~K-matrix and (bottom)~QMI models and the data for (left)~$B^+$ and (right)~$B^-$ decays.
    }
    \label{fig:app:gof}
\end{figure}

As an indication of fit quality, $\chi^2$ goodness-of-fit indicators in the square Dalitz plot are produced with an adaptive binning procedure that requires set minimum numbers of signal candidates per bin when integrated over charge. The uncertainty is statistical only, assuming the number of entries per bin is Poisson distributed. 
A comparison between the three S-wave approaches can be found in Table~\ref{tab:app:chi2ndf}, in which the number of bins $N_{\rm Bin}$ and free parameters $N_{\rm Par}$ are also recorded. 
Signed $\chi^2$ distributions for each S-wave approach with at least 30 signal candidates per bin can be seen in Fig.~\ref{fig:app:gof}.

\begin{table}[b]
    \centering
    \caption{\label{tab:app:chi2ndf}
        Goodness-of-fit indicators for each S-wave model with various adaptive binning schemes.
    }

    \renewcommand{\arraystretch}{1.1}
    \begin{tabular}
    {l@{\hspace{0.5cm}}
     @{\hspace{0.25cm}}c@{\hspace{0.25cm}}
     @{\hspace{0.25cm}}r@{\hspace{0.25cm}}
     @{\hspace{0.25cm}}r@{\hspace{0.25cm}}
     @{\hspace{0.5cm}}c}\hline

    Model & Minimum & $B^+$: $\chi^2$ & $B^-$: $\chi^2$ & $N_{\rm Bin} - N_{\rm Par} - 1$ \\
    & entries/bin  & & & \\
    \hline
    Isobar & 15 & 5221.4 & 5760.6 & $3600 - \phantom{0}98 - 1$\\
    & 30 & 2879.4 & 2923.1 & $1764 - \phantom{0}98 - 1$\\
    & 45 & 2194.5 & 2074.0 & $1225 - \phantom{0}98 - 1$\\\hline

    K-matrix & 15 & 5163.7  & 5772.6 & $3600 - 107 - 1$\\
    & 30 & 2836.3 & 2945.0 & $1764 - 107 - 1$\\
    & 45 & 2156.0 & 2098.3 & $1225 - 107 - 1$\\\hline

    QMI & 15 & 5026.2 & 5621.4 & $3600 - 445 - 1$\\
    & 30 & 2702.4 & 2853.2 & $1764 - 445 - 1$\\
    & 45 & 2025.0 & 2040.8 & $1225 - 445 - 1$ \\
    \hline
    \end{tabular}
\end{table}

\clearpage

\section{Additional Isobar results}
\label{app:isobar}

\begin{figure}[!b]
    \centering
    \includegraphics[width=0.2\linewidth]{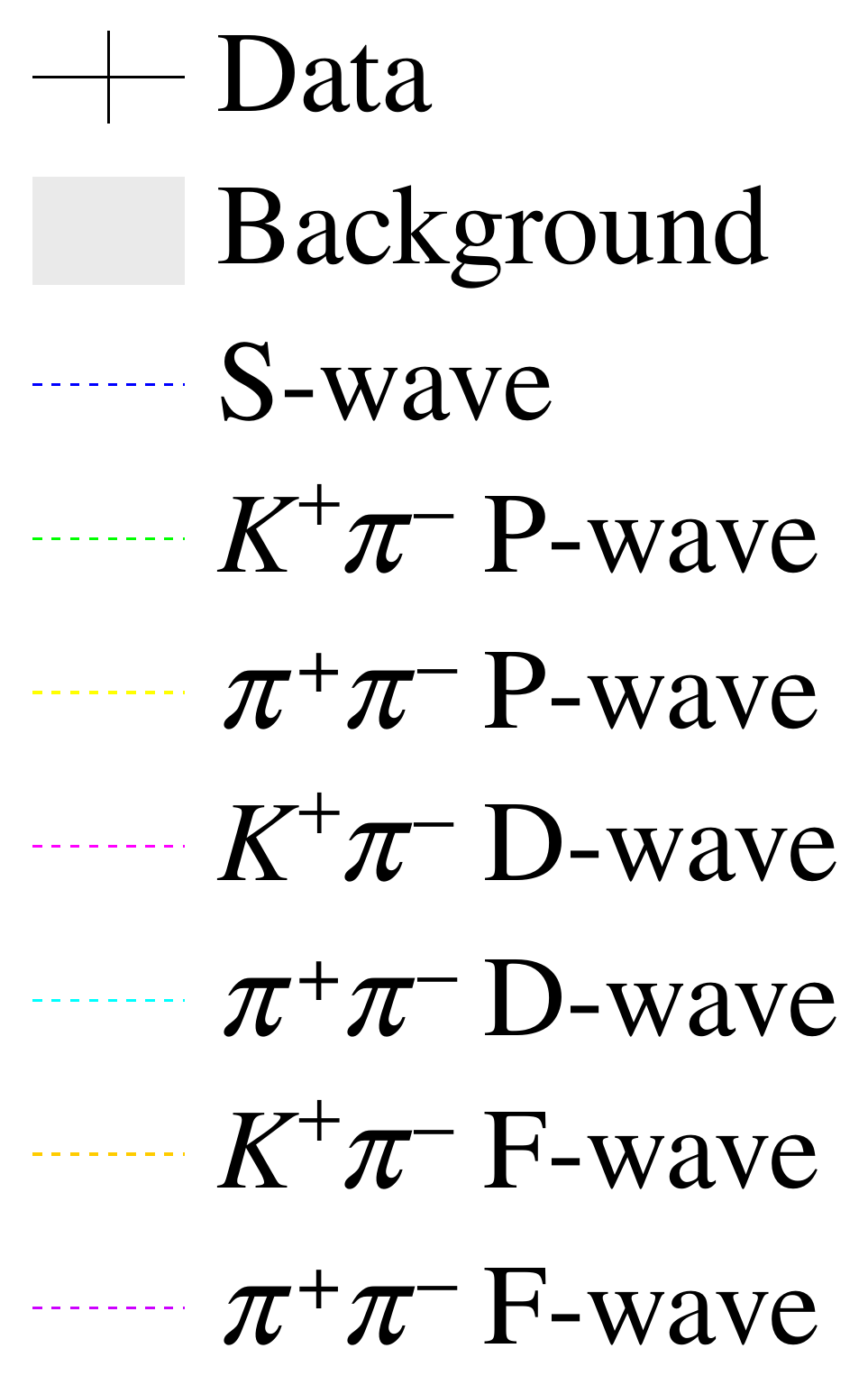}

    \caption{
        Colour legend of each partial wave shown in the subsequent figures of this section.
    }
    \label{fig:app:legiso}
\end{figure}

\begin{figure}[tb]
    \centering
    \includegraphics[width=0.5\linewidth]{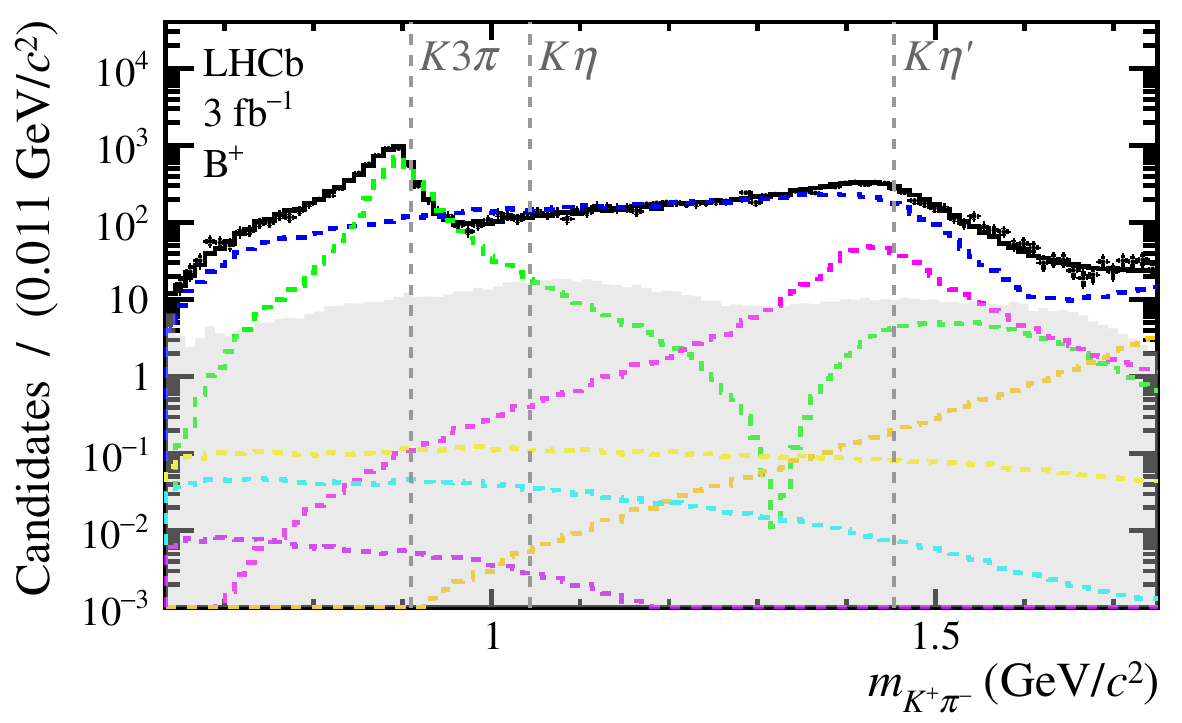}%
    \includegraphics[width=0.5\linewidth]{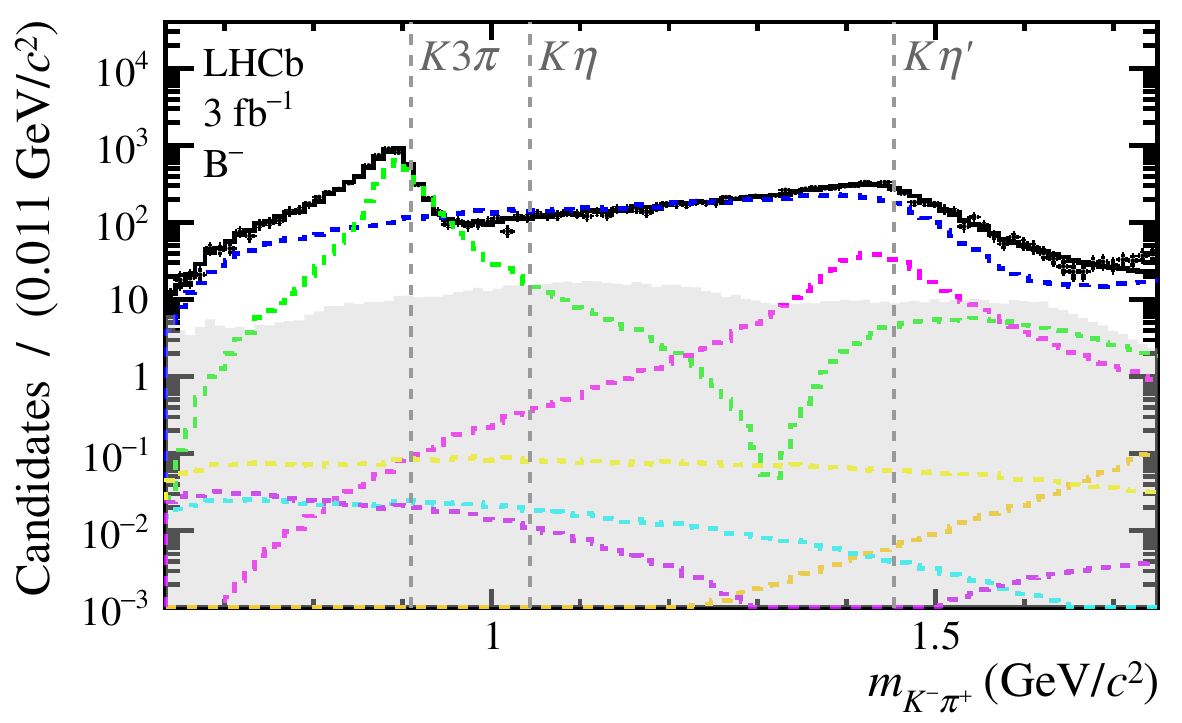}

    \includegraphics[width=0.5\linewidth]{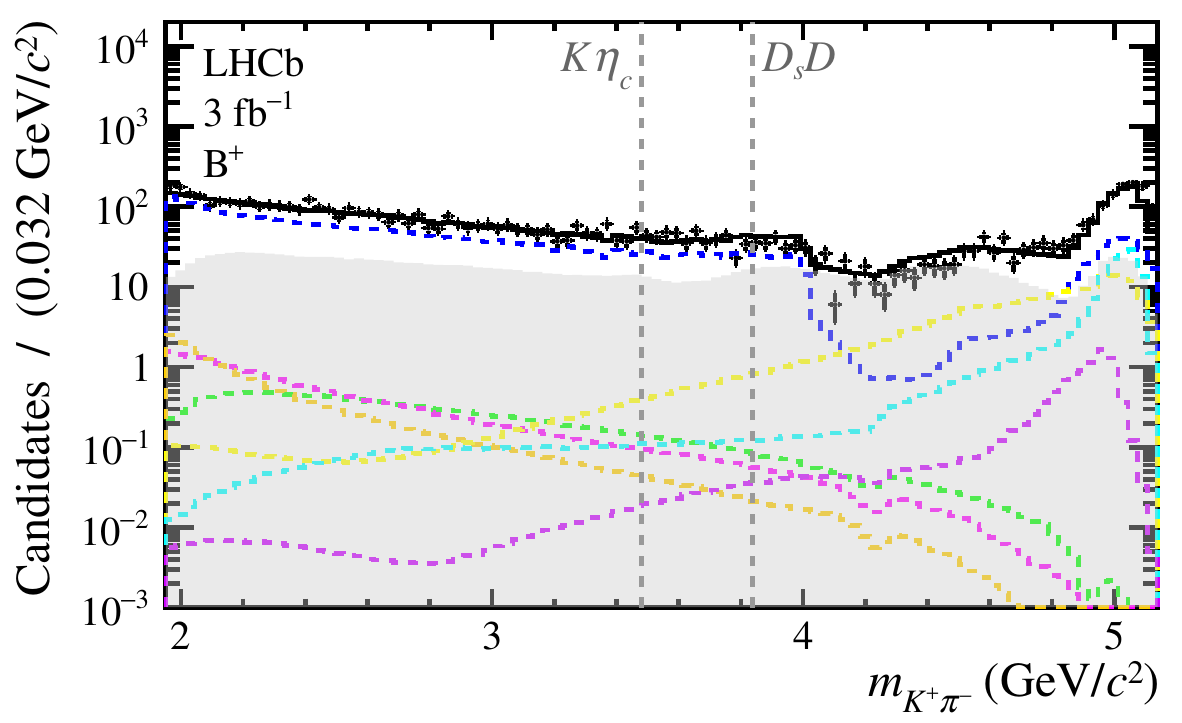}%
    \includegraphics[width=0.5\linewidth]{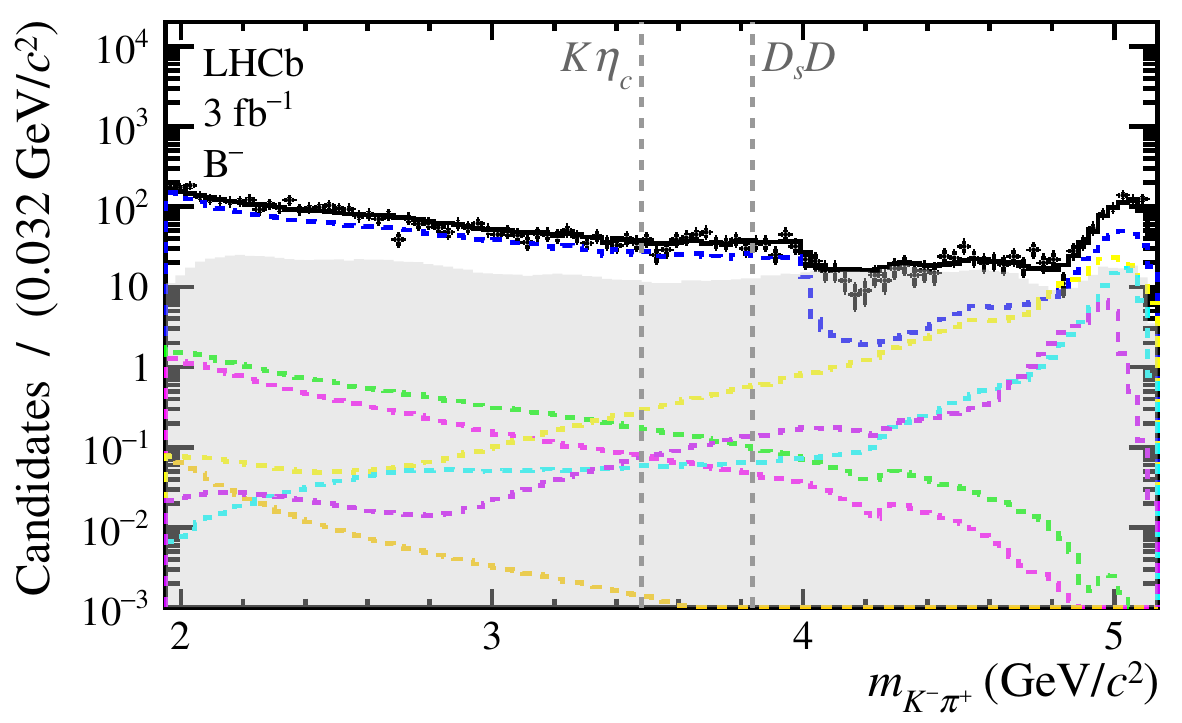}

    \caption{
        Fit projections of $m_{K^+ \pi^-}$ for (left)~$B^+$ and (right)~$B^-$ (top)~below and (bottom)~above the open charm threshold, enhanced with $\cos\theta_{K^+ \pi^-} < 0$. 
        Vertical dashed lines correspond to the opening thresholds of the indicated coupled channels.
    }
    \label{fig:app:mkpiiso}
\end{figure}

\begin{figure}[tb]
    \centering
    \includegraphics[width=0.5\linewidth]{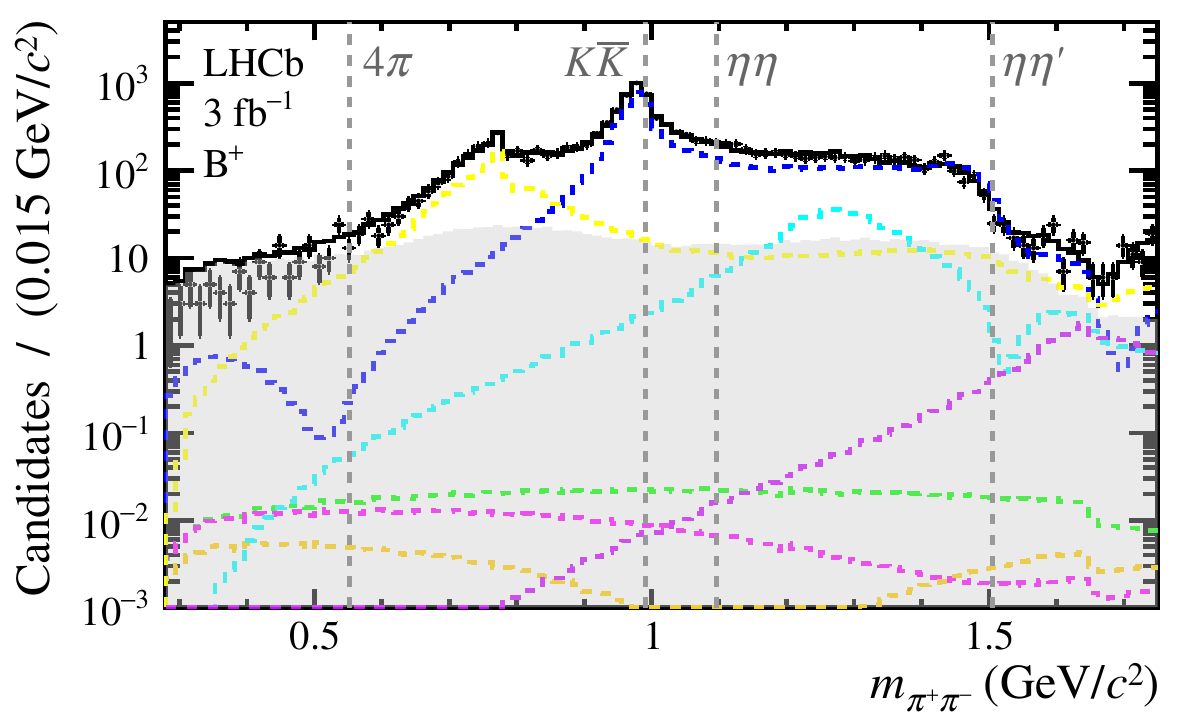}%
    \includegraphics[width=0.5\linewidth]{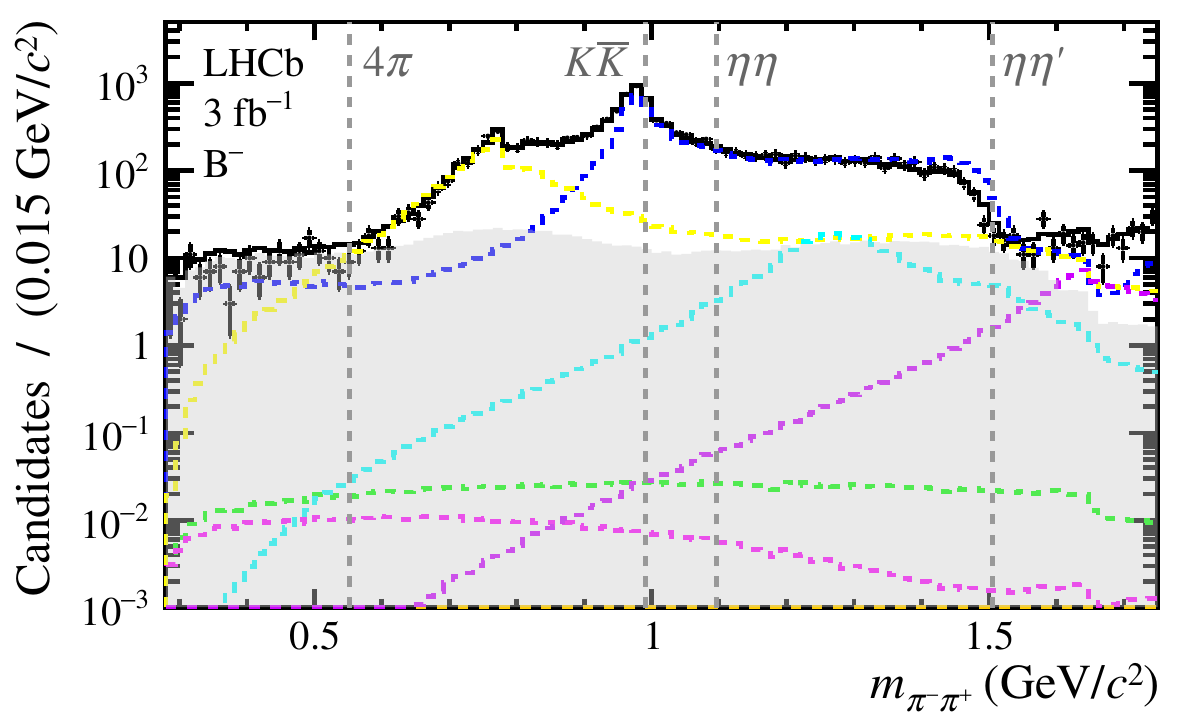}
    
    \includegraphics[width=0.5\linewidth]{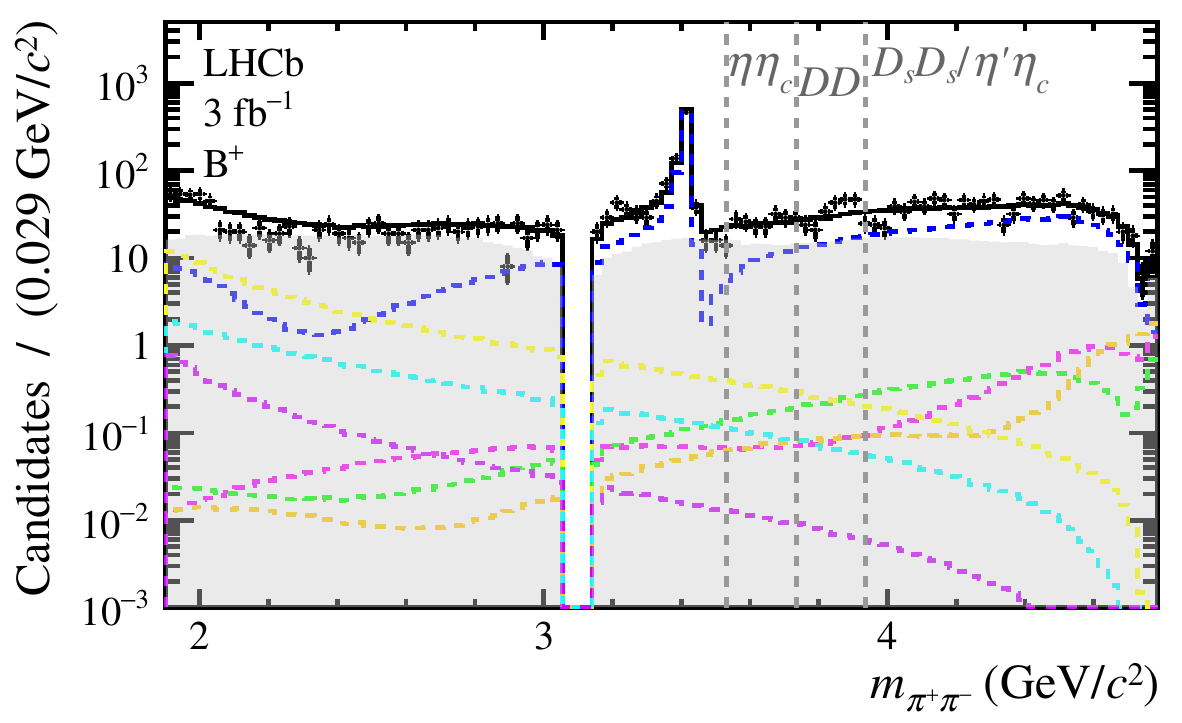}%
    \includegraphics[width=0.5\linewidth]{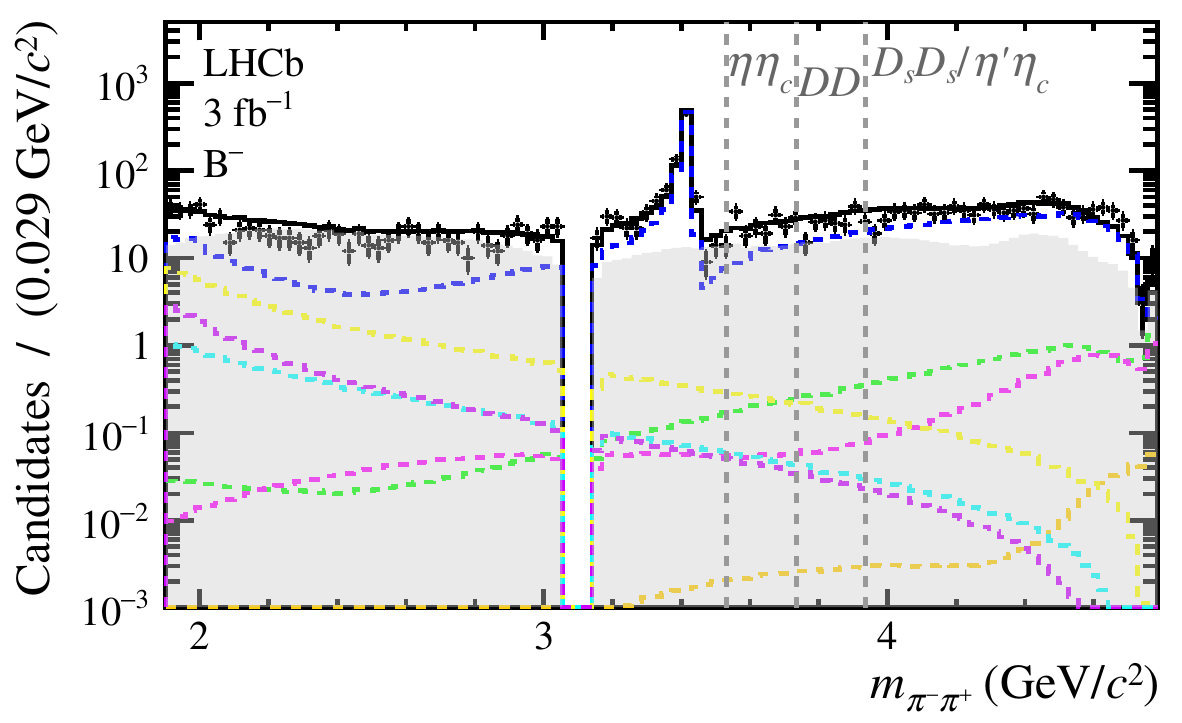}

    \caption{
        Fit projections of $m_{\pi^+ \pi^-}$ for (left)~$B^+$ and (right)~$B^-$ (top)~below and (bottom)~above the open charm threshold, enhanced with $\cos\theta_{\pi^+ \pi^-} < 0$. 
        Vertical dashed lines correspond to the opening thresholds of the indicated coupled channels.
    }
    \label{fig:app:mpipiiso}
\end{figure}

\begin{figure}[tb]
    \centering
    \includegraphics[width=0.5\linewidth]{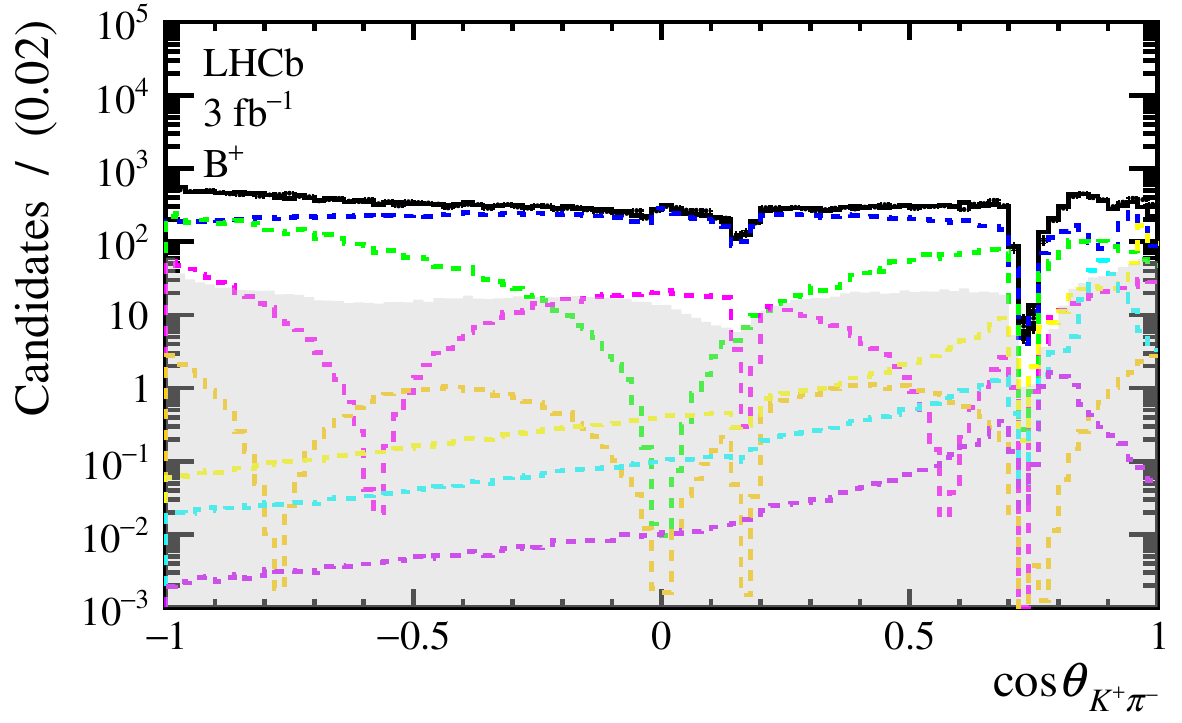}%
    \includegraphics[width=0.5\linewidth]{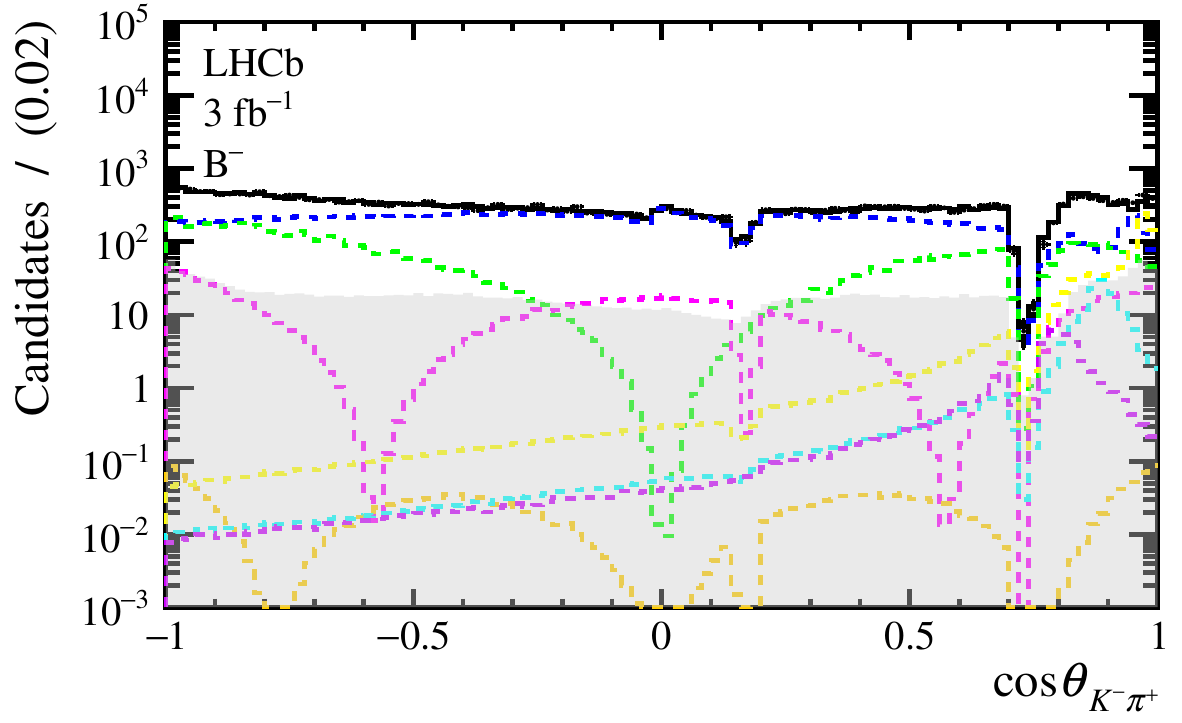}

    \includegraphics[width=0.5\linewidth]{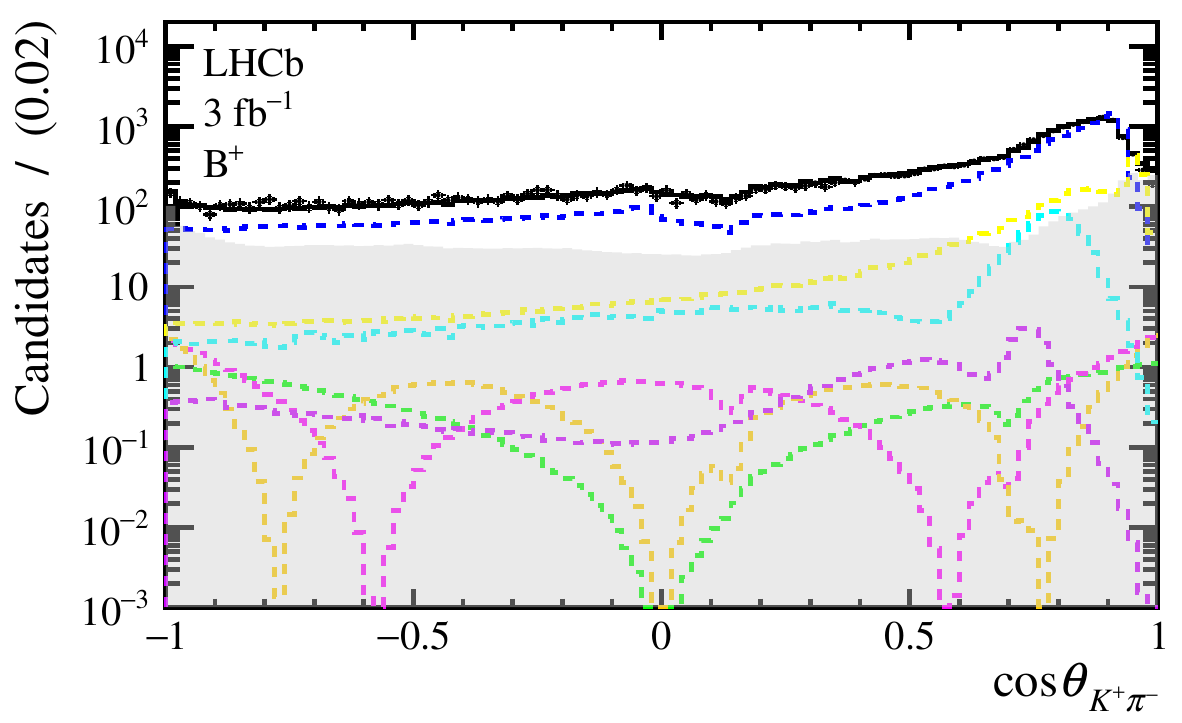}%
    \includegraphics[width=0.5\linewidth]{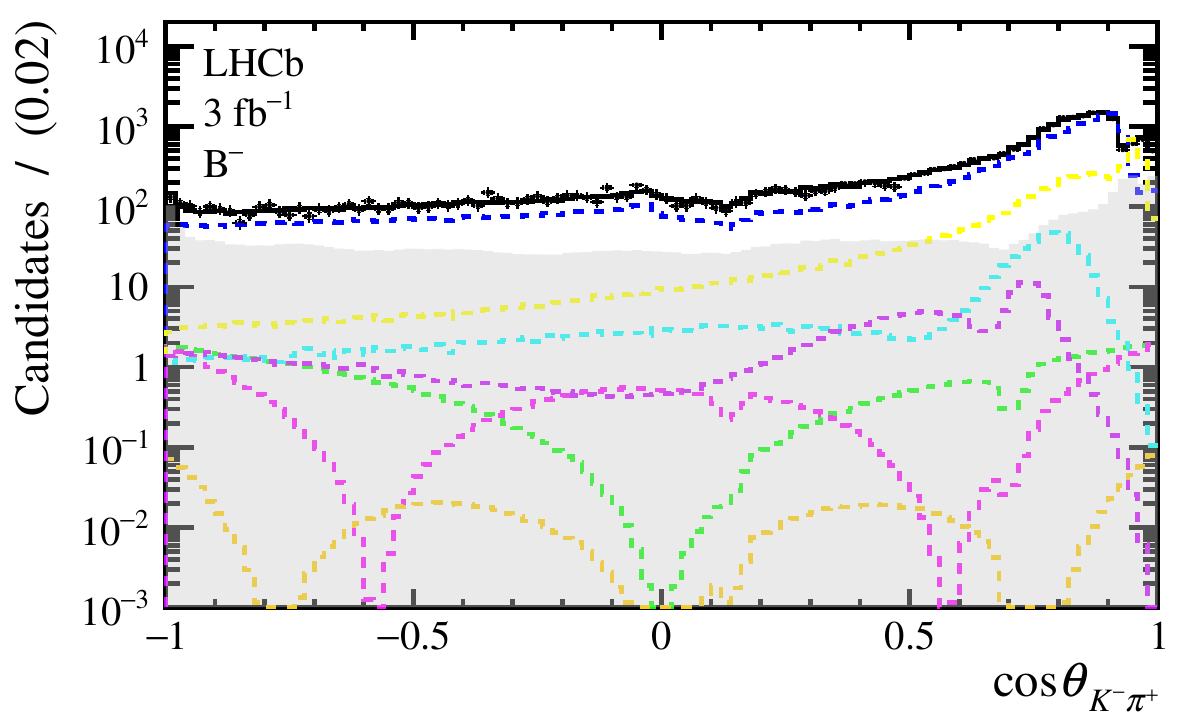}

    \caption{
        Fit projections of $\cos\theta_{K^+ \pi^-}$ for (left)~$B^+$ and (right)~$B^-$ (top)~below and (bottom)~above the open charm threshold in $m_{K^+ \pi^-}$.
    }
    \label{fig:app:ckpiiso}
\end{figure}

\begin{figure}[tb]
    \centering
    \includegraphics[width=0.5\linewidth]{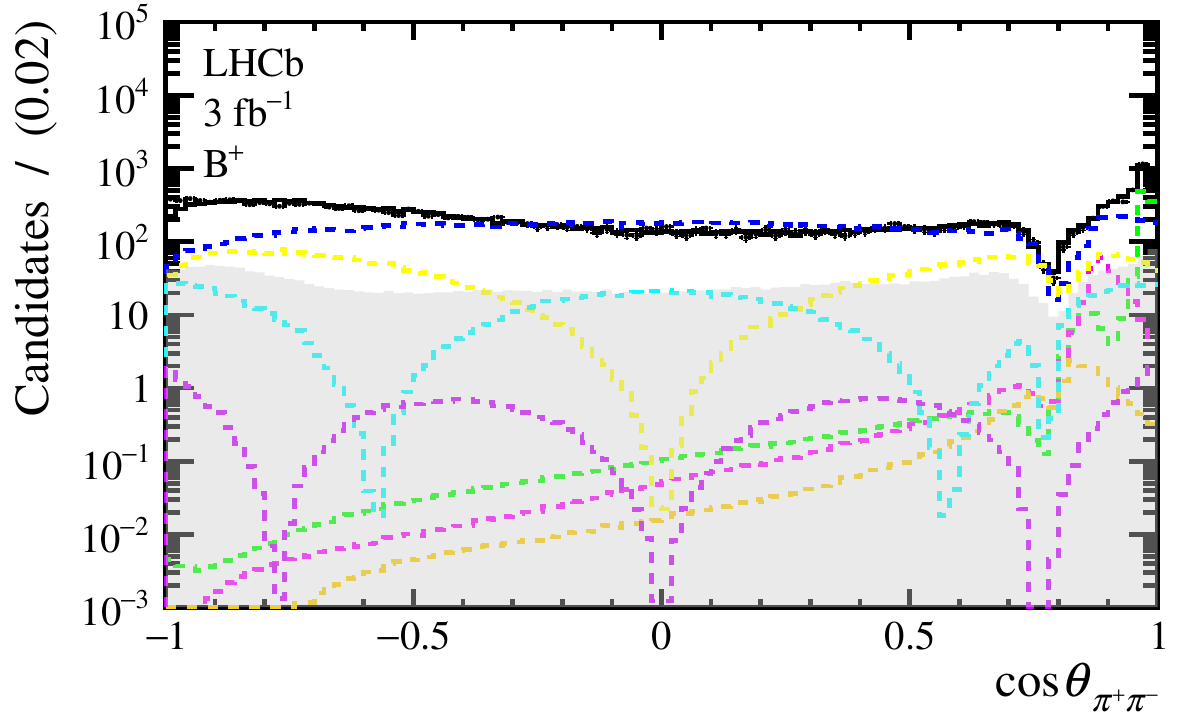}%
    \includegraphics[width=0.5\linewidth]{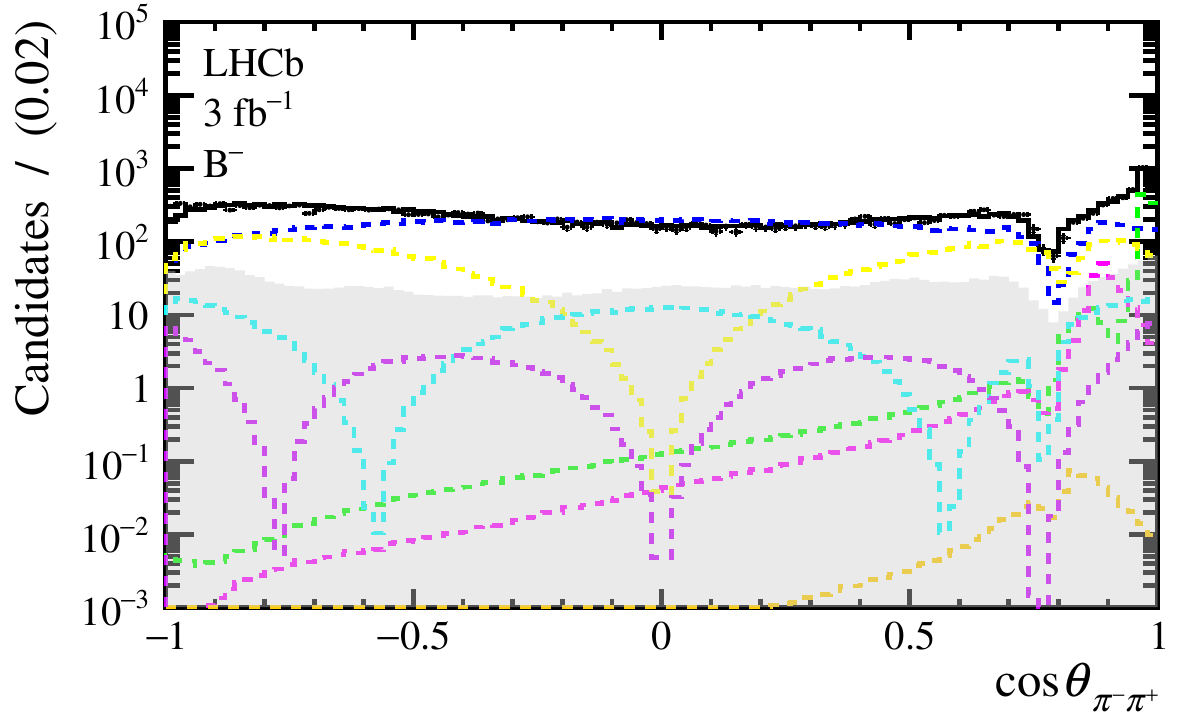}

    \includegraphics[width=0.5\linewidth]{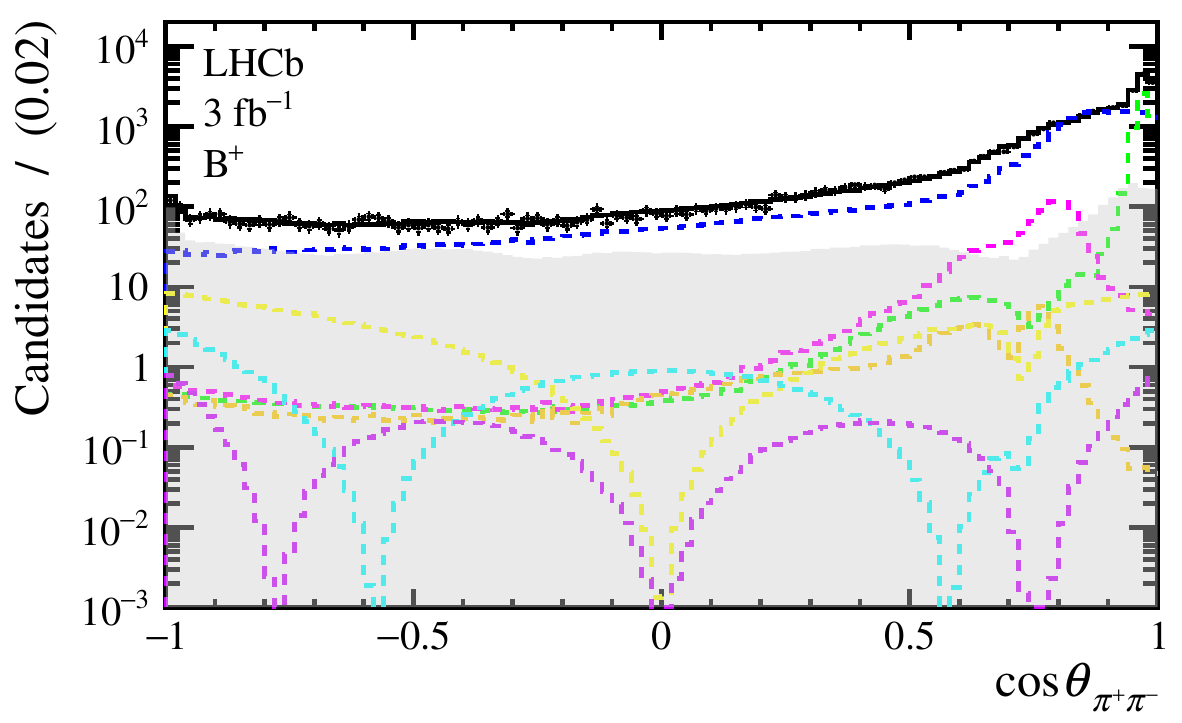}%
    \includegraphics[width=0.5\linewidth]{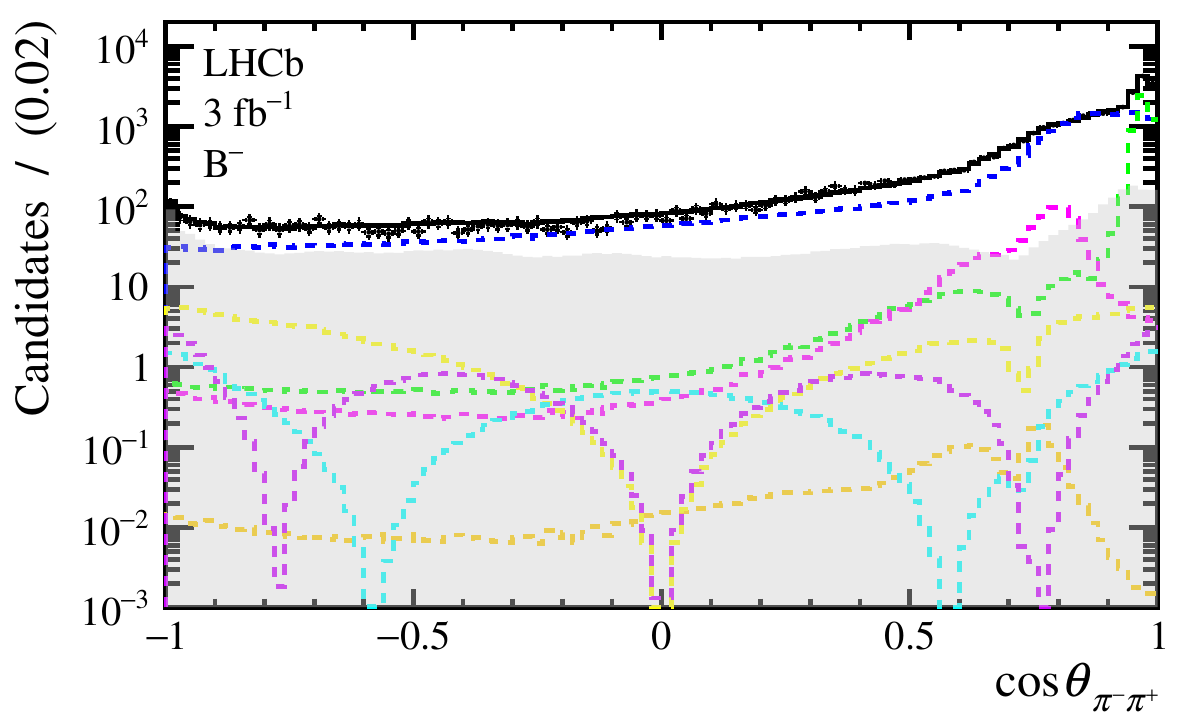}

     \caption{
        Fit projections of $\cos\theta_{\pi^+ \pi^-}$ for (left)~$B^+$ and (right)~$B^-$ (top)~below and (bottom)~above the open charm threshold in $m_{\pi^+ \pi^-}$.
    }
    \label{fig:app:cpipiiso}
\end{figure}

The lineshape parameters determined in the baseline model are given in Table~\ref{tab:app:iso}. 
Various projections of the data and fit results that serve to highlight the individual contributions of each partial wave are shown here with the colour legend displayed in Fig.~\ref{fig:app:legiso}. Mass projections with negative associated values of the cosine of the helicity angle, which removes the bulk of the reflections, are shown in Figs.~\ref{fig:app:mkpiiso} and~\ref{fig:app:mpipiiso}. The helicity structure enhanced in the regions below and above the open-charm threshold are given in Figs.~\ref{fig:app:ckpiiso} and~\ref{fig:app:cpipiiso} for the $\Kp\pim$ and $\pip\pim$ systems, respectively.

\begin{table}[!b]
    \centering
    \caption{\label{tab:app:iso}
        Lineshape parameters determined in the Isobar S-wave approach, where the uncertainties are statistical, systematic and model-induced, respectively.
    }

    \renewcommand{\arraystretch}{1.1}
    \begin{tabular}
    {l@{\hspace{0.25cm}}
     @{\hspace{0.25cm}}l@{\hspace{0.25cm}}
     @{\hspace{0.25cm}}r@{\hspace{0.1cm}}l
     @{\hspace{0.125cm}}l}\hline
    Component & Parameter & \multicolumn{2}{c}{Value} & \\ \hline

GLASS & $m_0$ & $1.461$ & $ \pm\,0.002 \pm 0.001 \pm 0.001$ & \!\gevcc \\
& $\Gamma_0$ & $0.216$ & $ \pm \,0.004 \pm 0.004 \pm 0.002$ & \!\gev \\
& $a$ & $6.900$ & $ \pm \,0.244 \pm \,0.771 \pm 0.700$ & $c/\!\gev$ \\
& $r$ & ${-}2.199$ & $ \pm \,0.051 \pm 0.172 \pm 0.082$ & $c/\!\gev$ \\
& $\phi_{\rm Nonres}$ & $3.373$ & $ \pm \,0.011 \pm 0.015 \pm 0.022$ & rad \\
& $\phi_{\rm ER}$ & $0.386$ & $ \pm\, 0.014 \pm 0.040 \pm 0.025$ & rad \\\hline
$f_0(980)$ & $m_0$ & $0.961$ & $ \pm \,0.002 \pm 0.002 \pm 0.002$ & \!\gevcc \\
& $g_{\pi\pi}$ & $0.117$ & $ \pm \,0.003 \pm 0.005 \pm 0.003$ & \!\gevgevcccc \\
& $g_{K\bar{K}/\pi\pi}$ & $2.571$ & $ \pm 0.123 \pm 0.182 \pm 0.178$ & \\

    \hline
    \end{tabular}
\end{table}

\clearpage

\section{Additional K-matrix results}
\label{app:kmatrix}

\begin{figure}[!b]
    \centering
    \includegraphics[width=0.2\linewidth]{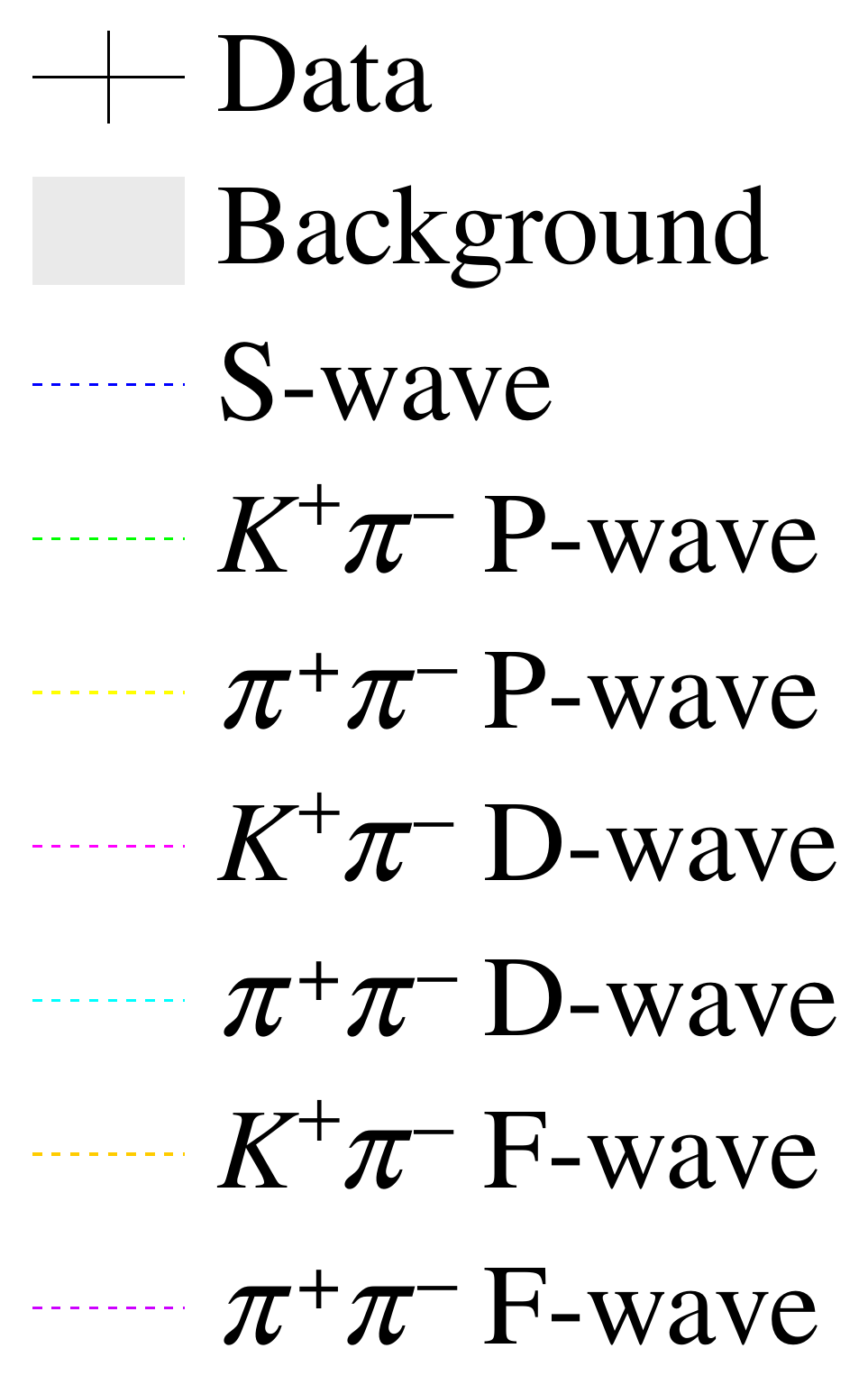}

    \caption{
        Colour legend of each partial wave shown in the subsequent figures of this section.
    }
    \label{fig:app:legkm}
\end{figure}

\begin{figure}[tb]
    \centering
    \includegraphics[width=0.5\linewidth]{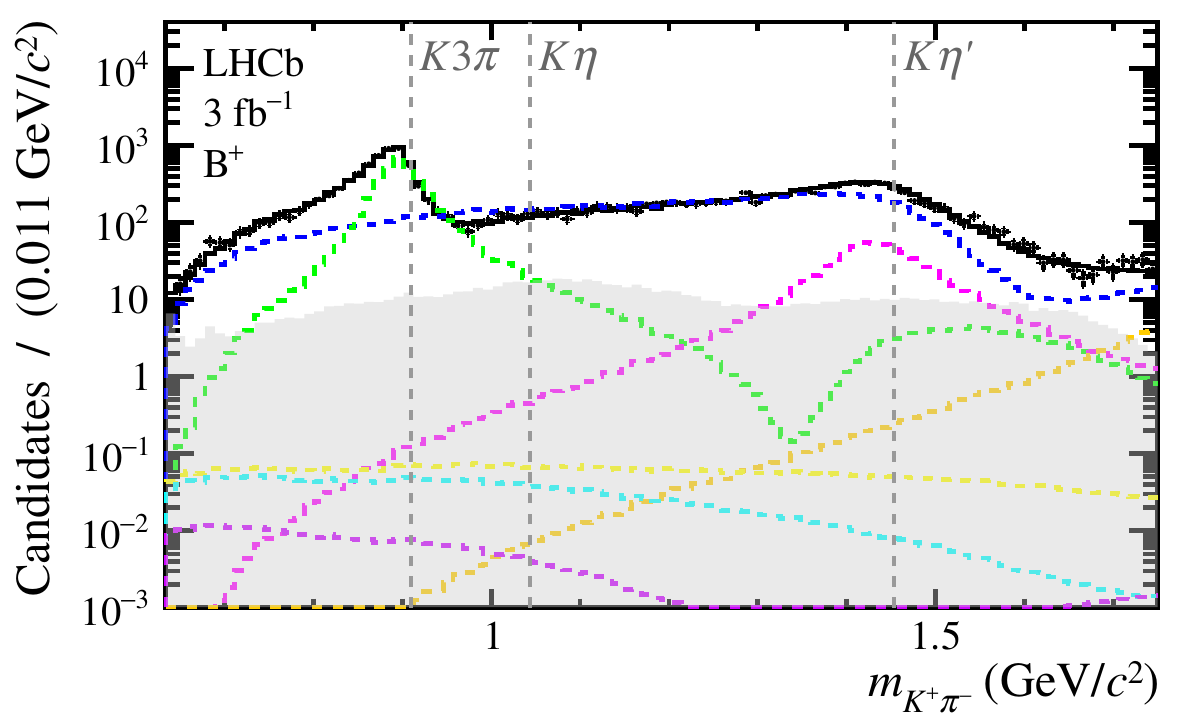}%
    \includegraphics[width=0.5\linewidth]{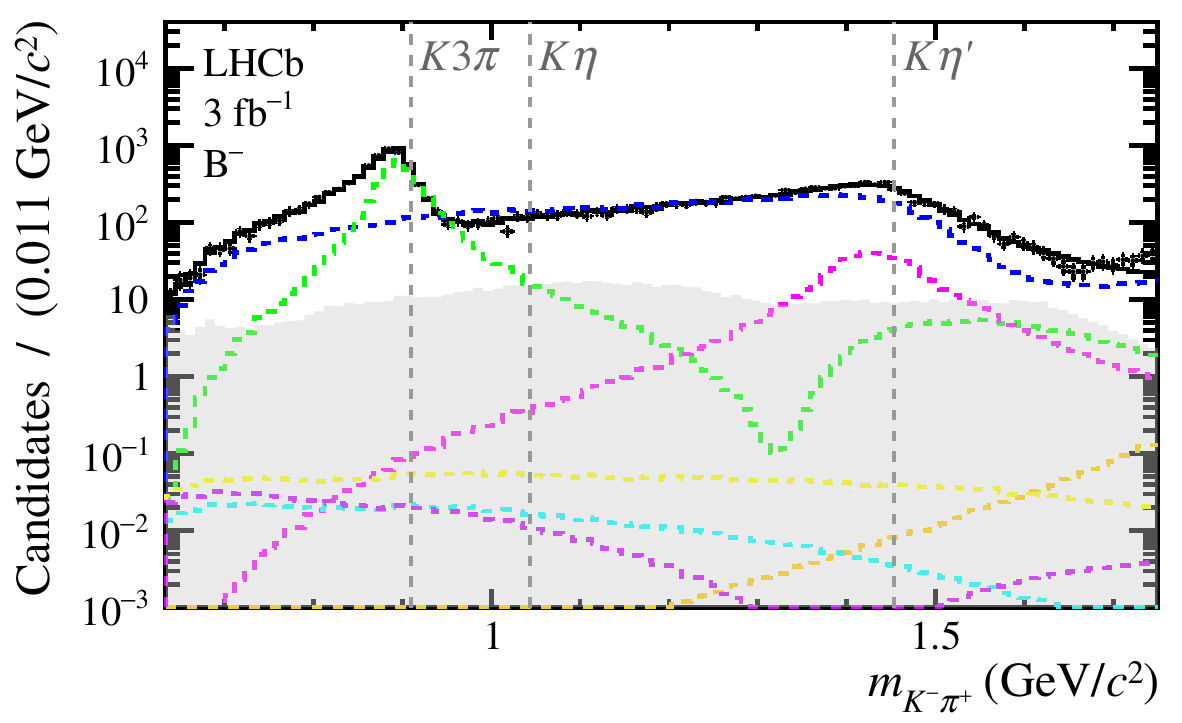}

    \includegraphics[width=0.5\linewidth]{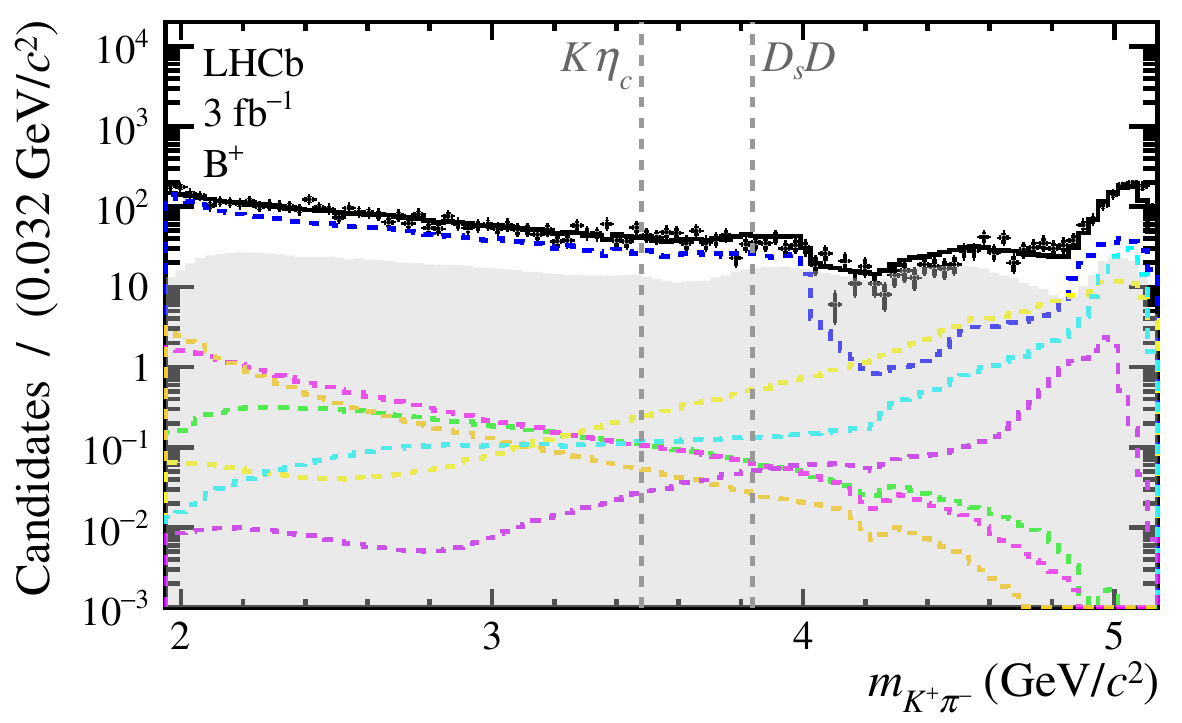}%
    \includegraphics[width=0.5\linewidth]{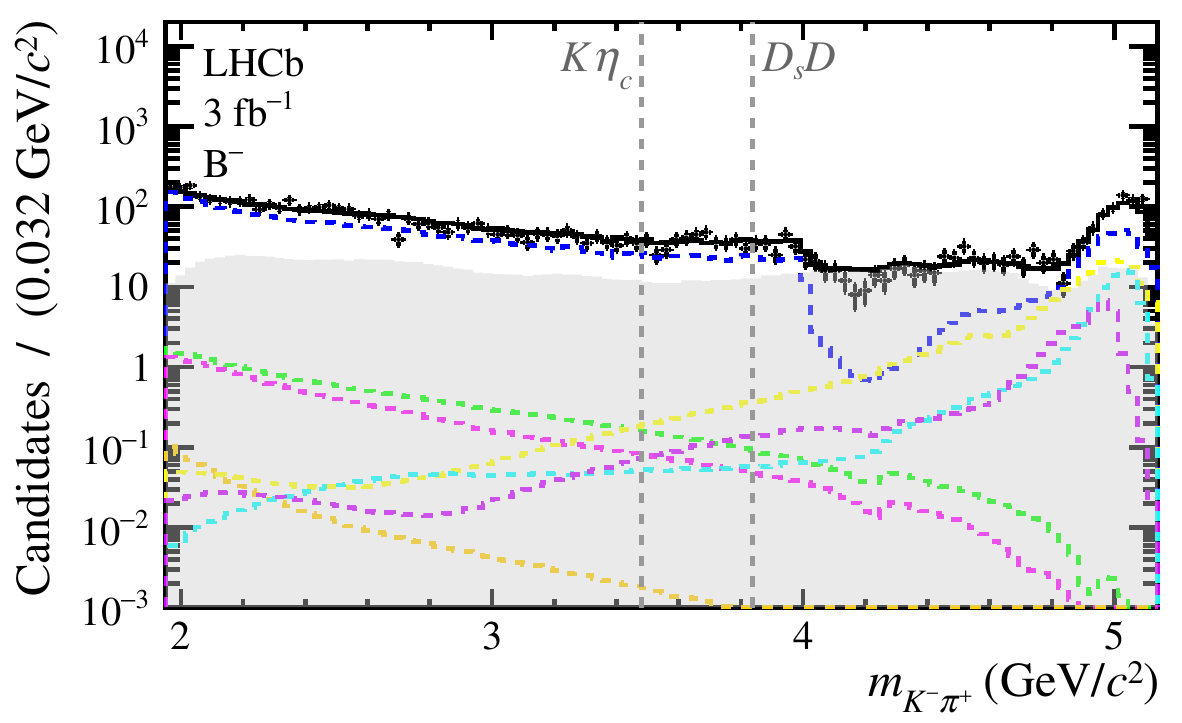}

    \caption{
        Fit projections of $m_{K^+ \pi^-}$ for (left)~$B^+$ and (right)~$B^-$ (top)~below and (bottom)~above the open charm threshold, enhanced with $\cos\theta_{K^+ \pi^-} < 0$. 
        Vertical dashed lines correspond to the opening thresholds of the indicated coupled channels.
    }
    \label{fig:app:mkpikm}
\end{figure}

\begin{figure}[tb]
    \centering
    \includegraphics[width=0.5\linewidth]{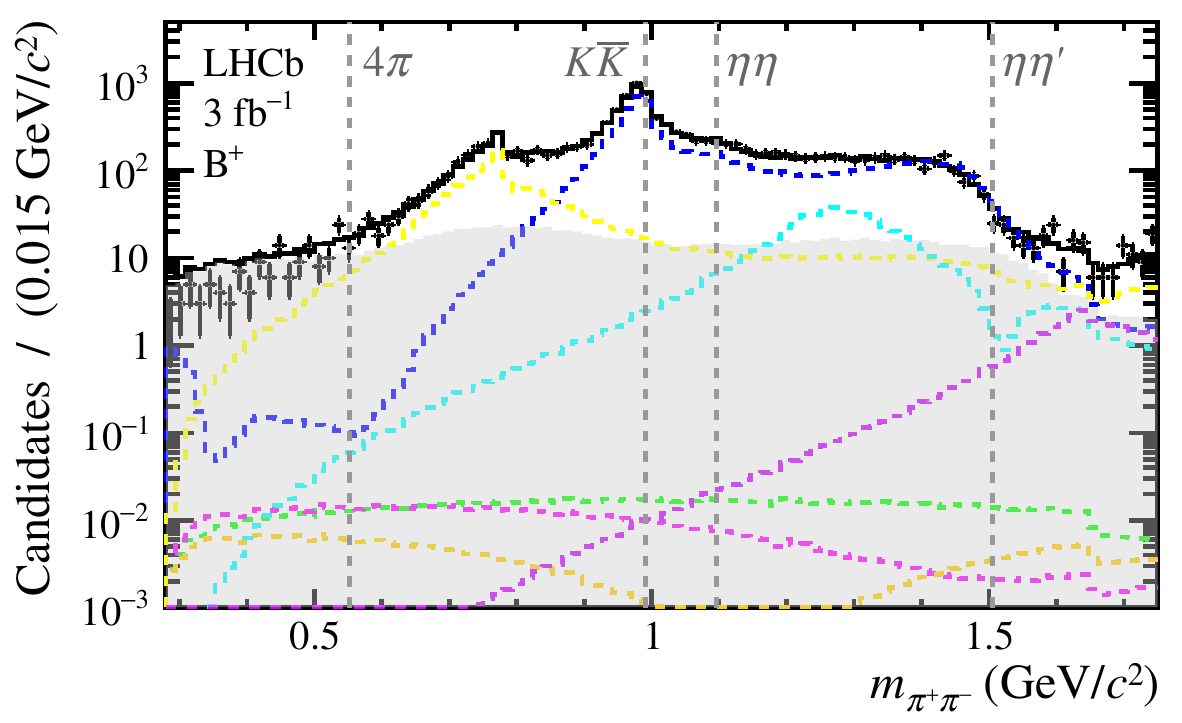}%
    \includegraphics[width=0.5\linewidth]{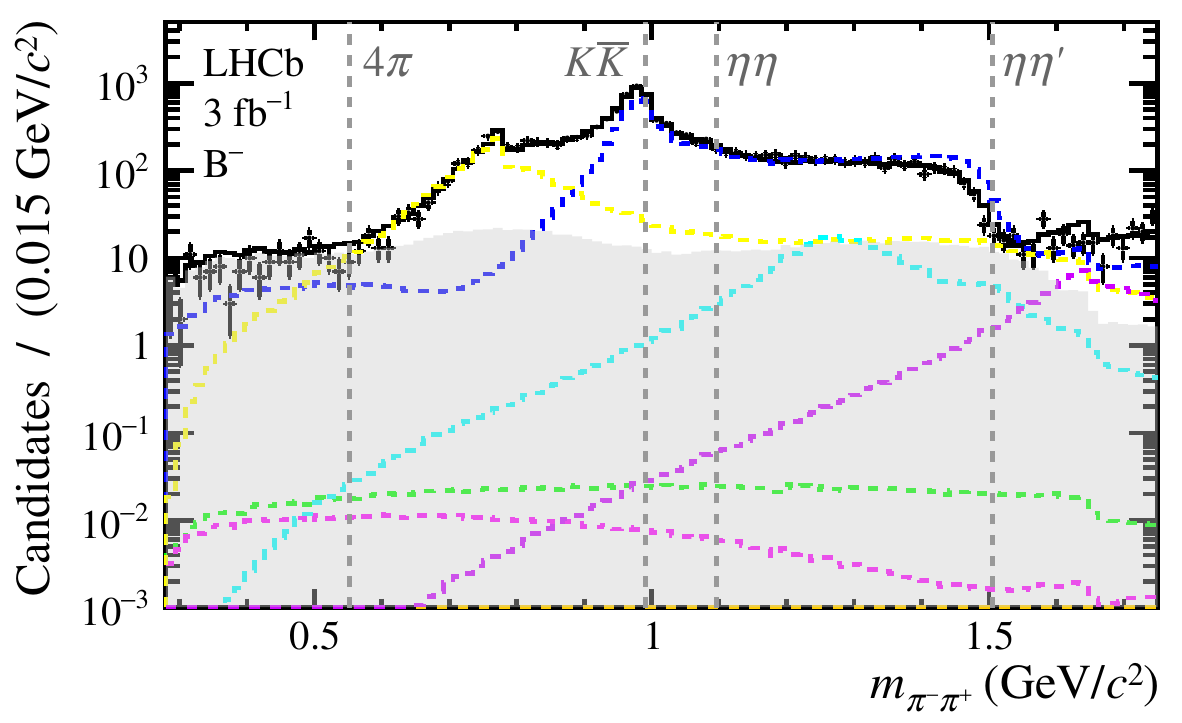}
    
    \includegraphics[width=0.5\linewidth]{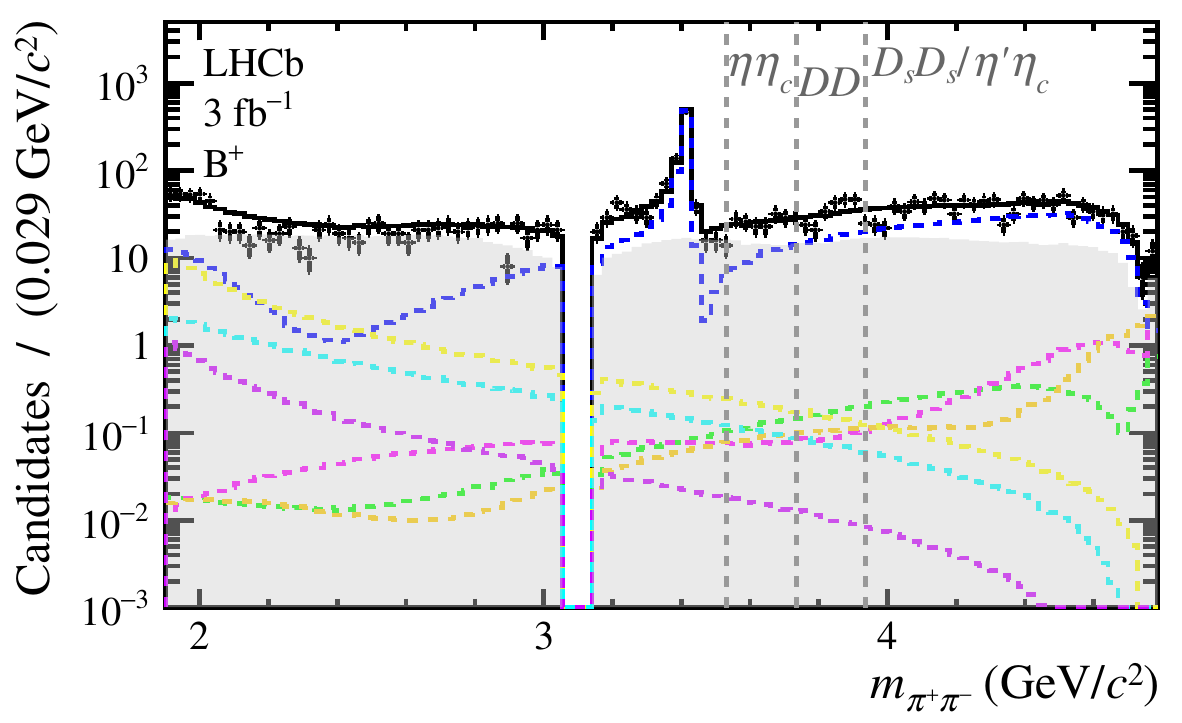}%
    \includegraphics[width=0.5\linewidth]{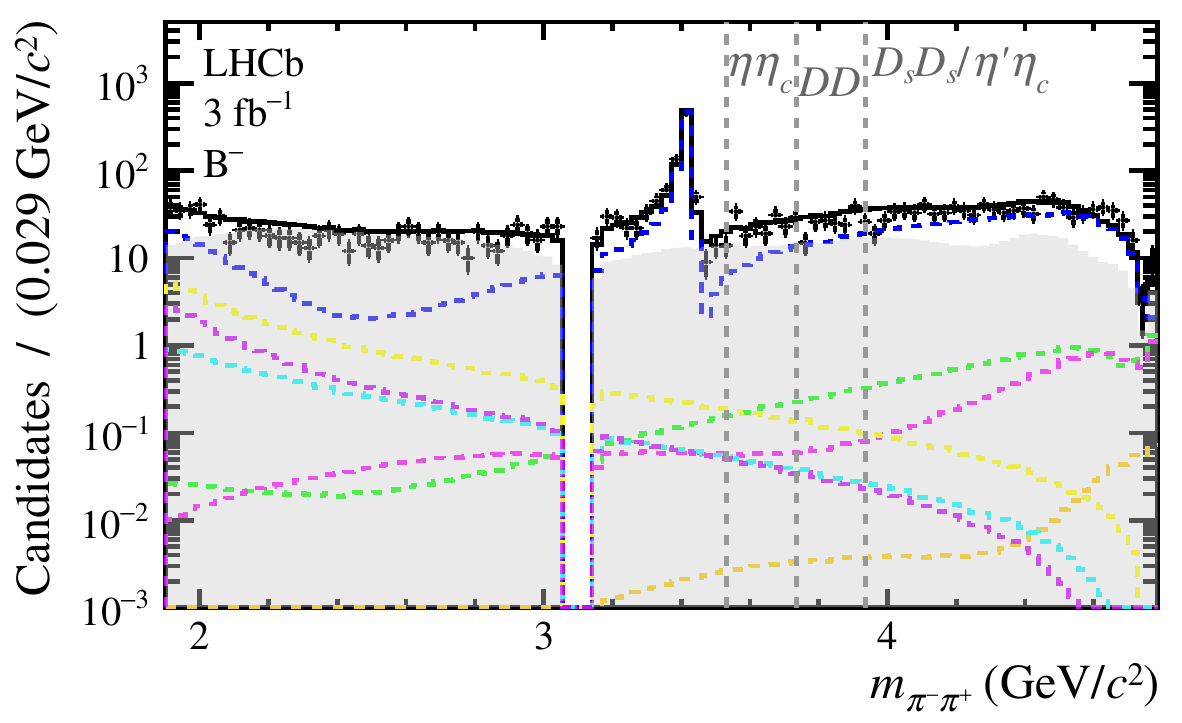}

    \caption{
        Fit projections of $m_{\pi^+ \pi^-}$ for (left)~$B^+$ and (right)~$B^-$ (top)~below and (bottom)~above the open charm threshold, enhanced with $\cos\theta_{\pi^+ \pi^-} < 0$. 
        Vertical dashed lines correspond to the opening thresholds of the indicated coupled channels.
    }
    \label{fig:app:mpipikm}
\end{figure}

\begin{figure}[tb]
    \centering
    \includegraphics[width=0.5\linewidth]{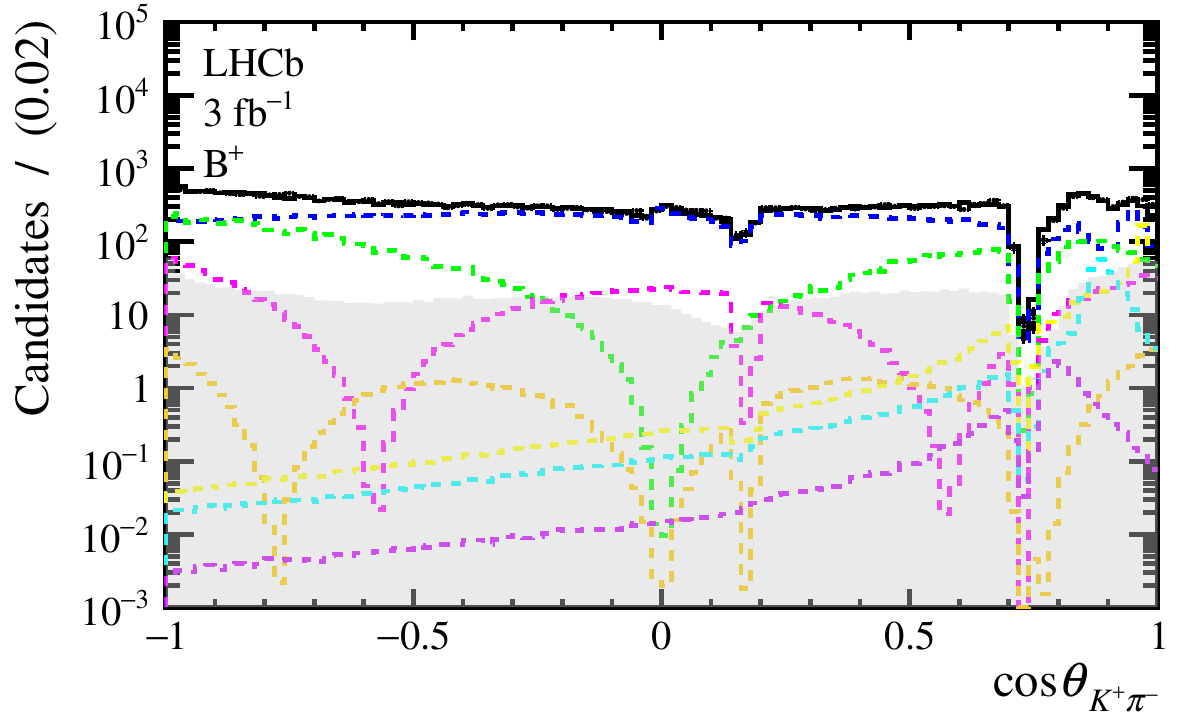}%
    \includegraphics[width=0.5\linewidth]{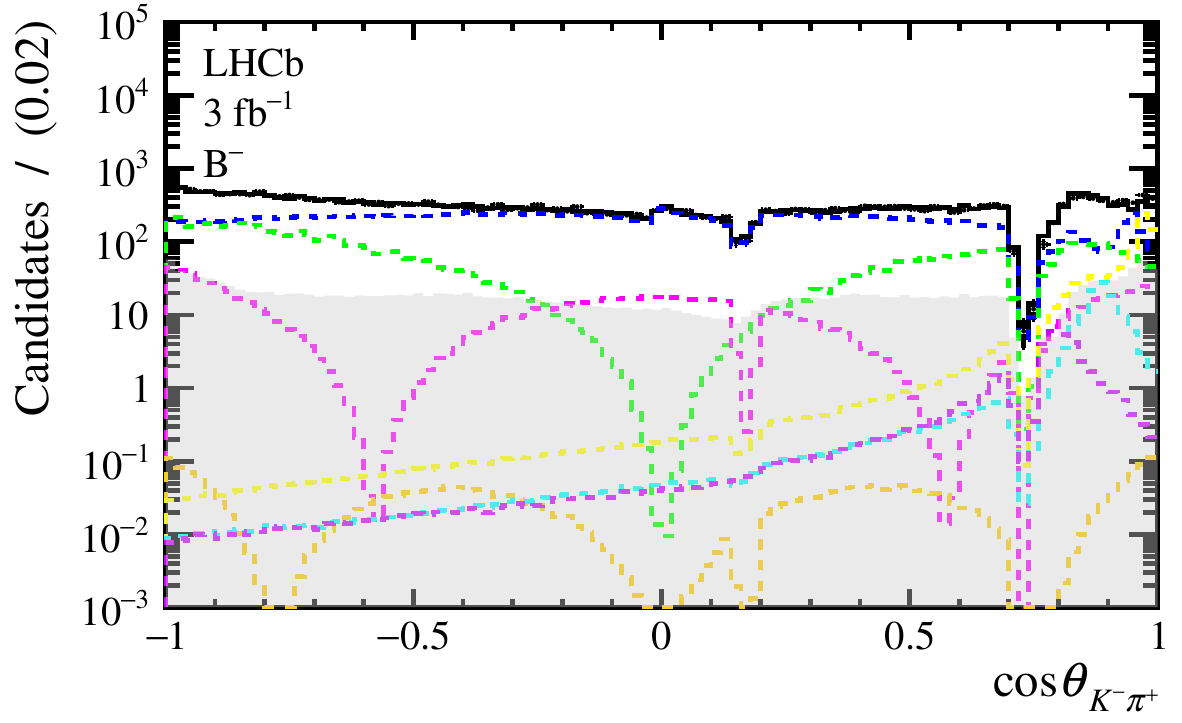}

    \includegraphics[width=0.5\linewidth]{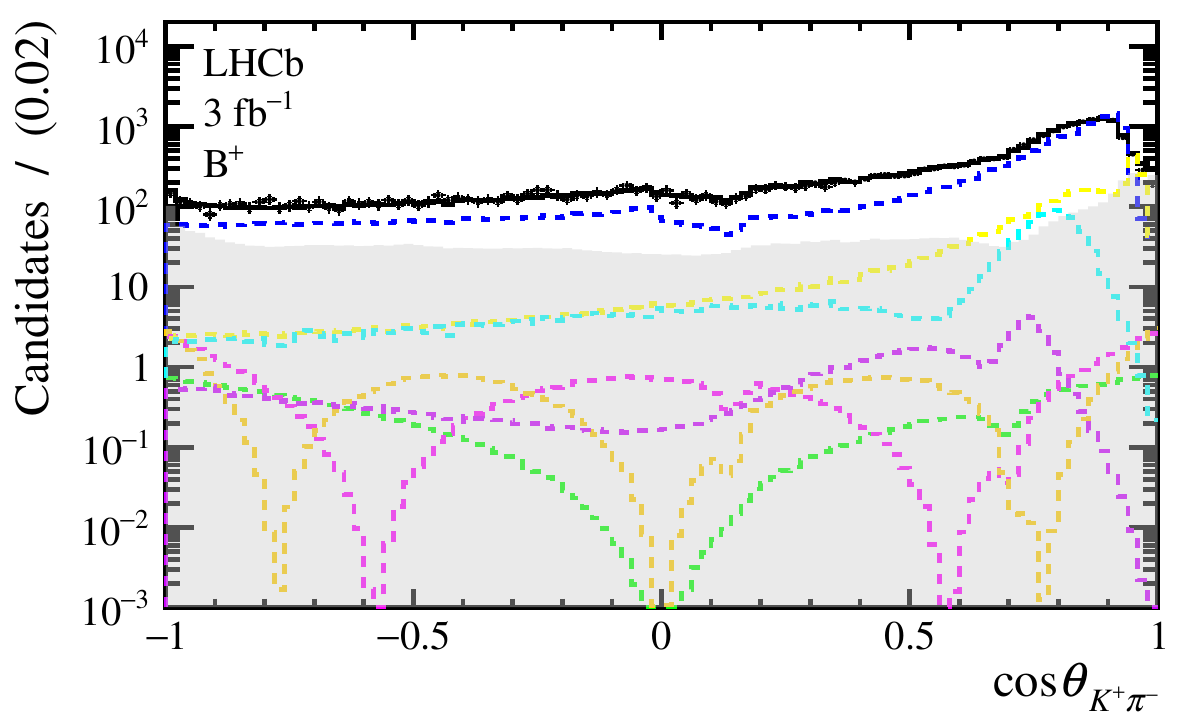}%
    \includegraphics[width=0.5\linewidth]{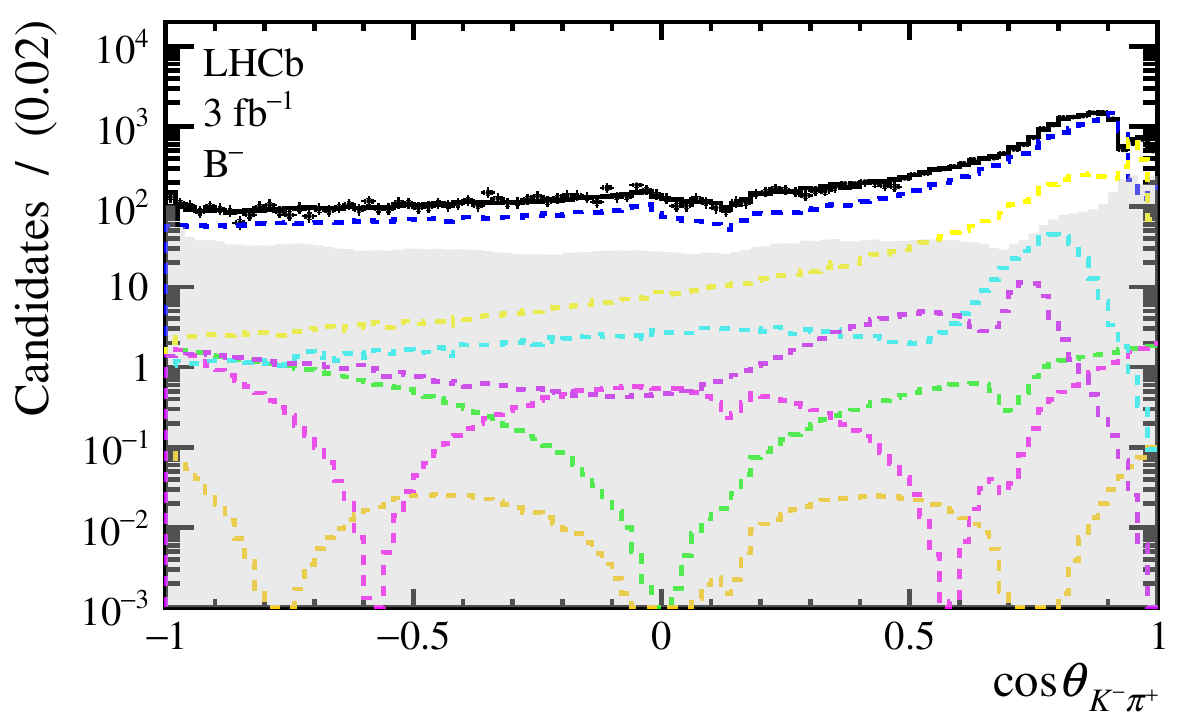}

    \caption{
        Fit projections of $\cos\theta_{K^+ \pi^-}$ for (left)~$B^+$ and (right)~$B^-$ (top)~below and (bottom)~above the open charm threshold in $m_{K^+ \pi^-}$.
    }
    \label{fig:app:ckpikm}
\end{figure}

\begin{figure}[tb]
    \centering
    \includegraphics[width=0.5\linewidth]{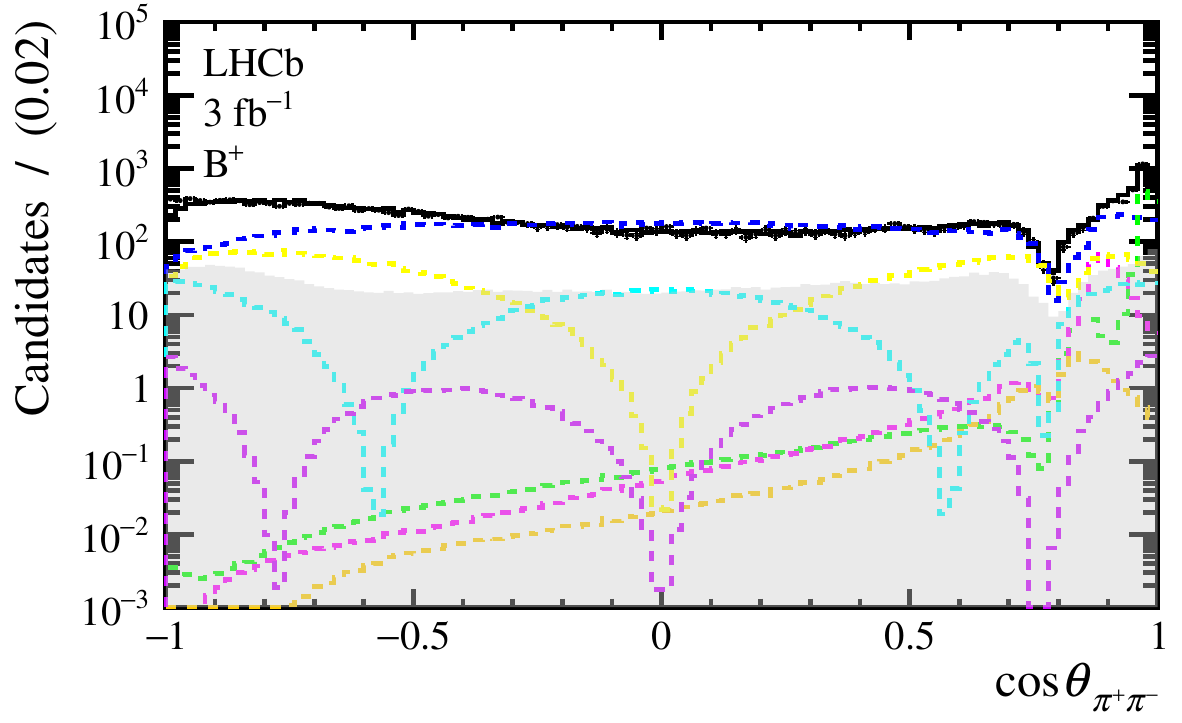}%
    \includegraphics[width=0.5\linewidth]{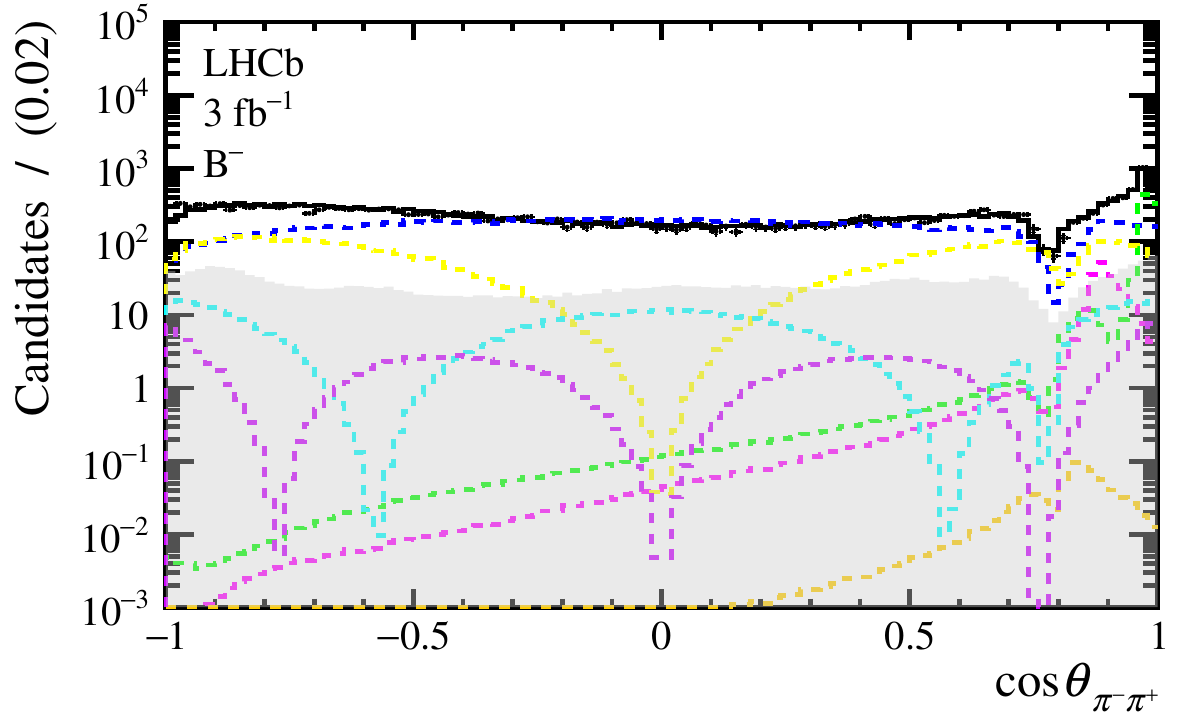}

    \includegraphics[width=0.5\linewidth]{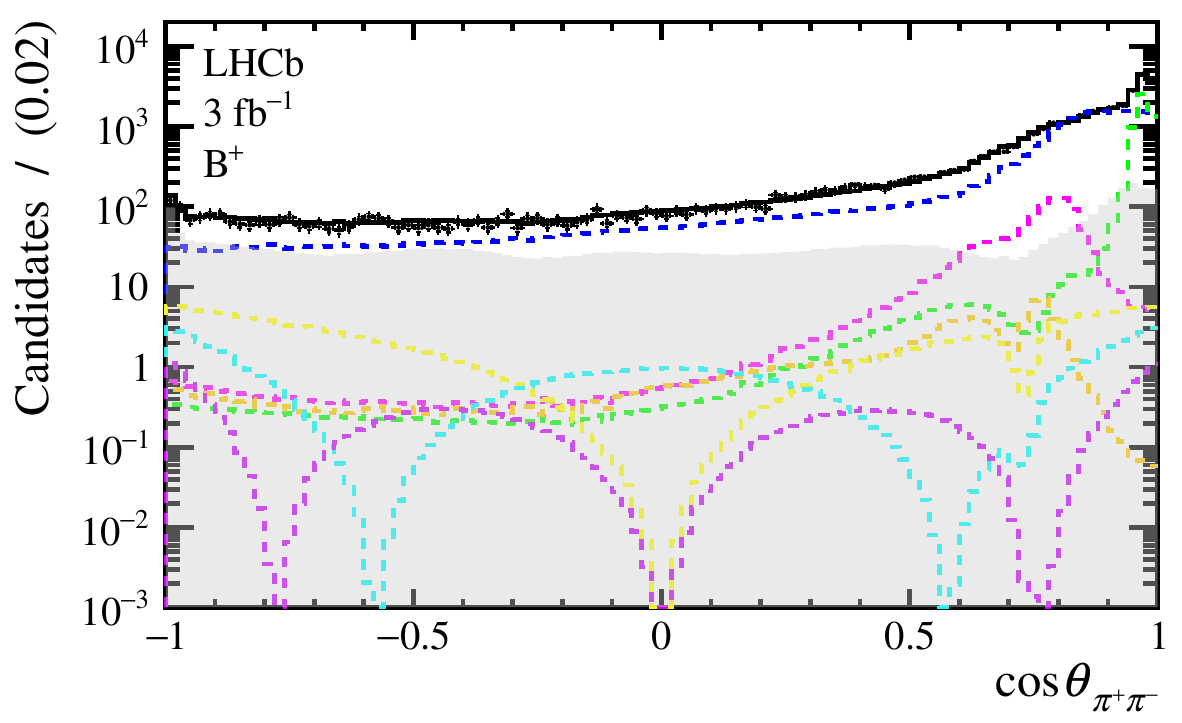}%
    \includegraphics[width=0.5\linewidth]{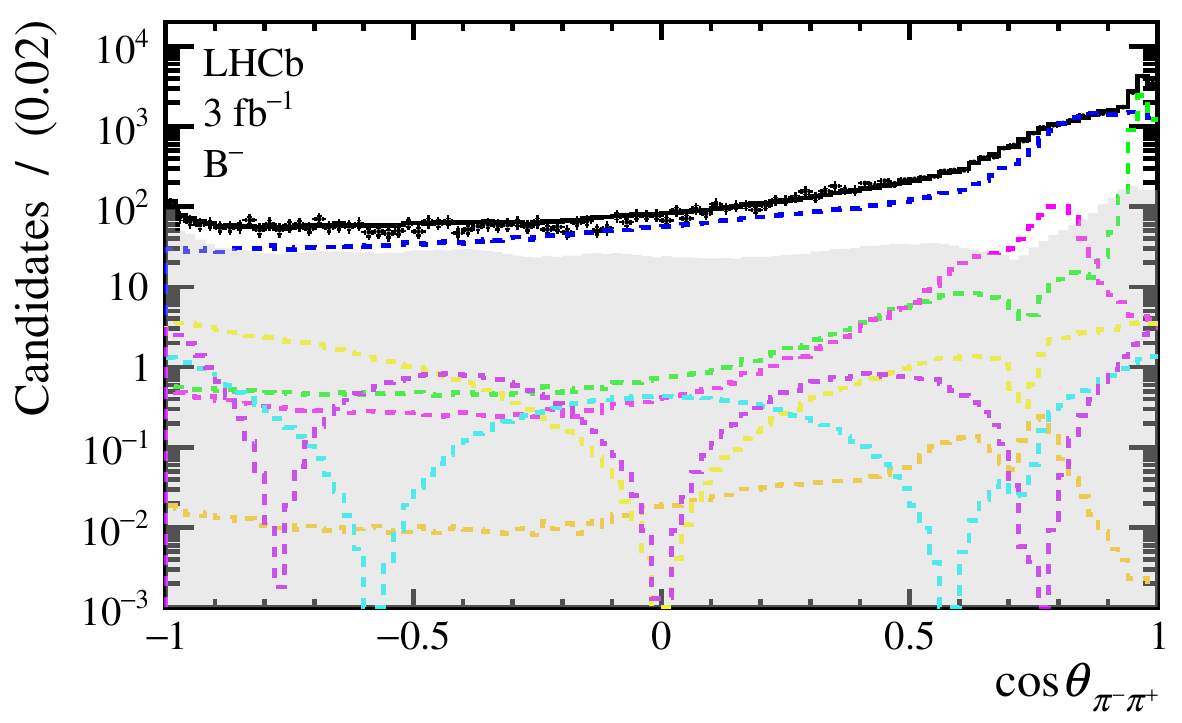}

    \caption{
        Fit projections of $\cos\theta_{\pi^+ \pi^-}$ for (left)~$B^+$ and (right)~$B^-$ (top)~below and (bottom)~above the open charm threshold in $m_{\pi^+ \pi^-}$.
    }
    \label{fig:app:cpipikm}
\end{figure}

The lineshape parameters determined in the baseline model are given in Table~\ref{tab:app:km}. 
Various projections of the data and fit results that serve to highlight the individual contributions of each partial wave are shown here with the colour legend displayed in Fig.~\ref{fig:app:legkm}. Mass projections with negative associated values of the cosine of the helicity angle, which removes the bulk of the reflections, are shown in Figs.~\ref{fig:app:mkpikm} and~\ref{fig:app:mpipikm}.
The helicity structure enhanced in the regions below and above the open-charm threshold are given in Figs.~\ref{fig:app:ckpikm} and~\ref{fig:app:cpipikm} for the $\Kp\pim$ and $\pip\pim$ systems, respectively.

\begin{table}[b]
    \centering
    \caption{\label{tab:app:km}
        Lineshape parameters determined in the K-matrix S-wave approach, where the uncertainties are statistical, systematic and model-induced, respectively.
    }

    \renewcommand{\arraystretch}{1.1}
    \begin{tabular}
    {l@{\hspace{0.25cm}}
     @{\hspace{0.25cm}}l@{\hspace{0.25cm}}
     @{\hspace{0.25cm}}r@{\hspace{0.1cm}}l
     @{\hspace{0.125cm}}l}\hline
    Component & Parameter & \multicolumn{2}{c}{Value} & \\ \hline
    
GLASS & $m_0$ & 1.460 & $\pm \,0.002 \pm 0.002 \pm 0.001$ & \!\gevcc \\
& $\Gamma_0$ & 0.217 & $\pm\, 0.004 \pm 0.004 \pm 0.001$ & \!\gev \\
& $a$ & 6.967 & $\pm\, 0.234 \pm 0.722 \pm 0.135$ & $c/\!\gev$ \\
& $r$ & $-2.180$ & $\pm\, 0.036 \pm 0.118 \pm 0.077$ & $c/\!\gev$ \\
& $\phi_{\rm Nonres}$ & 3.483 & $\pm\, 0.008 \pm 0.036 \pm 0.006$ & rad \\
& $\phi_{\rm ER}$ & 0.592 & $\pm \,0.009 \pm 0.034 \pm 0.015$ & rad \\

    \hline
    \end{tabular}
\end{table}

\clearpage

\section{Additional QMI results}
\label{app:qmi}

\begin{figure}[b]
    \centering
    \includegraphics[width=0.2\linewidth]{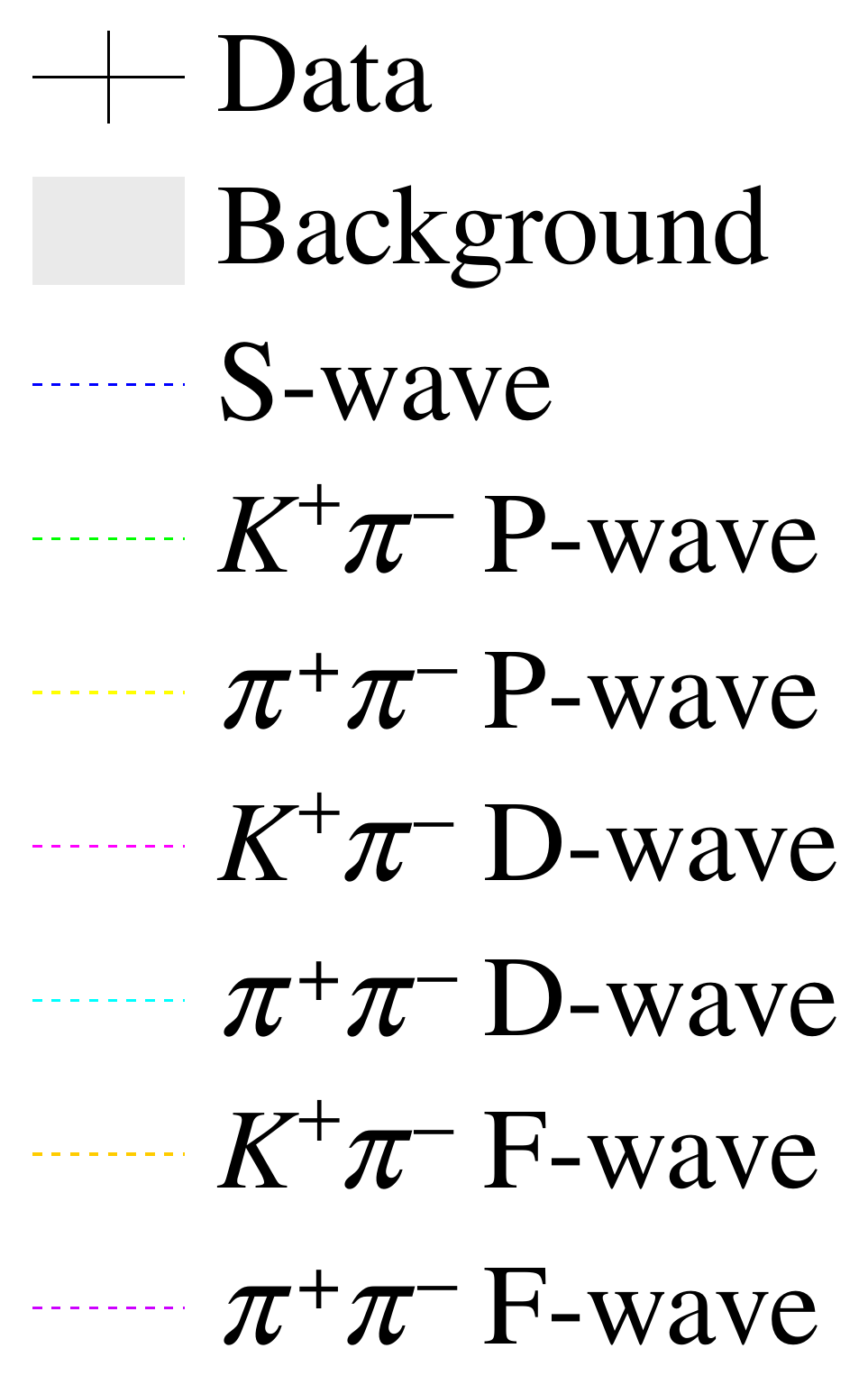}

    \caption{
        Colour legend of each partial wave shown in the subsequent figures of this section.
    }
    \label{fig:app:legqmi}
\end{figure}

\begin{figure}[tb]
    \centering
    \includegraphics[width=0.5\linewidth]{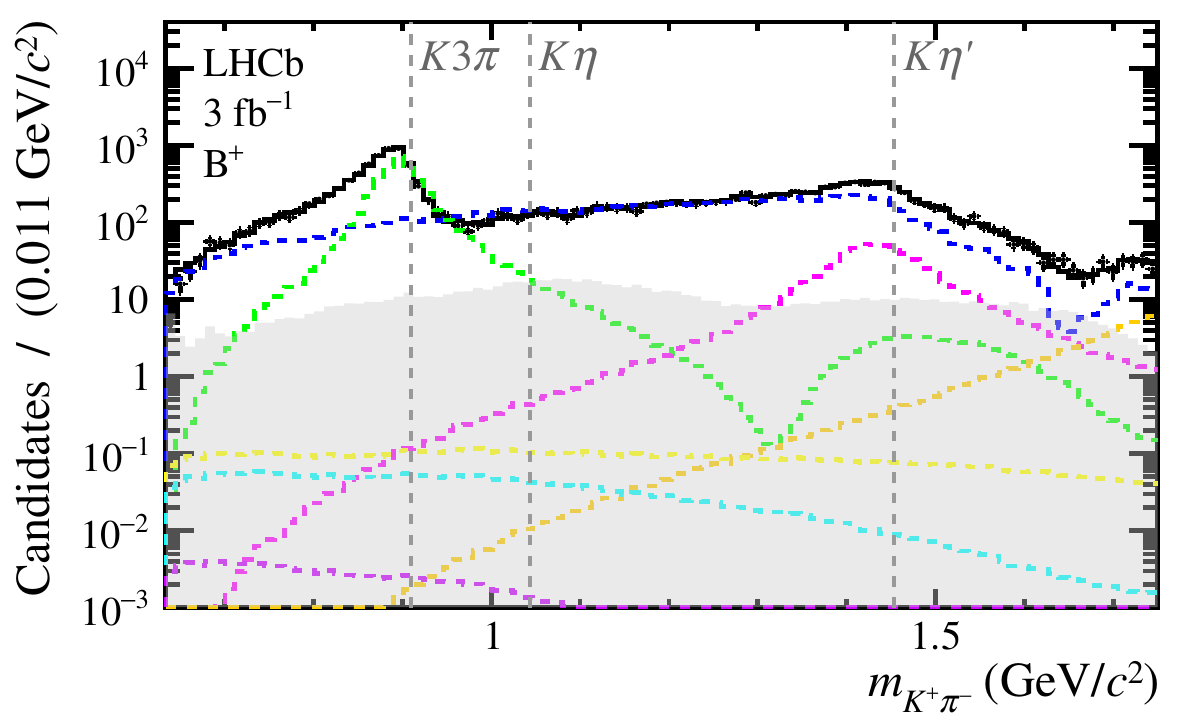}%
    \includegraphics[width=0.5\linewidth]{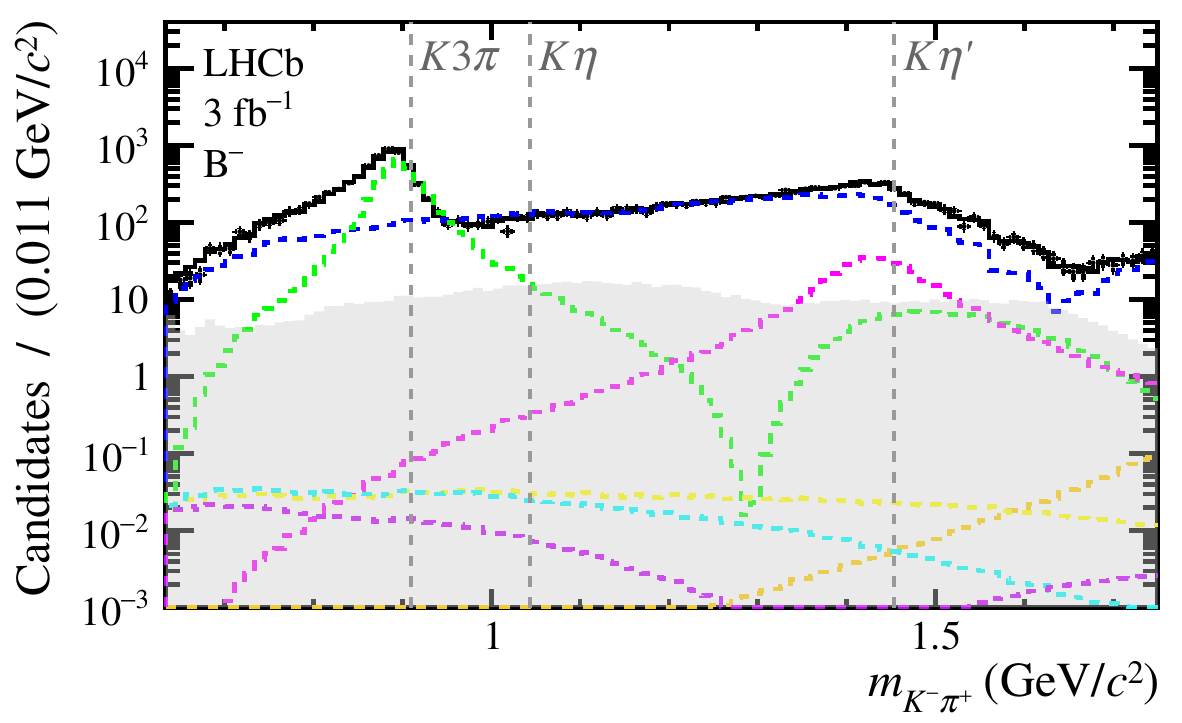}

    \includegraphics[width=0.5\linewidth]{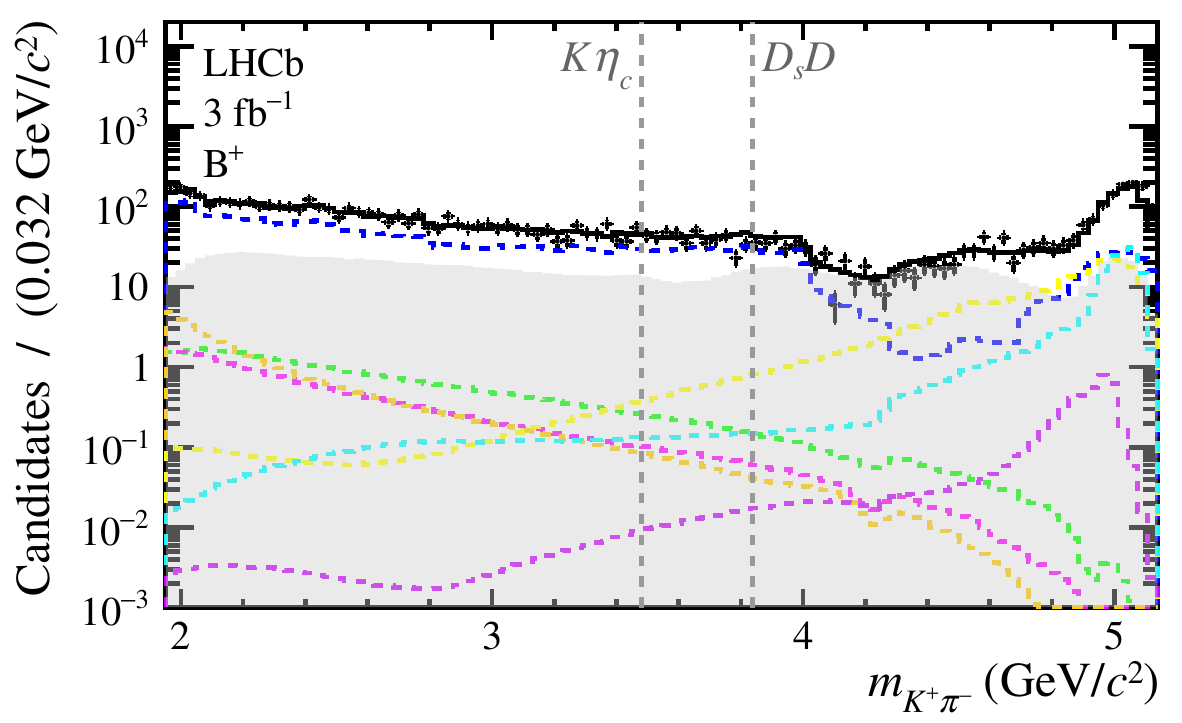}%
    \includegraphics[width=0.5\linewidth]{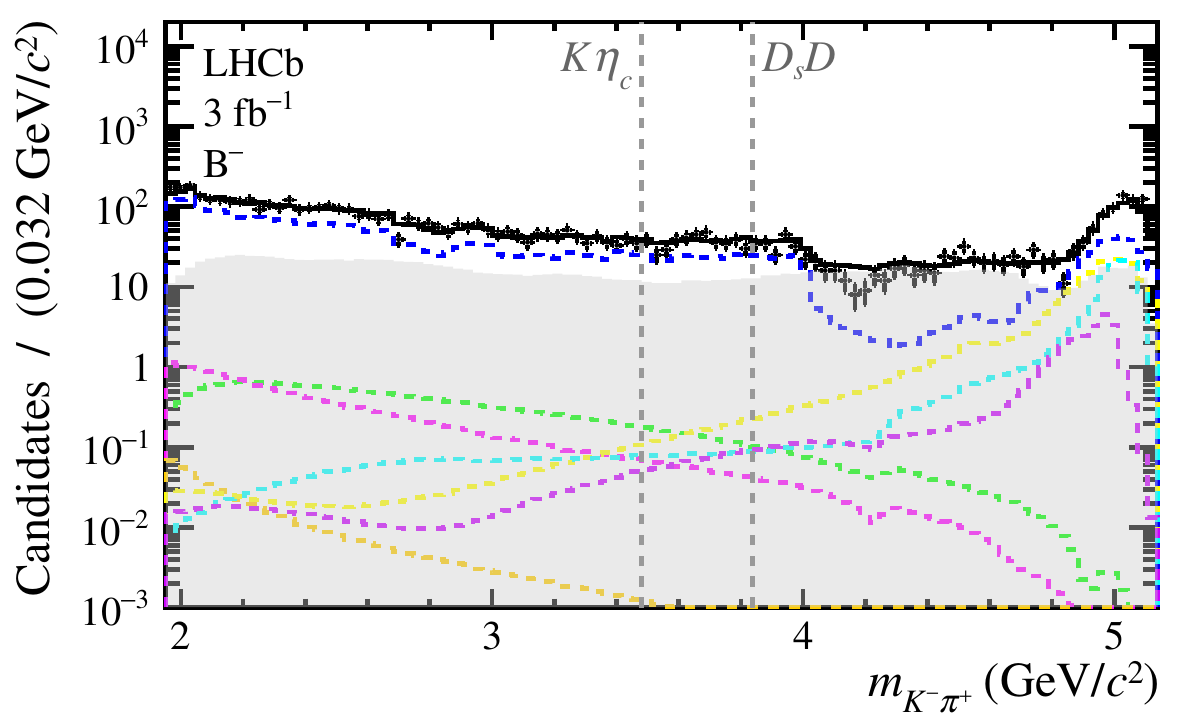}

    \caption{
        Fit projections of $m_{K^+ \pi^-}$ for (left)~$B^+$ and (right)~$B^-$ (top)~below and (bottom)~above the open charm threshold, enhanced with $\cos\theta_{K^+ \pi^-} < 0$. 
        Vertical dashed lines correspond to the opening thresholds of the indicated coupled channels.
    }
    \label{fig:app:mkpiqmi}
\end{figure}

\begin{figure}[tb]
    \centering
    \includegraphics[width=0.5\linewidth]{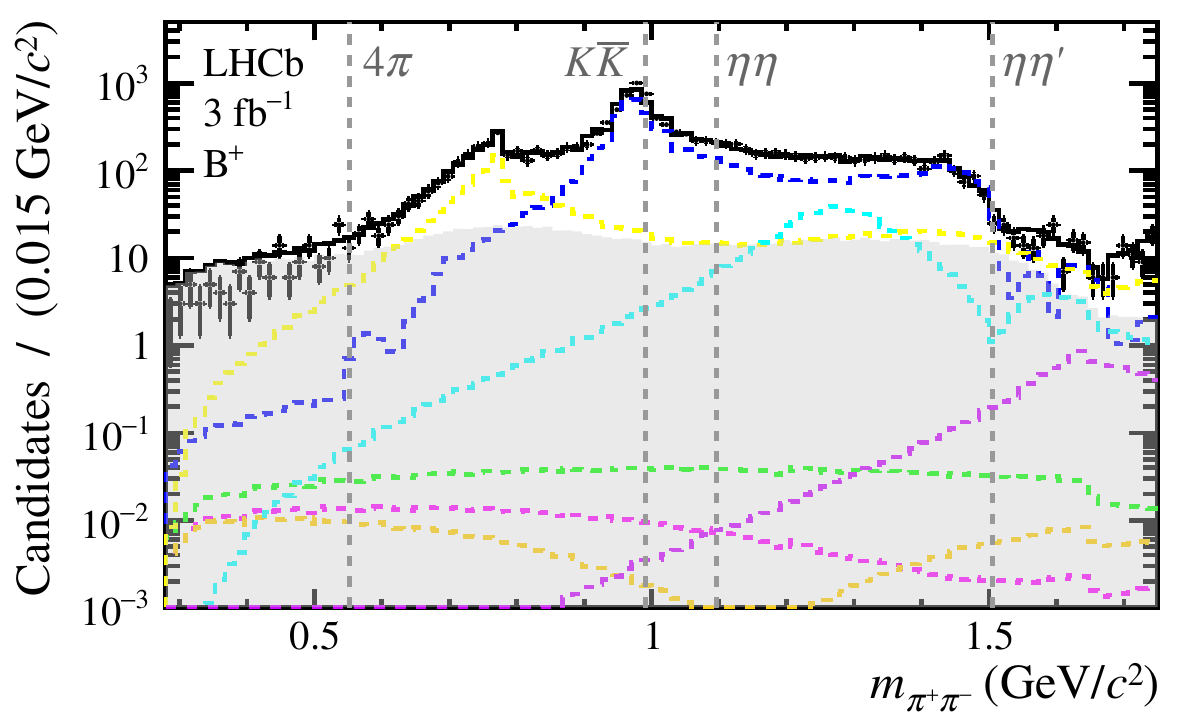}%
    \includegraphics[width=0.5\linewidth]{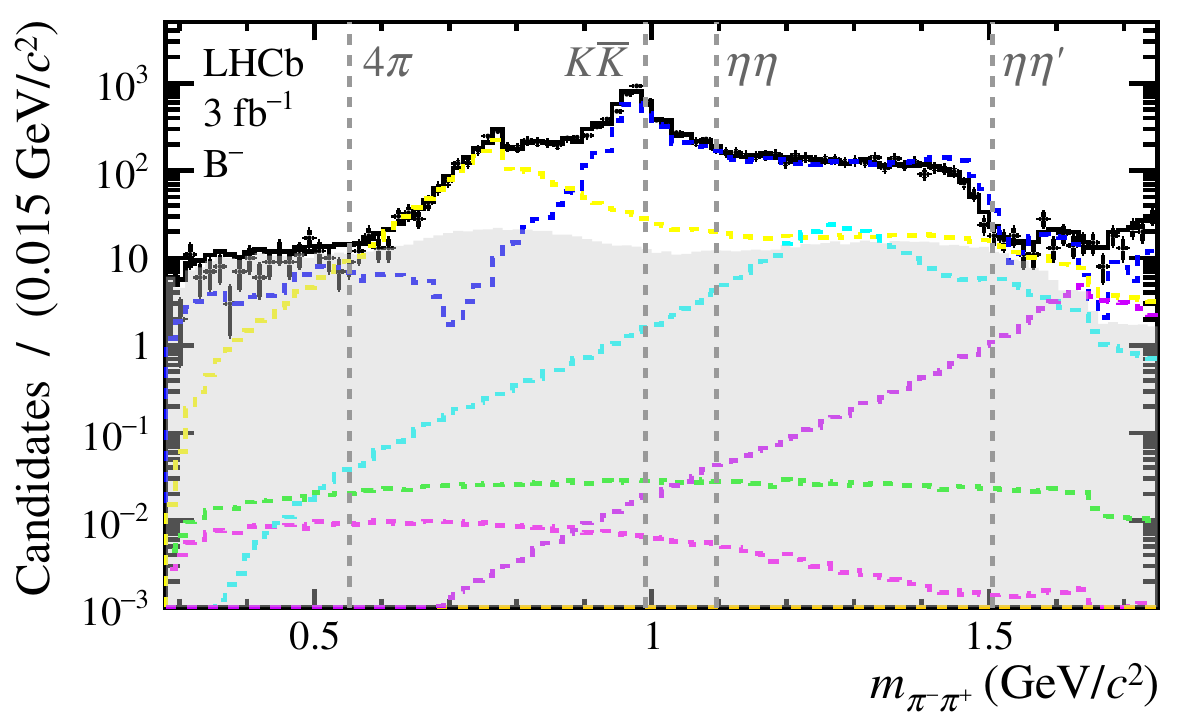}
    
    \includegraphics[width=0.5\linewidth]{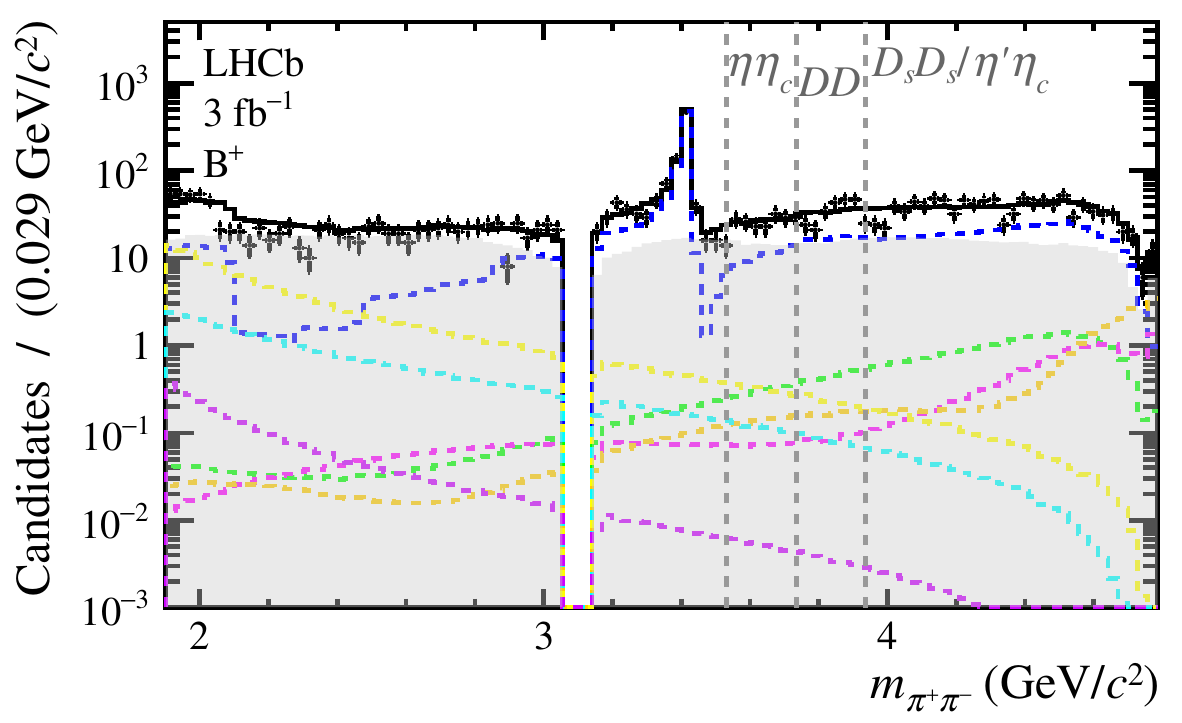}%
    \includegraphics[width=0.5\linewidth]{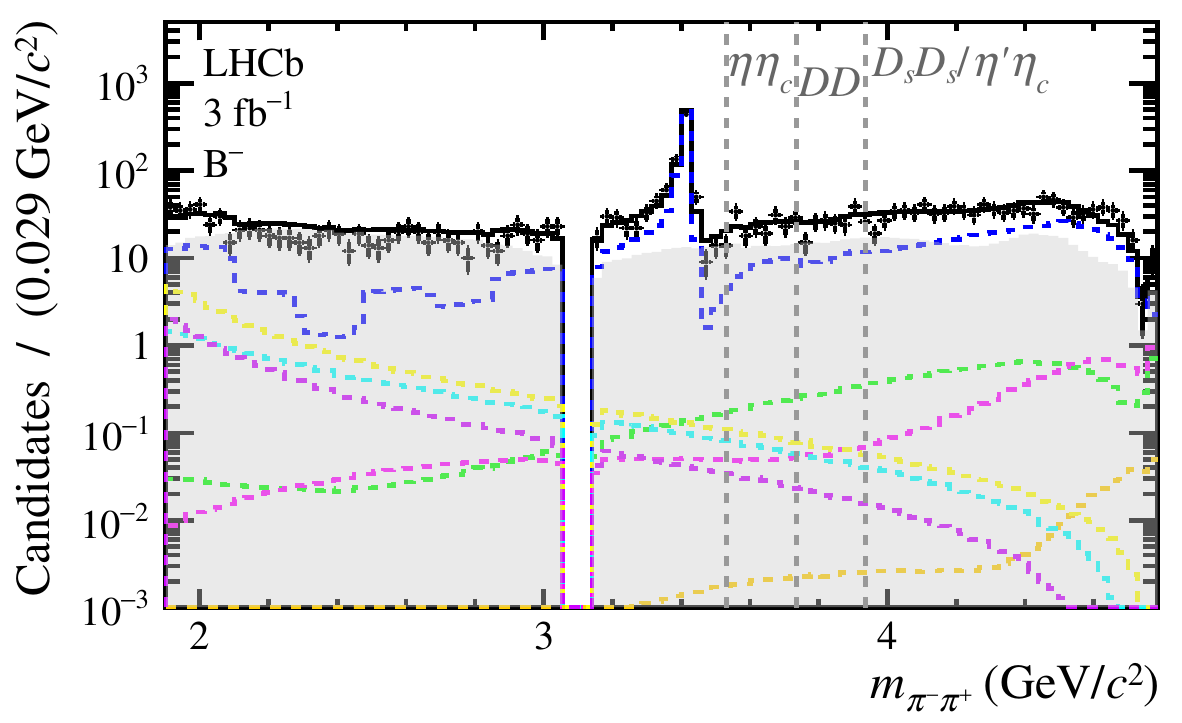}

    \caption{
        Fit projections of $m_{\pi^+ \pi^-}$ for (left)~$B^+$ and (right)~$B^-$ (top)~below and (bottom)~above the open charm threshold, enhanced with $\cos\theta_{\pi^+ \pi^-} < 0$. 
        Vertical dashed lines correspond to the opening thresholds of the indicated coupled channels.
    }
    \label{fig:app:mpipiqmi}
\end{figure}

\begin{figure}[tb]
    \centering
    \includegraphics[width=0.5\linewidth]{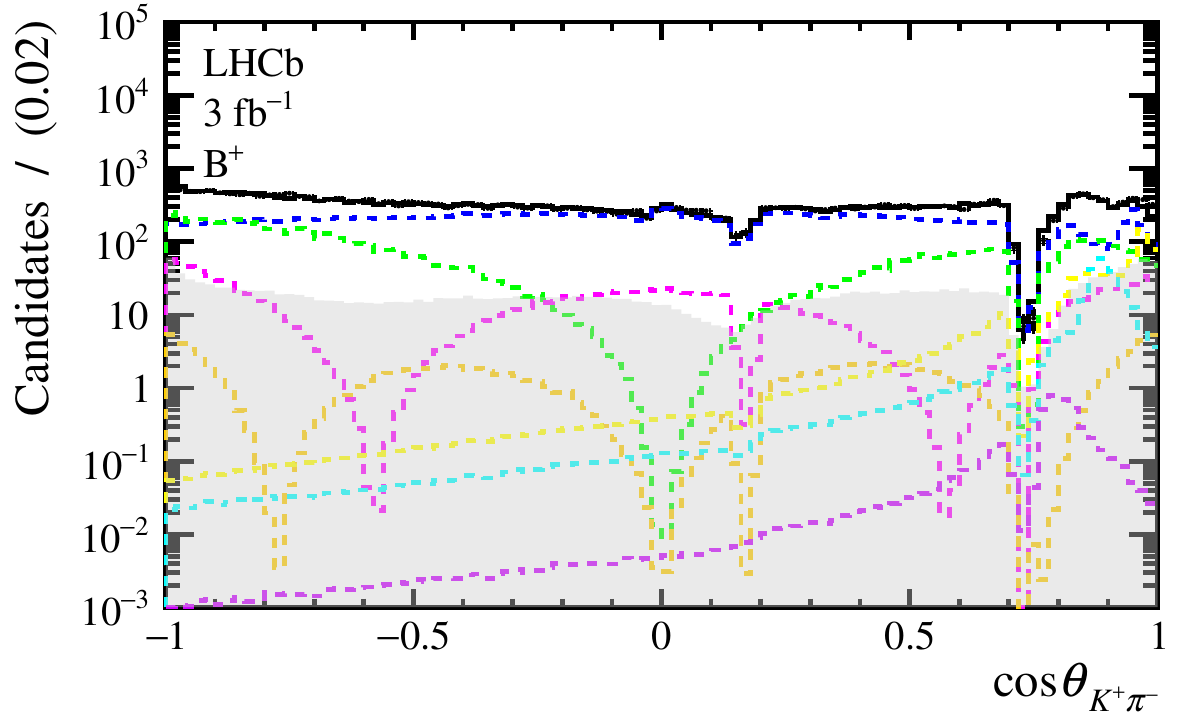}%
    \includegraphics[width=0.5\linewidth]{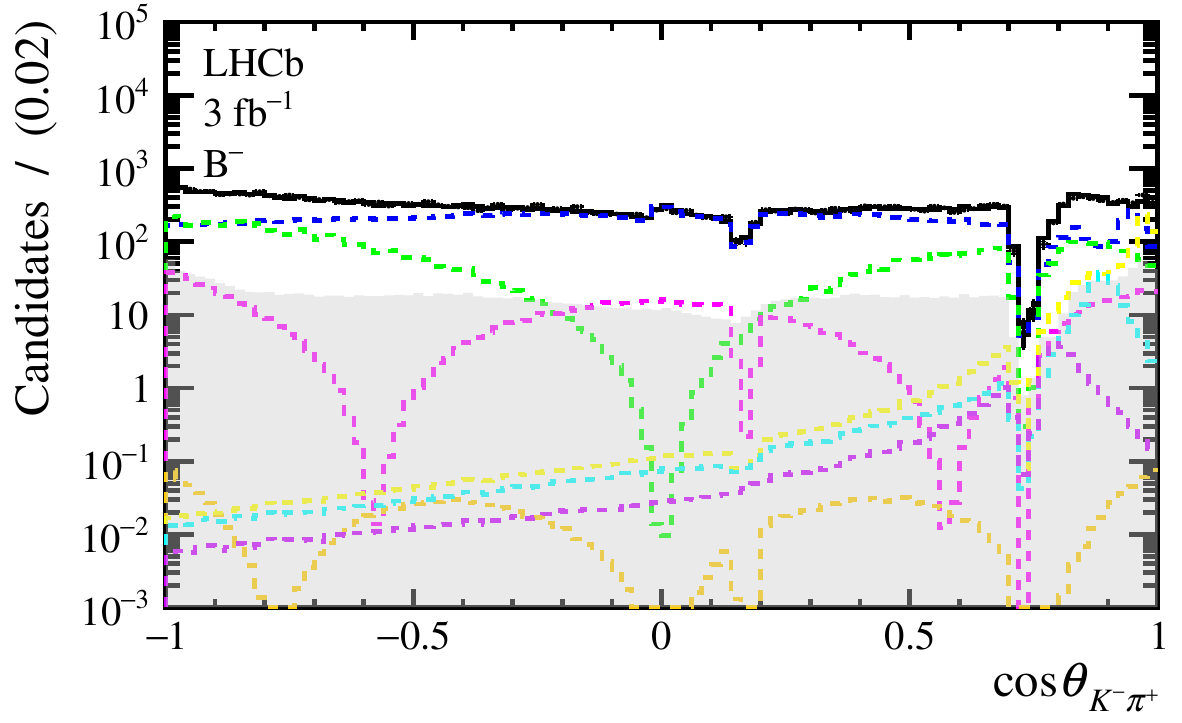}

    \includegraphics[width=0.5\linewidth]{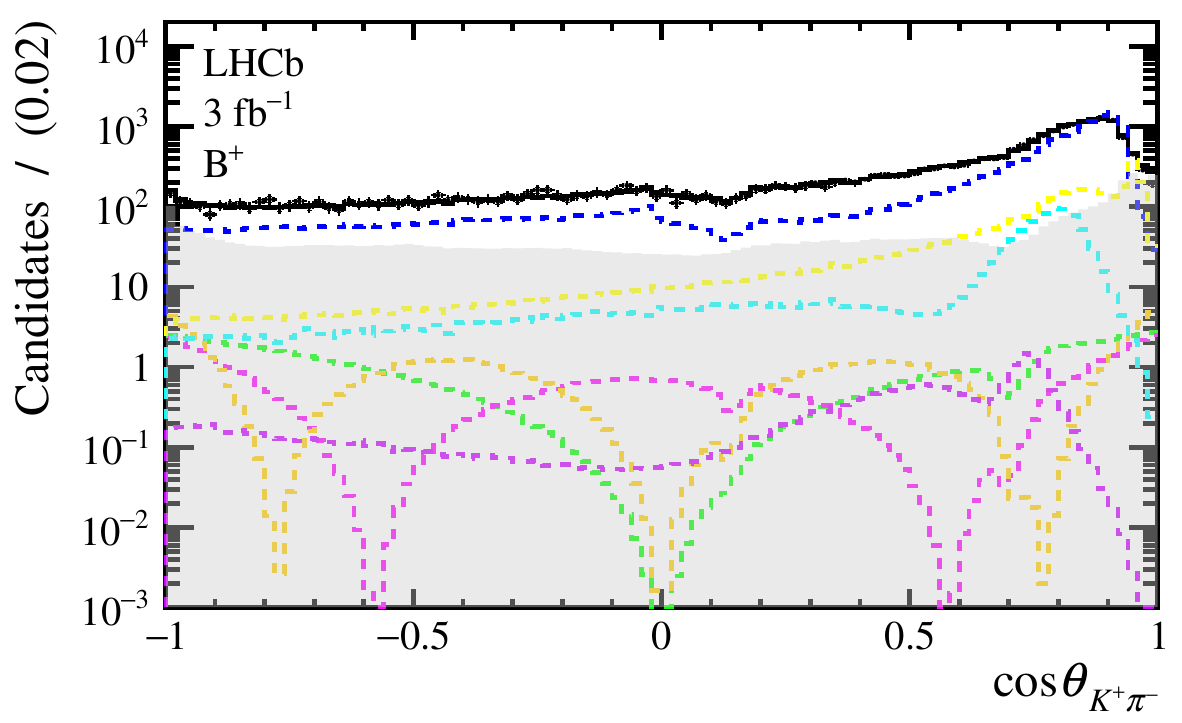}%
    \includegraphics[width=0.5\linewidth]{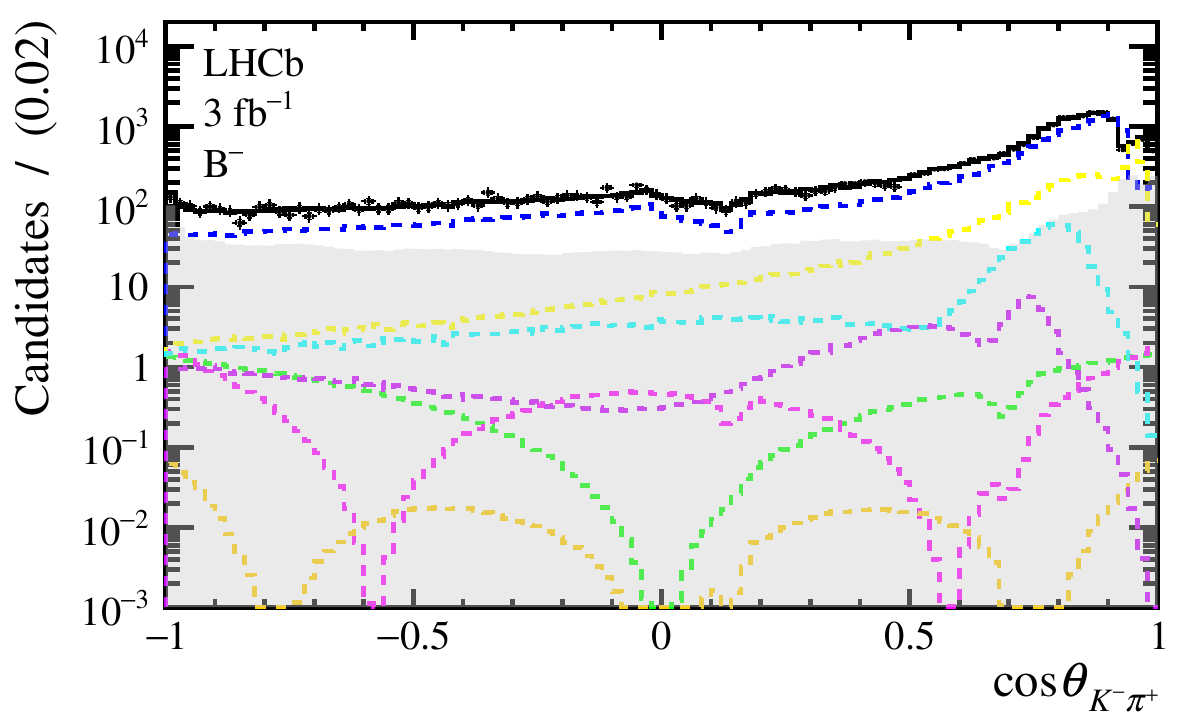}

    \caption{
        Fit projections of $\cos\theta_{K^+ \pi^-}$ for (left)~$B^+$ and (right)~$B^-$ (top)~below and (bottom)~above the open charm threshold in $m_{K^+ \pi^-}$.
    }
    \label{fig:app:ckpiqmi}
\end{figure}

\begin{figure}[tb]
    \centering
    \includegraphics[width=0.5\linewidth]{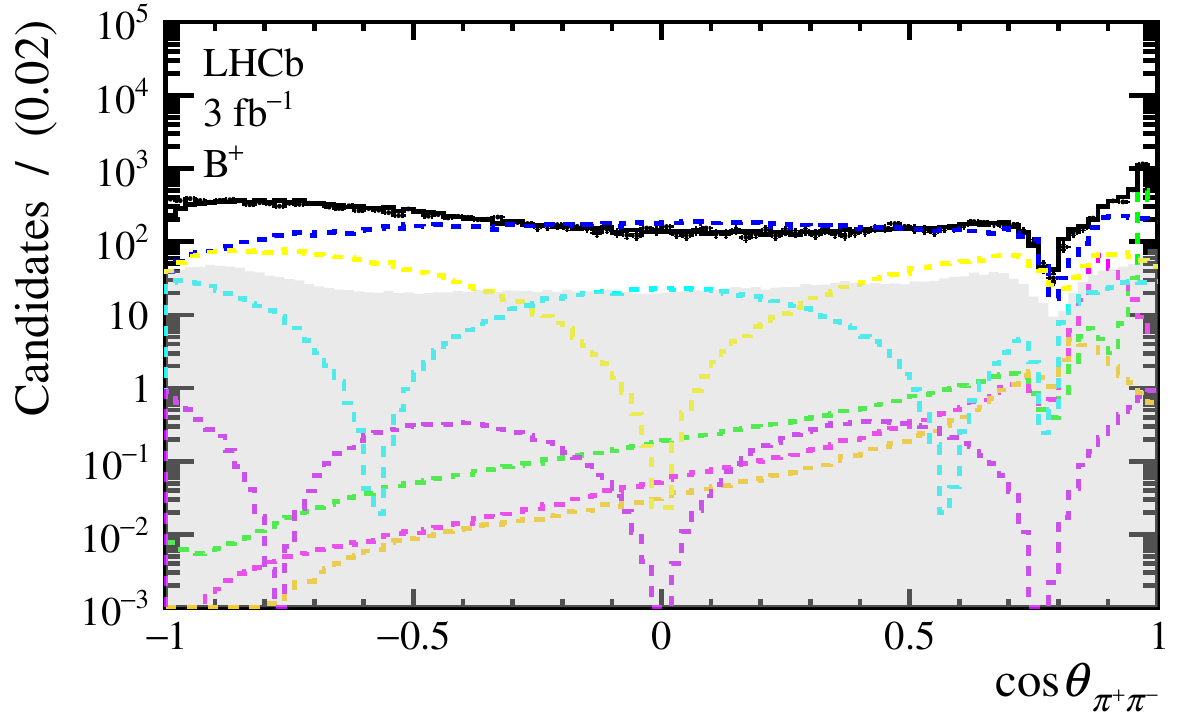}%
    \includegraphics[width=0.5\linewidth]{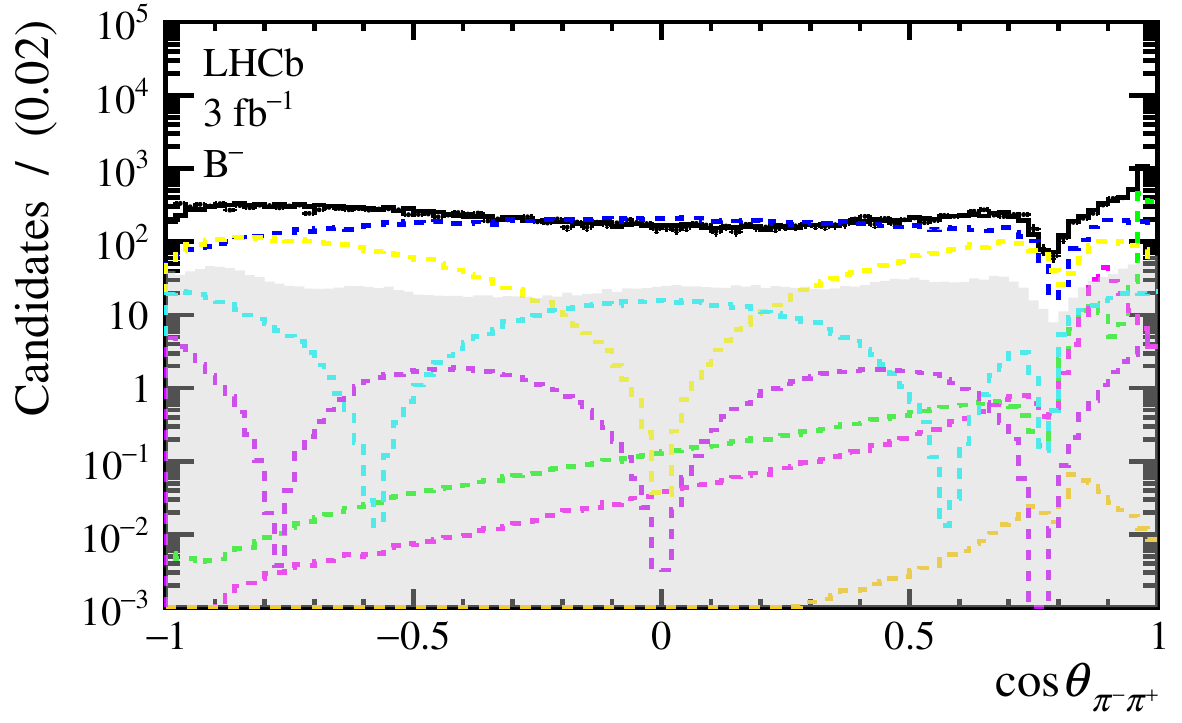}

    \includegraphics[width=0.5\linewidth]{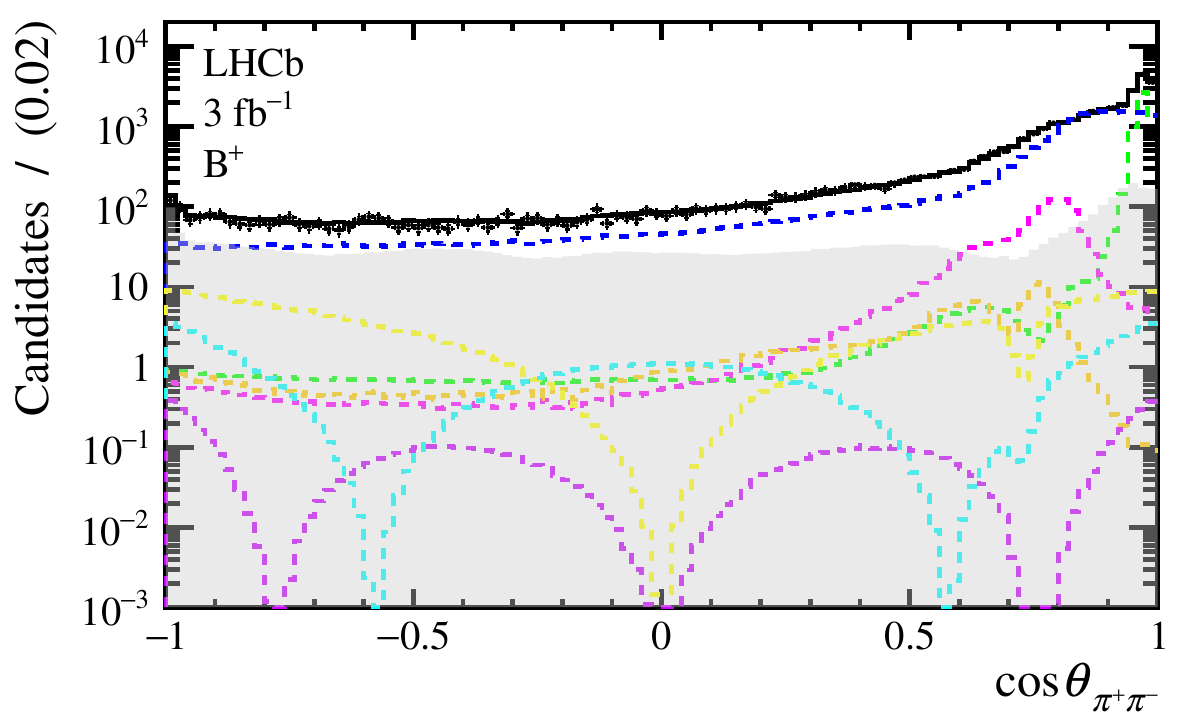}%
    \includegraphics[width=0.5\linewidth]{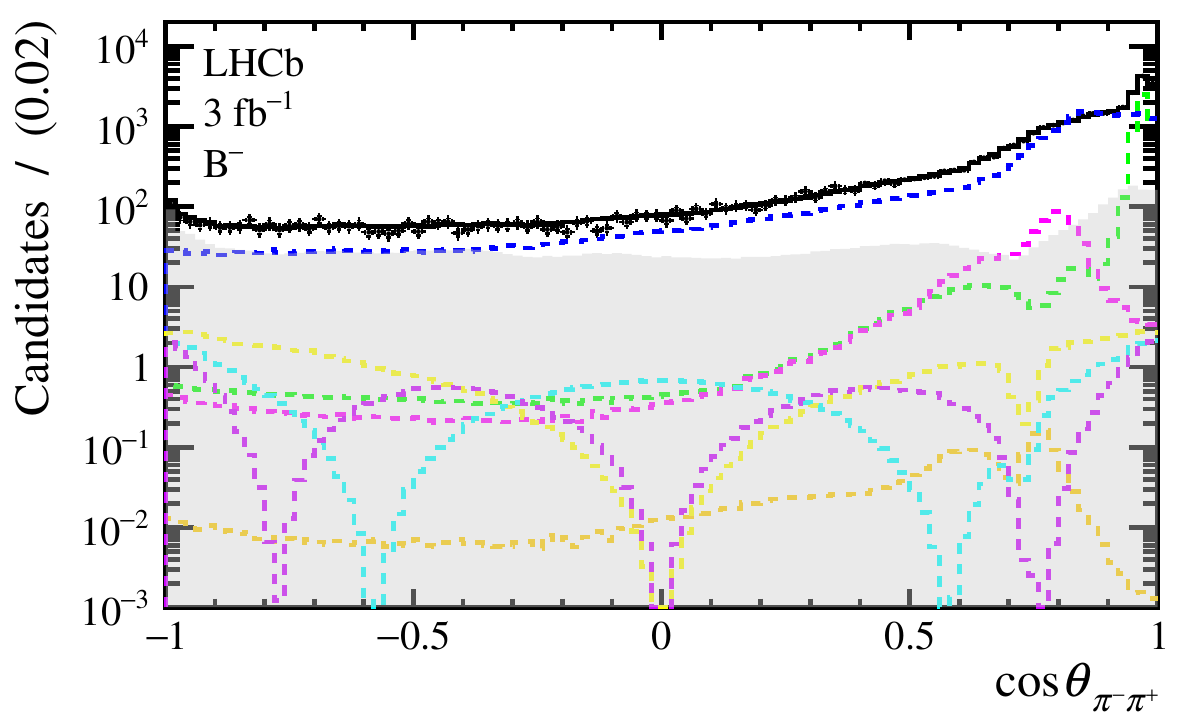}

    \caption{
        Fit projections of $\cos\theta_{\pi^+ \pi^-}$ for (left)~$B^+$ and (right)~$B^-$ (top)~below and (bottom)~above the open charm threshold in $m_{\pi^+ \pi^-}$.
    }
    \label{fig:app:cpipiqmi}
\end{figure}

Various projections of the data and fit results that serve to highlight the individual contributions of each partial wave are shown here with the colour legend displayed in Fig.~\ref{fig:app:legqmi}. Mass projections with negative associated values of the cosine of the helicity angle, which removes the bulk of the reflections, are shown in Figs.~\ref{fig:app:mkpiqmi} and~\ref{fig:app:mpipiqmi}. The helicity structure enhanced in the regions below and above the open-charm threshold are given in Figs.~\ref{fig:app:ckpiqmi} and~\ref{fig:app:cpipiqmi} for the $\Kp\pim$ and $\pip\pim$ systems, respectively.

\clearpage

\section{Additional S-wave comparisons}
\label{app:swave}

\begin{figure}[!b]
    \centering
    \includegraphics[width=0.5\linewidth]{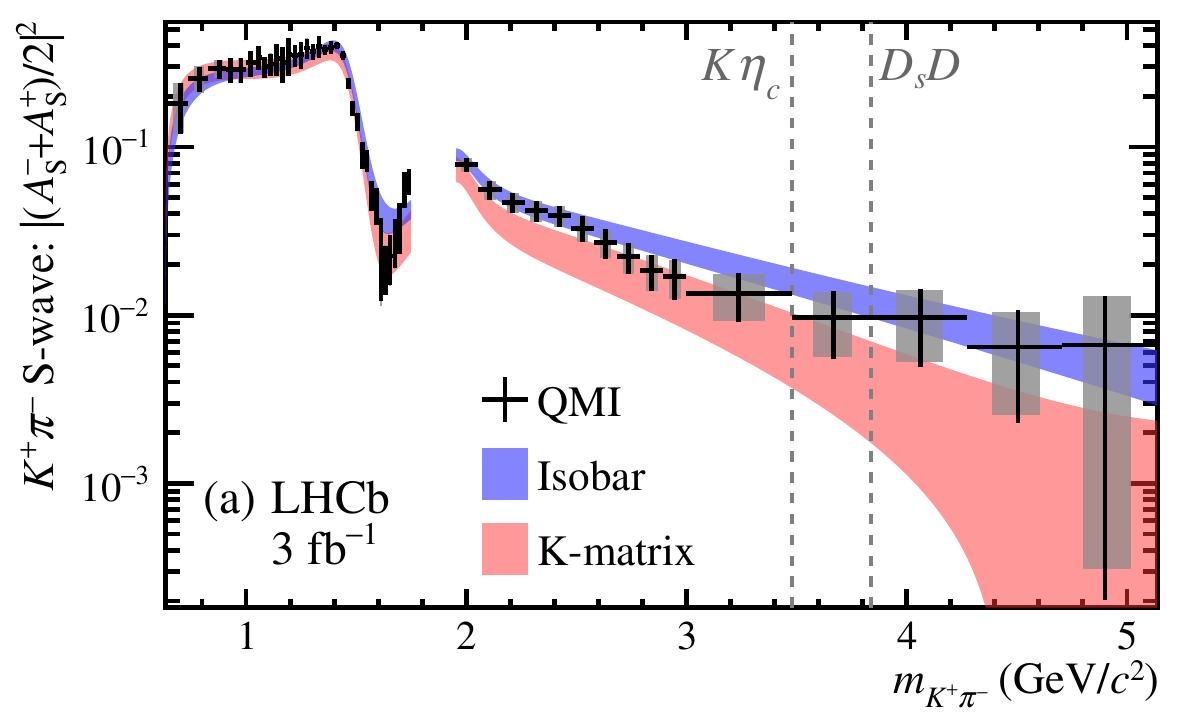}%
    \includegraphics[width=0.5\linewidth]{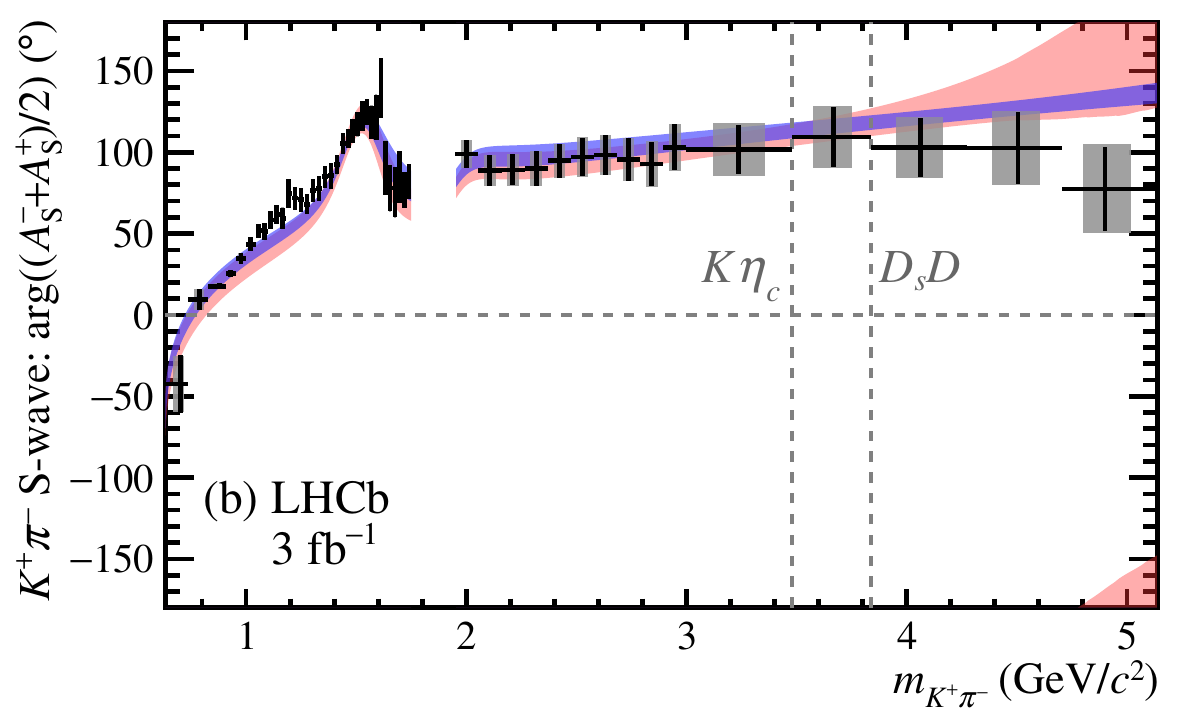}

    \includegraphics[width=0.5\linewidth]{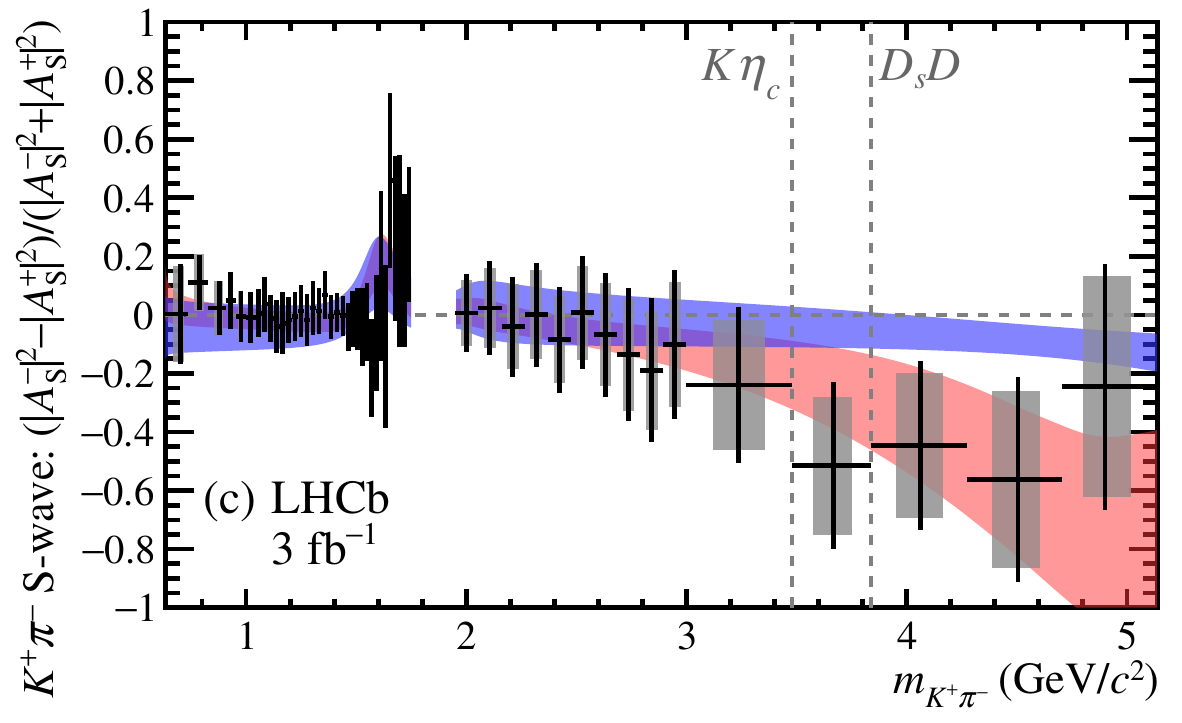}%
    \includegraphics[width=0.5\linewidth]{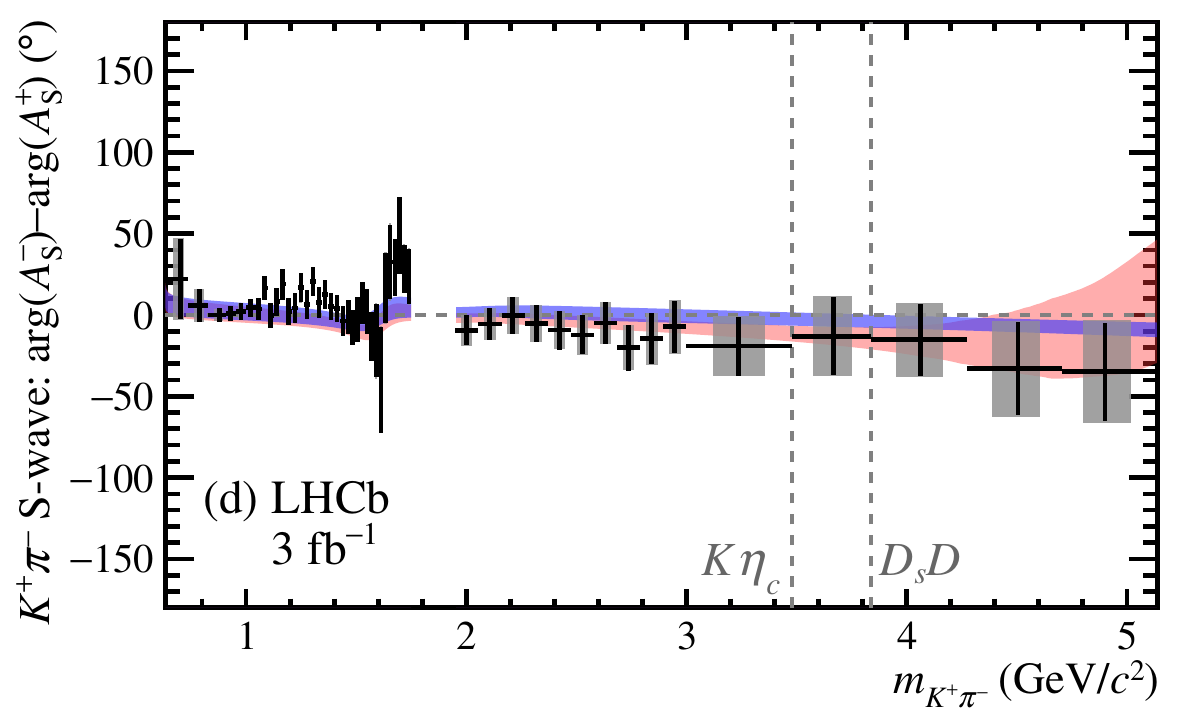}

    \caption{
        Comparison of the $K^+ \pi^-$ S-wave results for the (a)~$C\!P$-averaged magnitude squared, (b)~$C\!P$-averaged phase, (c)~asymmetry of the magnitude-squared and (d)~phase difference highlighting the charm region, where the vertical grey bands show the total QMI uncertainty without accounting for an additional systematic effect arising from the fixed-bin values set from the Isobar result. 
        Vertical dashed lines correspond to the opening thresholds of the indicated coupled channels.
    }
    \label{fig:app:swavekpi}
\end{figure}

\begin{figure}[!tb]
    \centering
    \includegraphics[width=0.5\linewidth]{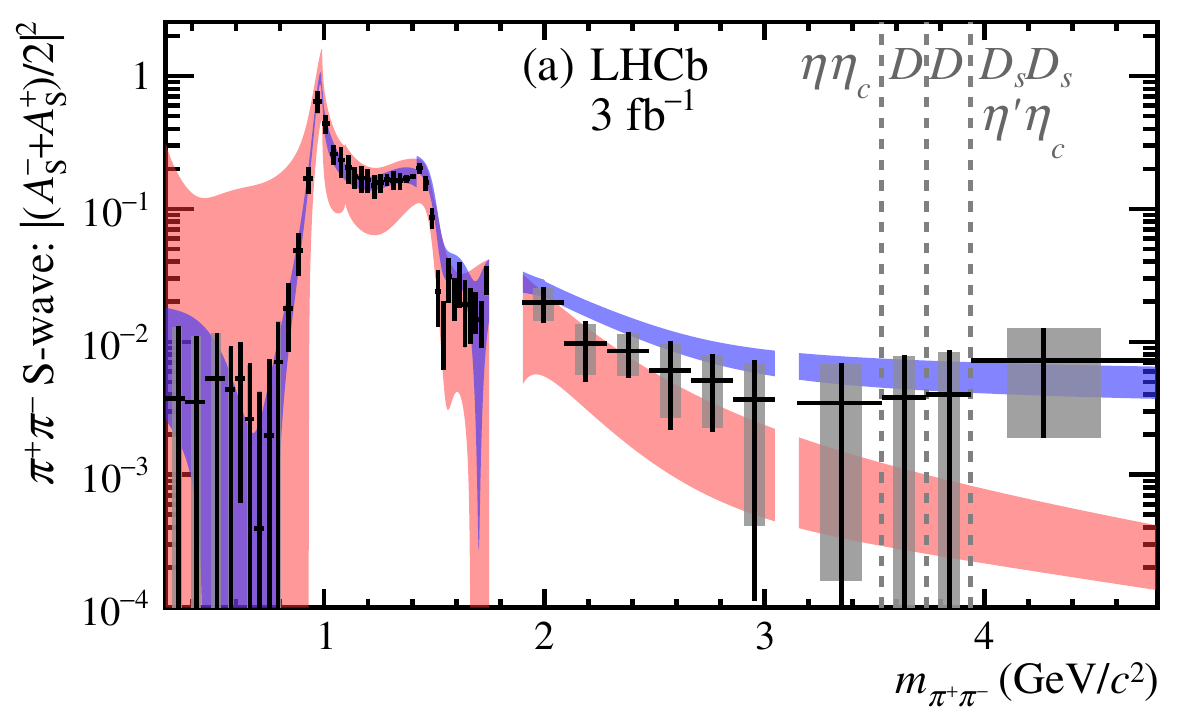}%
    \includegraphics[width=0.5\linewidth]{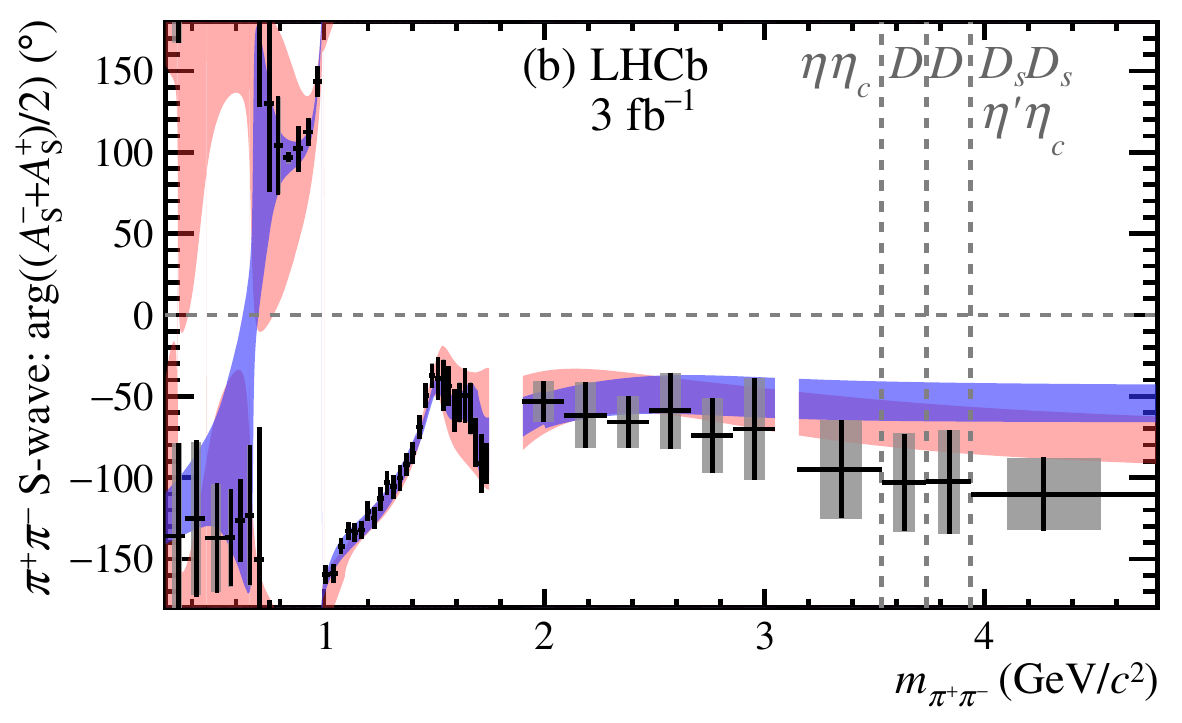}

    \includegraphics[width=0.5\linewidth]{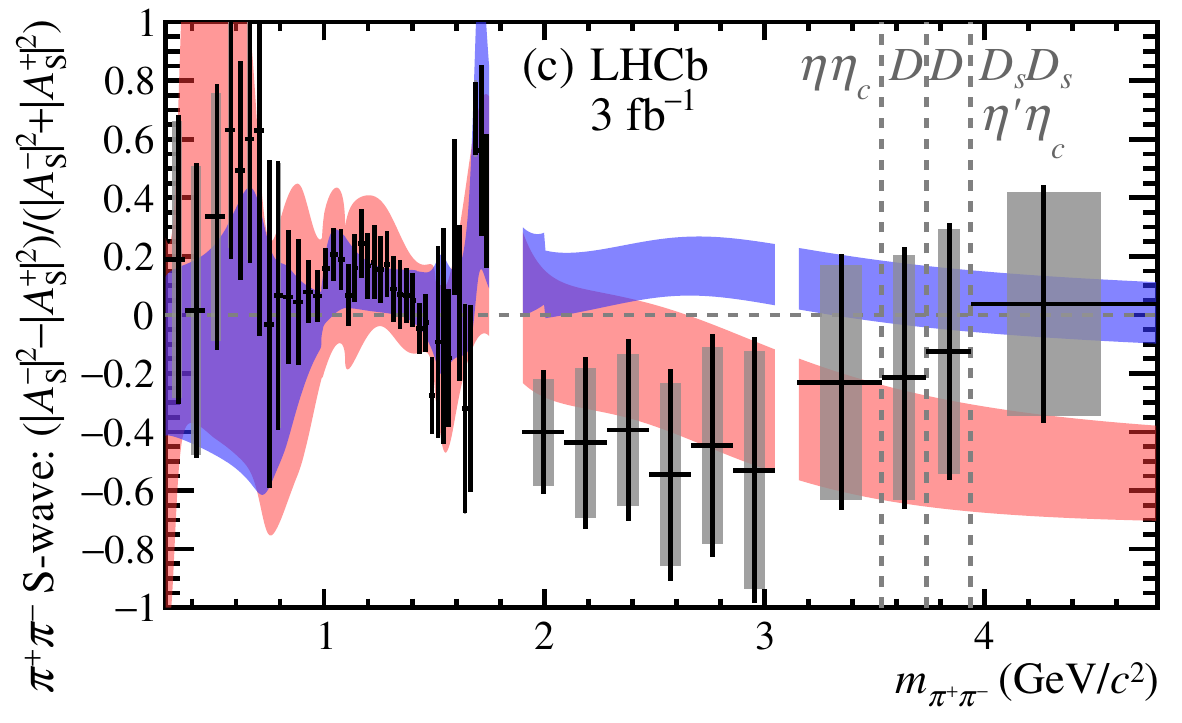}%
    \includegraphics[width=0.5\linewidth]{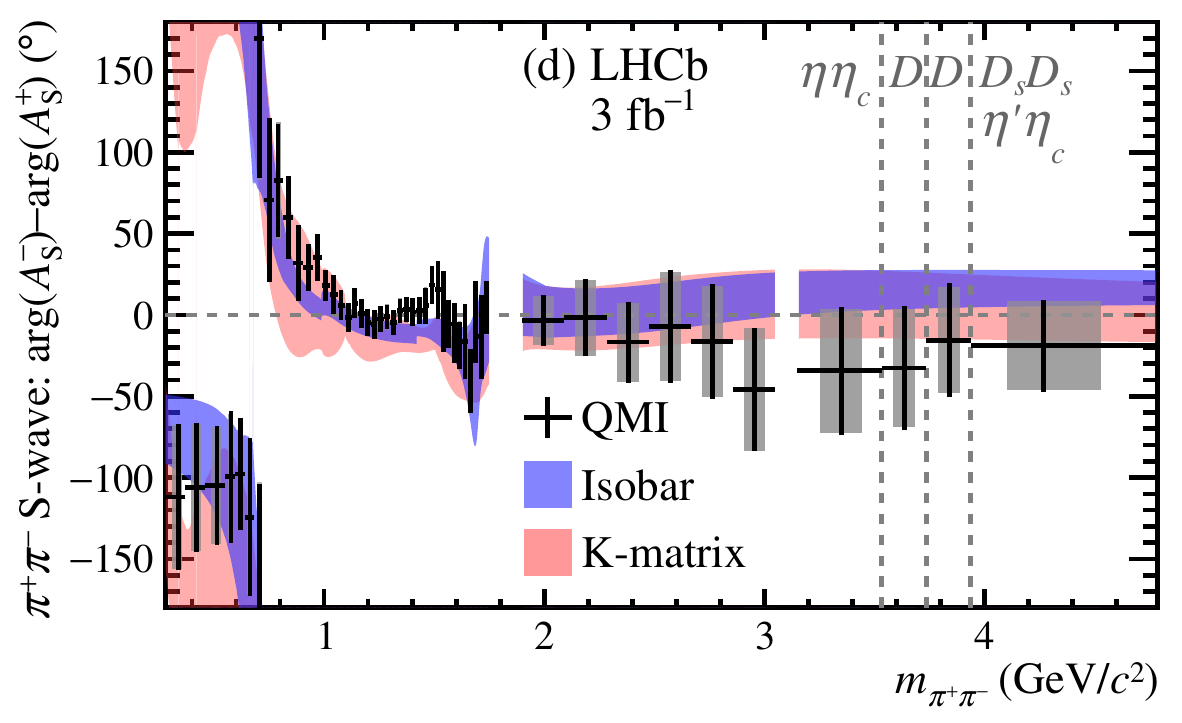}

    \caption{
        Comparison of the $\pi^+ \pi^-$ S-wave results for the (a)~$C\!P$-averaged magnitude squared, (b)~$C\!P$-averaged phase, (c)~asymmetry of the magnitude-squared and (d)~phase difference highlighting the charm region, where the vertical grey bands show the total QMI uncertainty without accounting for an additional systematic effect arising from the fixed-bin values set from the Isobar result. 
        Vertical dashed lines correspond to the opening thresholds of the indicated coupled channels.
    }
    \label{fig:app:swavepipi}
\end{figure}

\begin{figure}[!b]
    \centering
    \includegraphics[width=0.33\linewidth]{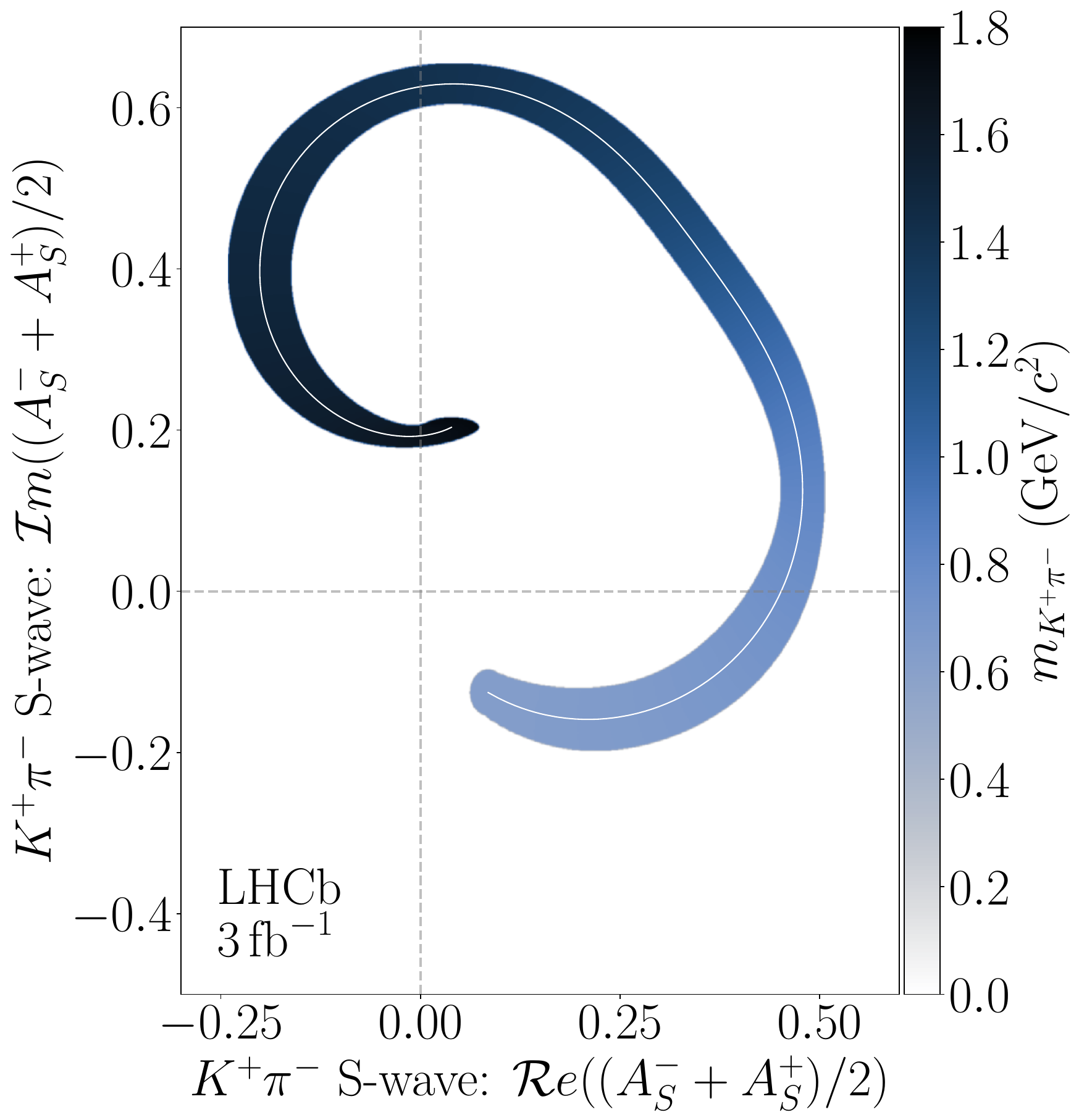}%
    \includegraphics[width=0.33\linewidth]{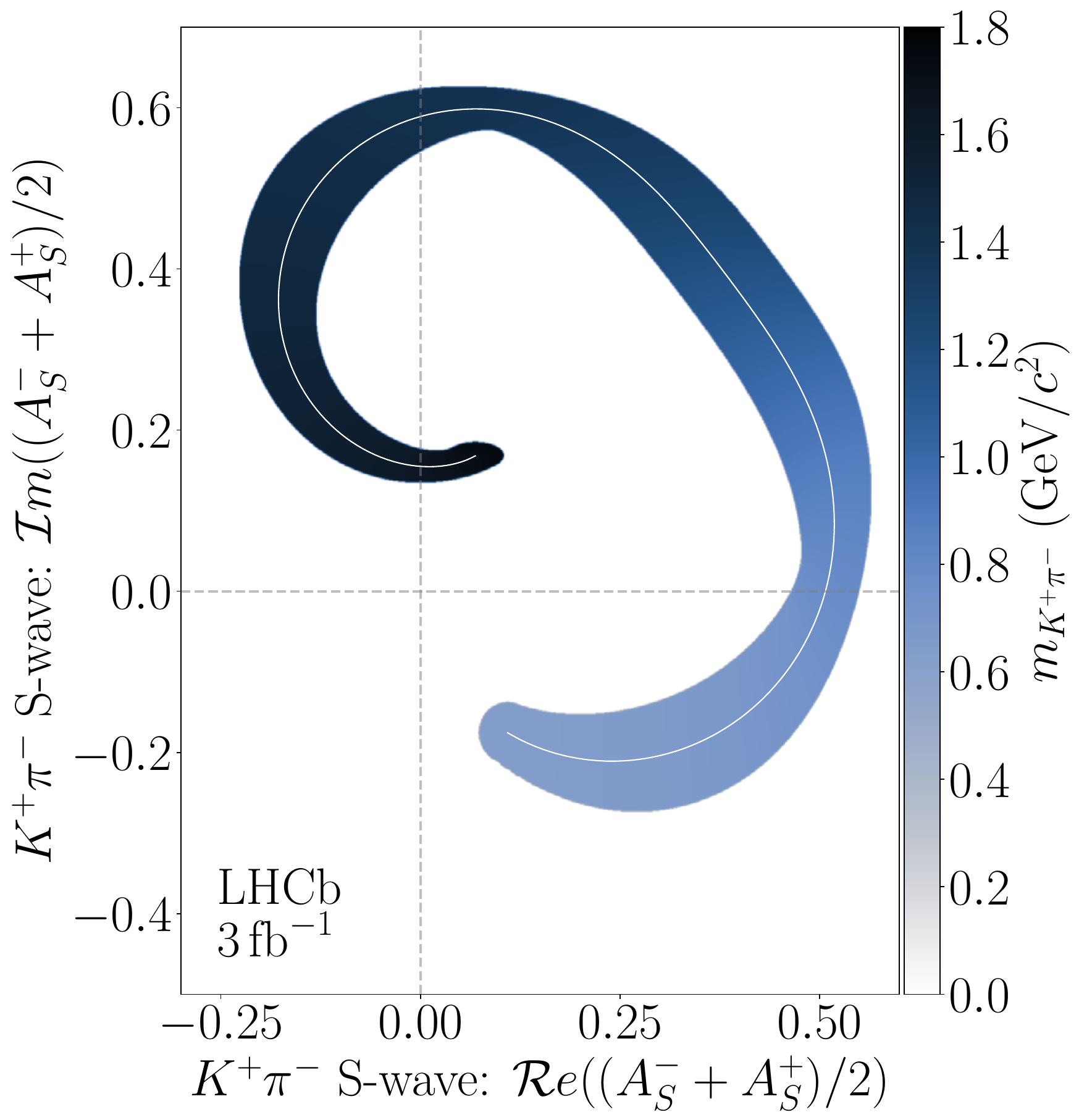}%
    \includegraphics[width=0.33\linewidth]{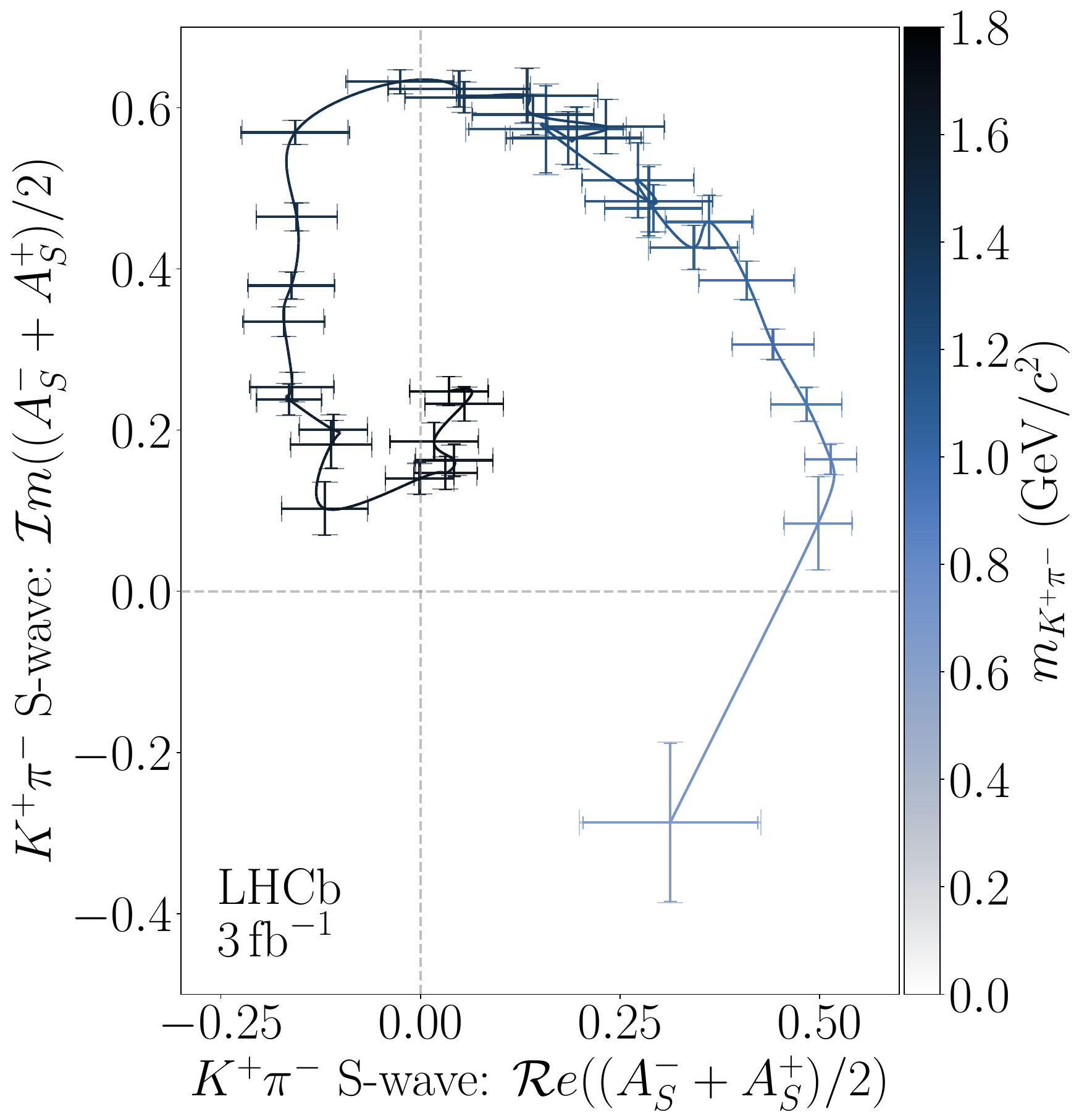}

    \includegraphics[width=0.33\linewidth]{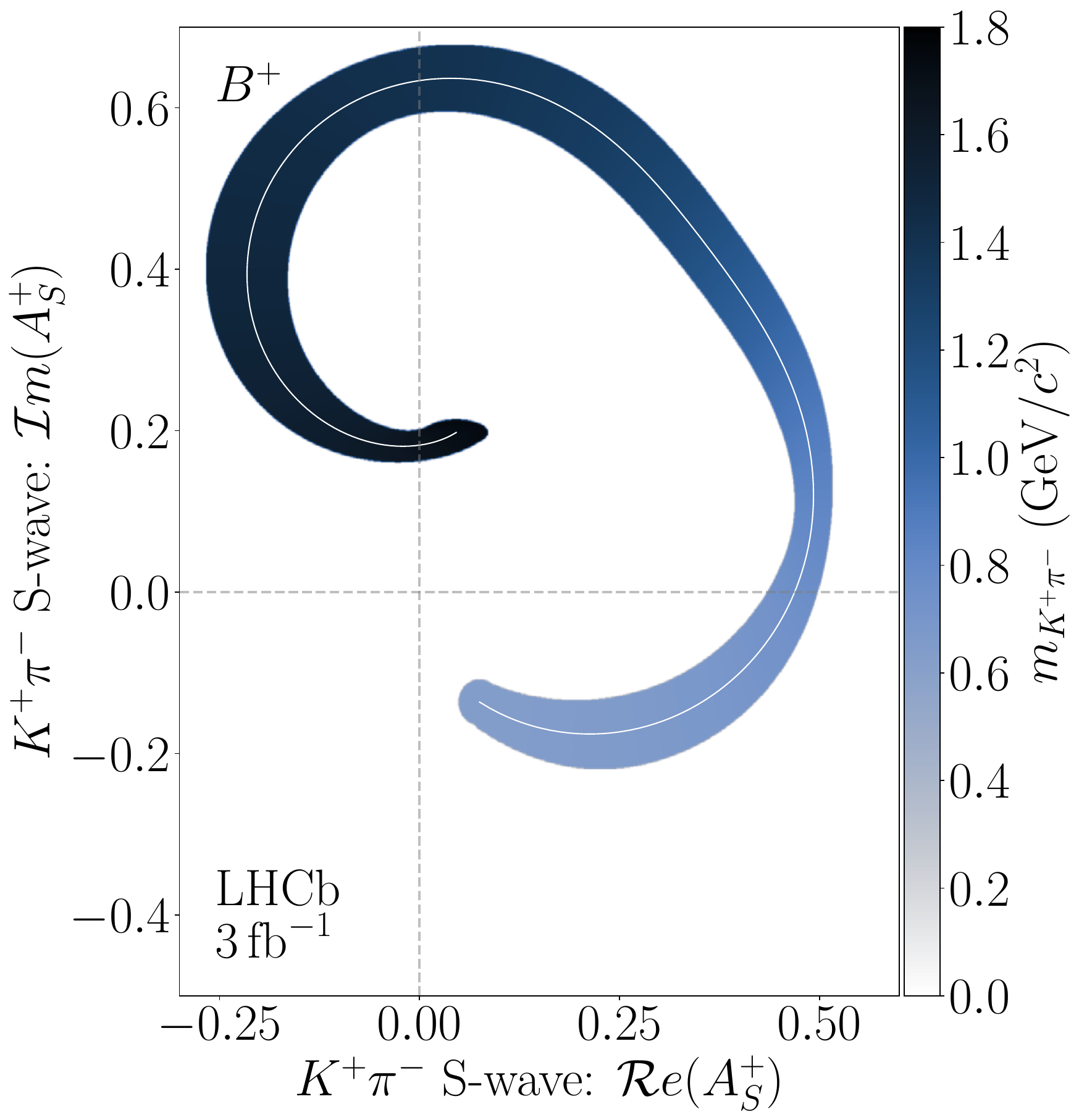}%
    \includegraphics[width=0.33\linewidth]{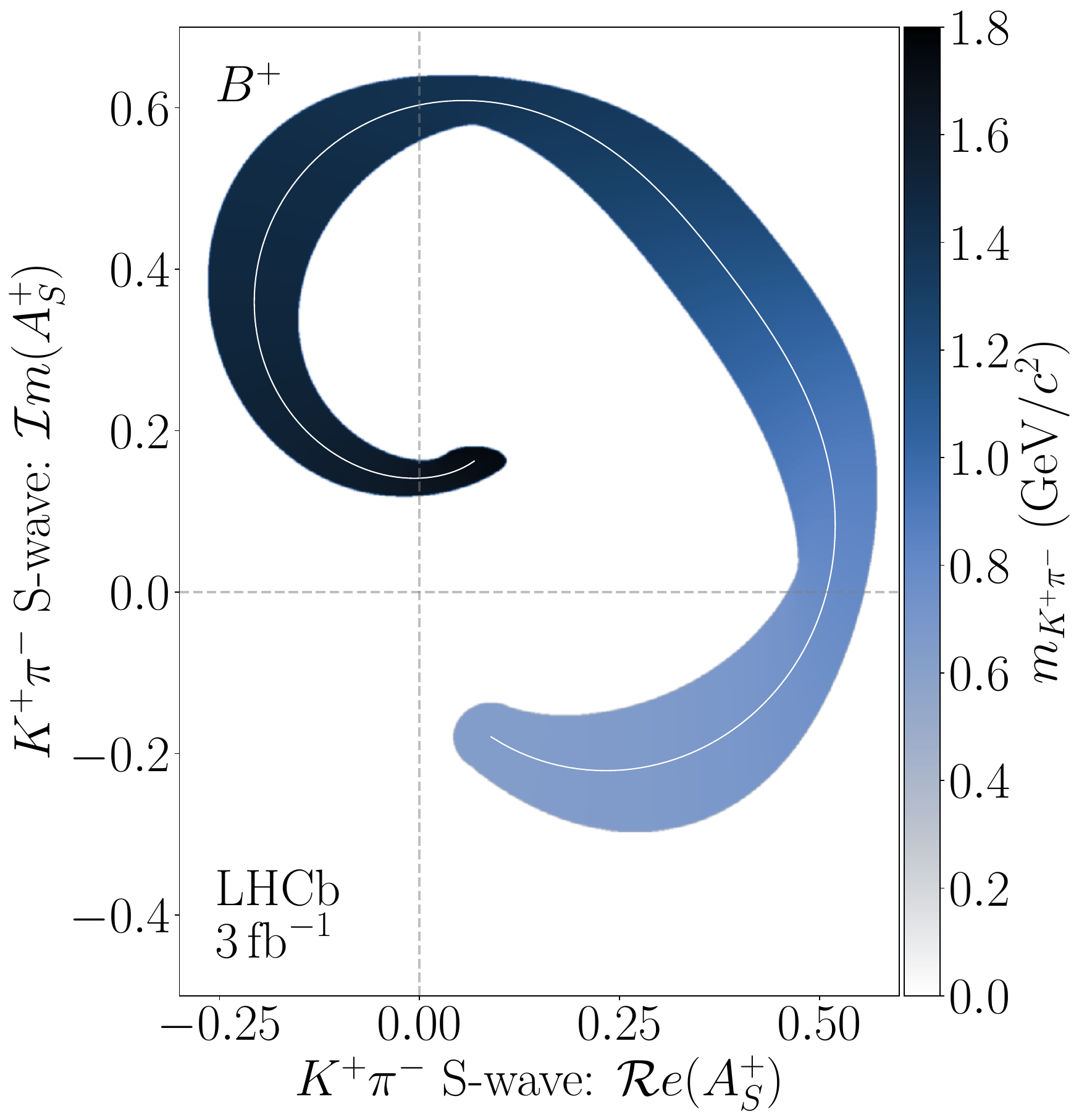}%
    \includegraphics[width=0.33\linewidth]{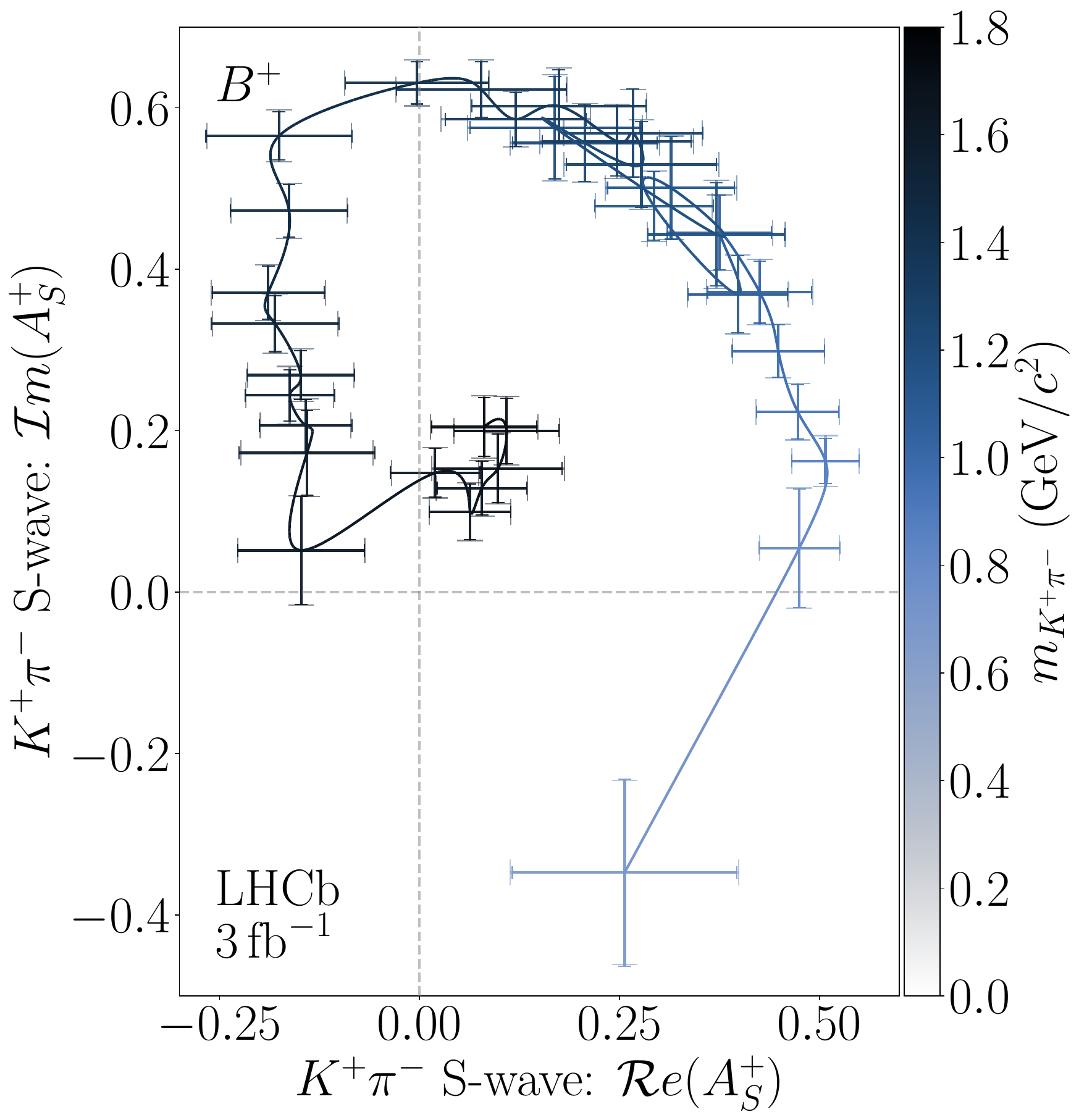}

    \includegraphics[width=0.33\linewidth]{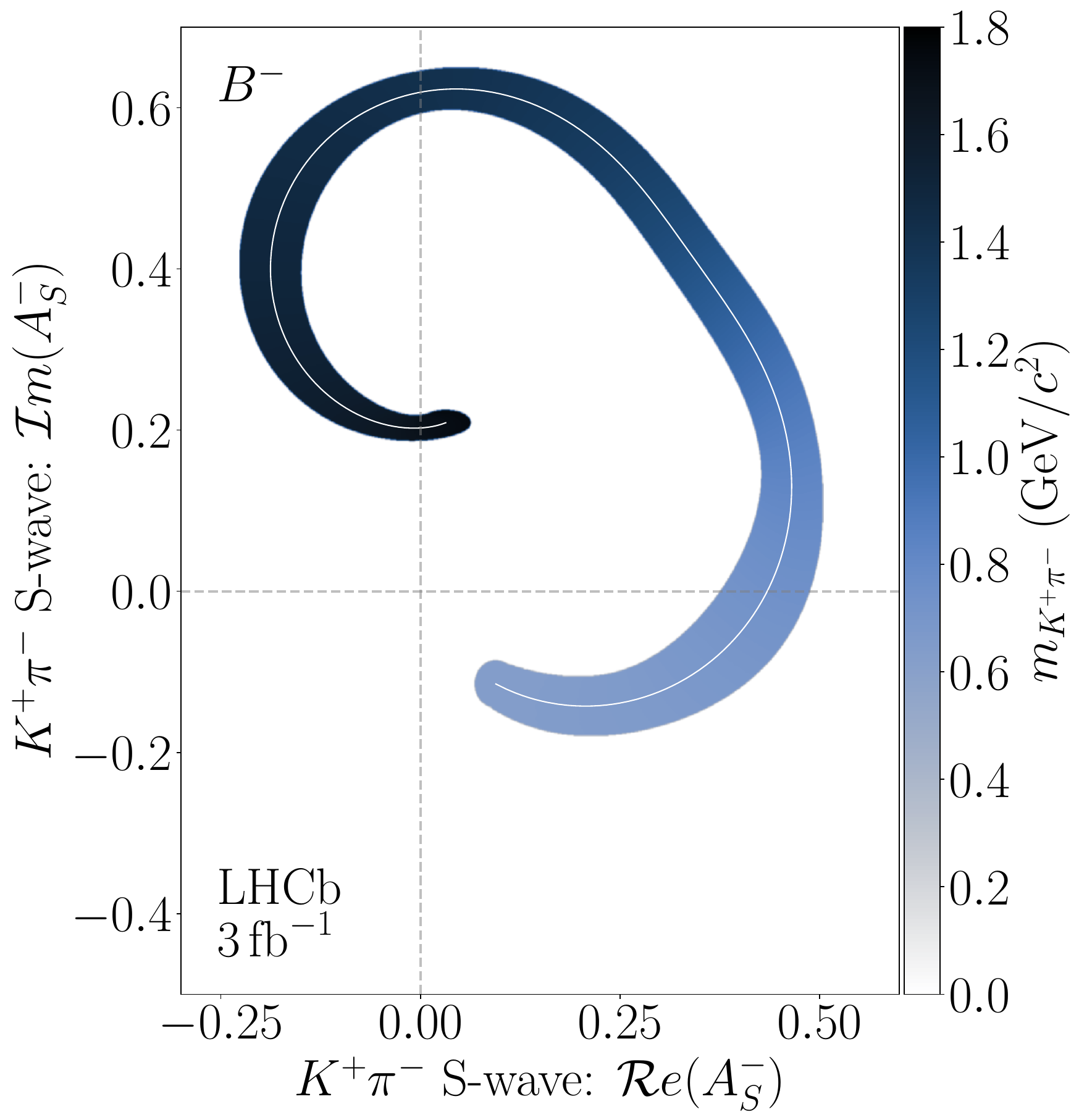}%
    \includegraphics[width=0.33\linewidth]{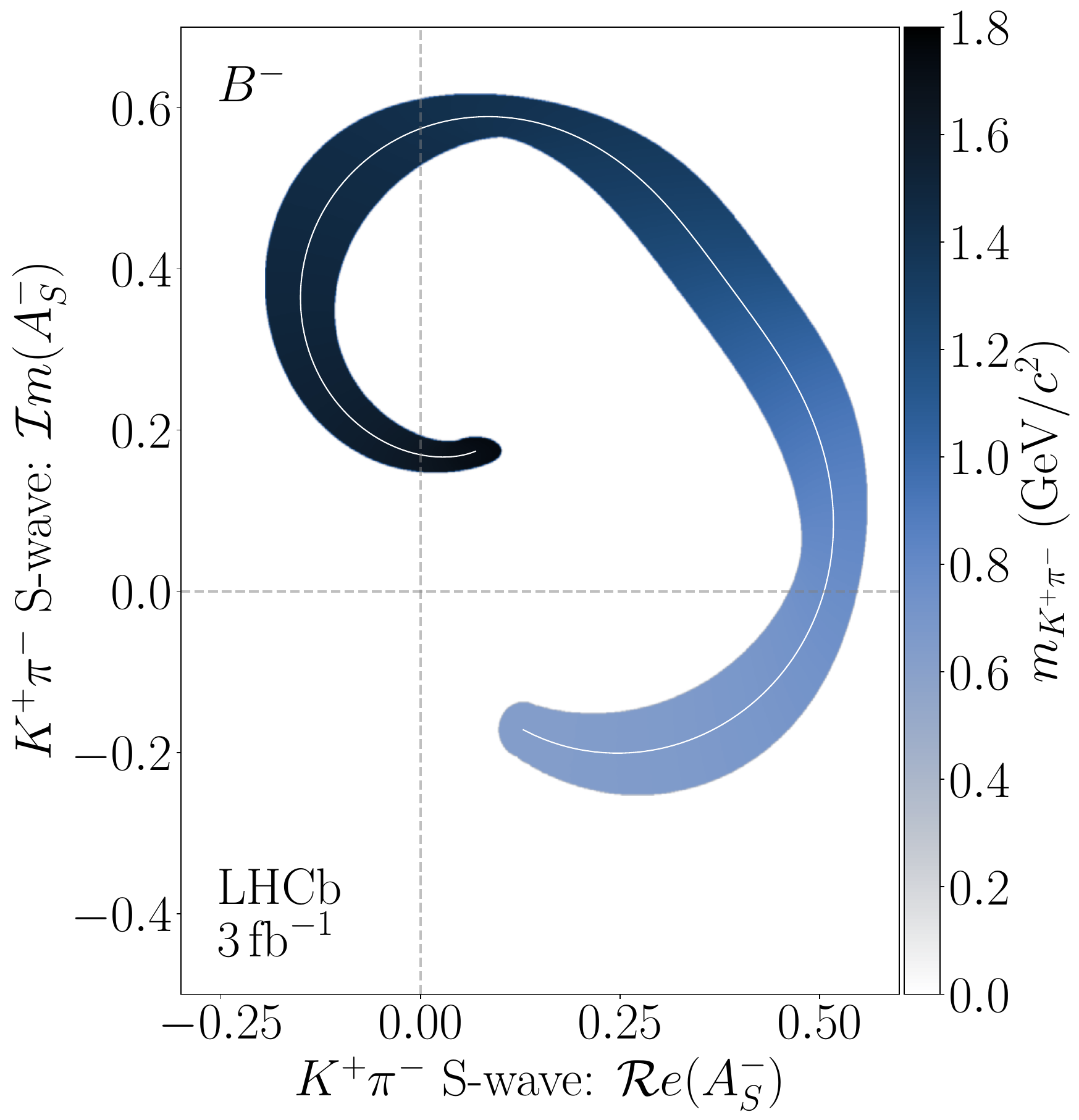}%
    \includegraphics[width=0.33\linewidth]{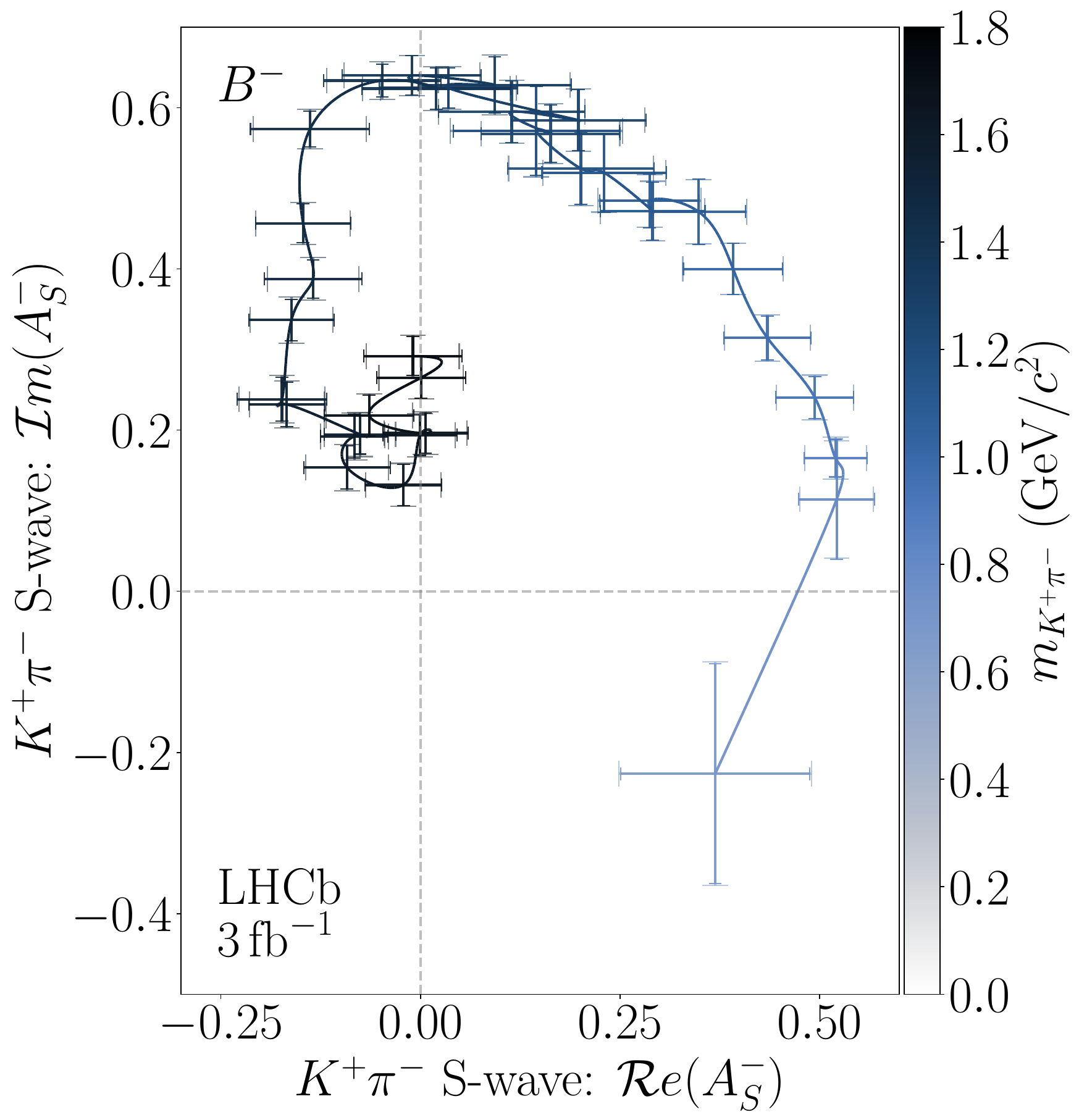}

    \caption{
        Argand diagrams of the (top)~$C\!P$-averaged, (middle)~$B^+$ and (bottom)~$B^-$ S-wave results in $K^+ \pi^-$ below the open-charm threshold for the (left)~Isobar, (middle)~K-matrix and (right)~QMI approaches. White curves indicate central values while smoothed bands cover the total uncertainty at $1\sigma$. In the QMI plots, cubic splines are provided purely for visual guidance, and the shorter error bars indicate the total QMI uncertainty without accounting for an additional systematic effect arising from the fixed-bin values set from the Isobar result.
    }
    \label{fig:app:argandkpi}
\end{figure}

\begin{figure}[!b]
    \centering
    \includegraphics[width=0.33\linewidth]{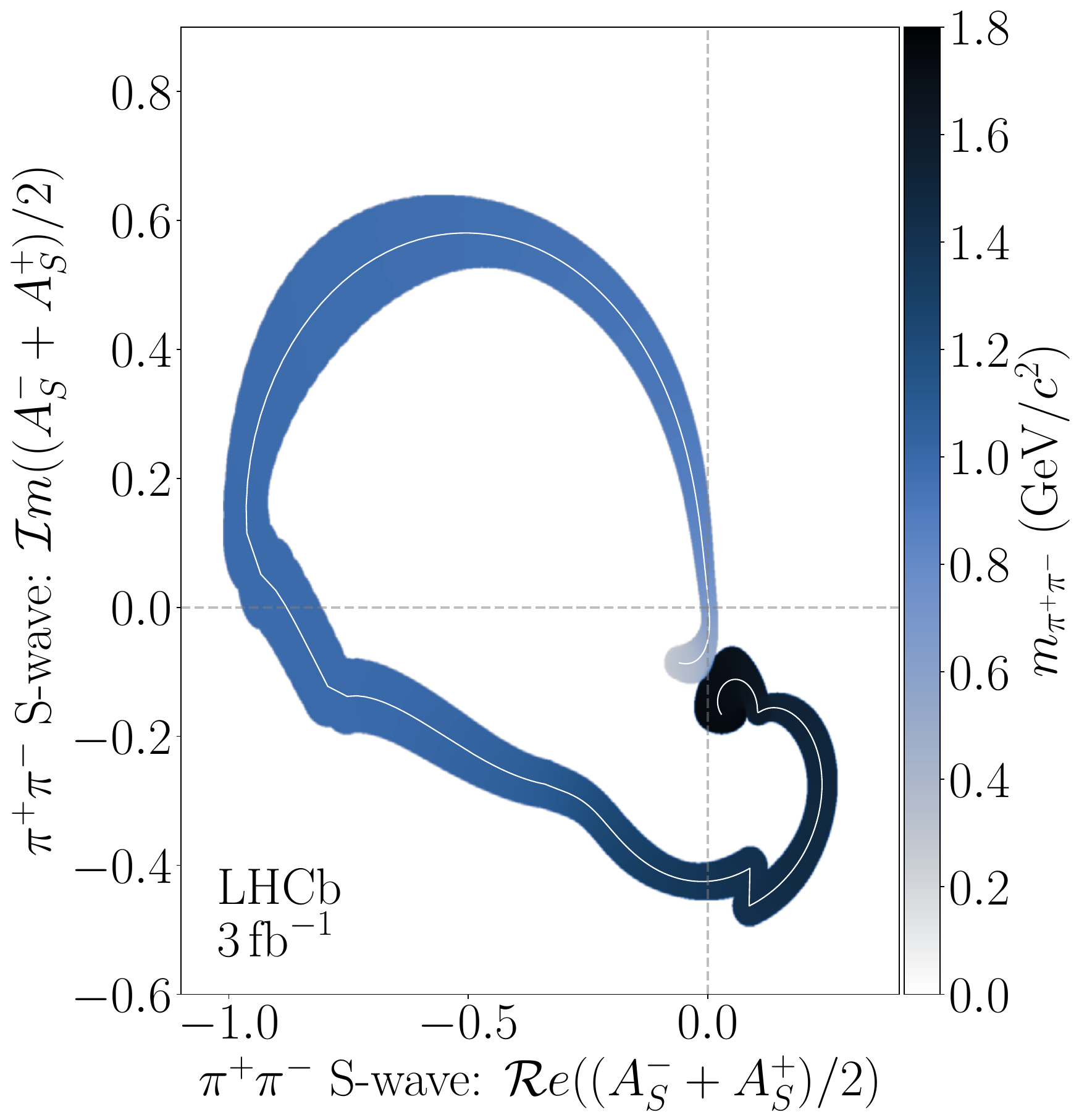}%
    \includegraphics[width=0.33\linewidth]{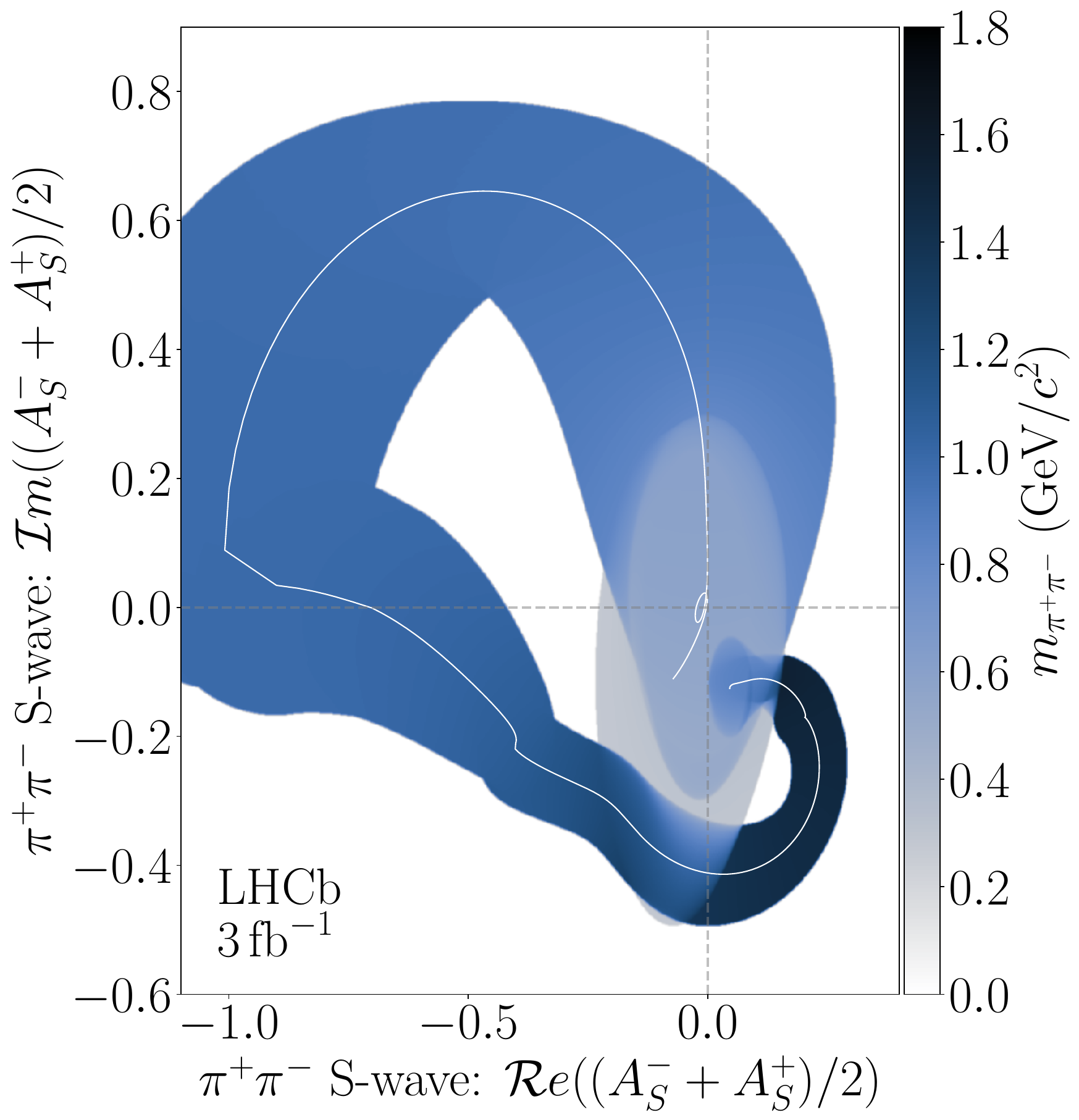}%
    \includegraphics[width=0.33\linewidth]{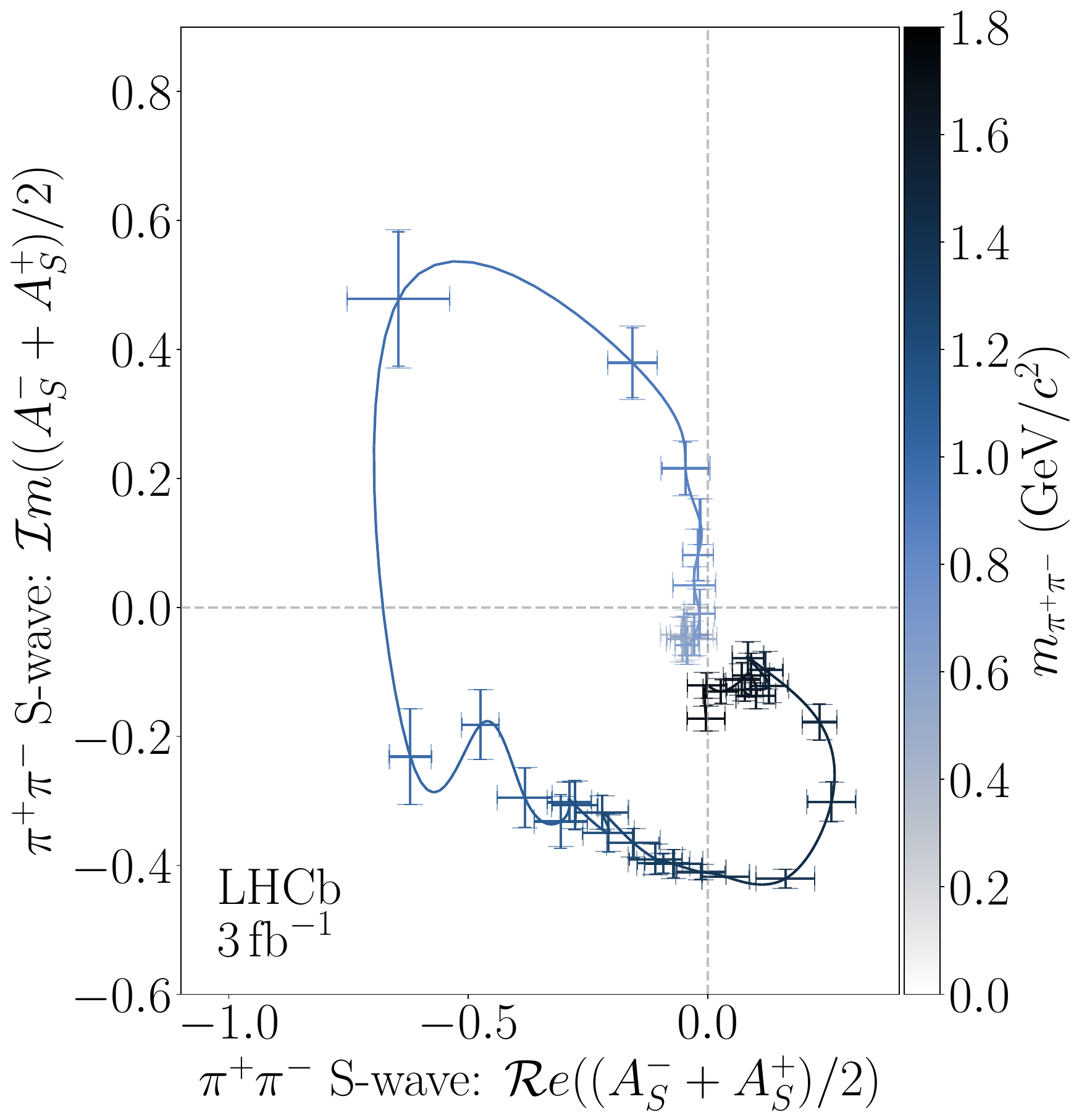}

    \includegraphics[width=0.33\linewidth]{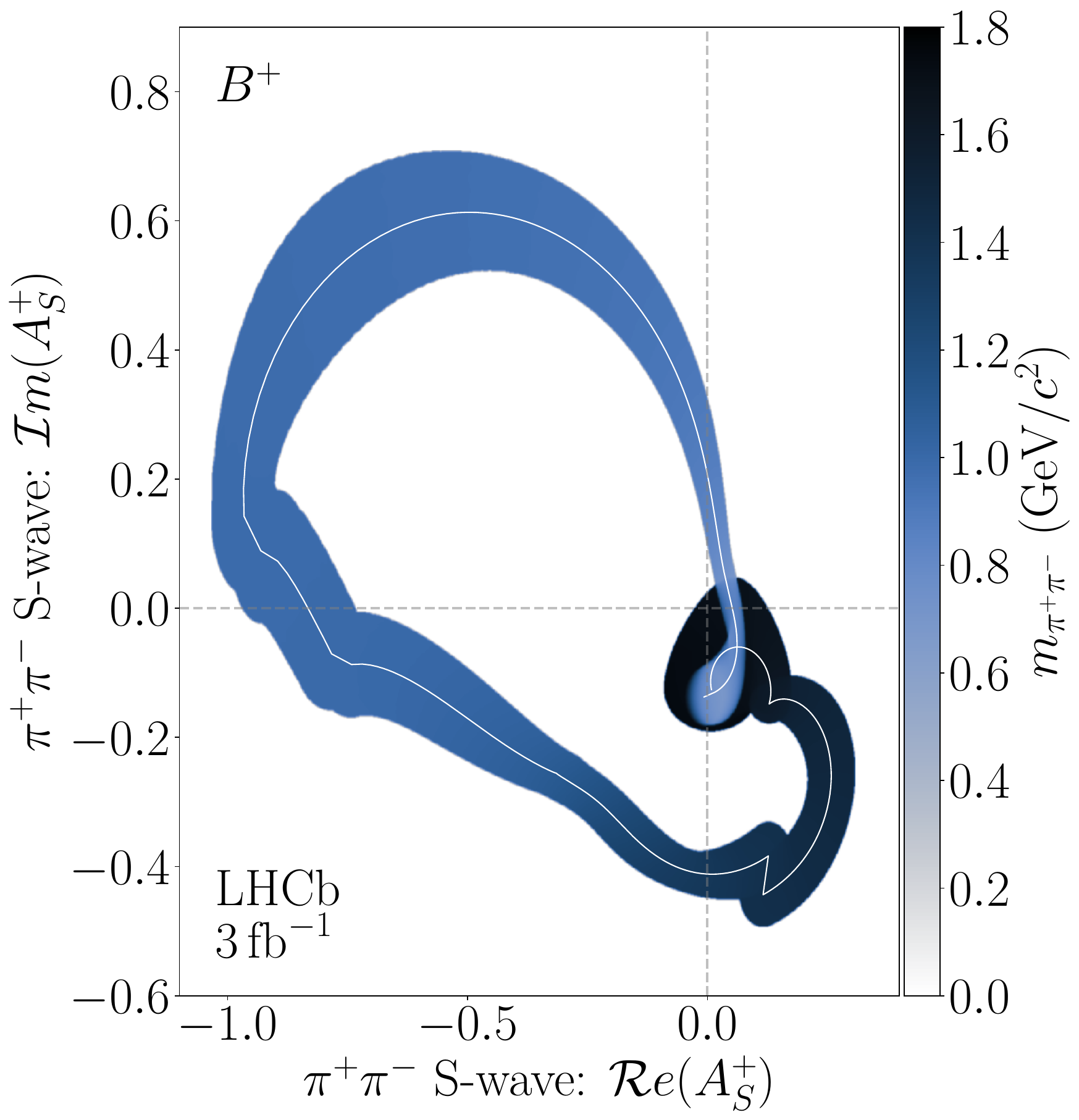}%
    \includegraphics[width=0.33\linewidth]{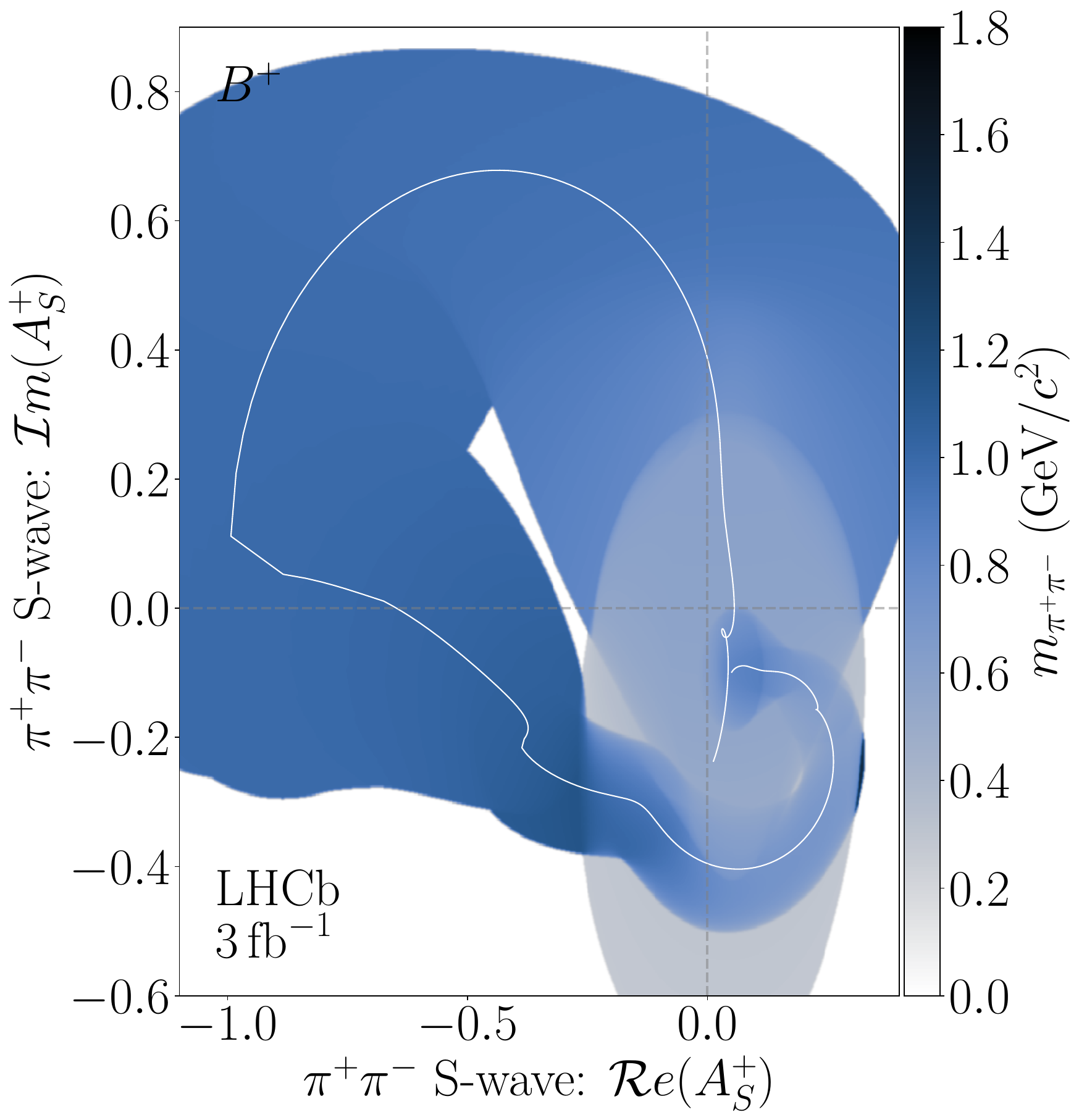}%
    \includegraphics[width=0.33\linewidth]{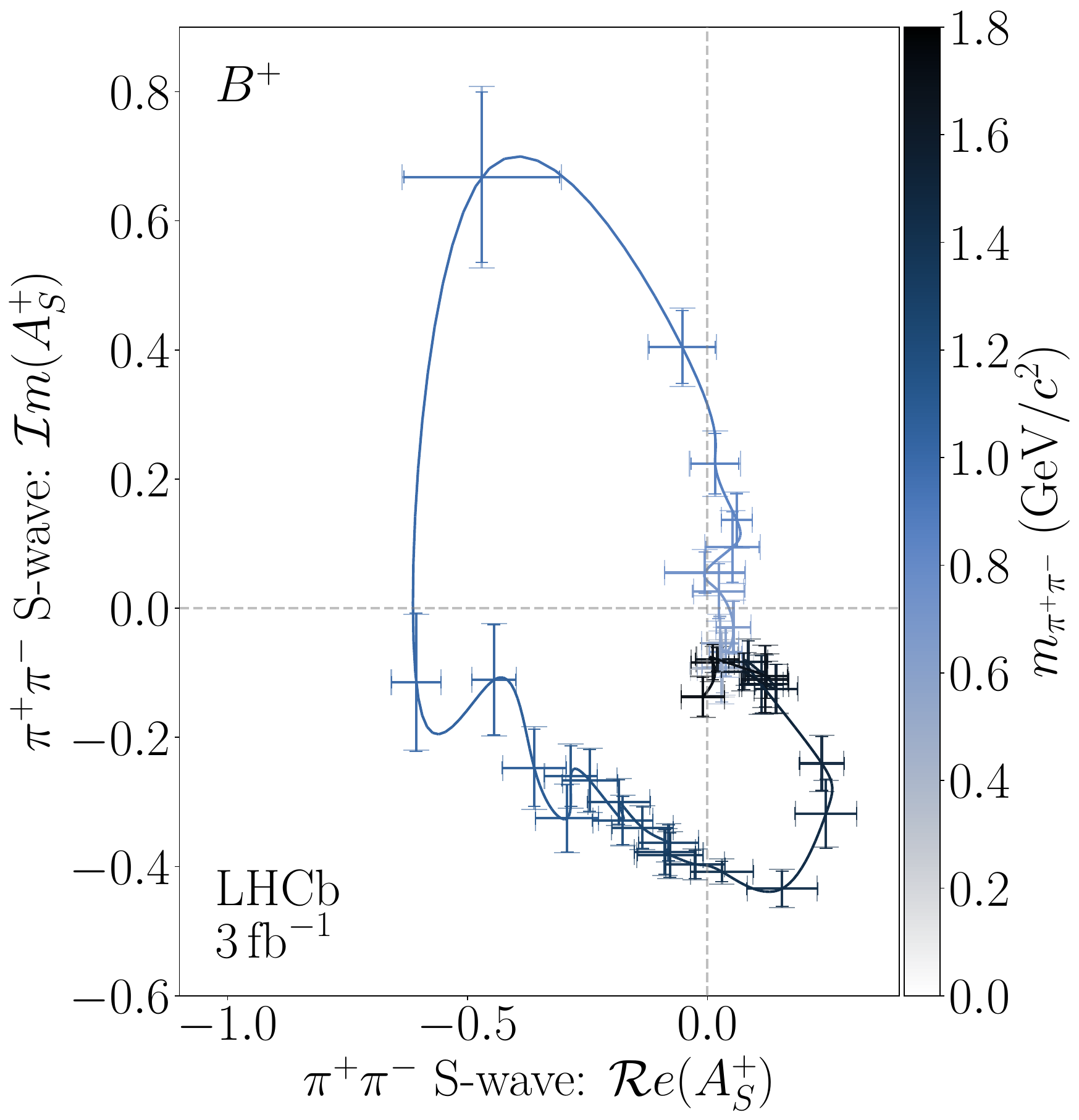}

    \includegraphics[width=0.33\linewidth]{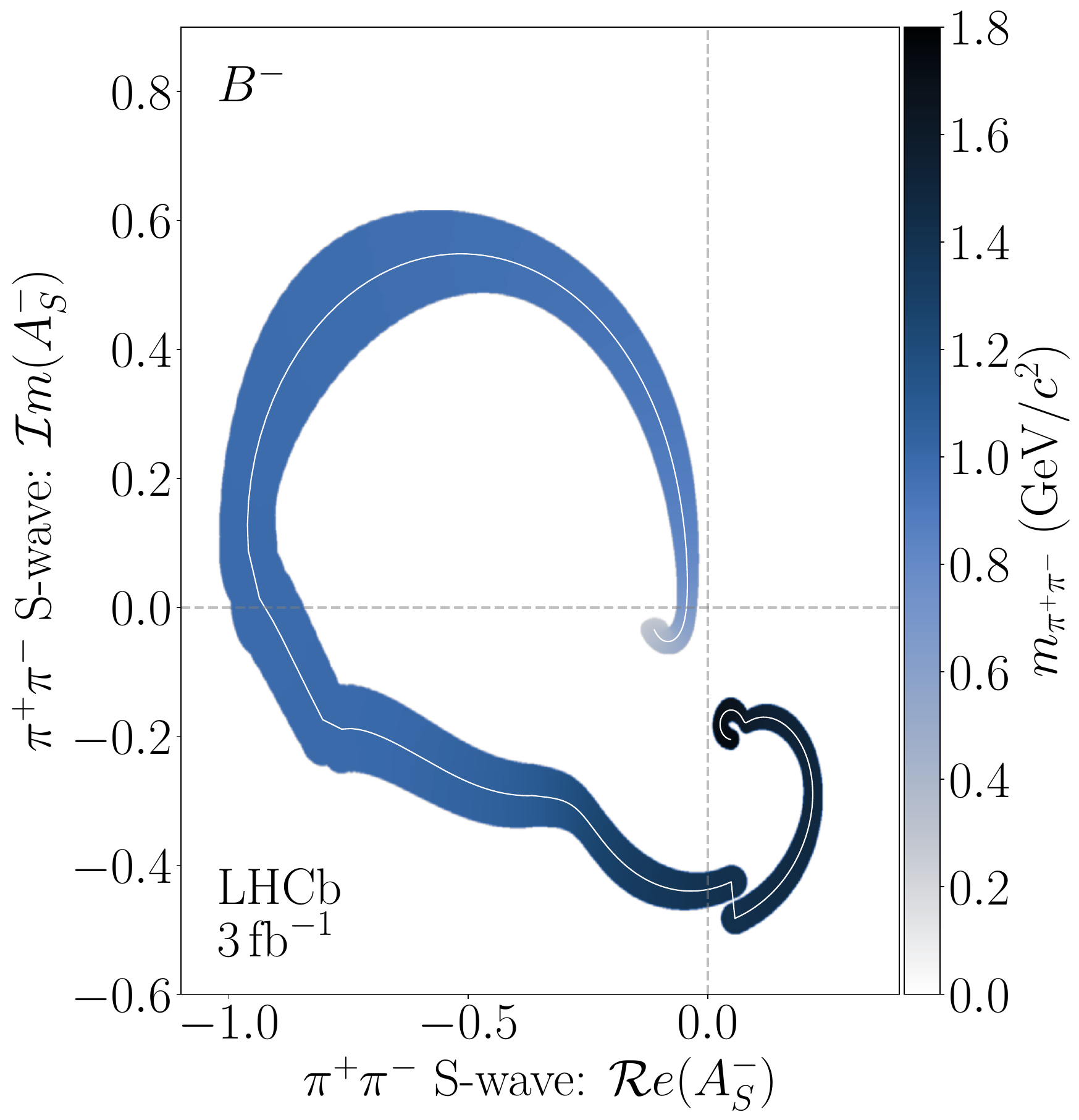}%
    \includegraphics[width=0.33\linewidth]{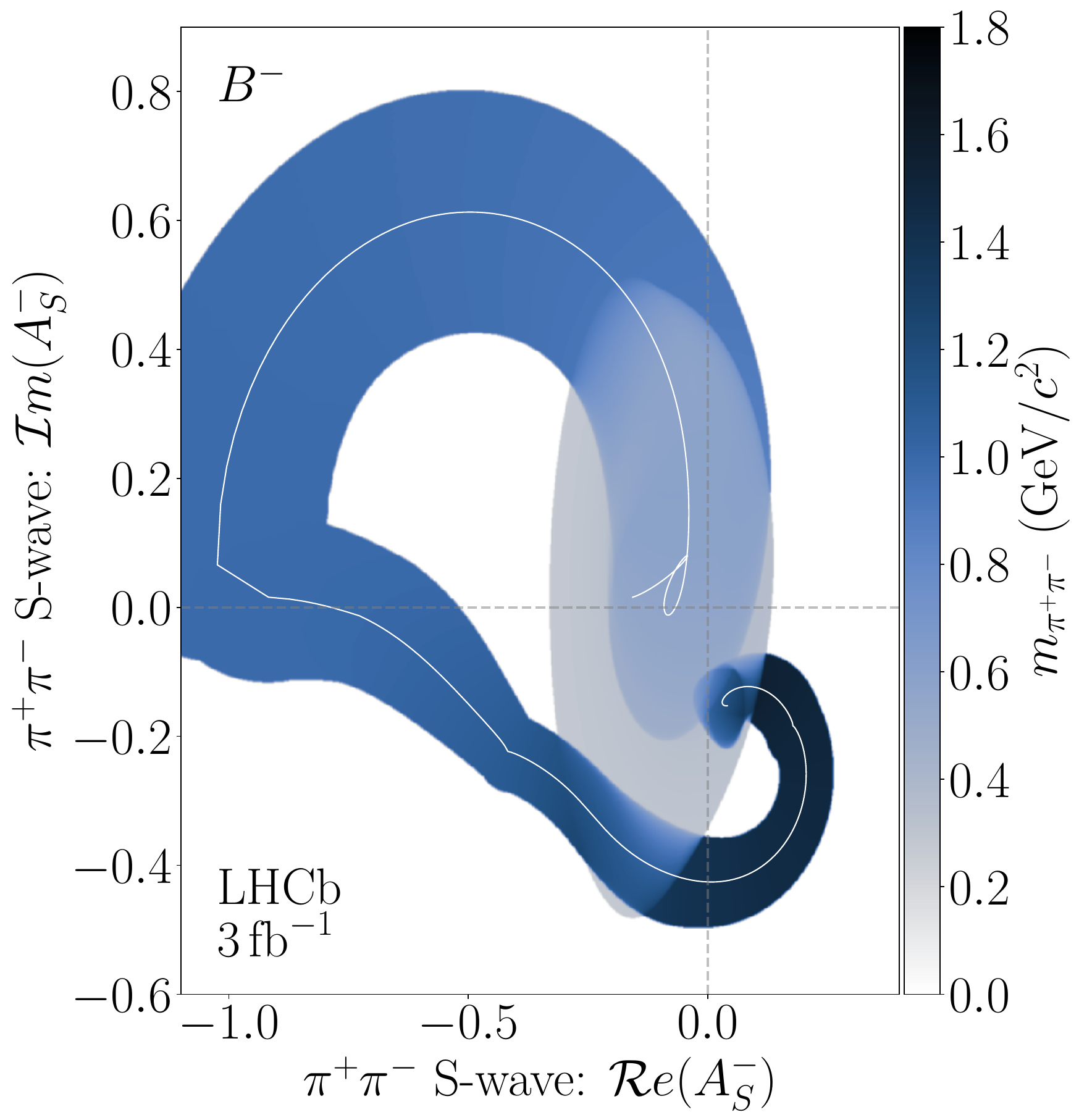}%
    \includegraphics[width=0.33\linewidth]{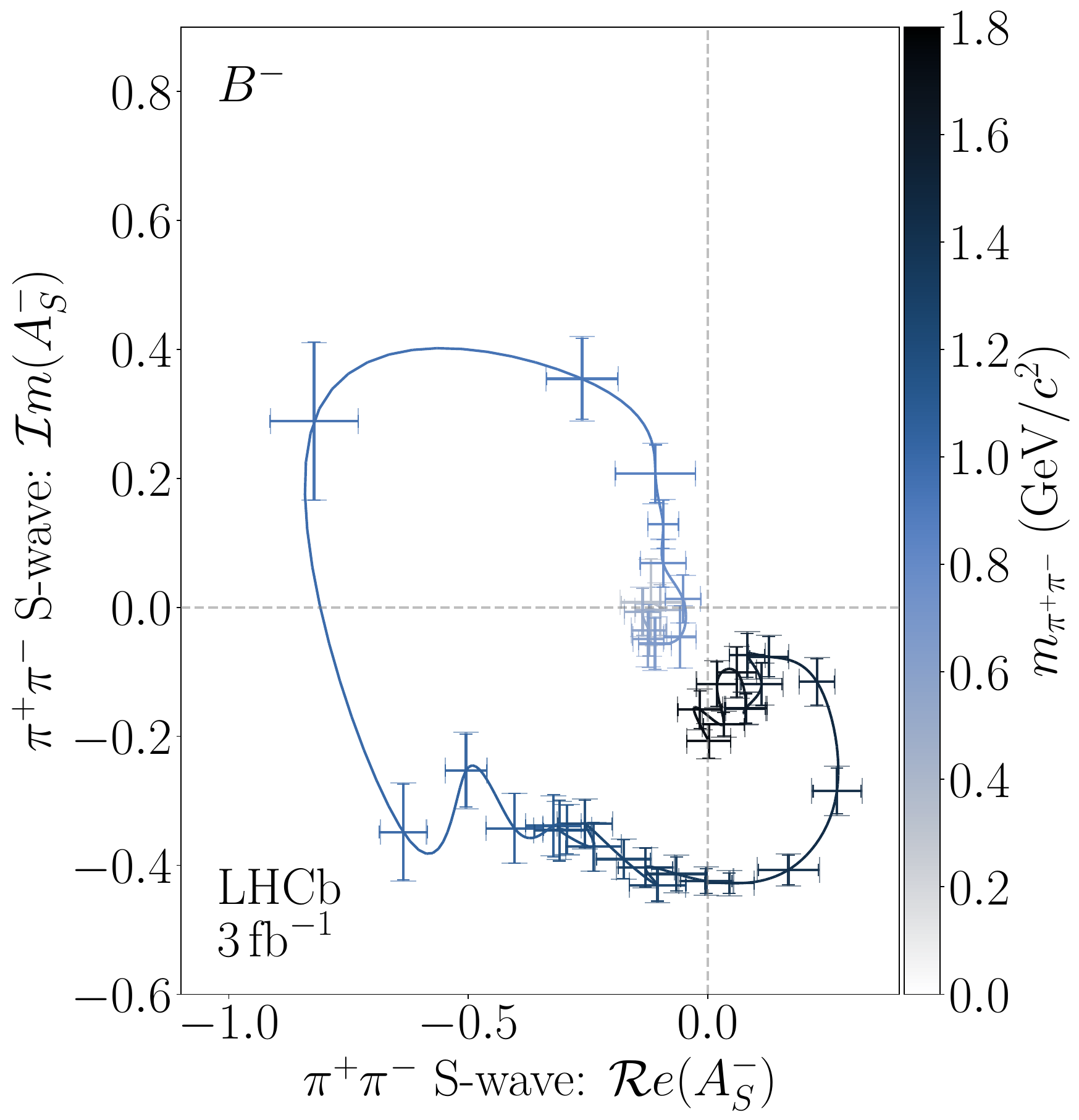}

    \caption{
        Argand diagrams of the (top)~$C\!P$-averaged, (middle)~$B^+$ and (bottom)~$B^-$ S-wave results in $\pi^+ \pi^-$ below the open-charm threshold  for the (left)~Isobar, (middle)~K-matrix and (right)~QMI approaches. White curves indicate central values while smoothed bands cover the total uncertainty at $1\sigma$. In the QMI plots, cubic splines are provided purely for visual guidance, and the shorter error bars indicate the total QMI uncertainty without accounting for an additional systematic effect arising from the fixed-bin values set from the Isobar result.
    }
    \label{fig:app:argandpipi}
\end{figure}

Isobar and K-matrix S-wave results projected onto the one-dimensional QMI S-waves across the full \Kp\pim and \pip\pim mass ranges are shown in Figs.~\ref{fig:app:swavekpi} and~\ref{fig:app:swavepipi}, respectively. These can alternatively be expressed in two dimensions on the Argand plane as shown in Figs.~\ref{fig:app:argandkpi} and~\ref{fig:app:argandpipi}.

\clearpage

\section{Legendre-weighted mass projections}
\label{app:moments}

\begin{figure}[!htb]
    \centering
    \includegraphics[width=0.5\linewidth]{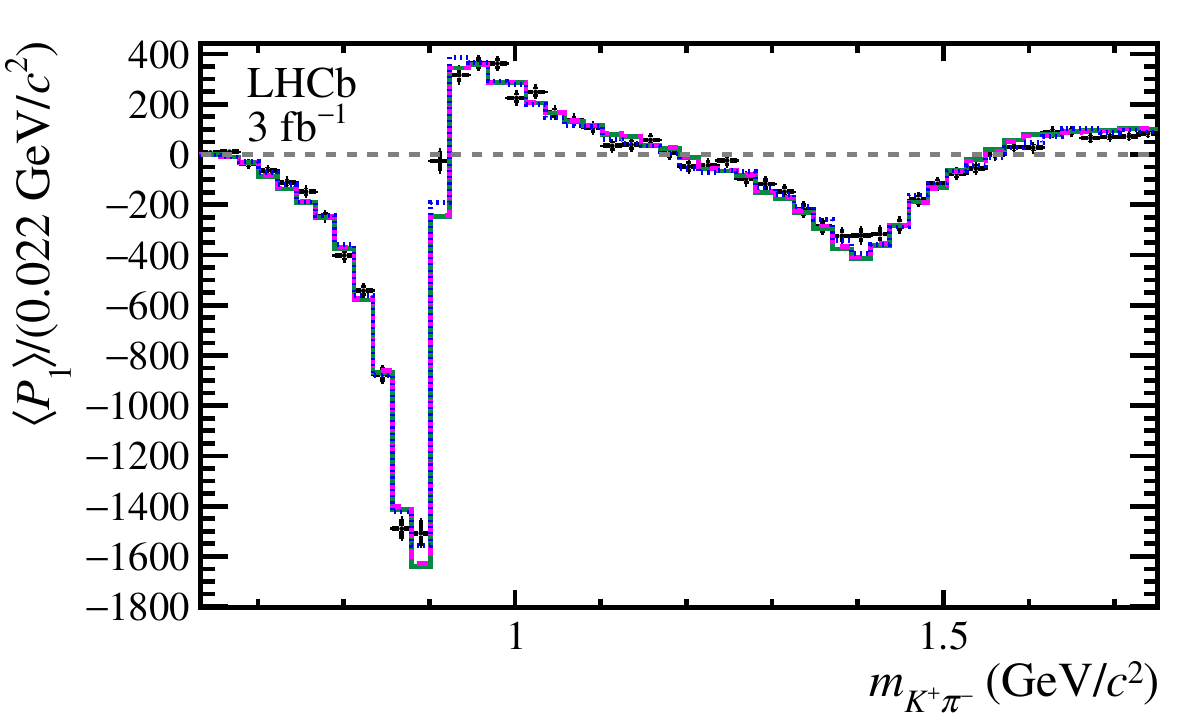}%
    \includegraphics[width=0.5\linewidth]{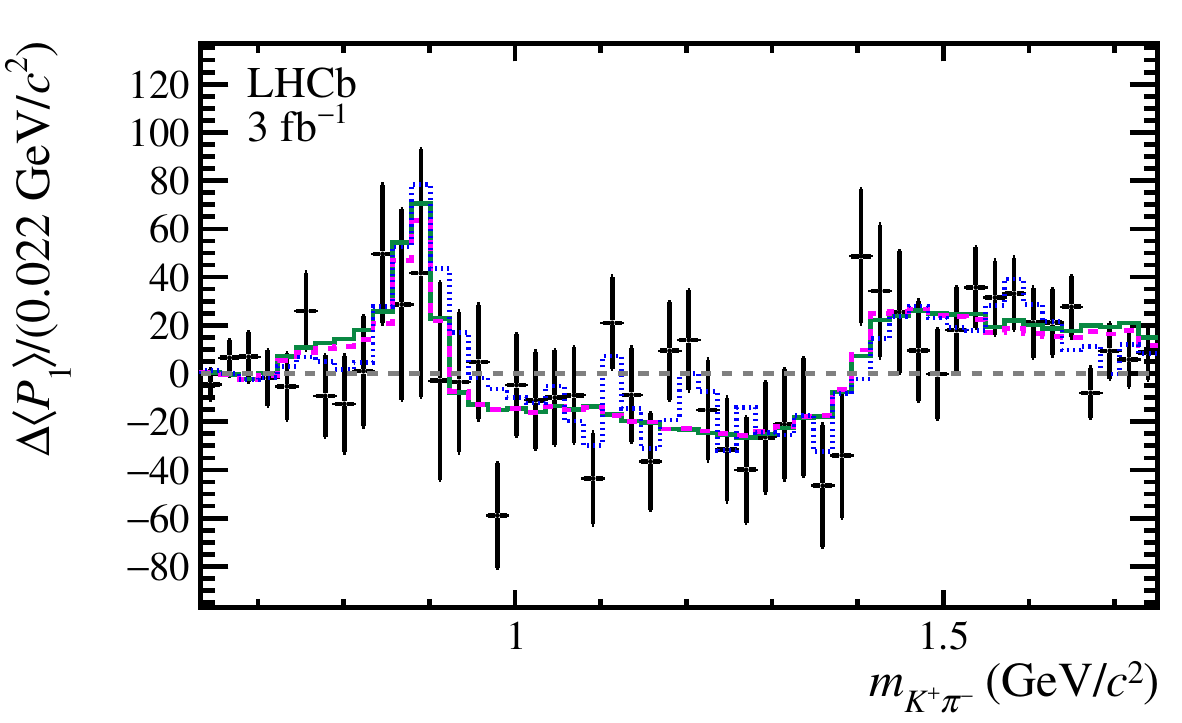}

    \includegraphics[width=0.5\linewidth]{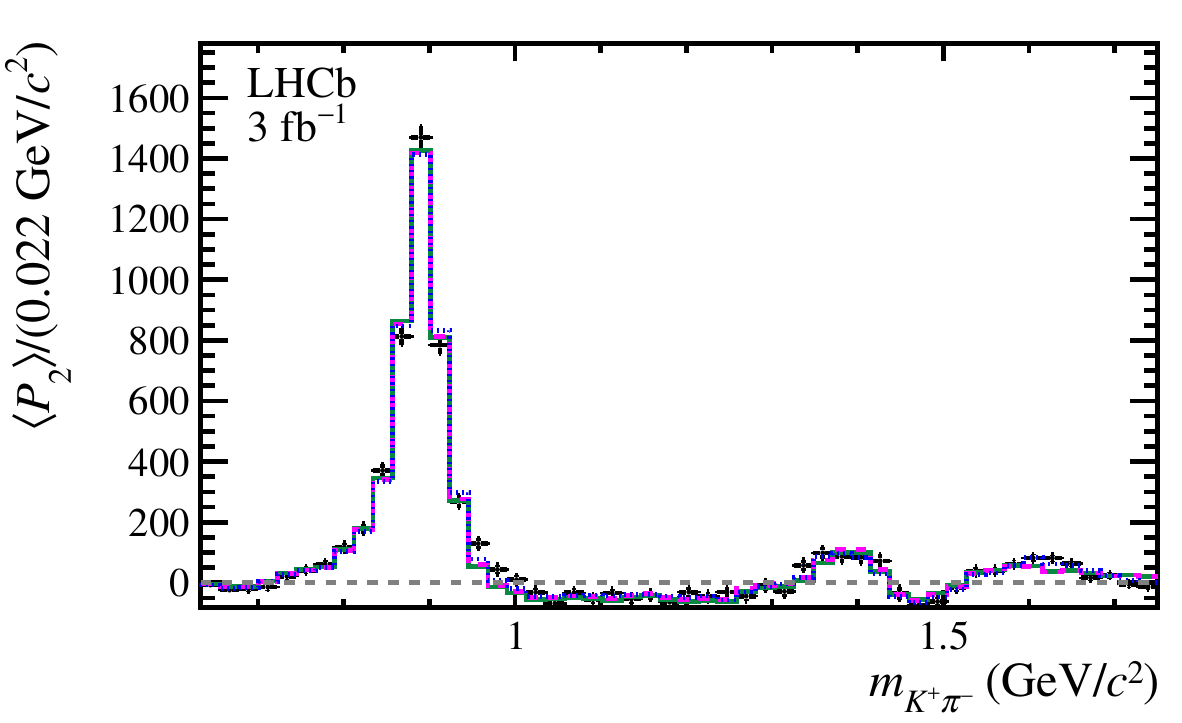}%
    \includegraphics[width=0.5\linewidth]{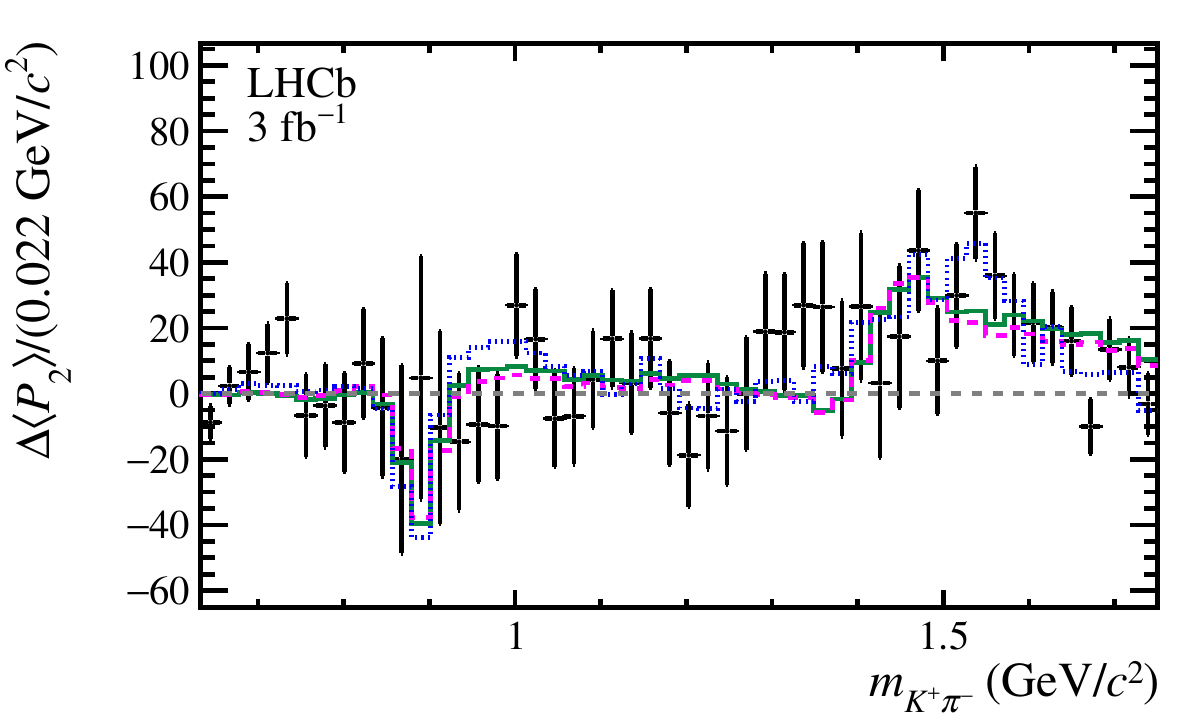}

    \includegraphics[width=0.5\linewidth]{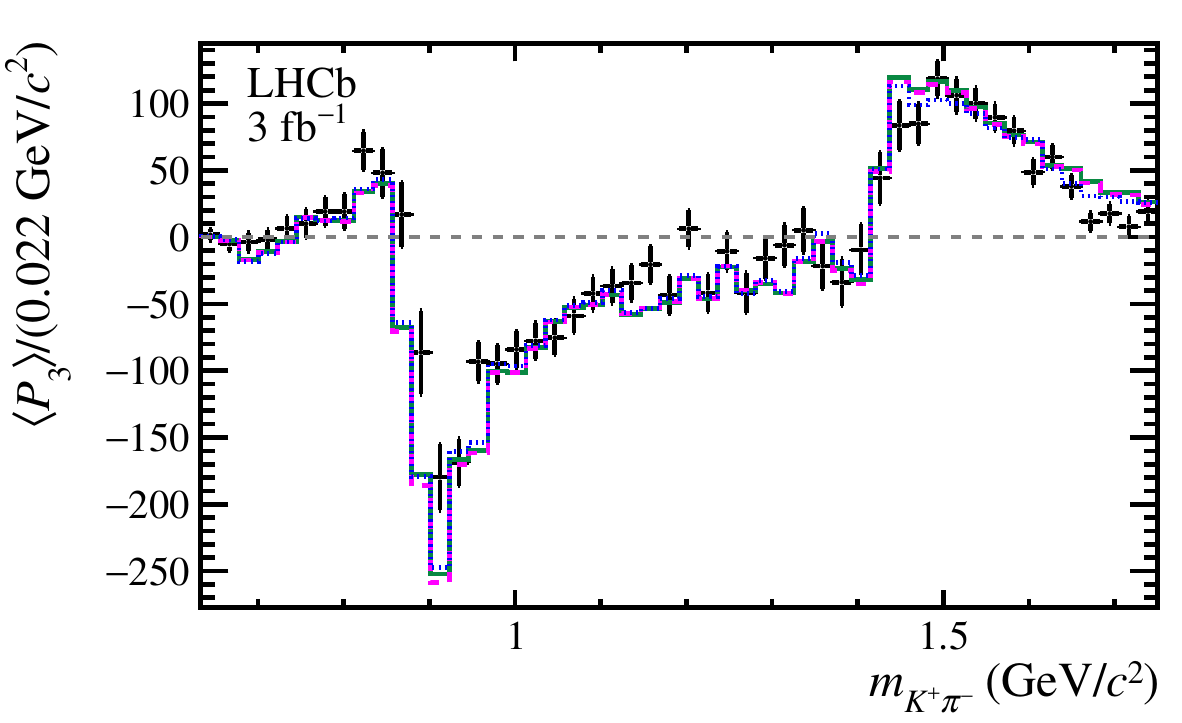}%
    \includegraphics[width=0.5\linewidth]{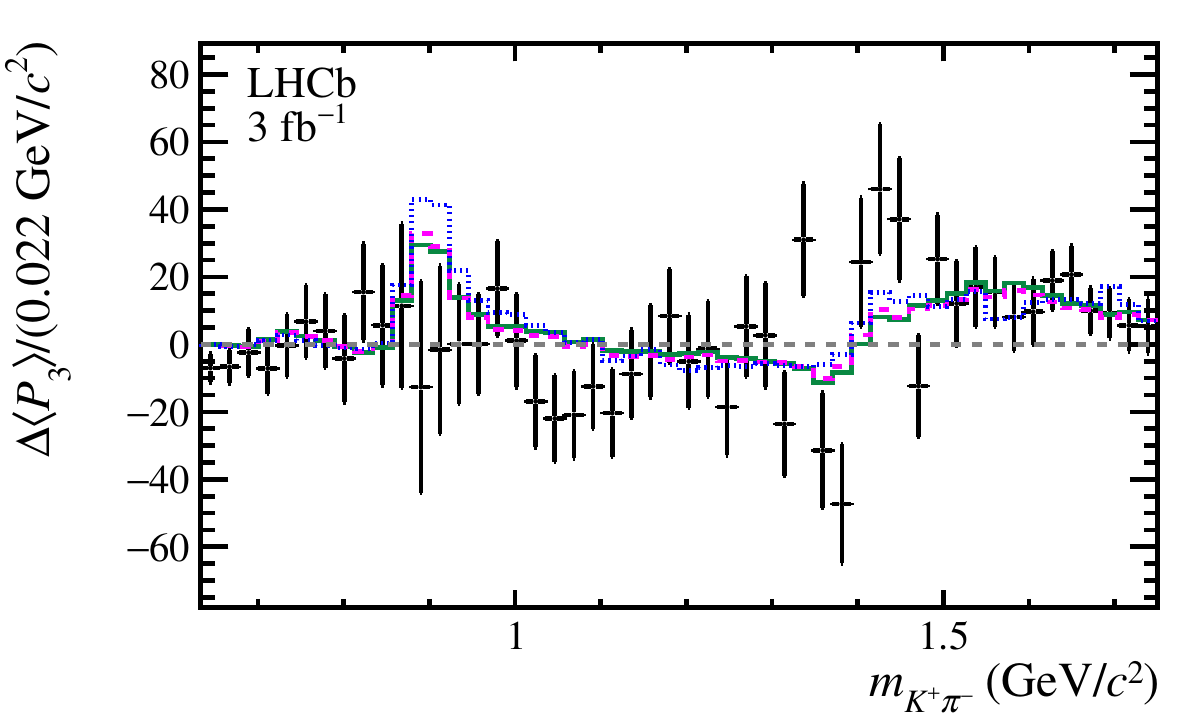}

    \includegraphics[width=0.5\linewidth]{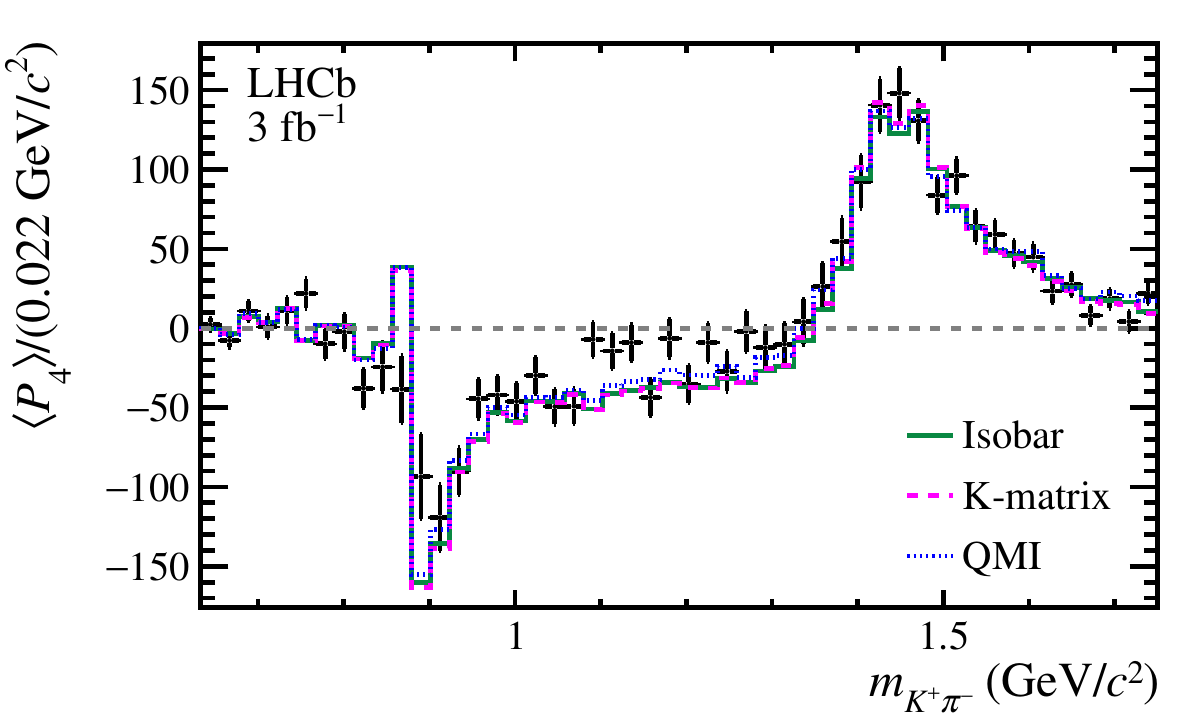}%
    \includegraphics[width=0.5\linewidth]{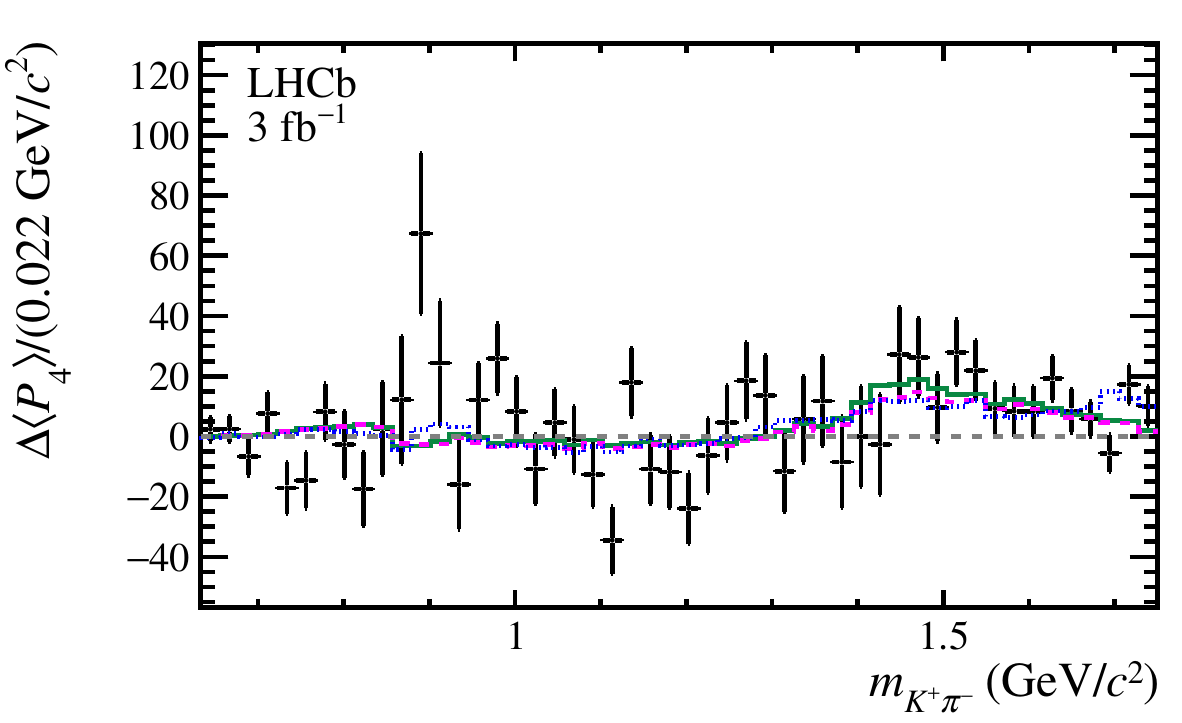}

    \caption{
        (left)~Background-subtracted data weighted by the first four Legendre polynomials in $\cos\theta_{K^+ \pi^-}$ below the open charm threshold and integrated over the $B$ charge, along with each signal model. (right)~Differentials of these distributions between $B^-$ and $B^+$.
    }
    \label{fig:app:pwakpi1}
\end{figure}

\begin{figure}[!htb]
    \centering
    \includegraphics[width=0.5\linewidth]{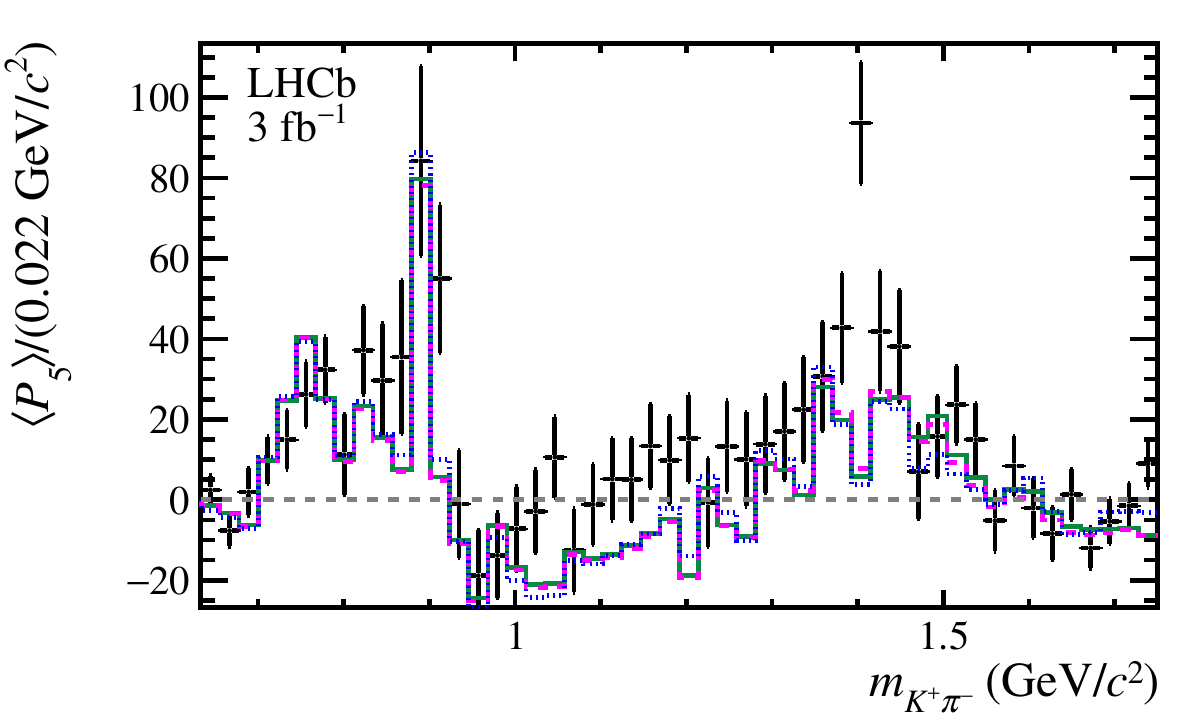}%
    \includegraphics[width=0.5\linewidth]{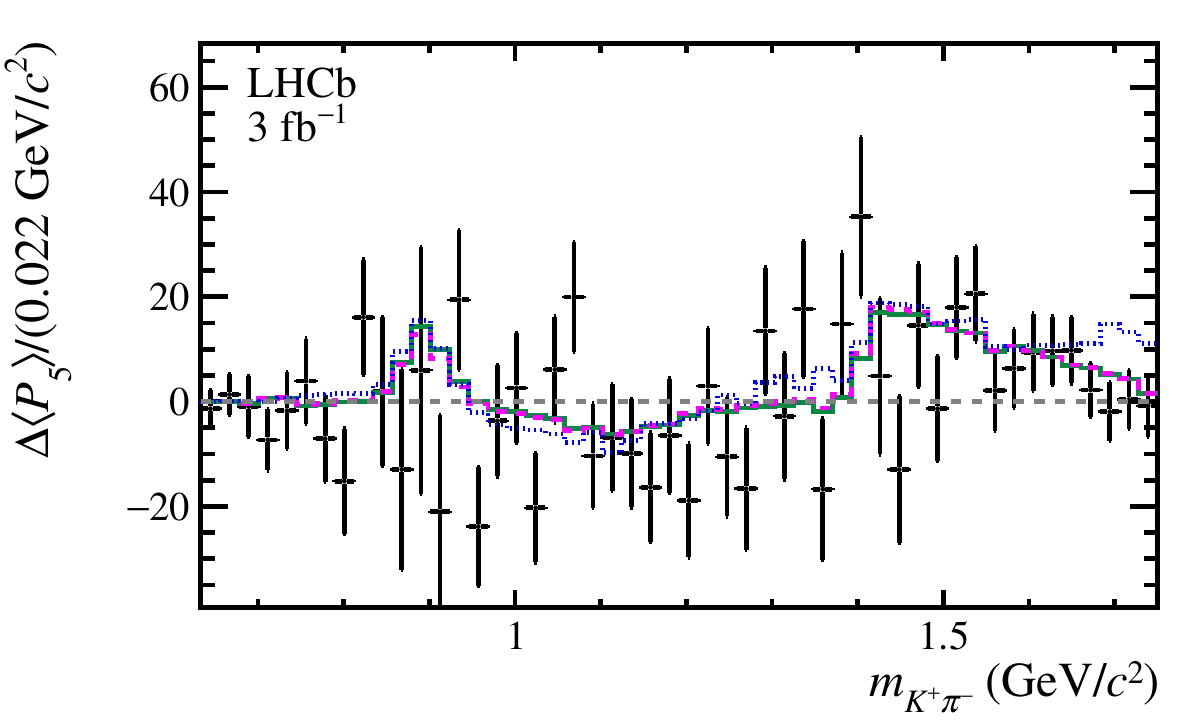}

    \includegraphics[width=0.5\linewidth]{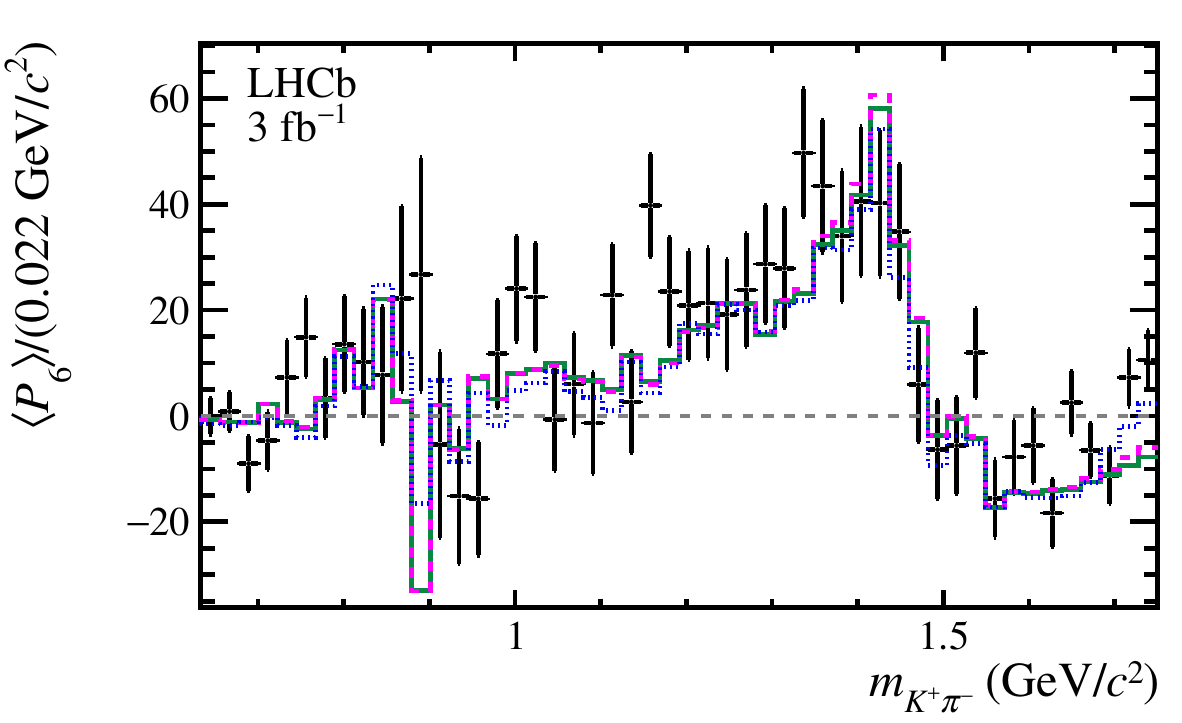}%
    \includegraphics[width=0.5\linewidth]{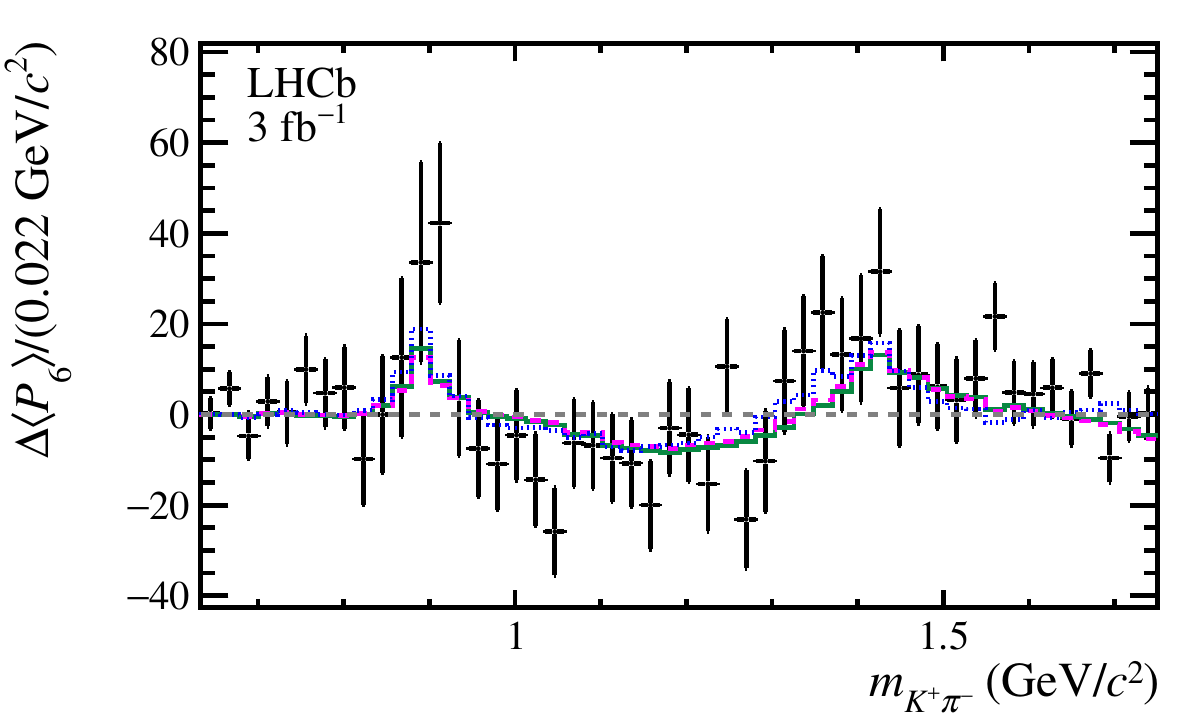}

    \includegraphics[width=0.5\linewidth]{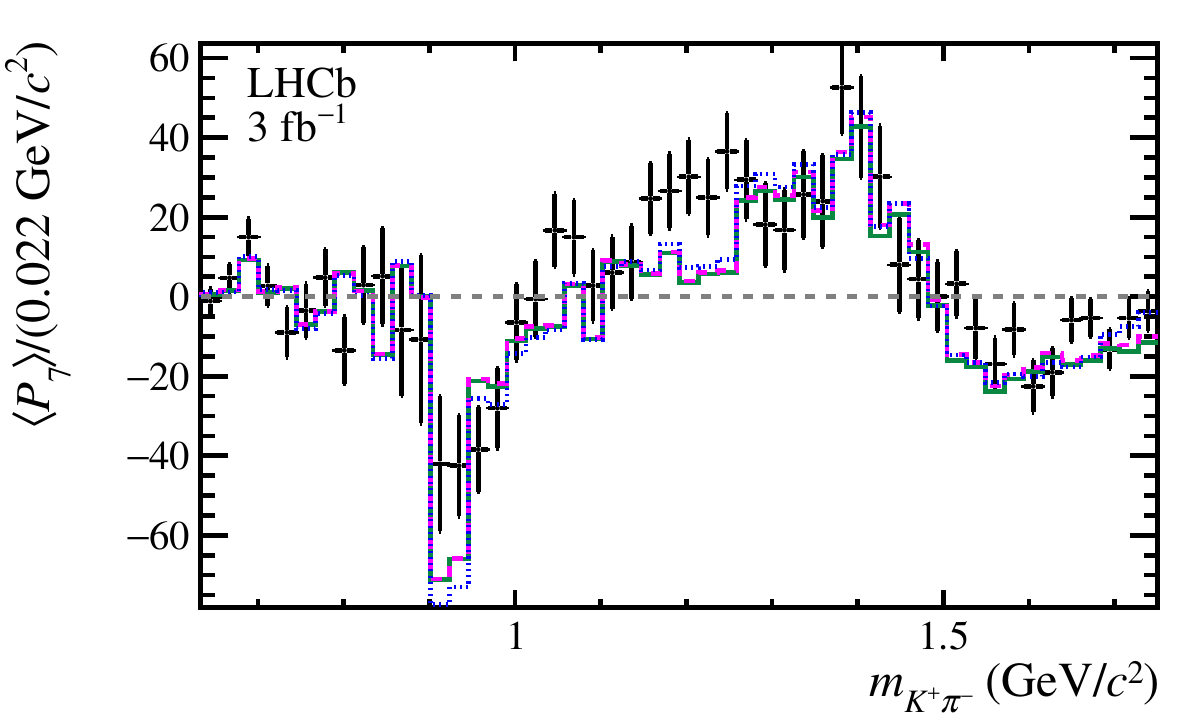}%
    \includegraphics[width=0.5\linewidth]{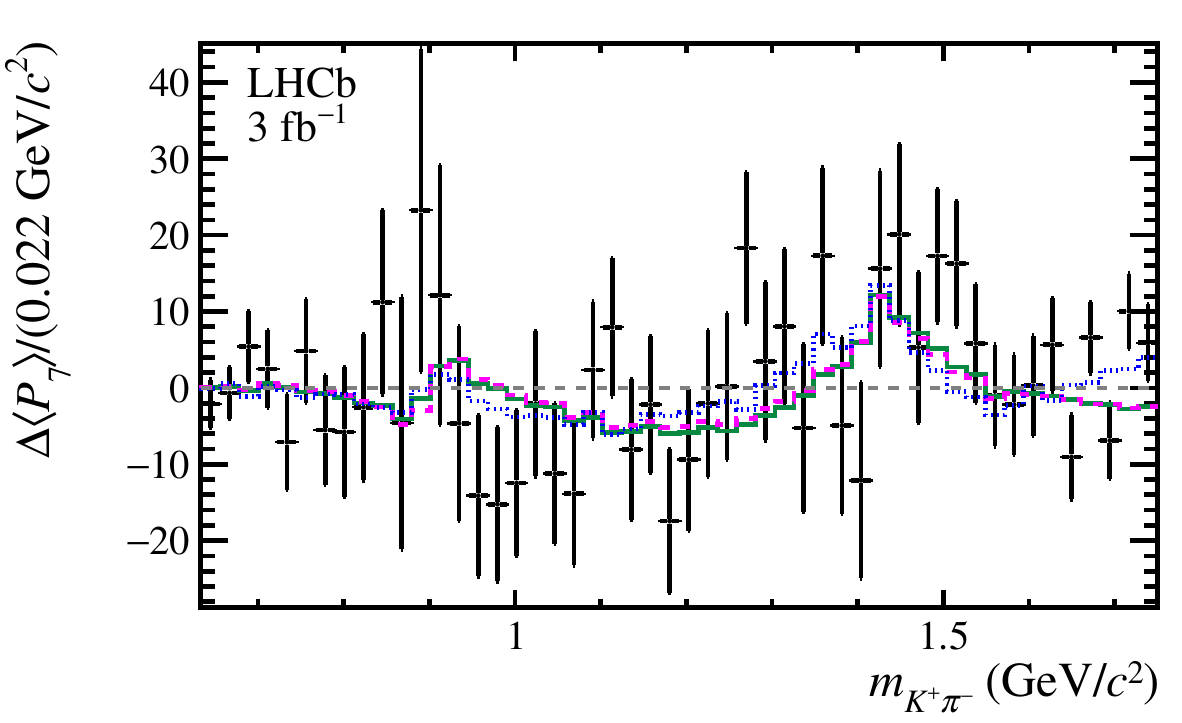}

    \includegraphics[width=0.5\linewidth]{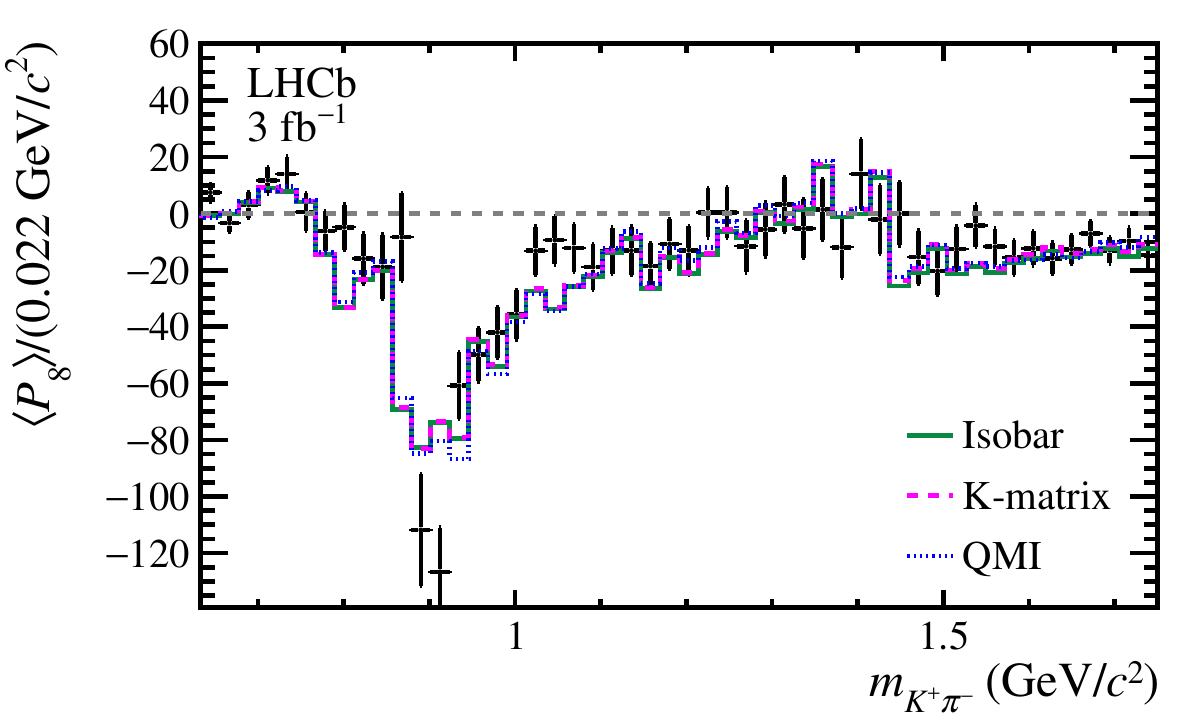}%
    \includegraphics[width=0.5\linewidth]{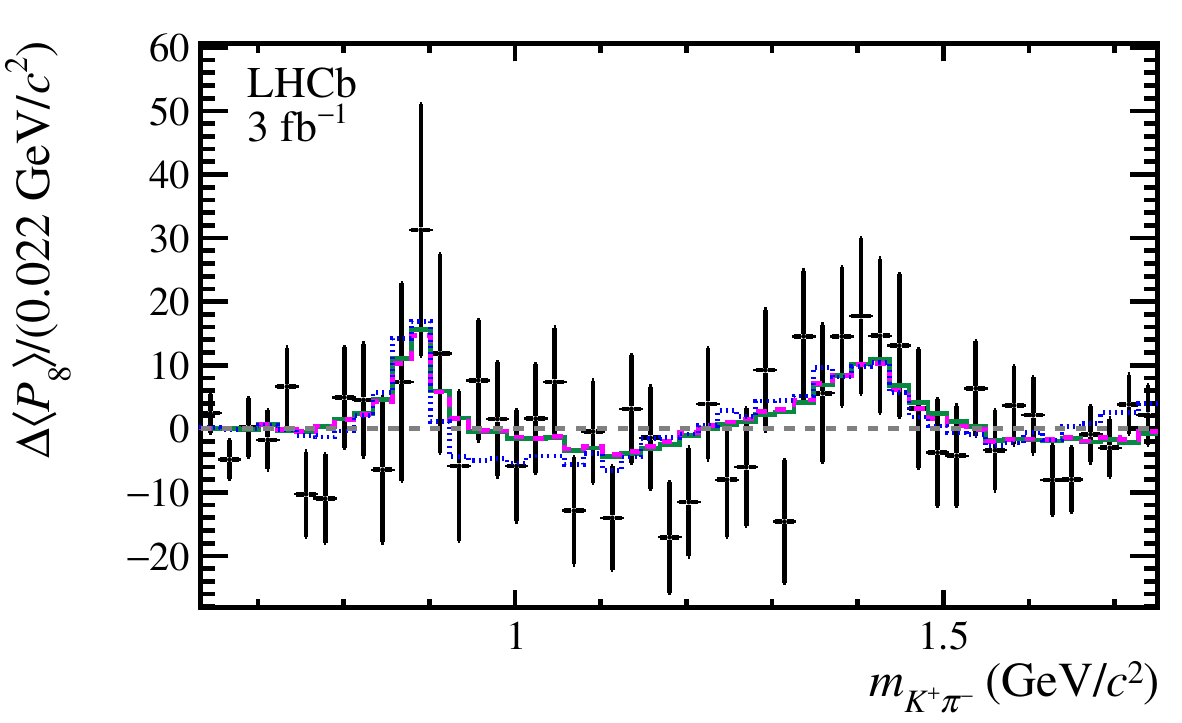}

    \caption{
        (left)~Background-subtracted data weighted by the next four Legendre polynomials in $\cos\theta_{K^+ \pi^-}$ below the open charm threshold and integrated over the $B$ charge, along with each signal model. (right)~Differentials of these distributions between $B^-$ and $B^+$.
    }
    \label{fig:app:pwakpi2}
\end{figure}

\begin{figure}[!htb]
    \centering
    \includegraphics[width=0.5\linewidth]{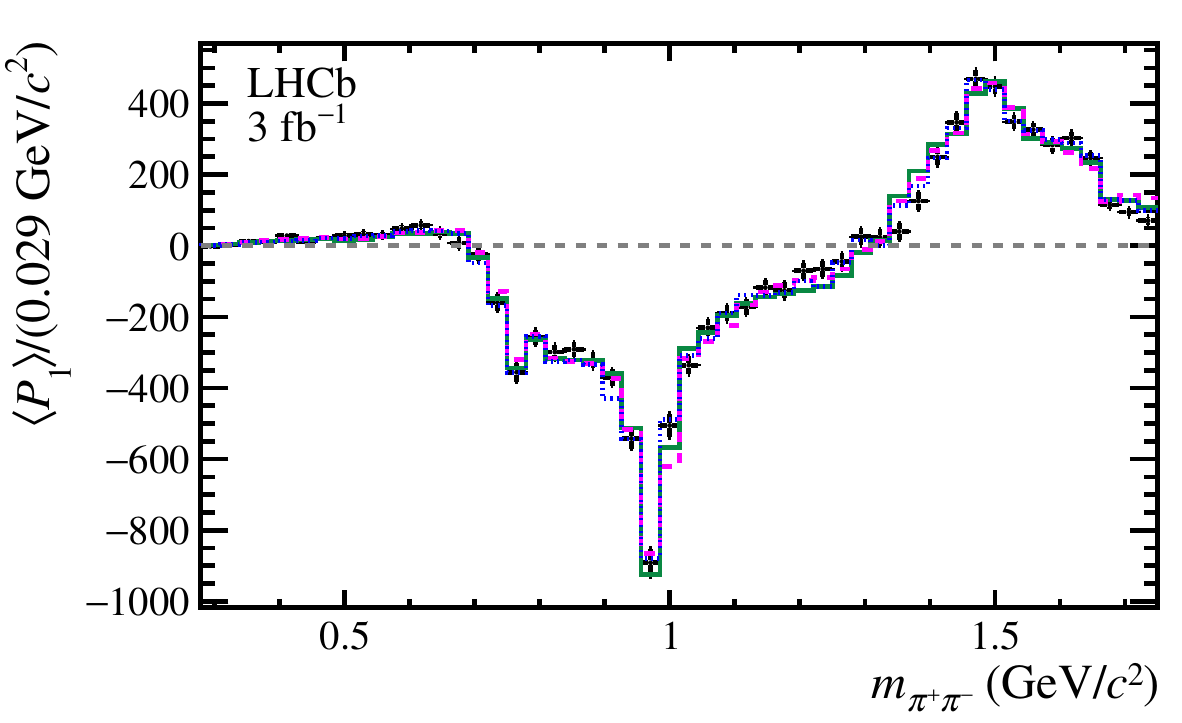}%
    \includegraphics[width=0.5\linewidth]{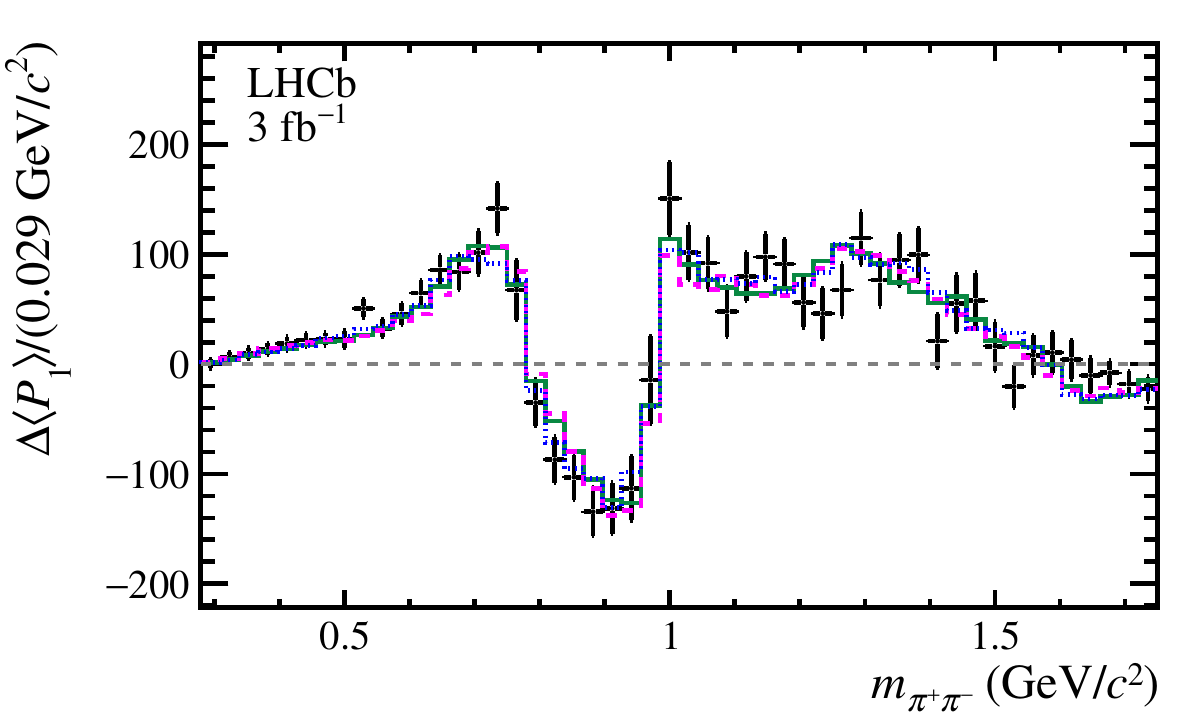}

    \includegraphics[width=0.5\linewidth]{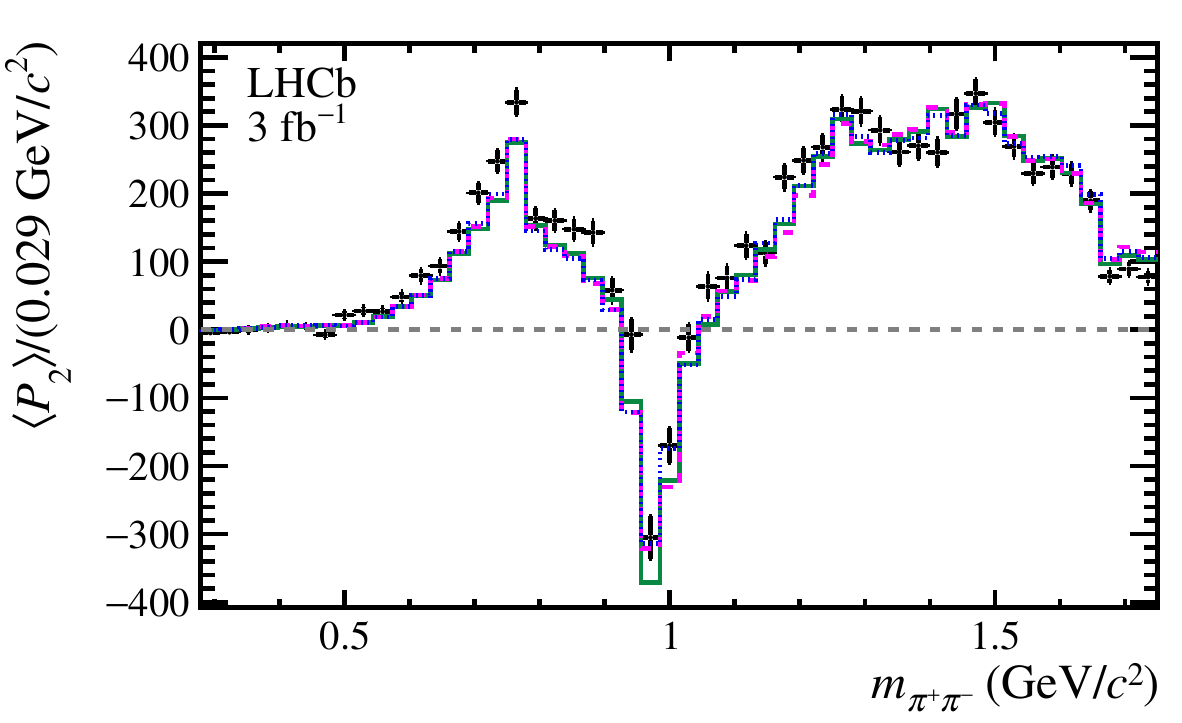}%
    \includegraphics[width=0.5\linewidth]{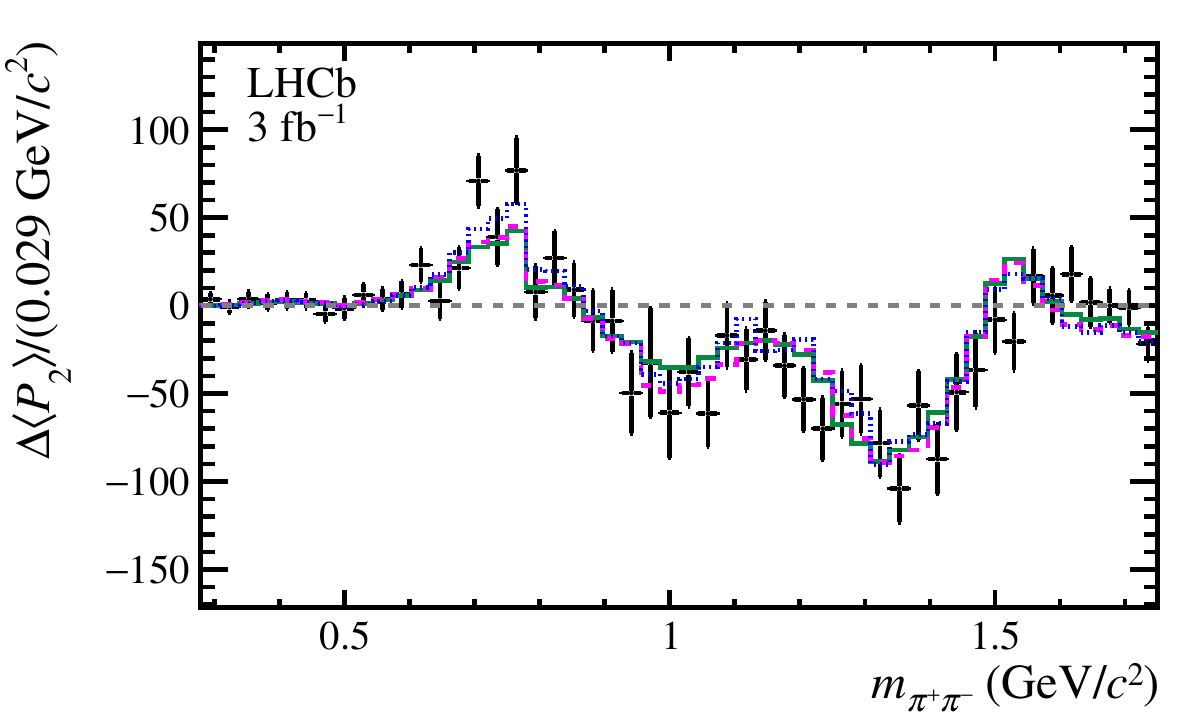}

    \includegraphics[width=0.5\linewidth]{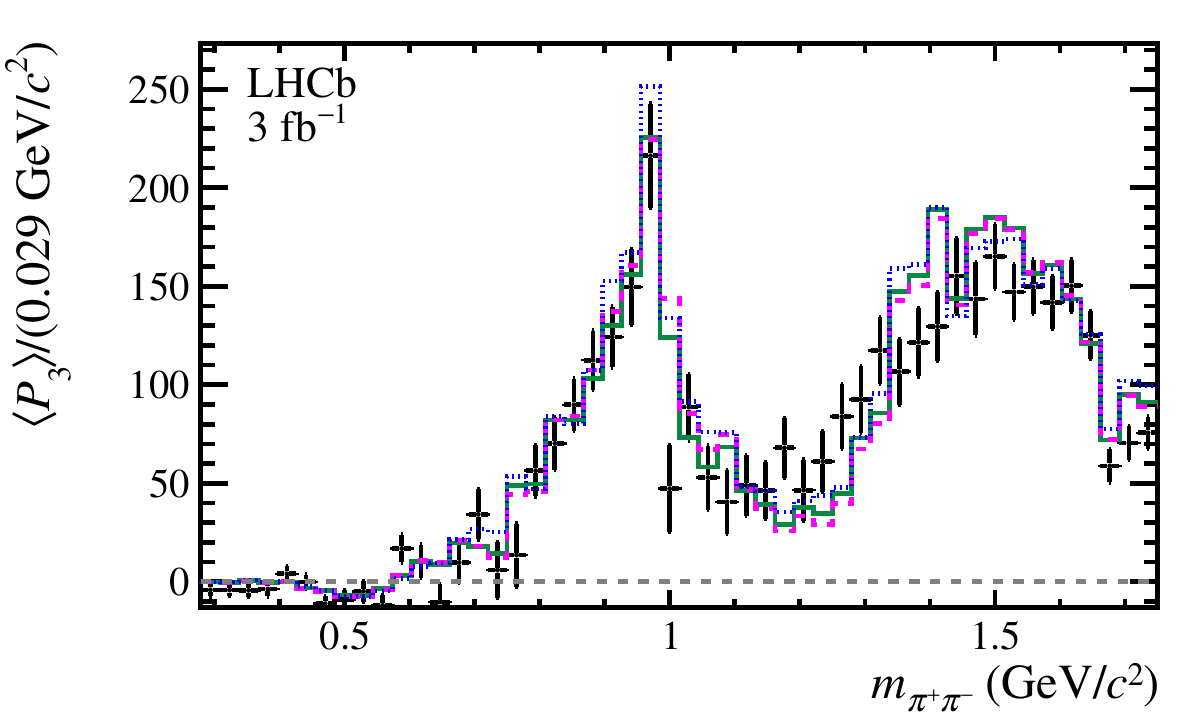}%
    \includegraphics[width=0.5\linewidth]{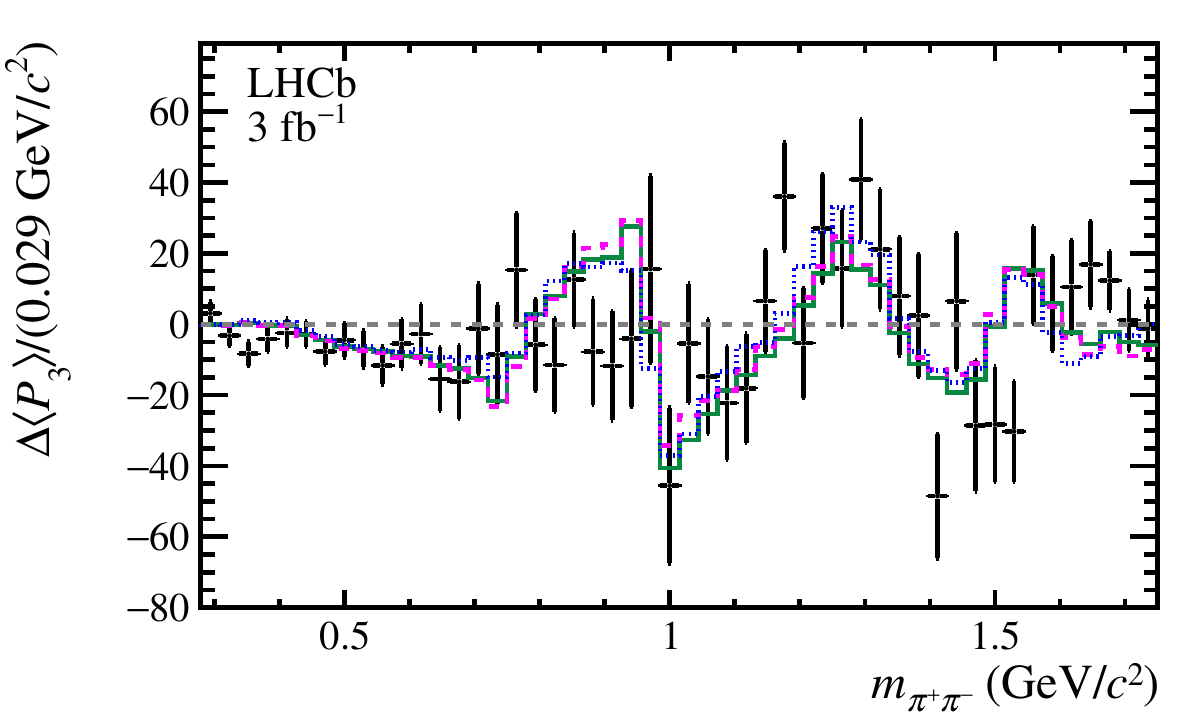}

    \includegraphics[width=0.5\linewidth]{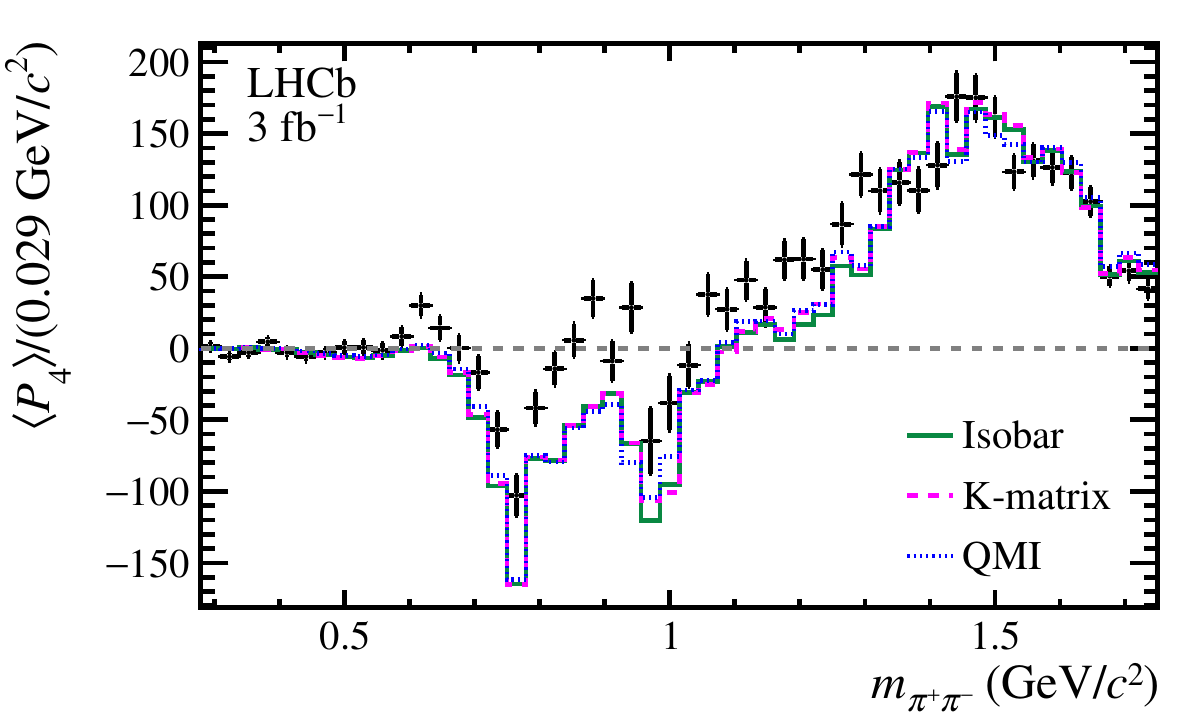}%
    \includegraphics[width=0.5\linewidth]{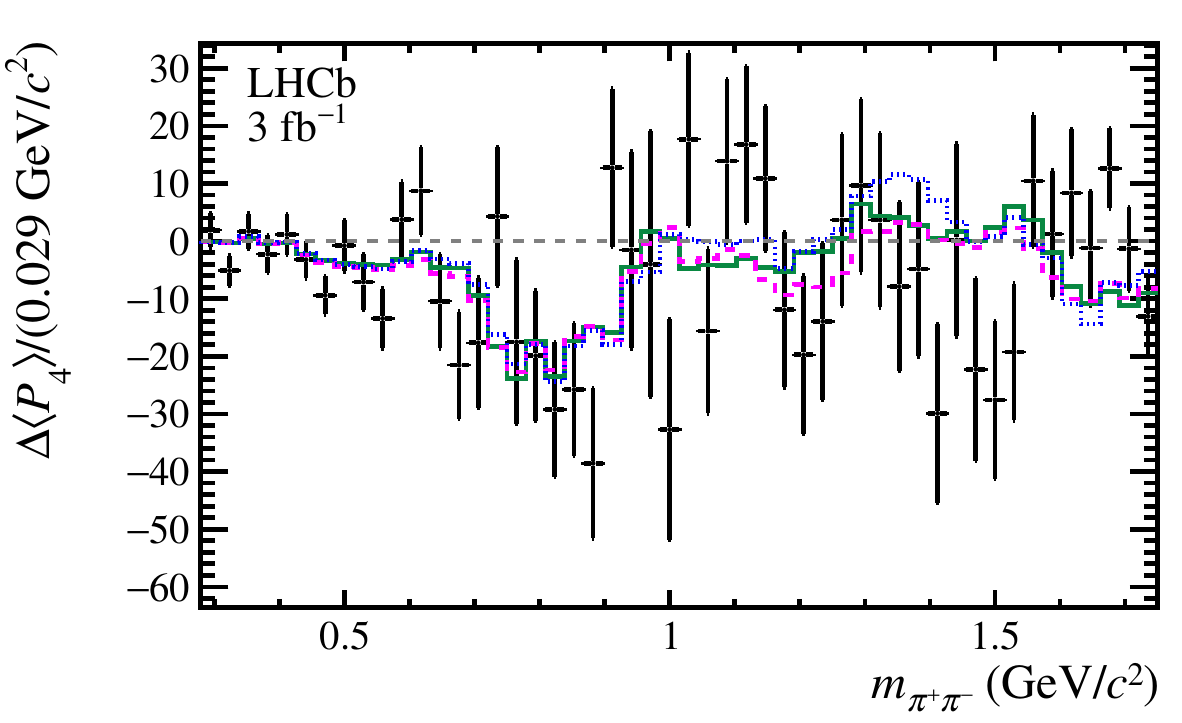}

    \caption{
        (left)~Background-subtracted data weighted by the first four Legendre polynomials in $\cos\theta_{\pi^+ \pi^-}$ below the open charm threshold and integrated over the $B$ charge, along with each signal model. (right)~Differentials of these distributions between $B^-$ and $B^+$.
    }
    \label{fig:app:pwapipi1}
\end{figure}

\begin{figure}[!htb]
    \centering
    \includegraphics[width=0.5\linewidth]{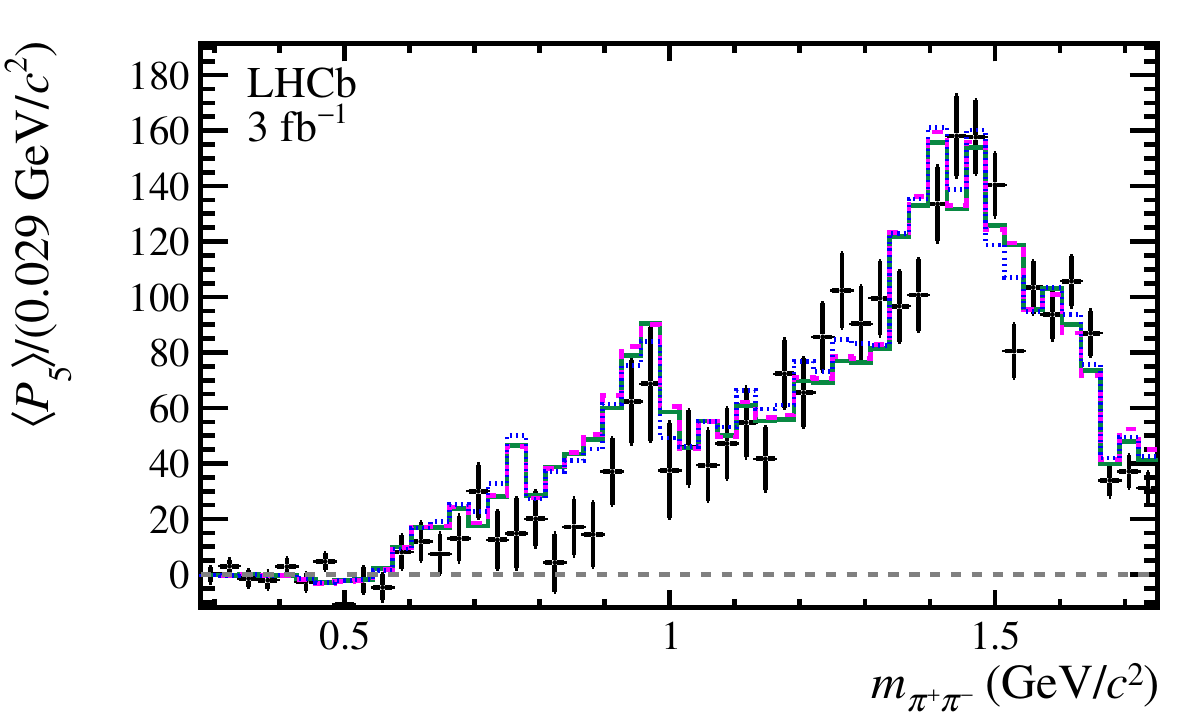}%
    \includegraphics[width=0.5\linewidth]{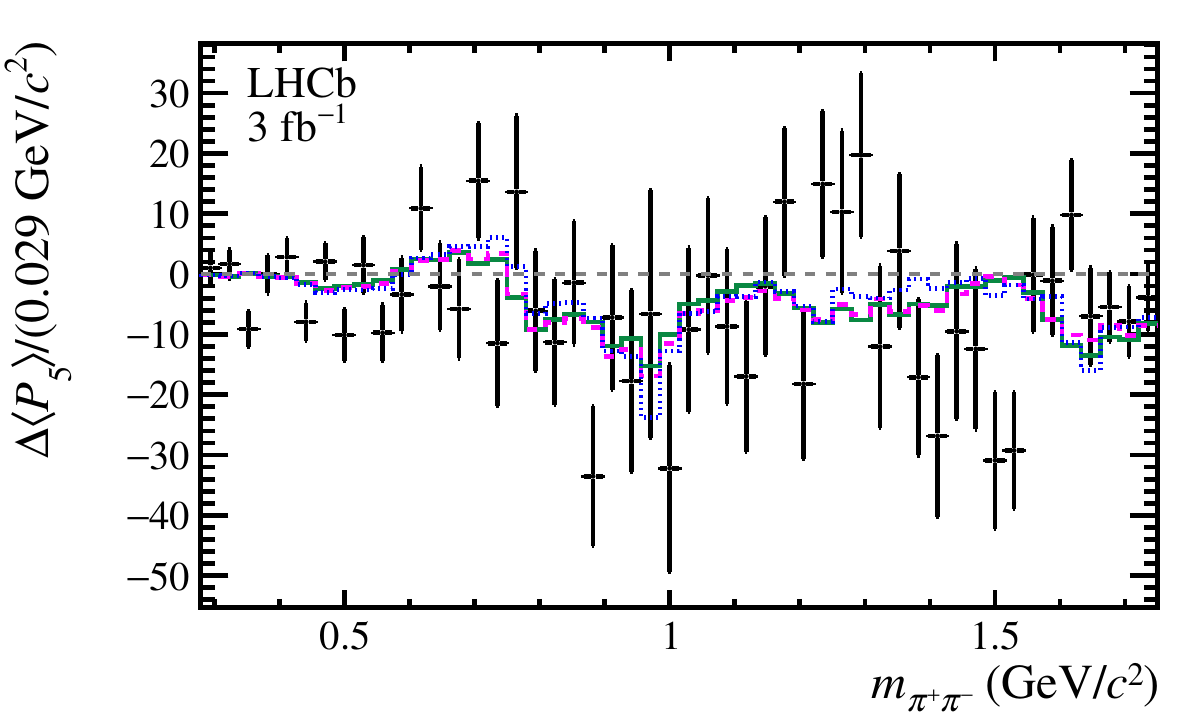}

    \includegraphics[width=0.5\linewidth]{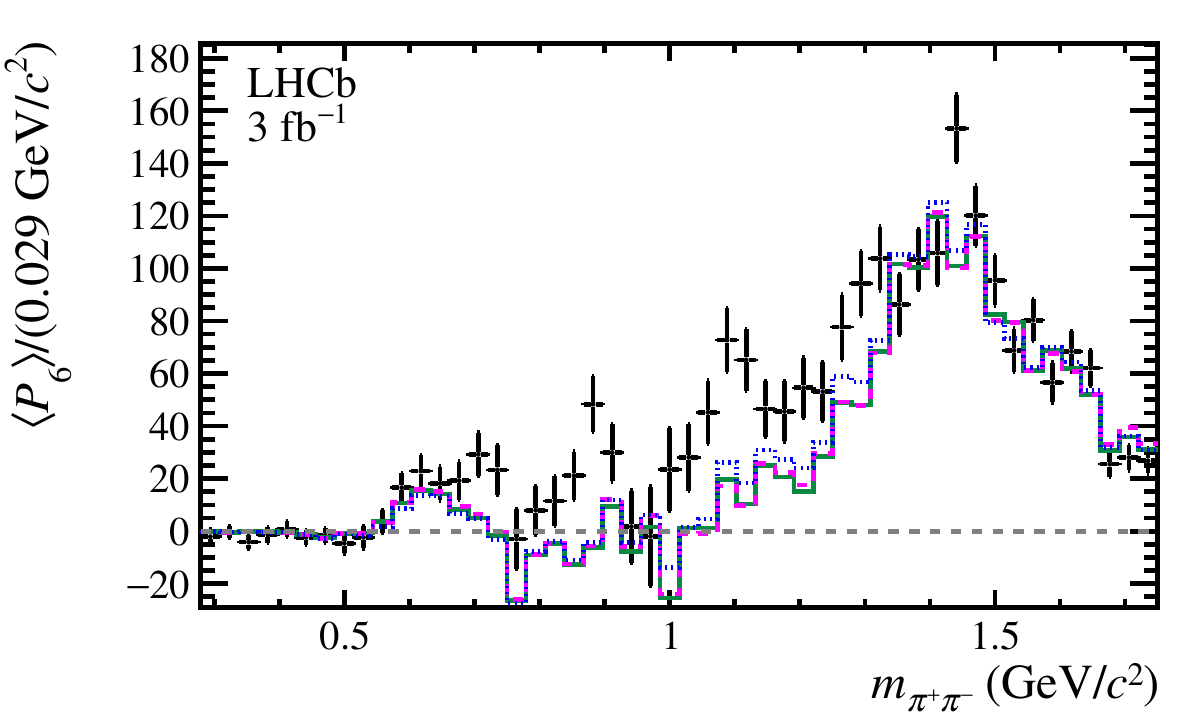}%
    \includegraphics[width=0.5\linewidth]{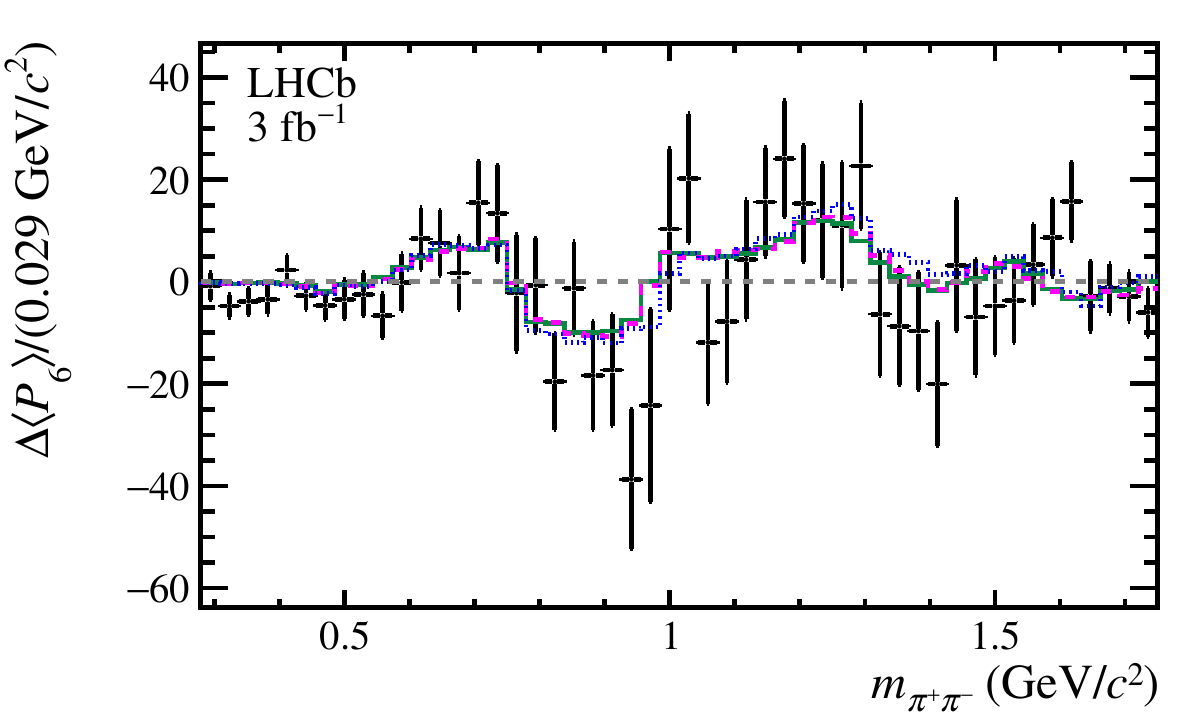}

    \includegraphics[width=0.5\linewidth]{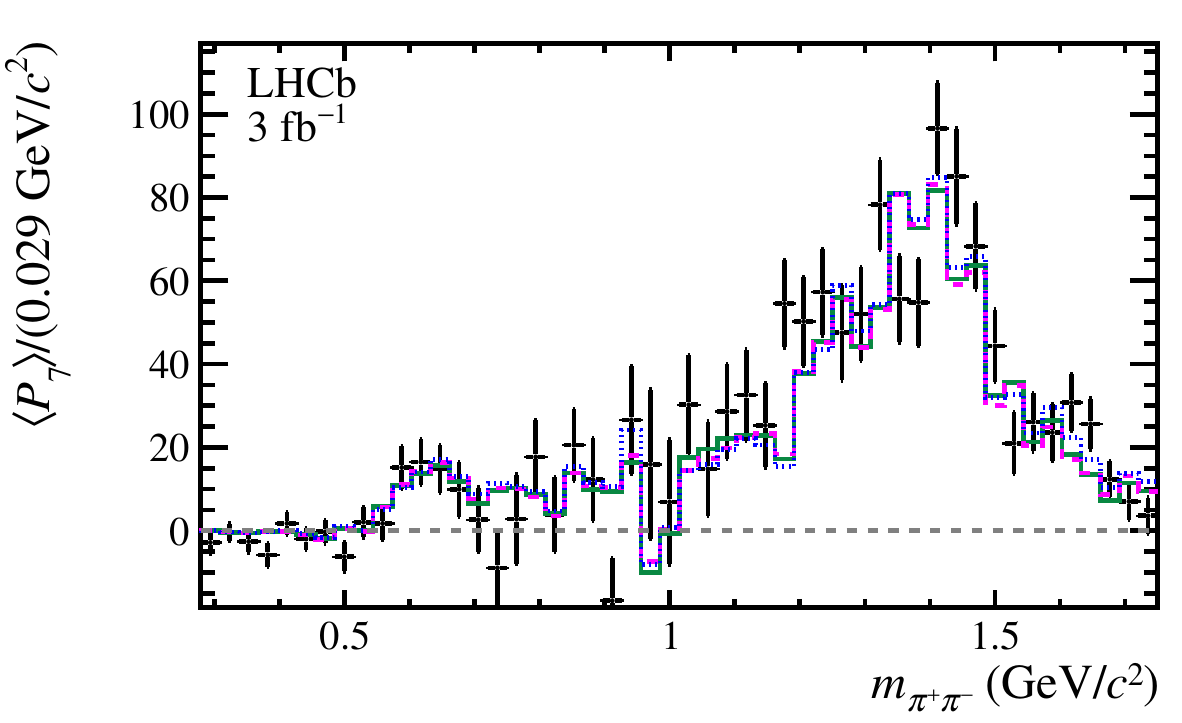}%
    \includegraphics[width=0.5\linewidth]{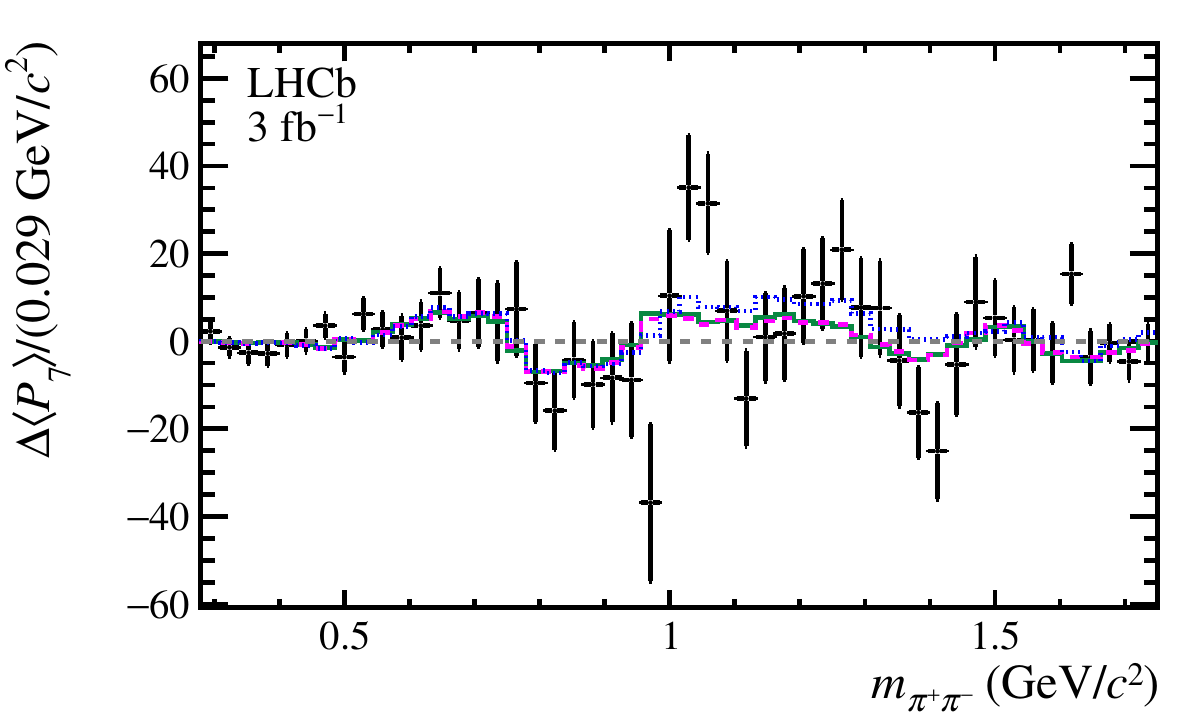}

    \includegraphics[width=0.5\linewidth]{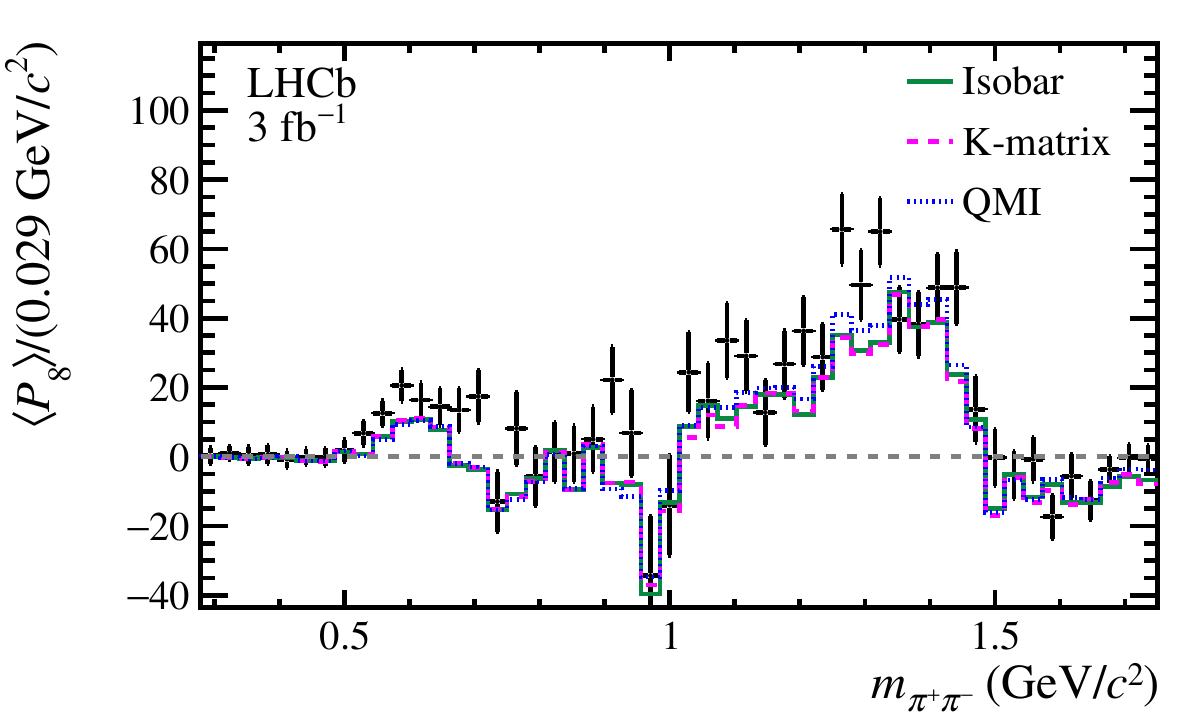}%
    \includegraphics[width=0.5\linewidth]{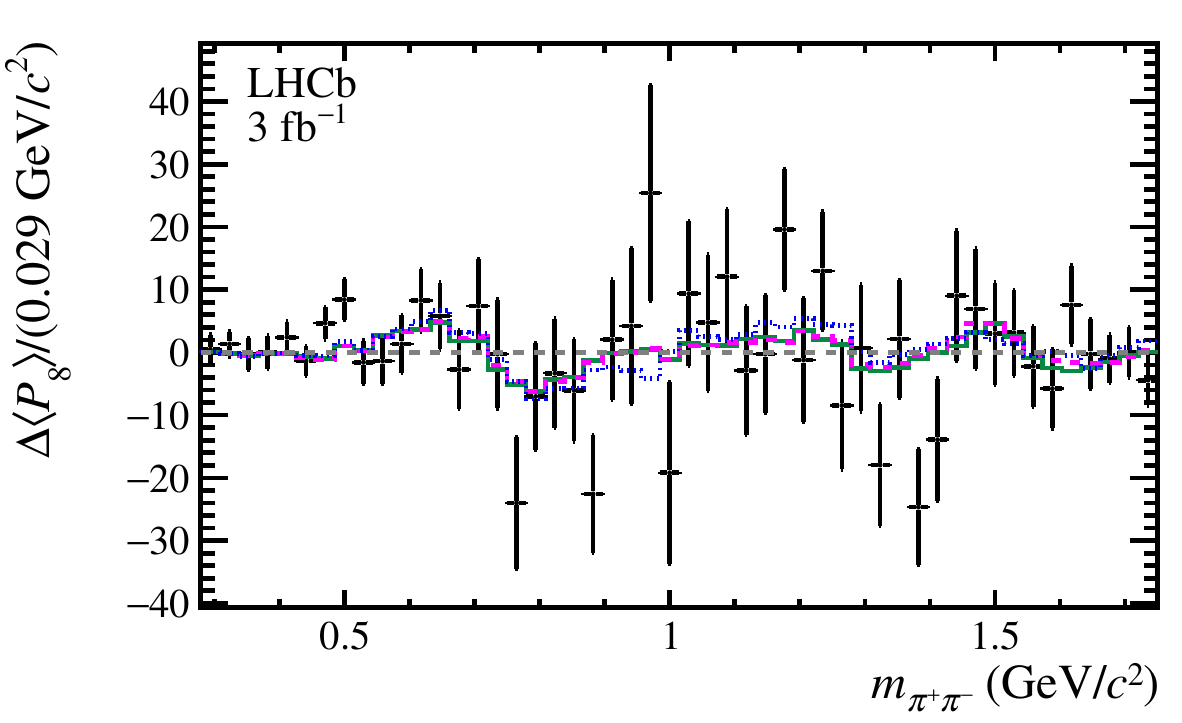}

    \caption{
        (left)~Background-subtracted data weighted by the next four Legendre polynomials in $\cos\theta_{\pi^+ \pi^-}$ below the open charm threshold and integrated over the $B$ charge, along with each signal model. (right)~Differentials of these distributions between $B^-$ and $B^+$.
    }
    \label{fig:app:pwapipi2}
\end{figure}

Contributions from different partial waves and their interference can be visualised through weighting data in the mass by Legendre polynomials $\left< P_n \right>$, of the cosine of the associated helicity angle. In this way, \CP violation mediated by the interference between partial waves can also be seen by weighting the \Bm and \Bp samples separately, and then taking their difference, $\Delta\!\left< P_n \right>$. These are shown in Figs.~\ref{fig:app:pwakpi1} and~\ref{fig:app:pwakpi2} for \mKpi and Figs.~\ref{fig:app:pwapipi1} and~\ref{fig:app:pwapipi2} for \mpipi, both below the open charm threshold.

As all known resonances decaying to either the \Kp\pim or \pip\pim final state in the charmless region are already included, areas in which all models disagree with the data hint at where non-S-wave amplitude models could be improved in future analyses. This could occur either at theoretical level to incorporate rescattering effects in higher partial waves or through further experimental innovation again with quasi-model-independent solutions. The inclusion of \cquark\cquarkbar loop contributions may also provide an explanation~\cite{El-Bennich:2009gqk,Heuser:2025mnk}.

\clearpage

%% file: supplemental.tex
\addcontentsline{toc}{section}{Supplemental material}
\section*{Supplemental material}
\label{sec:supplemental}

In addition to the results presented in the main body, a supplementary collection of files recording the complete expressions of the amplitude models in each S-wave approach is provided as \verb|*.json.gz|, in both their raw and convention-independent forms as they are impractical to publish either in the main text or appendices. These files are given in the JavaScript Object Notation~(JSON) format, which is both machine and human readable, containing parameter names, central values, total uncertainties and their corresponding two-dimensional arrays representing correlations between parameters. Additional steering files that drive the \textsc{Laura++} Dalitz-plot software package to generate pseudoexperiments based on the Isobar and K-matrix approaches are also included as \verb|LauraIso.cc| and \verb|LauraKM.cc|, respectively.

%% file: Authorship_LHCb-PAPER-2025-067.tex
\centerline
{\large\bf LHCb collaboration}
\begin
{flushleft}
\small
R.~Aaij$^{38}$\lhcborcid{0000-0003-0533-1952},
M.~Abdelfatah$^{69}$,
A.S.W.~Abdelmotteleb$^{57}$\lhcborcid{0000-0001-7905-0542},
C.~Abellan~Beteta$^{51}$\lhcborcid{0009-0009-0869-6798},
F.~Abudin\'en$^{59}$\lhcborcid{0000-0002-6737-3528},
T.~Ackernley$^{61}$\lhcborcid{0000-0002-5951-3498},
A.A.~Adefisoye$^{69}$\lhcborcid{0000-0003-2448-1550},
B.~Adeva$^{47}$\lhcborcid{0000-0001-9756-3712},
M.~Adinolfi$^{55}$\lhcborcid{0000-0002-1326-1264},
P.~Adlarson$^{87}$\lhcborcid{0000-0001-6280-3851}
% ...

{\footnotesize \it

% ...
$^{38}$Nikhef National Institute for Subatomic Physics, Amsterdam, Netherlands\\
% ...
$^{47}$Instituto Galego de F{\'\i}sica de Altas Enerx{\'\i}as (IGFAE), Universidade de Santiago de Compostela, Santiago de Compostela, Spain\\
% ...
$^{51}$Physik-Institut, Universit{\"a}t Z{\"u}rich, Z{\"u}rich, Switzerland\\
% ...
$^{55}$H.H. Wills Physics Laboratory, University of Bristol, Bristol, United Kingdom\\
% ...
$^{57}$Department of Physics, University of Warwick, Coventry, United Kingdom\\
% ...
$^{59}$School of Physics and Astronomy, University of Edinburgh, Edinburgh, United Kingdom\\
$^{60}$School of Physics and Astronomy, University of Glasgow, Glasgow, United Kingdom\\
$^{61}$Oliver Lodge Laboratory, University of Liverpool, Liverpool, United Kingdom\\
% ...
$^{69}$Syracuse University, Syracuse, NY, United States\\
% ...
$^{87}$Department of Physics and Astronomy, Uppsala University, Uppsala, Sweden, associated to $^{60}$\\
% ...
}
\end{flushleft}